\documentclass[aps,prd,floatfix,showpacs,superscriptaddress,nofootinbib,notitlepage]{report}
\usepackage{geometry}
\usepackage[alf]{abntex2cite}
\DeclareOldFontCommand{\it}{\normalfont\itshape}{\mathit}
\usepackage[colorlinks=true, pdfstartview=FitV, linkcolor=blue, citecolor=red, urlcolor=magenta]{hyperref}
\usepackage[utf8]{inputenc}
\usepackage{cite}
\DeclareOldFontCommand{\rm}{\normalfont\rmfamily}{\mathrm}
\usepackage{graphicx}
\usepackage{latexsym}
\usepackage{amsmath}
\usepackage{amsfonts}
\usepackage{amssymb}
\usepackage{braket}
\usepackage{float}
\def\a{\alpha}
\def\b{\beta}
\def\d{\delta}

\def\m{\mu}
\def\n{\nu}

\def\e{\xi}

\def\l{\lambda}
\def\w{\omega}

\def\r{\rho}
\def\s{\sigma}

\def\o{\theta}

\def\Y{\Psi}

\def\vf{\varphi}

\newcommand{\be}{\begin{equation}}
\newcommand{\ee}{\end{equation}}
\newcommand{\ha}{\frac{1}{2}}
\newcommand{\pa}{\partial}

\def\mc{\mathcal}
\newcommand{\beq}{\begin{eqnarray}}
\newcommand{\eeq}{\end{eqnarray}}

\def\a{\alpha}

\def\b{\beta}
\def\m{\mu}
\def\s{\sigma}
\def\n{\nu}
\def\r{\rho}
\def\l{\lambda}

\def\m{\mu}
\def\o{\omega}

\def\r{\rho}
\def\s{\sigma}

\def\d{\partial}

\def\vf{\varphi}

\def\e{\epsilon}

\def\dd #1 #2{{\delta #1\over \delta #2}}

\usepackage[colorlinks=true, pdfstartview=FitV, linkcolor=blue, citecolor=red, urlcolor=magenta]{hyperref}
\usepackage[utf8]{inputenc}
\usepackage{graphicx}
\usepackage{latexsym}
\usepackage{amsmath}
\usepackage{amsfonts}
\usepackage{amssymb}
\usepackage{bbm}
\usepackage{verbatim}
\usepackage{appendix}
\usepackage{float}
\usepackage{braket}
\def\a{\alpha}

\def\b{\beta}
\def\m{\mu}
\def\s{\sigma}
\def\n{v}
\def\r{\rho}
\def\l{\lambda}

\def\o{\omega}

\def\r{\rho}
\def\s{\sigma}

\def\d{\partial}

\def\vf{\varphi}

\def\e{\epsilon}

\def\dd #1 #2{{\delta #1\over \delta #2}}

\def\ha{\frac{1}{2}}

\def\As{\mathcal{A}\!\!\!/}

\def\n{v}
\def\m{\mu}
\def\n{v}

\begin{document}
\title{Working towards a gauge-invariant description of the Higgs model: from local composite operators
	to spectral density functions.}	
\begin{figure}[H]
	\center
	\includegraphics[width=18cm]{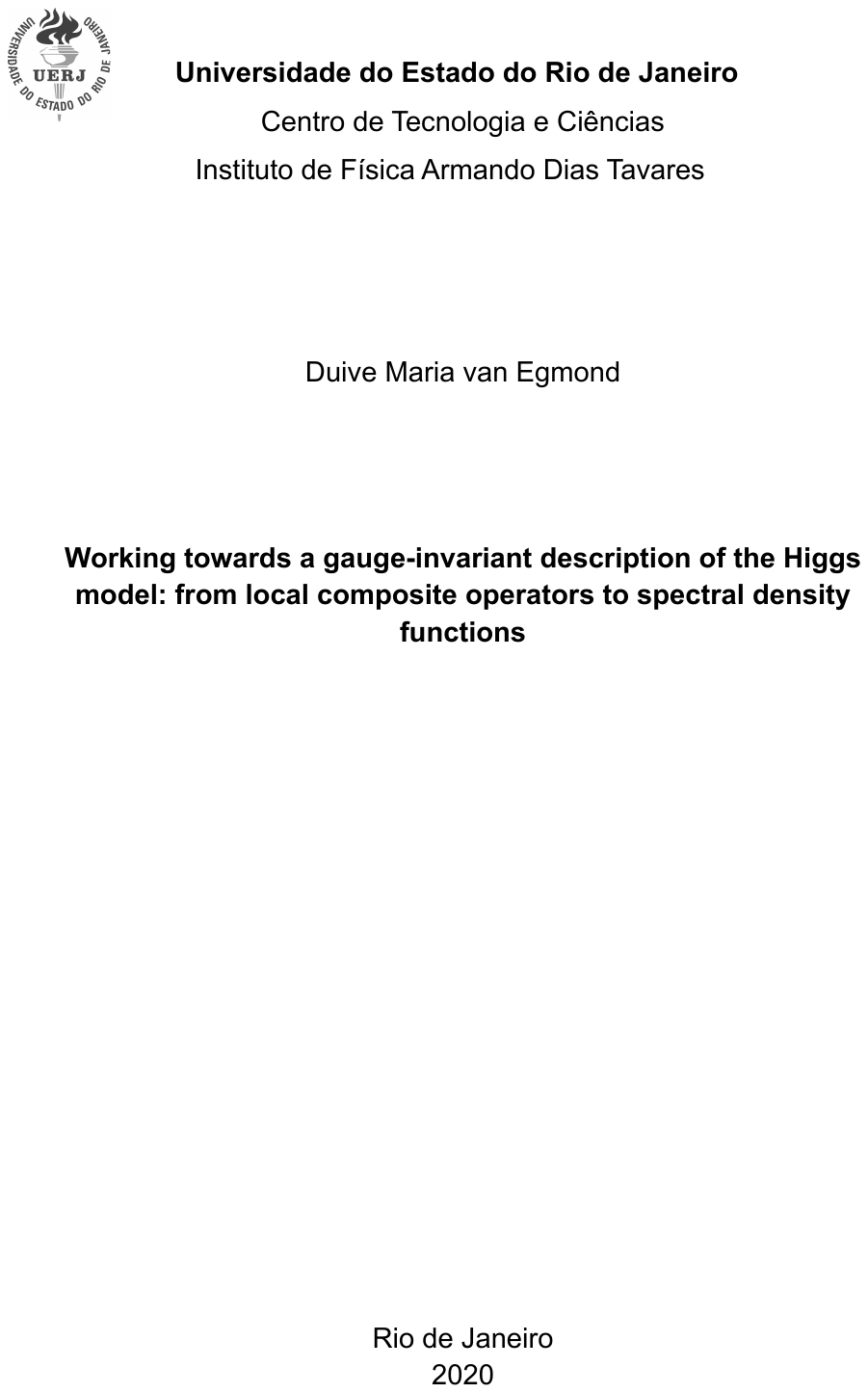}
\end{figure}
\thispagestyle{empty}
\newpage
\begin{figure}[H]
	\center
	\includegraphics[width=18cm]{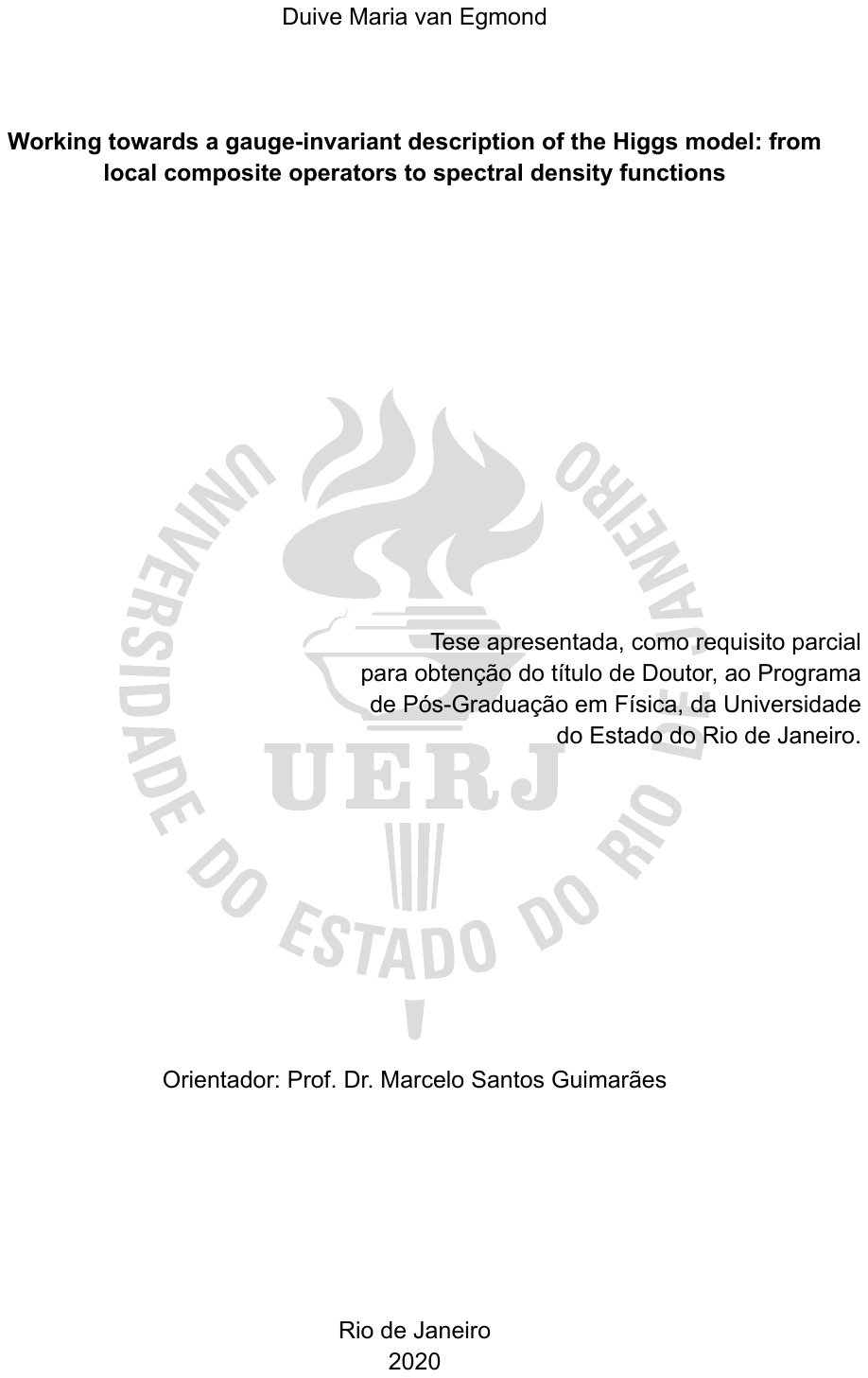}
\end{figure}
\thispagestyle{empty}
\newpage
\begin{figure}[H]
	\center
	\includegraphics[width=18cm]{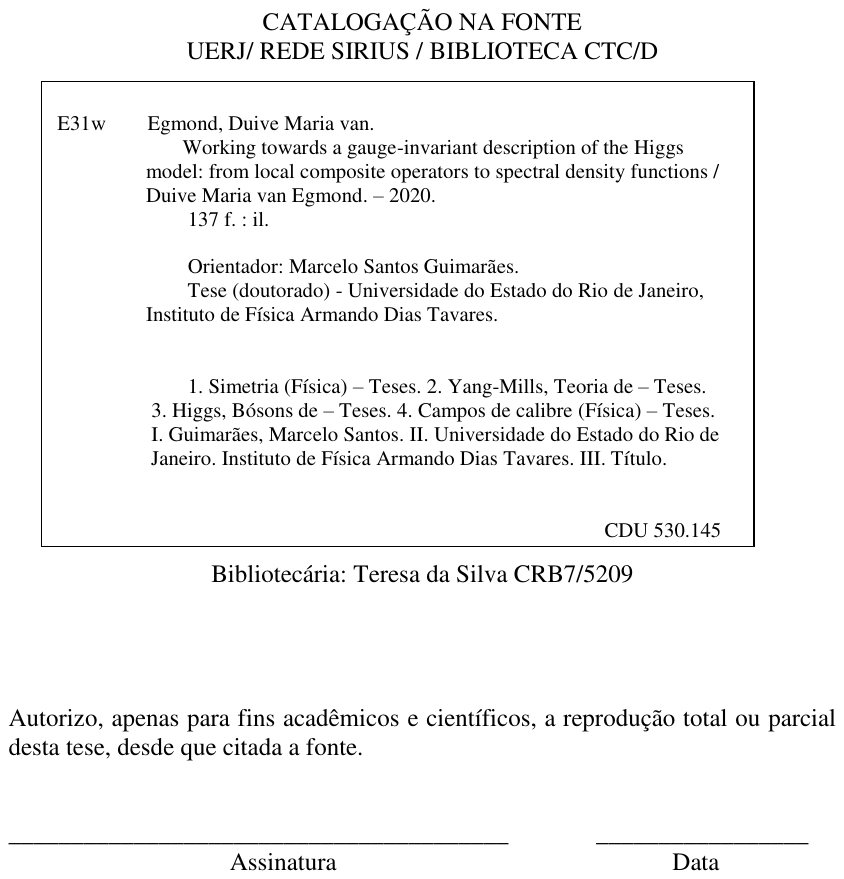}
\end{figure}
\thispagestyle{empty}
\newpage
\begin{figure}[H]
	\center
	\includegraphics[width=18cm]{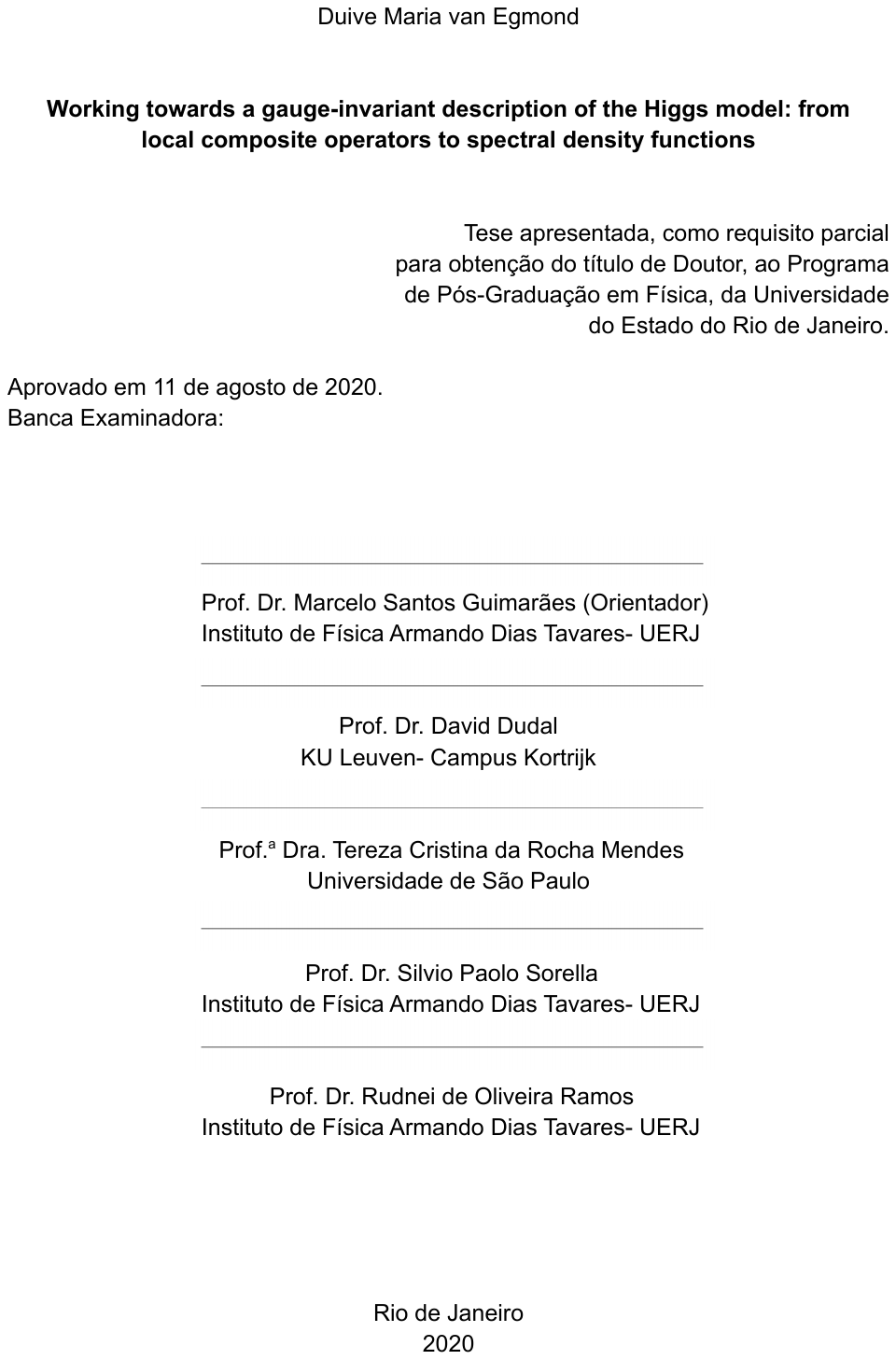}
\end{figure}
\thispagestyle{empty}
\newpage
\begin{center}
This thesis is dedicated to my father, Dr. Jacob Jan van Egmond (1949-2017). He saw this work before I did.
\end{center}
\thispagestyle{empty}

\newpage
\newpage
\section*{\center A word of thanks}
This thesis was supposed to be written during my last months as a PhD student in Rio de Janeiro. Unfortunately, the COVID-19 pandemic kept me in my hometown in the Netherlands, where I have been reminiscing about my time in Brazil while writing this thesis. All of this under the fantastic care of my mother Marieke, and for this and many more reasons I want to thank her first and foremost.
\vspace{.5cm}\\
I want to thank both my parents for providing a home in which education was valued above everything else. I also want to thank my brothers Isaac and Jac and my sister Marie Beth for always, and tirelessly, challenging me with their great intelligence.
\vspace{.5cm}\\
Since I went to Rio de Janeiro four years ago, I have had nothing but support from my friends and family in the Netherlands. I want to thank them all for their love and friendship. All the visits and phone calls have made me feel connected with you throughout this time. I would also like to thank my friends and colleagues at UERJ, with whom I have shared a lot of nice moments, during and after work. A special thanks to Oz\'orio, for all the great discussions, for his encouragement and for supporting me during difficult moments. The way you do not let your love for physics, or in fact, anything you love, be disturbed by anything or anyone, has inspired me a great deal. To have found a partner with whom I can share so many aspects of life is something I could not have dreamt of until we were sitting in a caf\'e together, writing equations on napkins.
\vspace{.5cm}\\
In the four years of my PhD at UERJ, I have learned what it means to be a researcher. There are four people I have to especially thank for that. First of all, prof. dr. Marcelo Guimarães, who taught me to trust myself and follow my curiosity. Then, prof. dr. Silvio Sorella, from whom I learned the art of writing an article, and whose phrase “fecha os olhos e vai” (close your eyes and go) has helped me in many moments during my research. Prof. dr. David Dudal, with whom I share a passion for the connection between small details and the greater picture. Prof. dr. Urko Reinosa, who taught me that you can never be sure enough about a calculation, and that there is sheer pleasure in that.
\vspace{.5cm}\\
Finally, I would like to thank prof. dr. Leticia Palhares and prof. dr. Marcela Peláez for showing me what women in physics can achieve.
\vspace{.5cm}\\
The present work was carried out with the support of the Coordination for the Improvement of Higher Education Personnel - Brazil (CAPES) - Financing Code 001.

\thispagestyle{empty}
\newpage
\begin{flushright}
	\vfill
``All language is but a poor translation"- \textit{Franz Kafka} 
\end{flushright}
\thispagestyle{empty}
\newpage
\begin{abstract}

We analyze different BRST invariant solutions for the introduction of a mass term in Yang-Mills (YM) theories. First, we analyze the non-local composite gauge-invariant field $A^h_{\mu}(x)$, which can be localized by the Stueckelberg-like field $\xi^a(x)$. This enables us to introduce a mass term in the $SU(N)$ YM model, a feature that has been indicated at a non-perturbative level by both analytical and numerical studies. The configuration of $A^h(x)$ , obtained through the minimization of $\int d^4 x A^2$ along the gauge orbit, gives rise to an all orders renormalizable action,
a feature which will be illustrated by means of a class of covariant gauge fixings which, as
much as 't Hooft's $R_{\xi}$-gauge of spontaneously broken gauge theories, provide a mass for the Stueckelberg-like field. \\
Then, we consider the unitary Abelian Higgs model and investigate its spectral functions at one-loop order. This analysis allows to disentangle what is physical and what is not at the level of the elementary particle propagators, in conjunction with the Nielsen identities. We highlight the role of the tadpole graphs and the gauge choices to get sensible results. We also introduce an Abelian Curci-Ferrari action coupled to a scalar field to model a massive photon which, like the non-Abelian Curci-Ferarri model, is left invariant by a modified non-nilpotent BRST symmetry. We clearly illustrate its non-unitary nature directly from the spectral function viewpoint. This provides a functional analogue of the Ojima observation in the canonical formalism: there are ghost states with nonzero norm in the BRST-invariant states of the Curci-Ferrari model.\\
Finally, the spectral properties of a set of local gauge-invariant composite operators are investigated in the $U(1)$ and $SU(2)$ Higgs model quantized in the 't Hooft $R_{\xi}$ gauge. These operators enable us to give a gauge-invariant description of the spectrum of the theory, thereby surpassing certain incommodities when using the standard elementary fields. The corresponding two-point correlation functions are evaluated at one-loop order and their spectral functions are obtained explicitly. It is shown that the spectral functions of the elementary fields suffer from a strong unphysical dependence from  the gauge parameter $\xi$, and can even exhibit positivity violating behaviour. In contrast, the BRST invariant local operators exhibit a well defined positive spectral density.
\end{abstract}
\thispagestyle{empty}
\tableofcontents	
\thispagestyle{empty}
\newpage

\newpage

\newpage
\chapter*{Introduction}
\addcontentsline{toc}{chapter}{Introduction}

\section*{Fundamental forces and field theory: a historical overview}
\addcontentsline{toc}{section}{Fundamental forces and field theory: a historical oversight}
Four fundamental forces seem to constitute nature: Electromagnetic force, Gravitational force and the Strong and Weak nuclear forces.\\
\\
The electromagnetic force was at the heart of the development of modern physics. Classically, it is described by Maxwell's equations from 1862 \cite{maxwell1865viii} in terms of electric charges that generate an electromagnetic field, which in turn exerts a force on other electric charges within that field. At the end of the nineteenth century, this picture was challenged by several experimental observations such as blackbody radiation and the photoelectric effect, which did not have an explanation within the Maxwell equations. In 1900,  Max Planck introduced the idea that energy is quantized in order to derive, heuristically, a formula for the energy emitted by a black body. Planck's law \cite{planck2013theory} marked the beginning of quantum physics.  Subsequently, Einstein explained the photoelectric effect by stating that the energy of an electromagnetic field is quantized in discrete units that were called $photons$. The photon is a $\textit{force carrier}$ or $\textit{messenger particle}$, neutrally charged and massless, that mediates the electromagnetic force between electrically charged matter particles. The phenomenological theories of Planck and Einstein were followed by more rigorous 
descriptions of quantum-mechanical sytems, with the Schr\"odinger wave equation (1925) as the main postulate of modern quantum physics \cite{schrodinger1926undulatory}.\\
\\
 Simultaneously to the emergence of quantum physics, Einstein's theory of special relativity \cite{einstein1905electrodynamics}, built on Maxwell's equations, meant a revolution in the perception of space and time. However, the concepts of relativity seemed incompatible with quantum physics: relativity theory treats time and space on an equal footing, while in quantum physics spatial coordinates are promoted to operators, with time a label. In 1928, Dirac succeeded in writing a relativistic wave equation for the electron \cite{dirac1928quantum}, which however hypothesized negative energy solutions, pointing to the existence of anti-particles (we now know the antiparticle of the electron to be the positron). This eventually led to the development of the first quantum field theory, dubbed Quantum Electrodynamics (QED) by Dirac. In QED, electrons and other matter particles are pictured as excited states of an underlying field, just as photons are excited states of the electromagnetic field. In the associated field operator $\varphi(x,t)$, position and time are both labels, in line with relativity theory. Moreover, QED naturally incorporates negative energy states. \\
 \\
In order to understand the classical electromagnetic field in the light of quantum field theory, we take the classical magnetic vector potential $A$, whose curl is equal to the magnetic field. This quantity had been known from Maxwell's equations but within the Schr\"odinger picture it overtook the electromagnetic field as the fundamental quantity. By construction, the vector potential can be changed by terms that have a vanishing curl without changing the magnetic field. This ``gauge symmetry" became central to QED, where the potential was promoted to a relativistic four-vector $A_{\mu}(x,t)$, the photon field, and interactions of fields were summarized in a Lagrangian that is invariant under transformations of the local $U(1)$ group, called gauge transformations. \\
\\
QED gives the most accurate predictions quantum physics currently has to offer. For example, the calculation of the electron's magnetic moment in QED agrees with experimental data up to ten digits. The success of QED can be attributed to the fact that many of the phenomena in our visible (low-energy) world can be approximated by perturbing the electromagnetic quantum vacuum, known as perturbation theory. Feynman diagrams are the visual representation of the perturbative contributions. However, QED is not well-defined as a quantum field theory to arbitrarily high energies, because the coupling constant (interaction strength) runs to infinity at finite energy. This divergence at high energy scale is known as the Landau pole \cite{landau1955niels}. Other questions concerning the electromagnetic force, such as the existence of a magnetic monopole, still remain open problems today.
\\
\\
 QED has served as the model and template for quantum field theories that try to describe the other three fundamental forces. Attempts to fit Einstein's gravity theory of curved spacetime \cite{einstein2019relativity} into the concepts of quantum field theory have not been an unqualified success. For example, the hypothetical force carrier of the gravitational force, the $\textit{graviton}$, has never been detected. Moreover, the description of gravity as a field theory has been shown to fail at Planck length. Most modern research in quantum gravity is conducted in the framework of String Theory, which takes a different approach to unite quantum phyics and relativity: instead of making time and space both labels, they are both operators. Today, it is even disputed if gravitational force is truly a fundamental force, or an emergent effect of deeper quantum mechanical processes \cite{Verlinde:2016toy}. \\
 \\
 In 1954, Yang and Mills extended the abelian $U(1)$ gauge group of QED to non-abelian gauge groups \cite{Yang:1954ek}. Their goal was to find an explanation for the strong interaction which holds subatomic particles together. Today, we know the Yang-Mills (YM) $SU(3)$ theories as the quantum field theory describing the strong interactions. The matter particles for this force are $\textit{quarks}$, while the strong interaction is mediated by force carriers called $\textit{gluons}$. Particles that interact under the strong force carry a $\textit{color charge}$, and the YM theory of strong interactions is therefore also called Quantum Chromodynamics (QCD). Different from QED, the gauge fields themselves carry color and can thus interact with eachother. Therefore the gauge boson $A_{\mu}$ is displayed as $A_{\mu}^a T^a$, with $a$ the color charge and $T^a$ the $SU(3)$ generators. There are as many $A_{\mu}^a$'s as there are generators, and because the number of generators for an $SU(N)$ group is $N^2-1$, we have eight differently colored gluons.\\
 \\
  The fact that gluons carry color has dramatic consequences for the relation between QCD interactions and energy scale. In QED, the fact that charged particles interact less when they get further away is explained by $\textit{charge screening}$. A charged particle like the electron is surrounded by the vacuum, a cloud of virtual photons and electron-positron pairs continuously popping in and out of existence. Because of the attraction between opposite charges, the virtual positrons tend to be closer to the electron and screen the electron charge. Thus, the effective charge becomes smaller at large distances (low energies, also called the infrared (IR) region), and grow stronger at small distances (high energies, also called the ultraviolet (UV) region). This is completely intuitive to us: the further away, the less interaction. It is also how Feynman diagrams are designed: a free particle comes in from the distance, interacts, and vanishes into the distance again. \\
  \\
 In QCD, quark-antiquark pairs screen the color charged particles in the same way as the electron-positron pairs screen the electrically charged particles. However, the QCD vacuum also contains pairs of the charged gluons, which do not only carry color but also an anti-color magnetic moment. The net effect is not the screening of the color charged particle, but the augmentation of its charge. This means that the particles will interact more at larger distances, or lower energies, and vice versa. This was also observed in experiments: when high energies were applied to quarks, they hardly interacted with eachother. This means a theoretical model should have a coupling constant descreasing to zero in the UV limit. In 1973, Gross, Wilczek and Politzer proved that this is indeed the case for YM theories \cite{gross1973ultraviolet, politzer1973reliable}. This behaviour, called $\textit{asymptotic freedom}$, is seemingly counterintuitive and can be best explained by imagining particles inside an elastic rod: in rest, there is no force between them, but the further you try to pull it apart, the more they will try to get back together. It also reminds of a superconductor: if a magnetic field is forced to run through the superconductor, the energy associated with the created flux tube grows with distance. This has led to some QCD models based on superconductivity, known as dual superconductor models \cite{Ripka:2003vv}. \\
 \\
 Asymptotic freedom means that for QCD, the relation between energy scale and interaction strength is inverted with reference to QED. This means that it should be possible to use perturbation theory in the UV limit. Indeed, for high energy phenomena the perturbative QCD models are in excellent agreement with the experiments \cite{Jegerlehner:2001fb,Jegerlehner:2002em,Martin:2015lxa,Martin:2015rea}. In order to be quantized, these perturbative  models demand a gauge-fixing, or choosing a gauge, in the Faddeev-Popov (FP) procedure. This introduces in YM theories so-called $\textit{ghost fields}$, which are Grassmann fields that violate spin-statistics and can therefore not be physical. After gauge-fixing there is no longer a local gauge-invariance, but there is still a residual symmetry, known as Becchi-Rouet-Stora-Tyutin (BRST) symmetry. BRST symmetry guarantees the unitarity of the quantized YM theories and can be used to derive the Slavnov-Taylor identities, which are fundamental to renormalization methods such as Algebraic Renormalization \cite{Piguet:1995er}.  Important is that the BRST symmetry is nilpotent, which means the BRST variation $s$ applied on another BRST variation is zero, $s^2=0$. Kugo and Ojima \cite{Kugo:1977yx, Kugo:1977mk} used this to distinguish between two types of states that are annihilated by a BRST transformation: those that trivially do so because they themselves are a BRST transformations of another state, and those that are not BRST transformations of another states. Physical states are of the second type and are said to be in the BRST $\textit{cohomology}$. Non-physical states, such as the ghost fields, are of the first type and are outside the BRST cohomology. BRST symmetry therefore is an important tool for the definition of physical space.
\\  
\\
The UV regime of QCD is well described by perturbative YM theories with a FP gauge fixing. However, the standard perturbation theory is unable to access the IR regime because it presents a Landau pole. This may be related to the fact that for large coupling constants the FP procedure is not valid because the gauge-fixing is no longer unique, leading to an infinite number of copies of the gauge field in the model. This was first observed by Gribov in 1978, and is called the $\textit{Gribov ambiguity}$. Over the years, various attempts have been made to deal with this
problem, see for example  \cite{zwanziger1989local,zwanziger1993renormalizability,Dudal:2008sp,Serreau:2012cg,Capri:2015nzw,Zwanziger:2001kw}.
  Until today, no coherent analytical model for the IR region in non-abelian gauge theories is available. Nonetheless, we have learned several things about the IR physics of QCD from lattice simulations. It was found \cite{bowman2007scaling, Cucchieri:2004mf, Strauss:2012dg, Dudal:2013yva, Dudal:2019gvn}, in lattice simulations for the minimal Landau gauge, that the spectral function of the gluon propagator is not non-negative everywhere, which means that in the IR there is no physical interpretation for this propagator like there is for the photon propagator in QED. This behaviour of the gluon
  spectral function is commonly associated with the concept of confinement \cite{cornwall2013positivity,Krein:1990sf,Roberts:1994dr,Lowdon:2017gpp}, which means that quarks and gluons clump together  under the strong force to form composite particles called $\textit{hadrons}$.  Because they are confined, we do not see isolated quarks and gluons in nature. The non-positivity of the spectral
  function then becomes a reflection of the inability of the gluon to exist as a free physical particle. A further important observation in lattice QCD is that the gluon propagator shows massive behaviour in the non-perturbative region. This has prompted research into massive analytical models for the QCD confinement region, which we will discuss in further detail in section \ref{massym}, as well as in chapter \ref{Amin}, where we will discuss a BRST invariant solution for the massive QCD model. 
 \\
\\
Through experiments in the 1950s the weak interaction, responsible for radioactive decay of the atoms, was predicted to be carried by three gauge bosons: two $W$ bosons and one $Z$ boson. This fits with the three generators of the $SU(2)$ YM model. Because the $W$ bosons carry electric charge, the weak and electromagnetic force are described together in an $SU(2) \times U(1)$ theory, called electroweak theory. However, the $W,Z$ bosons are massive, unlike the photon. In the Lagrangian, a mass term for the gauge boson would take the form 

\beq
\mathcal{L} \supset \frac{1}{2}m^2 A_{\mu}A_{\mu},
\eeq
which is not gauge-invariant and can not be implemented into the Lagrangian by hand. We now know that the solution for this problem is given by the spontaneous symmetry breaking (SSB) due to the non-zero vacuum value of the scalar $SU(2)$ Higgs field $\Phi(x)$, and the Higgs mechanism\footnote{A more accurate name is the Higgs-Brout-Englert-Guralnik-Hagen-Kibble mechanism \cite{higgs1964broken, englert1964broken, guralnik1964global}} that gives mass to the gauge bosons. The $SU(2)$ gauge theory with a scalar field is referred to as Yang-Mills-Higgs (YMH) theory. The Glashow-Weinberg-Salam (GWS) model of electroweak symmetry breaking \cite{Glashow:1961tr, Weinberg:1967tq, Salam:1968rm}, based on the Higgs mechanism, gives mass to the $W,Z$ bosons, while keeping the photon massless. Together with QCD, the electroweak theory constitutes the Standard Model (SM) of particle physics. \\
\\
The electroweak sector can be shown to be both unitary and renormalizable by employing a class of gauge-fixing called 't Hooft or $R_{\xi}$ gauge, which introduces the gauge parameter $\xi$. Different choices of $\xi$ highlight different properties of the model. In the formal limit $\xi \rightarrow \infty$, we end up in the unitary gauge, which is considered the physical gauge as it decouples the non-physical particles. However, this gauge is known to be  non-renormalizable \cite{Peskin:1995ev}.  For any finite choice of $\xi$, the model is renormalizable because the $R_{\xi}$ gauge cancels an unrenormalizable mixing term between the gauge boson and an unphysical Goldstone boson. For $\xi \rightarrow 0$, we end up in the Landau gauge, a very useful gauge choice that picks out the transverse part of the gauge boson. Of course, any physical process or quantity should be independent of the $\xi$ parameter. Therefore, the $R_{\xi}$ gauge provides a powerful check on practical calculations.

\section*{Higgs mechanism without spontaneous symmetry breaking}
\addcontentsline{toc}{section}{Higgs mechanism without spontaneous symmetry breaking}

In most textbook accounts, the mass generation of gauge bosons through the Higgs mechanism is displayed in terms of the notion of spontaneous symmetry breaking (SSB) of the local gauge symmetry. However, the meaning of SSB in connection to the Higgs mechanism is ambiguous, since it was established already in 1975 by Elitzur \cite{elitzur1975impossibility} that local gauge symmetries can never be spontaneously broken. In this section, I will try to make sense of these contradictory accounts and discuss the solution provided by the gauge-invariant composite local operators of 't Hooft \cite{tHooft:1980xss} and Fr\"ohlich, Morchio and Strocchi  \cite{Frohlich:1980gj, Frohlich:1981yi}, which are the central subject of this thesis. \\
\\
The term ``spontaneous symmetry breaking'' originated in the statistical physics of phase transitions. One of the best known examples is that
of SSB in a ferromagnet. Above the Curie temperature $T_C$, the ground state of the ferromagnet is rotationally symmetric because of the random spin orientation of the atoms. Below $T_C$ however, the ground state consists of spins which are aligned within a certain domain, thus breaking the rotational symmetry. The orientation of this alignment is random in the sense that each direction is equally likely to occur, but nevertheless one direction is chosen. So, even though the system still has a rotational symmetry, the ground state is not invariant under this symmetry and we say the symmetry is spontaneously broken.\\
Any situation in physics in which the ground state of
a system has less symmetry than the system itself, exhibits SSB. Two different configurations of the ground state are seperated by an energy barrier, and the phenomenon of SSB therefore goes accompanied by a discontinuous change of a physical quantity related to the free energy, called the order parameter, as a function of another quantity, the control parameter. In the case of a ferromagnet, the magnetic susceptibility (order parameter) goes to infinity at the critical temperature (control parameter).
\begin{figure}[H]
	\center
	\includegraphics[width=10cm]{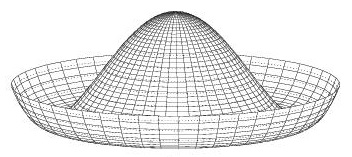}
	\caption{Representation of the potential for the global $U(1)$ symmetry}
	\label{mexhat}
\end{figure}
Symmetry breaking of the ground state is always connected with mass generation: movement of the ground state in direction of the energy barrier has an energy cost and as is therefore ``massive'', while a movement in the direction perpendicular to the energy barrier costs zero energy, i.e. it is a ``massless" mode. These concepts become more explicit in field theory, where the massive and massless modes correspond to genuine massive and massless bosons. This was first described by Nambu \cite{Nambu:1960tm} for the spontaneous breaking of the chiral symmetry of massless
fermions and was subsequently elucidated by Goldstone \cite{Goldstone:1961eq}. We can illustrate the occurrence of massless and massive SSB bosons in the simple model of a complex scalar field $\varphi(x)$ with $U(1)$ symmetry. The Lagrangian

\beq
\mathcal{L}&=&\partial_{\mu} \varphi^{\dagger}\partial_{\mu} \varphi - V(\varphi^{\dagger} \varphi) \,\,\,\,\,{\text{ with}} \,\,\,\ V(\varphi \varphi^{\dagger})=\frac{\lambda}{2} \left(\varphi^{\dagger} \varphi -\frac{v^2}{2}\right)^2, \,\,\,\lambda> 0
\label{11}
\eeq
is invariant under the global $U(1)$ transformation $\varphi \rightarrow e^{i \alpha} \varphi$. The minimum of the potential is degenerate, because any configuration of the vacuum expectation value (vev) $\braket{\varphi}=\frac{v}{\sqrt{2}} e^{i \theta}$, with $\theta$ an arbitrary phase, minimizes the potential. This is the famous ``Mexican hat'' potential depicted in Figure \ref{mexhat}. Any choice for $\theta$ will lead to the same $U(1)$ invariant Langrangian. However, after choosing a particular $\theta$, the vev $\braket{\varphi}$ is not $U(1)$ invariant since under a global $U(1)$ transformation $\theta \rightarrow \theta + \alpha$. Thus, the $U(1)$ symmetry is spontaneously broken by the vev of the scalar field. 
Let us now choose one vacuum orientation, e.g. $\theta=0$ so that $\braket{\varphi}=\frac{v}{\sqrt{2}}$. We can define two new real scalar field $\rho(x)$ and $\chi(x)$ with zero vev, $\braket{\rho}=\braket{\chi}=0$, and set $\varphi= \frac{1}{\sqrt{2}}\left(v+\rho\right)e^{-i \chi/v}$. Then, in the Lagrangian \eqref{11} the $\rho$ field acquires a mass $m_{\rho}= \lambda v$, while the field $\chi$ remains massless. The massless boson is called the Nambu-Goldstone (NG) boson, and the $U(1)$ case is the simplest example of the $\textit{Goldstone theorem}$, which states that for every broken generator there is one massless Goldstone boson. For the $U(1)$ model, the SSB breaks the complete symmetry, but for more complex systems, SSB can break the symmetry group down to a subgroup under which the ground state is invariant. For example, the SSB of the chiral symmetry in QCD breaks a global $SU(3)\times SU(3)$ chiral flavor symmetry down to the diagonal $SU(3)$ group, generating eight NG bosons.
\\
\\
As is well-known, the global $U(1)$ symmetry can be promoted to a local $U(1)$ gauge symmetry by changing the derivative $\partial_{\mu}\varphi$ in \eqref{11} to the covariant derivative $D_{\mu}\varphi=\partial_{\mu}\varphi - i e A_{\mu }\varphi$, with $A_{\mu}$ the gauge boson. The Lagrangian  

\beq
\mathcal{L}&=&F_{\mu\nu}F_{\mu \nu}+(D_{\mu} \varphi)^{\dagger} D_{\mu} \varphi - V(\varphi^{\dagger} \varphi)\,\,\, \,\text{with}\,\,\,\, V(\varphi \varphi^{\dagger})=\frac{\lambda}{2} \left(\varphi^{\dagger}\varphi -\frac{v^2}{2}\right)^2, \,\lambda> 0,
\label{121b}
\eeq
with $F_{\mu\nu}=\partial_{\mu}A_{\nu}-\partial_{\nu}A_{\mu}$, is invariant under the local $U(1)$ transformations $\varphi\rightarrow e^{i e \alpha(x)} \varphi$  and $A_{\mu} \rightarrow A_{\mu}+ \partial_{\mu}\alpha(x)$. Naturally, the global $U(1)$ transformations form a subgroup of the local $U(1)$ transformations, namely when $\alpha(x)$ is constant for every point in spacetime, $\alpha(x)=\alpha$. At first glance, it looks like we can repeat the SSB mechanism described for the global $U(1)$ symmetry by taking $\braket{\varphi}=\frac{v}{\sqrt{2}}$. The field configuration $\varphi= \frac{1}{\sqrt{2}}\left(v+h\right)e^{-i \chi/v}$ then will give a mass to the gauge boson, $m_A= e v$. While the Lagrangian is still invariant under a $U(1)$ gauge transformation, the vev of the scalar field is not; we seem to have a SSB of the local gauge symmetry. The $\chi(x)$ field would be the NG boson for the broken $U(1)$ gauge theory, but for gauge theories this is not a physical degree of freedom since we can use the gauge freedom to set $\chi(x)=0$; this gauge choice is called unitary gauge.\\
\\
The process described above is how the Higgs mechanism is displayed in most introductory text on the subject. It can easily be generalized to non-abelian gauge theories, and it is used to describe the $SU(2)_L\times U(1)_Y \rightarrow U(1)_{EM}$ local symmetry breaking which explains mass generation for the $W$ and $Z$ bosons, as well as for several fermion fields. In section \ref{hew} we will go further into the details of the Higgs mechanism in the electroweak sector.\\
\\
 Despite the phenomological success, however, there are some important objections to be made against the concept of SSB of a local gauge symmetry. The most important result in this context is Elitzur's theorem, which states that $\textit{local gauge symmetries can}$ $\textit{never be broken spontaneously}$. The theorem is rigorously proven on the lattice \cite{elitzur1975impossibility}, but can be understood by the 2-D Ising model, a simple lattice configuration with two symmetry-breaking ground states: all $\uparrow$ and all $\downarrow$. In this system, to go from one ground state to the other there is an extensive energy cost, i.e. there is an energy barrier to overcome, because a change in the ground state configuration encompasses the whole system. Therefore, at low temperature the system gets stuck in a particular spin alignment and the symmetry is broken. In constrast, for local symmetries this argument does not apply. Imagine a local transformation, only acting on a very small subset of the system: this will change a ground state configuration into another ground state configuration at zero energy cost. As a consequence, there is no energy barrier between any two ground states, because they are always related  by a sequence of local transformations. The system will be able to explore the entire space of ground states and there is no symmetry breaking. In the case of the local gauge symmetry, this means that $\phi (x)$, rather than being stuck in a certain gauge configuration, can move around freely around the ``gauge orbit". This means that when you compute the average value of the operator over the group it must be zero; Elitzur's theorem is also stated as that $\textit{all gauge non-invariant operators must have a vanishing vev}$. For a more formal treatment of Elitzur's theorem in lattice theory, see for example \cite{itzykson1991statistical}. 
\\
\\
Of course, local gauge symmetries can be broken $\textit{explicitly}$ by adding a symmetry breaking term to the Lagrangian. In fact, for the quantization of a field theory we have to break the local gauge symmetry by choosing a specific gauge configuration, i.e. by gauge fixing. It seems that gauge fixing offers a way out of Elitzur's theorem. Say, for example, that in the $U(1)$ model \eqref{121b} we fix the gauge in the Landau gauge, $\partial_{\mu}A_{\mu}=0$. Since the very purpose of a gauge fixing constraint is to restrict the domain of field configurations, the constraint itself is of course not invariant under a local gauge transformation. Exceptions are the so-called Gribov copies, which appear when the gauge fixing constraint is preserved under local gauge transformations. Gribov copies jeopardize the field description of non-abelian theories in the low-energy (non-perturbative) regime. For small coupling constants, however, the Gribov copies can be ignored and it can be shown that e.g. the Landau gauge does not allow for local gauge transformations in the perturbative region. Even so, it is easy to see that $\partial_{\mu}A_{\mu}=0$ is still invariant under the global subgroup of the $U(1)$ gauge transformation, $\alpha(x)=\alpha$. The residual global symmetry which remains after gauge fixing is called the $\textit{remnant global gauge symmetry}$ \cite{Caudy:2007sf}. There is no theorem that forbids the SSB of $\textit{this}$ symmetry. We can then state that the Higgs mechanism is in fact the breaking of the remnant global symmetry by setting $\braket{\phi}=\frac{v}{\sqrt{2}}$ after gauge fixing. This is in fact what is done in some modern takes on the Higgs mechanism \cite{englertorigin}.\\
\\
However, there are some important distinction to be made between the SSB of the global gauge symmetry and the original formulation of SSB. In the ferromagnet, the SSB of the spin alignment means that out of all $\textit{physically}$ distinct configurations, one is actually realised. In contrast, the choice of a particular configuration within the global gauge symmetry does not reflect any physical outcome, and we therefore do not expect an accompanying abrupt change in any physical properties, in particular those related to the free energy. This is in agreement with findings on the lattice \cite{Fradkin:1978dv} that there is no order parameter which varies discontinuously along the path going from the ``confinement phase'' (strong coupling) to the ``Higgs phase'' (weak coupling), at least in the fundamental representation of the Higgs field. This absence of a phase boundary is known as ``Higgs complementarity". \\
\\
A further observation about remnant global gauge symmetry is that it is ambiguous because different gauge choices will lead to different global subgroups after fixing the gauge. For example, the aforementioned unitary gauge leaves no global symmetry, while the Coulomb gauge $\partial_i A_i=0$ leaves a large invariance, $\alpha(x)=\alpha(t)$. The question is then to decide which, if any, of these gauge-symmetry breakings is associated with a transition between physically different phases. In \cite{Caudy:2007sf}, it was found on the lattice that for SSB of the global remnant symmetry, the Landau gauge and the Coulomb gauge show distinct transition lines. Moreover, they both show a SSB at points where there is no actual phase change, making their role in the confinement-Higgs transition highly doubtful.\\
\\
Another, often forgotten problem with the approach of gauge-fixing to surpass Elitzur's theorem is that choosing a gauge leaves a residual local symmetry: the BRST symmetry. The vacuum should, even after gauge-fixing, always be BRST invariant because it is a physical quantity. It thus seems there is no way around using gauge-invariant fields to represent physical particles. In non-abelian theories, these fields are necessarily composite. In \cite{tHooft:1980xss}, 't Hooft defined composite operators for the $SU(2)$ Higgs model. These operators are gauge-invariant combinations of the elementary YM fields, the gauge boson and the fermion fields, with the Higgs field. They are constructed in such a way that for a constant value of the Higgs field, we regain the elementary field. The gauge-invariant composite extension of the neutral $Z$-boson is defined as  $\Phi^{\dagger}D_{\mu}\Phi$, while that of the charged $W^{\pm}$ bosons is given by $\varepsilon_{ij}\Phi_i D_{\mu}\Phi_j$ and its complex conjugate.\footnote{While they do not play a role in this work, fermions can also be given a gauge-invariant extension, e.g. $\Phi^{\dagger}\Psi$ for the neutrino and $\varepsilon_{ij}\Phi_i \Psi_j$ for the electron.} The Higgs field itself can be obtained from $\Phi^{\dagger}\Phi$. 't Hooft's definitions incorporate the Higgs complementarity, because there is no fundamental difference between the particle in the Higgs phase and the confining phase. In the confinement phase, the composite operators are bound states of the fundamental field with extremely strong confining forces. In the Higgs phase, the perturbative expansion of these composite operators will lead to gauge-invariant descriptions of
the expected weakly-interacting degrees of freedom.\\
\\
In \cite{Frohlich:1980gj, Frohlich:1981yi}, Fr\"ohlich, Morchio and Strocchi (FMS) further formalized the role of gauge-invariant local composite operators in the Higgs mechanism. Because of their gauge invariance, the composite operators cover the whole gauge orbit, and a non-zero vev for these operators cannot break the local or global gauge invariance. Thus, taking for example the gauge-invariant extension of the Higgs field, $\mathcal{O}=\Phi^{\dagger}\Phi$, we can assign this field a vev $\braket{\mathcal{O}}=\frac{v^2}{2}$. Now, we can achieve the usual dynamics of the Higgs mechanism by taking $\Phi= \frac{1}{\sqrt{2}}\left(v+\rho\right)e^{-i \chi/v}$, with $\braket{\rho}=\braket{\chi}=0$, without having to impose $\braket{\Phi}=v$. This is the Higgs mechanism without a symmetry breaking order parameter. The FMS mechanism gives a simple relation between the correlation functions of the gauge-invariant fields $\tilde{\mathcal{O}}(x)$ and the corresponding gauge dependent ones $\varphi=(A_{\mu}^a, h)$ calculated in the standard perturbation expansion. Expanding the two-point function of the composite operator we find
\beq
\braket{\tilde{\mathcal{O}}(x)\tilde{\mathcal{O}}(y)}&\sim&\braket{\varphi(x)\varphi(y)}_{\text{1-loop}}+... \,,
\label{popo}
\eeq
where $...$ denote higher-order loop corrections, combinations of the elementary fields which make the sum of correlation functions gauge-invariant.
\\
\\
In QCD, the use of composite operators is common in the description of hypothetical composite bound states in the confinement (IR) region. For example, glueballs, bound states solely composed of gluons, have a field theoretical description as the gauge-invariant operator $F_{\mu\nu}^2$ \cite{v2011study}. There are however some important differences between configurations like glueballs and the local composite gauge-invariant operators. The local composite  gauge-invariant operators have a clear connection to elementary fields: the first transforms into the latter when the Higgs field acquires a constant value. As a consequence, these operators can be analyzed in the Higgs phase by perturbative loop calculations. \\
\\
The gauge-invariant FMS construction has gained little attention throughout the years. In most texts where it appears, it serves mainly as a justification of the Higgs mechanism, while for practical matters the SSB of the gauge symmetry is used. The question is whether the FMS mechanism serves a purpose outside this formal role. To try and answer that question, let us look at the gauge boson propagator

\beq
\braket{A_{\mu}(x)A_{\mu}(y)},
\label{prop}
\eeq
which plays a fundamental role in QED, describing the quantum dynamical path that a photon takes when it travels from $x$ to $y$, calculated by means of Feynman-diagrams. In $U(1)$ theory, the transversal part of the propagator is gauge-invariant and can therefore be associated with physically observable quantities. However, in $SU(N)$ theory the (transverse) propagator is not gauge-invariant (or BRST invariant after gauge fixing) and it is therefore impossible that this object would describe a physical quantity. Nevertheless, the propagator \eqref{prop} is a much used quantity in electroweak theory, and succesfully so: the  higher  order  calculations  of  the  pole  masses  and  cross  sections  derived from the propagator \eqref{prop} are in very accurate agreement with the experimental data \cite{Jegerlehner:2001fb,Jegerlehner:2002em,Martin:2015lxa,Martin:2015rea}. This apparent paradox can be explained by the FMS construction. Looking at eq. \eqref{popo}, we can see that the pole mass of the elementary field $\varphi(x)$ should coincide with the pole mass of the gauge-invariant composite operator $\tilde{\mathcal{O}}(x)$. This was confirmed in preliminary lattice result \cite{maas2014two, maas2015field}. This means the pole mass of the gauge-dependent propagator \eqref{prop} is gauge-invariant and can be interpreted as a physical quantity. 
The gauge-invariance of the pole mass of \eqref{prop} can also be proven by the so-called Nielsen
identities \cite{Nielsen:1975fs,gambino1999fermion,gambino2000nielsen,Grassi:2000dz} which follow from the Slavnov-Taylor identities. The Nielsen identities ensure that the pole masses of the propagators of the transverse gauge bosons and Higgs field
do not depend on the gauge parameter $\xi$ entering the $R_{\xi}$ gauge fixing condition.\\
 \\
 Nevertheless, as one can easily figure out, the use of the non-gauge-invariant fields has its own
limitations which show up in several ways. For example, the spectral density function of the elementary two-point correlation function $\braket{\varphi(x)\varphi(y)}$ in terms of the K\"all\'{e}n-Lehmann (KL) representation is not protected from gauge-dependence, because the higher-order loop correction in \eqref{popo} will contribute to the gauge-invariance of the $\tilde{\mathcal{O}}(x)$ spectral density function.  These higher-order loop corrections are the main subject of the present work.  We will show how the local gauge invariant composite operators can be introduced in the Higgs model and analyzed by perturbative loop calculations. The main goal is to gain insight in the spectral properties of the gauge bosons and Higgs fields.  Calculations of spectral properties of elementary fields are plagued by an unphysical gauge-dependency, but through the use of the gauge-invariant composite operators, these feautures can be made visible.

\section*{Outline of this thesis}

This thesis is organized as follows. In chapter \ref{I}, we will discuss existing massive solutions for YM theories. First, we will discuss in detail the GWS model of electroweak symmetry breaking of the Standard Model (SM) in section \ref{G}. We will also discuss other examples of the Higgs mechanism, that are not part of the SM, in \ref{beyond}. Massive solutions for the QCD sector are discussed in section \ref{massym}. Chapter \ref{Amin} is based on \cite{Capri_2018} and was also treated in \cite{ozthesis}. In this chapter, we will look at renormalizable gauge class for the non-local gauge invariant configuration $A_{\mu}^h(x)$, which can be localized by the dimensionless auxiliary Stueckelberg field $\xi$. In chapter \ref{VII}, which is based on \cite{Dudal:2019aew}, we will discuss the spectral properties of the $U(1)$ Higgs model. The main purpose of this chapter is to establish the dependence of the spectral properties for the elementary fields on the gauge parameter. This analysis allows to disentangle what is physical and what is not at the level of the
elementary particle propagators. In chapter \ref{VV}, the gauge-invariant spectral description of the $U(1)$ Higgs model from local gauge-invariant composite operators will be discussed, as was done in \cite{Dudal:2019pyg}. In \ref{VIII}, we will extend this analysis to the $SU(2)$ Higgs model, as done in \cite{Dudal:2019pyg2}. Chapter \ref{con} summarizes our conclusions and discusses some ideas for future projects, based on the results of this work. \\
\\
Throughout this thesis, we shall work in Euclidean four-dimensional space-time, unless otherwise indicated.
\newpage
\chapter{Mass generation within and beyond the Standard Model\label{I}}
\section{The Higgs mechanism in the Standard Model \label{hew}}

The SM describes the strong interactions, weak interactions and hypercharge in an $SU(3)_c \times SU(2)_L \times U(1)_Y$ gauge theory, where $c$ stands for color, $L$ for the left-handed fermions it couples to and $Y$ for the hypercharge. Here, we will discuss the Georgi-Weinberg-Salam (GWS) model, where $SU(2)_{L}\times U(1)_Y \rightarrow U(1)_{\text{EM}}$. Before engaging in the Higgs mechanism for this model, we will first discuss the gauge sector and the Higgs sector for the SM. We will not discuss the fermion sector here, but it is important to realize that quark and lepton masses are also generated through coupling with the Higgs field, known as Yukawa coupling. For a nice complete overview, see \cite{logan2014tasi}.

\subsubsection{The SM gauge sector \label{G}}
The four-dimensional action of the SM is described in terms of field strength tensors by:

\beq
S_{\textit{gauge}}&=&\int d^4 x \frac{1}{4} G^a_{\mu \nu}G^a_{\mu \nu}+ \frac{1}{4}W^a_{\mu \nu}W^a_{\mu \nu}+ \frac{1}{4} B_{\mu \nu} B_{\mu \nu},
\label{gauge}
\eeq
with repeated indices taken as summed. \\\\
For $SU(3)_c$ we have the following field strength tensor

\beq
G^a_{\mu \nu}&=& \partial_{\mu} G_{\nu}^a - \partial_{\nu} G_{\mu}^a+g_s \,f^{abc} G_{\mu}^b G_{\nu}^c,
\eeq
 with $g_s$ the strong interaction coupling strength and $f^{abc}$ the antisymmetric structure constant. There are eight different color charges, corresponding to the eight generators $T^a$ of $SU(3)$, given in matrix representation by $\lambda^a/2$, with $\lambda^a$ the Gell-Mann matrices. \\\\
For the $SU(2)_L$ interaction, the field strength tensor is given by 
\beq
W^a_{\mu \nu}&=& \partial_{\mu} W_{\nu}^a - \partial_{\nu} W_{\mu}^a+g\, \epsilon^{abc} W_{\mu}^b W_{\nu}^c,
\eeq
with $g$ the weak interaction coupling strength. There are three different charges, so the structure constant is equal to the Levi-Civita tensor $f^{abc}=\epsilon^{abc}$. The generators $t^a$ are given in matrix representation by $\tau^a/2$, with $\tau^a$ the Pauli matrices. \\\\
For both $SU(3)_c$ and $SU(2)_L$, and any non-abelian group, the relation between the group generators is given by 
\beq
[t^a, t^b]= i f^{abc} t^c.
\label{tta}
\eeq
Finally, we have the abelian field strength tensor for the $U(1)_Y$ interaction
\beq
B_{\mu \nu}&=& \partial_{\mu}B_{\nu}-\partial_{\nu}B_{\mu}.
\eeq
\\
The gauge transformations which leave the Lagrangian \eqref{gauge} invariant are 

\beq
SU(3)_c: \,\, G_{\mu}& \rightarrow& U_c(x) G_{\mu} U_c^{-1}(x)+ \frac{i}{g_c}U_c(x) \partial_{\mu} U_c^{-1}(x),  \nonumber\\
U_c &=& \exp (-i g_c \o_c^a(x)T^a)
\nonumber\\
\nonumber \\
SU(2)_L:\,\, W_{\mu}& \rightarrow& U_L(x) W_{\mu} U_L^{-1}(x)+ \frac{i}{g}U_L(x) \partial_{\mu} U_L^{-1}(x), \nonumber\\
  U_c &=& \exp (-i g \o_L^a(x)t^a) \nonumber\\
 \nonumber\\
U(1)_Y: \,\,B_{\mu}& \rightarrow &U_Y(x) B_{\mu} U_Y^{-1}(x)+ \frac{i}{
	g' Y}U_Y(x) \partial_{\mu} U_Y^{-1}(x), \nonumber\\
 U_{Y}(x)&=& \exp \left(-i g'\o_Y (x)Y\right)  \label{P5}
\eeq
with $G_{\mu}= G^a_{\mu}T^a$, $W_{\mu}= W^a_{\mu}t^a$ and $g'$ the coupling strength of the hypercharge interaction. In infinitesimal form, this gives 

\beq
G^a_{\mu}& \rightarrow& G^a_{\mu}-  \partial_{\mu}  \omega_c^a(x)-g f^{abc}G_{\mu}^b \o^c_c(x)
\nonumber\\
\nonumber \\
W^a_{\mu}& \rightarrow& W^a_{\mu}-  \partial_{\mu}  \o_{L}^a(x)-g \epsilon^{abc}W_{\mu}^b \o^c_L(x) \nonumber\\
\nonumber \\
B_{\mu}&\rightarrow & B_{\mu}-\partial_{\mu}\omega_Y(x)
  \label{P6}
\eeq

\subsection{The SM Higgs sector}

Clearly, the gauge sector does not allow for a massive term of the gauge bosons to be inserted in by hand, since this would violate gauge-invariance. To explain the experimentally-observed nonzero mass of the $SU(2)_L$ gauge bosons, the SM requires a new ingredient. We therefore introduce a single $SU(2)_L$-doublet scalar field, which will lead to a mass generation of the gauge bosons by means of the Higgs mechanism. \\\\
The Higgs field $\Phi$ is given by an $SU(2)_{L}$-doublet of complex scalar fields that can be written as 

\beq
\Phi&=&\frac{1}{\sqrt{2}}\begin{pmatrix} 
	\phi^+ \\
	\phi^0
\end{pmatrix}=\frac{1}{\sqrt{2}}\begin{pmatrix} 
	\phi_1+i \phi_2 \\
	\phi_3+i \phi_4
\end{pmatrix},
\label{uuz}
\eeq
with $\phi_i$ properly normalized real scalar fields. The Lagrangian for the Higgs sector is given by

\beq
\mathcal{L}_{\Phi}&=& \left(\mathcal{D}_{\mu} \Phi\right)^{\dagger}\left(\mathcal{D}_{\mu} \Phi\right)-V(\Phi),
\label{Lphi}
\eeq
with the covariant derivative 

\beq
\mathcal{D}_{\mu}&=& \partial_{\mu}- i g' B_{\mu}Y-i g W_{\mu}.
\label{3}
\eeq
The Lagrangian \eqref{Lphi} is invariant under the gauge transformations \eqref{P5} in combination with 

\beq
 &\Phi& \rightarrow U_{Y}(x) \Phi \nonumber\\
 &\Phi& \rightarrow U_{L}(x) \Phi\label{P},
\eeq
with $U_L(x)$ and $U_Y(x)$ as defined in \eqref{P5}. We assign a hypercharge $Y=1/2$ to the Higgs field, and make it a color singlet.\\\\
The most general gauge invariant potential energy function is given by

\beq
V(\Phi)=\frac{\lambda}{2} \left( \Phi^{\dagger} \Phi - \frac{v^2}{2} \right)^2,
\label{pott}
\eeq
where we will choose $\lambda$ and $v$ to be real and positive numbers.

\subsection{The SM Higgs mechanism}

In the introductory chapter, we have seen that the Higgs mechanism cannot be caused by a non-zero vev of the Higgs field $\Phi(x)$, because this field is not gauge-invariant. However, this is not contradictory with the statement that the potential \eqref{pott} is minimized by the configuration

\beq
\Phi_0= \frac{1}{\sqrt{2}} \begin{pmatrix} 
	0 \\
	v
\end{pmatrix},
\label{nietvev}
\eeq
as long as we do not attribute it any physical meaning, such as vacuum expectation values. In fact, eq. \eqref{nietvev} is effectively expressing the attribution of a non-zero vev to the composite local gauge-invariant operator, namely $\braket{\Phi^{\dagger}\Phi}=\Phi_0^{\dagger}\Phi_0=\frac{v^2}{2}$.
\\\\
We can establish the configuration \eqref{nietvev} by choosing

\beq
\phi_{3,0}=v, \,\,\,\,\,\,\,\,\,\,\,\,\,\,\, \phi_{1,0}=\phi_{2,0}=\phi_{4,0}=0.
\eeq
We can also define a new scalar field $h$ defined by

\beq
\phi_3=h+ v
\eeq
so that $h_0=0$, while renaming 

\beq
\phi_1&=& \rho_2, \,\,\,\, 
\phi_2 = \rho_1 ,\,\,\,\, 
\phi_4= -\rho_3.
\eeq
The Higgs field then becomes

\beq
\Phi = \frac{1}{\sqrt{2}} \begin{pmatrix} 
	i\rho_1 +  \rho_2 \\
	v + h  - i \rho_3
\end{pmatrix}= \frac{1}{\sqrt{2}}\left(\left(v+h\right)\mathbbm{1}+ i \rho^a \tau^a\right) \cdot \begin{pmatrix} 0 \\ 1 \end{pmatrix}.
\label{23}
\eeq
Putting this configuration into the potential function \eqref{pott}, we find that $h$ gains a mass $m_h=\sqrt{ \lambda} v$, while the $\rho^a$ fields are massless. We can rewrite \eqref{23} up to first order in the fields as

\beq
\Phi=  \frac{1}{\sqrt{2}} \exp \left( \frac{i \rho^a \tau^a}{v}\right) \begin{pmatrix}0 \\ v+h \end{pmatrix},
\eeq
and ``gauge away" the fields $\rho^a$ by making the appropriate $SU(2)$ gauge transformation. This means the $\rho^a$ fields are no physical fields; they are ``would-be" Goldstone bosons. This gauge choice is the unitary gauge, and the Higgs field after gauge fixing becomes 

\beq
\Phi=  \frac{1}{\sqrt{2}}  \begin{pmatrix}0 \\ v+h \end{pmatrix}.
\eeq
\\
The minimizing configuration \eqref{nietvev} is not invariant under the gauge transformations \eqref{P}, because $\tau^a \Phi_0 \neq0$ and $\mathbbm{1}\Phi_0 \neq 0$. Thus, all the generators of $SU(2)_L \times U(1)_Y$ are broken by the configuration \eqref{nietvev}. However, one linear combination of these generators remains unbroken

\beq
(t^3+ Y)\Phi_0 &=& \frac{1}{2 \sqrt{2}} \left(t^3+\mathbbm{1} \right)\begin{pmatrix} 
	0 \\
	v
\end{pmatrix}=0.
\eeq
This is the electric charge operator

\beq
Q= t^3+ Y = \begin{pmatrix} 
	1/2 & 0\\
	0 & -1/2
\end{pmatrix}
+
\begin{pmatrix} 
	1/2 & 0\\
	0 & 1/2
\end{pmatrix}
= 
\begin{pmatrix} 
	1 & 0\\
	0 & 0
\end{pmatrix}
\eeq
and we see from \eqref{uuz} that $\phi^+$ is charged, while $\phi^0$ is uncharged. Thus, the electromagnetic $U(1)$ symmetry group is unbroken, and the Higgs mechanism for the GWS model gives $SU(2)_L \times U(1)_Y \rightarrow U(1)_{EM}$.
\\
\subsection{Mass generation for gauge bosons}

Let us look at the gauge-kinetic term of the Lagrangian \eqref{Lphi}. The terms quadratic in the gauge fields are 
\beq
\left(\mathcal{D}_{\mu} \Phi\right)^{\dagger}\left(\mathcal{D}_{\mu} \Phi\right) \supset \frac{g^2 v^2}{8} W_{\mu}^a W_{\mu}^a+ \frac{g'^2 v^2}{8} B_{\mu}B_{\mu}-\frac{g g' v^2}{4} B_{\mu}W^{3 \mu},
\label{124}
\eeq
and we can write this in matrix form in the basis $(W^1, W^2, W^3,B)$:

\beq
M^2= \frac{v^2}{4} \begin{pmatrix}
	g^2 && 0 && 0 && 0 \\
	0 && g^2 && 0 && 0 \\
	0 && 0 && g^2 && -g g' \\
	0 && 0 && - g g' && g'^2
\end{pmatrix},
\eeq
so that for the pure $SU(2)$ theory, $g'=0$, there is a symmetry under the rotation $W^1 \leftrightarrow W^2 \leftrightarrow W^3$, i.e. it is invariant under $W^a_{\mu} \rightarrow W^a_{\mu}+ \varepsilon^{abc}\omega^b W^c_{\mu}$. This symmetry is called the $\textit{custodial symmetry}$, and we will discuss its origin in section \ref{cus}. \\
\\
We can diagonalize the mass matrix by defining

\beq
\begin{pmatrix}
	Z_{\mu}\\
	A_{\mu}
\end{pmatrix}
= \begin{pmatrix}
	\cos \theta_{\o} & -\sin \theta_{\o}\\
	\sin \theta_{\o} & \cos \theta_{\o}
\end{pmatrix}
\begin{pmatrix}
W_{\mu}^3\\
B_{\mu}
\end{pmatrix},
\label{wma}
\eeq
with $\cos \theta_{\o}= \frac{g}{\sqrt{g^2+g'^2}}$ and $\sin \theta_{\o}= \frac{g'}{\sqrt{g^2+g'^2}}$. $\theta_{\omega}$ is called weak mixing angle. In the basis $(W^1, W^2, Z, A)$, we have the mass matrix

\beq
M^2= \frac{v^2}{4} \begin{pmatrix}
	g^2 && 0 && 0 && 0 \\
	0 && g^2 && 0 && 0 \\
	0 && 0 && g^2+g'^2 && 0 \\
	0 && 0 && 0&& 0
\end{pmatrix}.
\label{mm}
\eeq
We identify $A_{\mu}$ as the photon and $Z$ as the neutral weak boson, while the charged weak bosons are given by the combinations

\beq
W^{\pm}= W_1 \mp i W_2.
\eeq
The masses of the $W$ and $Z$ bosons are related by

\beq
\frac{M_W}{M_Z}= \cos \theta_{\o},
\eeq
and for $g' \rightarrow 0$, we have $\cos \theta_{\o}=1$ and the custodial symmetry is restored.
\\
\\
The mass matrix \eqref{mm} allows for the masses of $W$ boson and $Z$ boson to be determined in terms of three experimentally well
known quantities. First, we have the Fermi coupling constant $G=1.16 \times 10^{-5} GeV^-2$, which gives the minimizing constant $v= \left( G \sqrt{2}\right)^{-1/2}$. Then, the parameters $g$ and
$g'$ can be expressed in terms of electric charge and weak mixing angle as $g \sin \theta_{\o} = g \cos \theta_{\o} = e$, where $e$ is related to the fine-structure constant as $\alpha= \frac{e^2}{4 \pi}= \frac{1}{137.04}$. The weak mixing angle, finally, has been determined from neutrino scattering experiments to be $\sin^2\theta_{\o}=0.235 \pm 0.005$. We have 

\beq
M_W = \left( \frac{\alpha \pi}{G \sqrt{2}}\right)^{1/2} \frac{1}{\sin \theta_{\o}},\,\,\,\,\,\,\,\, M_Z = \left( \frac{\alpha \pi}{G \sqrt{2}}\right)^{1/2} \frac{2}{\sin 2\theta_{\o}}
\eeq
which gives

\beq
M_W \approx 77 \, \text{GeV}, \,\,\,\,\,\,\,\, M_Z \approx 88\, \text{GeV}
\eeq
which is a reasonable, but not perfect, approximation of the experimentally determined values

\beq
M_W=80.22 \pm 0.26 \, \text{GeV}, \,\,\,\,\,\,\,\, M_Z =91.17 \pm 0.02 \, \text{GeV}.
\eeq
The approximation can be further improved by taking into account renormalization corrections.
\subsection{ Custodial symmetry \label{cus}}
We can see the origin of the custodial symmetry by writing out the potential energy function 

\beq
V(\phi)=\lambda \left(\phi_1^2+\phi_2^2+\phi_3^2+\phi_4^2  - \frac{v^2}{2} \right)^2.
\eeq
This potential is cleary invariant under a larger symmetry than $SU(2)_L \times U(1)_Y$: it preserves a global $O(4)$ symmetry under which the vector $(\phi_1, \phi_2, \phi_3, \phi_4)$ transforms. The global $O(4)$ corresponds to a global $SU(2)_L \times SU(2)_R$ symmetry, as can be seen by writing $\Phi$ in the form of a bidoublet 

\beq
\Phi=\frac{1}{\sqrt{2}} \begin{pmatrix} 
	\phi^{0*} && \phi^+ \\
	-\phi^{+*} && \phi^0
\end{pmatrix}
\label{bid}
\eeq
so that

\beq
V(\phi)= \frac{\lambda}{2} \left(\frac{1}{2} \text{Tr} \,\Phi^{\dagger} \Phi - \frac{v^2}{2} \right)^2,
\eeq
is invariant under the global $SU(2)_L \times SU(2)_R$ symmetry

\beq
\Phi &\rightarrow& M_L \Phi M^{-1}_R \nonumber\\
W_{\mu}&\rightarrow& M_L W_{\mu} M^{-1}_R
\label{MM}
\eeq
with $M_{R,L}= \exp\left(i a^a_{R,L} \frac{\sigma^a}{2} \right)$ two independent global $SU(2)$ matrices. 
The minimizing configuration for $\Phi$  is the bidoublet 

\beq
\Phi_0= \frac{1}{\sqrt{2}}\begin{pmatrix} 
	v && 0 \\
	0 &&v
\end{pmatrix},
\label{ve}
\eeq
which breaks the global $SU(2)_L \times SU(2)_R$ symmetry down to subgroup where $M_L=M_R$. This is the $\textit{diagonal subgroup}$ $SU(2)_{\text {diag}}$, and it is preserved thanks to the fact that we were able to write $\Phi_0$ proportional to the unit matrix. \\
\\
In order to have the dynamical term in the Lagrangian invariant under $SU(2)_L \times SU(2)_R$, we need a covariant derivative such that

\beq
\mathcal{D}_{\mu} \Phi \rightarrow  M_L \mathcal{D}_{\mu}\Phi M^{-L}_R,
\label{18}
\eeq
while preserving the form of the dynamical part of the Lagrangian around the vev eq. \eqref{124}. Let us start by switching off the hypercharge gauge interaction $U(1)_Y$, i.e. $g'=0$. In this case, we can easily meet the requirements \eqref{124} and \eqref{18} by promoting $SU(2)_L$ to a local gauge symmetry, $\alpha_L^a \rightarrow \omega^a_L(x)$. Thus, the Lagrangian

\beq
\mathcal{L}_{\Phi}&=& \text{Tr}\left(\mathcal{D}_{\mu} \Phi\right)^{\dagger}\left(\mathcal{D}_{\mu} \Phi\right)-\lambda \left( \text{Tr} \,\Phi^{\dagger} \Phi - \frac{v^2}{2} \right)^2,
\label{LL}
\eeq
with
\beq
\mathcal{D}_{\mu} &=& \partial_{\mu}- i \frac{g}{2}W^a_{\mu}\sigma^a,
\eeq
has an $SU(2)_{L, local} \times SU(2)_{R, global}$ symmetry. The minimizing configuration \eqref{ve} breaks this symmetry down to the global symmetry of the diagonal subgroup $SU(2)_{\text{diag}}$, defined by $M_L=M_R$ in eq. \eqref{MM}. Under this symmetry, the gauge field transforms in infinitesimal form as $W^a_{\mu} \rightarrow W^a_{\mu}+ \varepsilon^{abc}\omega^b W^c_{\mu}$. The diagonal subgroup $SU(2)_{\text{diag}}$ is the custodial symmetry.\\

Switching on $g'$, we see from \eqref{124} that the custodial symmetry is broken. We can see this in the Lagrangian when we try to add the $U(1)_Y$ gauge symmetry, while preserving the $SU(2)_{L, local} \times SU(2)_{R, global}$ symmetry. If we change the covariant derivative to the full $SU(2)_L \times U(1)_Y$ covariant derivative in eq. \eqref{3} while using the bioublet Higgs field \eqref{bid} , we would get an erroneous, `custodial' result 

\beq
\text{Tr}\left(\mathcal{D}_{\mu} \Phi\right)^{\dagger}\left(\mathcal{D}_{\mu} \Phi\right) \supset \frac{g^2 v^2}{8} W_{\mu}^a W_{\mu}^a+ \frac{g'^2 v^2}{8} B_{\mu}B_{\mu}.
\eeq
We could also define a new covariant derivative

\beq
\mathcal{D}_{\mu} &=& \partial_{\mu}- i \frac{g}{2}\left(W^a_{\mu}\sigma^a\right)_L+i\frac{g'}{2}\left( B_{\mu}\sigma^3\right)_R ,
\eeq
which gives the right result \eqref{124} at $\Phi=\Phi_0$.
We then have to gauge the third generator of $M_R$ and identify it with the hypercharge generator, $\alpha^3_R \rightarrow \o_Y(x)$. The generator of the custodial symmetry would be the electric charge $Q=t^3+Y$, which remains unbroken because $\Phi_0$ is not charged and therefore $Q=0$ at this point. However, gauging only one component of a global symmetry violates the global symmetry. Thus, the custodial symmetry is only an approximate global symmetry of the SM, violated by hypercharge gauge interactions. \\
\\
The custodial symmetry is important because it rules out some possible scalar field configurations. The $\rho$ parameter

\beq
\rho  \equiv \frac{M_W^2}{\cos^2 \theta_{\o} M_Z^2}
\eeq
was experimentally measured to be very close to its `custodial' value $\rho=1$. Now imagine a complex triplet $X$ with $Y=1$ and a minimizing configuration 

\beq
X_0= \begin{pmatrix} 0 \\
	 0 \\v_{X} \end{pmatrix}
\eeq
so that 

\beq
\left( \mathcal{D}_{\mu} X\right)^{\dagger} \left( \mathcal{D}_{\mu} X\right) \supset g^2 v_{X}^2 W_{\mu}^+ W_{\mu}^- + g^2 v_{X}^2 W_{\mu}^3 W_{\mu}^3+ g'^2 v_{X}^2 B_{\mu}B_{\mu}- 2 g g' v_{X}^2 B_{\mu}W^{3 \mu}.
\eeq
The mass matrix is therefore, in the basis $(W^1, W^2, W^3, B )$

\beq
M_X^2= v_{X}^2 \begin{pmatrix}
	g^2 && 0 && 0 && 0 \\
	0 && g^2 && 0 && 0 \\
	0 && 0 && 2 g^2 && -2 g g' \\
	0 && 0 && -2 g g'&& 2 g'^2
\end{pmatrix},
\label{mmm}
\eeq
which we can diagonalize this in the same way as the $\Phi$ mass matrix by using the weak mixing angle matrix \eqref{wma}, so that in the basis $(W^1, W^2, Z, A )$

\beq
M_X^2= v_{X}^2 \begin{pmatrix}
	g^2 && 0 && 0 && 0 \\
	0 && g^2 && 0 && 0 \\
	0 && 0 && 2 (g^2+g'^2) && 0 \\
	0 && 0 && 0&& 0
\end{pmatrix}.
\label{mmmm}
\eeq
The photon is still massless, but the custodial symmetry for $g' \rightarrow 0$ is no longer present. \\
\\
In the presence of both the $\Phi$ doublet and the $X$ triplet we have

\beq
M_W^2 = \frac{g^2}{4}(v^2+4v_X^2), \,\,\,\,\,\,\,\,\, M_Z^2 = \frac{g^2+g'^2}{4}(v^2+8 v_X^2)
\eeq
so that 

\beq
\rho = \frac{v^2+4 v_X^2}{v^2+8 v_X^2},
\eeq
so that $v_X$ has to be very small compared to $v$ to meet the experimental value of $\rho$. Thus, the existence of a scalar boson $X$ with a significant minimizing value $X_0$ is ruled out by the custodial symmetry. However, combinations of $X$ with other triplets can restore the custodial symmetry, and make for some interesting beyond-the-SM phenomenology \cite{logan2014tasi}. 

\subsection{$R_{\xi}$ gauge class \label{rxi}}

In the previous sections, we have restricted ourselves to a specific gauge choice, the unitary gauge. Even though the gauge choice should never affect any physical outcome, different gauge choices help us to understand different aspects of our model. For example, in the unitary gauge all fields are physical, which proves the unitarity of the $S$-matrix. On the other hand, using e.g. the Landau gauge, we can prove the renormalizability of the Higgs model. \\
\\
Instead of using different gauge fixing models, we can combine several gauge choices into a $\textit{gauge class}$. Gauge classes can go through different gauge choices by means of an unphysical $\textit{gauge parameter}$.
For example, the Linear Covariant Gauges (LCG) is a gauge class given by $\partial_{\mu}A_{\mu}= \alpha \,b$, with $b$ an auxiliary field and $\alpha$ the gauge parameter. For $\alpha=0$, we end up in the Landau gauge, while for finite $\alpha$ the gauge field acquires a longitudinal component. Of course, physical quantities should never depend on $\alpha$; the gauge parameter is therefore also an important check of the physicality of our outcome.\\
\\
In this section, we will discuss the $R_{\xi}$ gauge class, introduced by 't Hooft to prove the renormalizability of the Higgs model. It is therefore an important gauge class for the GWS model, but can be understood in the $U(1)$ Higgs model. In the action

\beq
S&=&\int d^4 x\left\{\frac{1}{4}F_{\mu\nu}F_{\mu \nu}+(D_{\mu} \varphi)^{\dagger} D_{\mu} \varphi + \frac{\lambda}{2} \left(\varphi^{\dagger}\varphi -\frac{v^2}{2}\right)^2\right\},
\label{1213}
\eeq
we can parametrize $\varphi(x)= \frac{1}{\sqrt{2}}(v+h(x)) e^{-i \chi(x)/v}$, and then make the unitary gauge choice $\chi(x)=0$. This decouples the non-physical ``would-be" NG boson. However, expanding the Lagrangian we have for the squared gauge boson terms

\beq
\mathcal{L}&\supset& -\frac{1}{2}A_{\mu}\partial^2 A_{\mu}+\frac{1}{2}A_{\mu}\partial_{\mu}\partial_{\nu}A_{\nu}+\frac{1}{2}m^2 A_{\mu}A_{\mu},
\eeq
with $m= e v $. The two-point function for the gauge boson is then given by

\beq
\braket{A_{\mu}(p)A_{\nu}(-p)}&=& \frac{1}{p^2+m^2}\mathcal{P}_{\mu \nu}(p^2)+ \frac{1}{m^2}\mathcal{L}_{\mu \nu}(p^2),
\label{ppg}
\eeq
with $\mathcal{P}_{\mu \nu}(p^2)= \delta_{\mu \nu}- \frac{p_{\mu}p_{\nu}}{p^2}$ and $\mathcal{L}(p^2)=\frac{p_{\mu}p_{\nu}}{p^2}$ the transversal and longitudinal projectors. The longitudinal part does not vanish for large values of the momentum $p$, and we cannot prove renormalizability. \\
\\
The problem with renormalizability can be made more explicit by adapting another parametrization of the scalar field in eq. \eqref{1213}, namely $\varphi(x)=\frac{1}{\sqrt{2}}(v+h(x)+ i \rho(x))$. This will lead to a term in the Lagrangian

\beq
\mathcal{L} &\supset&  m\, A_{\mu}\partial_{\mu}\rho,
\label{min}
\eeq
and this mixing term between the gauge boson and the would-be NG boson will lead to unrenormalizable results. We therefore employ the gauge-fixing action

\beq
S_{gf}= \int d^4 x \left\{\frac{1}{2 \xi} \left(\partial_{\mu}A_{\mu}+\xi m \rho\right)^2 \right\},
\label{uur}
\eeq
known as the $R_{\xi}$ gauge class with the gauge parameter $\xi$. The mixing term in \eqref{uur} will cancel exactly the unwanted mixing term \eqref{min}. The two-point function for the gauge boson is now given by 

\beq
\braket{A_{\mu}(p)A_{\nu}(-p)}&=& \frac{1}{p^2+m^2}\mathcal{P}_{\mu\nu}(p^2)+ \frac{\xi}{p^2+\xi m^2}\mathcal{L}_{\mu \nu} (p^2),
\eeq
and we can show that this gauge is renormalizable for all finite values of $\xi$. Notice that for $\xi=0$ we end up in the Landau gauge, while in the limit $\xi \rightarrow \infty$ we find back eq. \eqref{ppg}, the unitary gauge. This means the $R_{\xi}$-gauge is both unitary and renormalizable. The would-be Goldstone remains in the Lagrangian and has a two-point function 

\beq
\braket{\rho(p)\rho(-p)}= \frac{1}{p^2+\xi m^2},
\eeq
but the very fact that its mass depends on the $\xi$-parameter brands it as unphysical. The same thing happens for the ghost field.\\
\\
Another important feature of the $R_{\xi}$-gauge is that we can show, in non-abelian and abelian gauge theories, that the pole masses of the two-point function of both the gauge boson $A_{\mu}(x)$ and the Higgs field $h(x)$ do not depend on the gauge parameter $\xi$ \cite{gambino2000nielsen}. This property, contained in the Nielsen Identities \cite{Nielsen:1975fs}, is important because it allows to give a physical meaning to the polemass of the otherwise gauge-dependent (and therefore unphysical) correlation functions.\\
\\
Even though the $R_{\xi}$-gauge is unique to the Higgs model because it requires the Goldstone boson $\rho^a$, the concepts of this gauge class can be used to define a renormalizable class of gauge-fixing also outside of the Higgs environment, as we will see in chapter \ref{Amin}. \\

\section{Higgs beyond the Standard Model \label{beyond}}

The Higgs mechanism for the electroweak sector has been firmly established as the mechanism that gives mass to the $W,Z$ bosons, as well as some fermions, while leaving the photons massless. In 2012, the Higgs boson was discovered at the CERN Large Hadron Collidor (LHC) \cite{Aad:2012tfa, Chatrchyan:2012ufa}, in full agreement with the SM predictions. \\
\\
It is nonetheless important to realize that there is no $\textit{a priori}$ reason that the Higgs mechanism only occurs in the version $SU(2)\times U(1) \rightarrow U(1)$. The Higgs mechanism can be described for any gauge theory, such as the simple example $U(1) \rightarrow \text{nothing}$ that we discussed in the introduction. However, the photon is massless, so we know that nature did not provide for a $U(1)$ Higgs boson. Still, the $U(1)$ Higgs model is a useful toy model to study several properties of the Higgs model, as we will see in chapter \ref{VII} and \ref{VV} . Another useful model is $SU(2) \rightarrow \text{nothing}$, the Higgs-Yang-Mills (HYM) model which can be achieved by setting $g'=0$ in the electroweak model. This model will be central to chapter \ref{VIII}. As we have seen in the previous section, the pure $SU(2)$ model will exhibit a full custodial symmetry. 
\\
\\
Since the discovery of the Higgs mechanism, several symmetry breaking models have been proposed besides the GWS model. We will discuss two of them, both proposed by Georgi and Glashow. Then, we will also disuss how the Higgs mechanism would occur in QCD in the presence of an $SU(3)$ Higgs boson.
\\
\\
It is important to mention that while all the examples of the Higgs mechanism in this section are variations of the Higgs mechanism in the electroweak sector, the same mechanism of a non-zero minimizing value is also widely used in other areas such as condensed matter physics, where it was first established by Anderson \cite{anderson1962theory} as an analogy to the Landau-Ginzburg effective model of superconductivity.
\subsection{$SU(2) \rightarrow U(1)$}

This model, proposed in \cite{georgi1972unified}, provides an alternative Higgs model in the electroweak sector. Since $SU(2)$ is homeomorphic to $SO(3)$, we will use the Hermitian traceless generators

\beq
 \tau^1 =\begin{pmatrix} 
 	0 &&0 &&0\\
 	0 &&0 &&-i \\
 	 0&& i && 0
 \end{pmatrix}
\,\,\,
\tau^2 = \begin{pmatrix} 
	0 && 0 && i\\
	0 &&0 &&0 \\
	- i&&0 &&0
\end{pmatrix}
\,\,\,
 \tau^3 =\begin{pmatrix} 
	0 && -i&&0\\
	i && 0&&0 \\
	0&&0 &&0
\end{pmatrix},
\eeq
so that $(\tau^a)^{bc}=-i \epsilon^{abc}$. This is a real representation, so the Lagrangian is given by 

\beq
\mathcal{L}= -\frac{1}{4}F_{\mu \nu}^a F_{\mu \nu}^a + \frac{1}{2}(\mathcal{D}_{\mu} \phi)^2 - \frac{\lambda}{8} \left(\phi \phi^{\dagger}-v^2\right)^2,
\eeq
and the minimizing value of $\phi(x)$ is 

\beq
\phi = \begin{pmatrix} 
	0 \\
	0 \\
	v 
	\end{pmatrix}
\eeq
so that the mass matrix is given by

\beq
M^2 = \begin{pmatrix}
	m^2 && 0 && 0\\
	0 && m^2 && 0 \\
	0 && 0 && 0
	\end{pmatrix},
\eeq
which means two gauge bosons, $A_{\mu}^1$ and $A_{\mu}^2$, are massive, while $A_{\mu}^3$ is massless. \\
\\
The above model does not account for the neutral $Z$-boson and is therefore not the correct theory for the electroweak mass generation. However, as was shown in \cite{tHooft:1974kcl}, identifying the photon with the massless $A_{\mu}^3$ does provide for a radial magnetic field which indicates the existence of a magnetic monopole. The magnetic monopole, often hypothesized but never found, is  not present in the GWS model.

\subsection{$SU(5) \rightarrow  SU(3)\times SU(2)\times U(1)$ }
It would be aesthetically nice if the standard model was the low-energy phenomenology of a larger gauge theory. This is the idea of a Grand Unified Theory (GUT). Georgi and Glashow proposed a model to this effect in \cite{georgi1974unity}. The required Higgs field  transforms under the adjoint representation of SU(5) and can be represented by $5 \times 5$ hermitian traceless matrix. The Higgs field transforms under the adjoint representation as

\beq
\Phi^a \rightarrow \Phi^a + f^{abc}\omega^b \Phi^c
\eeq
or, adding a generator $t^a$ on both sides

\beq
\Phi \rightarrow \Phi -i \omega^a[t^a, \Phi],
\label{pphi}
\eeq
so that configurations of $\Phi$ that commute with a (sub)group with generators $t^a$ are invariant under the gauge transformation related to this (sub)group.\\
\\
 As for any non-abelian gauge theory, the covariant derivative in the adjoint representation is given by

\beq
D_{\mu}^{ab}= \delta^{ab}\partial_{\mu}-g f^{abc}A_{\mu}^c
\label{vk}
\eeq
with $N^2-1=24$ different color charges, corresponding to the 24 generators $t^a$ of $SU(5)$. We can use eq. \eqref{tta} and the normalization $[t^a,t^b]= \frac{\delta^{ab}}{2}$ to write the mass term of the Lagrangian as

\beq
\mathcal{L}_m = \frac{g^2}{2}( f^{abc}A_{\mu}^b \Phi_0^c)^2= -g^2 \text{Tr}[[t^a,\Phi_0][t^b,\Phi_0]]A_{\mu}^a A_{\mu}^b,
\label{iii}
\eeq 
which means, from eq. \eqref{pphi}, that gauge bosons corresponding to unbroken symmetries remain massless.\\
\\
If we choose

\beq
\Phi_0 = v \begin{pmatrix} 2 & 0 & 0 & 0 & 0\\
	0 & 2 & 0 & 0 & 0\\
	0 & 0 & 2 & 0 & 0 \\
	0 & 0 & 0 & -3 & 0 \\
	0 & 0 & 0 & 0 & -3
	\end{pmatrix},
		\eeq
this commutes with the $SU(3)$ subgroup $\begin{pmatrix} T^a & 0 \\ 0 & 0 \end{pmatrix}$, the $SU(2)$ subgroup $\begin{pmatrix} 0 & 0 \\ 0 & t^a \end{pmatrix}$ and the $U(1)$ 
subgroup corresponding to the generator proportional to $\Phi_0$. We then have twelve massless gauge bosons, corresponding to the unbroken generators of $SU(3) \times SU(2) \times U(1)$, and twelve massive gauge bosons. This adjoint Higgs particle would only accomplish the seperation of the different interactions. Another Higgs field, in the fundamental representation, would provide electroweak symmetry breaking.\\
\\
The attractiveness of the $SU(5)$ model, besides its simplicity, lies in the fact that matter particles fit neatly into representations of $SU(5)$. A single generation of the standard model fits perfectly into two irreducible representations of $SU(5)$, the $\overline{5}$ and $10$. However, the model predicts many unobserved phenomena such as proton decay, and quark-to-lepton mass ratios that don’t agree with experiment. It also predicts particles so heavy ($M_{GUT}\approx 10^{15}$ GeV) that they cannot be detected.

\subsection{$SU(3) \rightarrow SU(2) \times U(1)$}

The $SU(3)$ Higgs mechanism works in a similar way as that of $SU(5)$. In the adjoint representation, the minimizing configuration

\beq
\Phi_0 =v \begin{pmatrix}
	1 & 0 & 0\\
	0 & 1 & 0 \\
	0 & 0 & -2
	\end{pmatrix},
	\eeq
commutes with the $SU(2)$ subgroup $\begin{pmatrix} t^a & 0 \\ 0 & 0 \end{pmatrix}$, the Gell-Mann matrices $\lambda_{1,2,3}$, and the $U(1)$ subgroup corresponding to the generator proportional to $\Phi_0$, i.e. $\lambda_8$. This amounts to four massless gauge bosons, corresponding to the unbroken generators of $SU(2)\times U(1)$, and four massive gauge bosons, corresponding to $\lambda_{4,5,6,7}$. From \eqref{iii} one can then find the mass of the gauge bosons, $m= 3 g v$.\\
\\
It is well-known that gluons in the high-energy (perturbative) regime are massless. Therefore, we do not expect a Higgs boson for the strong interaction. However, in non-perturbative regimes there are indications of massive behaviour, as we will discuss in the next section.

\section{Massive solutions for the QCD sector\label{Ic}}

\begin{figure}[H]
	\center
	\includegraphics[width=8cm]{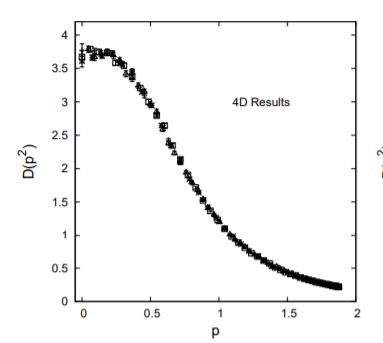}
	\caption{The gluon propagator $D(p^2)$ as a function of the lattice momenta $p$, with $p$ given in GeV. Figure from \cite{Cucchieri:2010xr}.}
	\label{pplr}
\end{figure}

In the high-energy regime, the FP procedure gives the gauge-fixed YM action for the massless QCD sector:
\beq
S&=& \int d^4 x \left\{\frac{1}{4} F^a_{\mu \nu} F^a_{\mu \nu} + b^a \partial_{\mu}A^a_{\mu}+ \overline{c}^a \partial_{\mu}D_{\mu}^{ab}c^b\right\},
\label{pp1}
\eeq
with the covariant derivative $D^{ab}_{\mu}$ in the adjoint representation of the gauge group as in eq. \eqref{vk}, the fields $(\overline{c}^a,c^a)$ denoting the FP ghosts and $b^a$ the Lagrange multiplier implementing the Landau gauge condition $\partial_{\mu} G^a_{\mu}=0$. The gauge-fixed Lagrangian is invariant under the nilpotent BRST symmetry 

\beq
s A_{\mu}^a&=&-D_{\mu}^{ab}c^b\nonumber\\
s c^a &=& \frac{1}{2}g f^{abc}c^b c^c\nonumber\\
s \overline{c}^a &=& b^a \nonumber \\
s b^a &=& 0,
\label{rapa}
\eeq 
with $s^2=0$. Notice that the Lagrangian in eq. \eqref{pp1} describes any $SU(N)$ theory. As we will see in what follows, in some cases an $SU(2)$ theory is used as a simplification of the $SU(3)$ theory to describe the gluon dynamics of the strong interaction. Notice also that we use the denotation for the gluon field strength tensor $F^a_{\mu\nu}$ and for the vector field $A_{\mu}^a$, instead of $G^a_{\mu\nu}$ and $G^a_{\mu}$ from section \ref{G} \footnote{This is in line with most articles written on the subject of YM theories, since for pure YM theories there is no danger of confusion with the photon field $A_{\mu}$.}.
\\
\\
The predictions in the QCD sector derived from the Lagrangian in eq. \eqref{pp} through perturbation theory agree with what has been observed in high-energy experiments. However, when extending this analysis to finite energies, the coupling constant $g_c$ diverges and hits a Landau pole. In early studies of QCD, this behaviour was seen as responsible for the IR counterpart of asymptotic freedom, the ``IR slavery" that keeps the gluons and quarks confined \cite{Alkofer_2001}. In other words, confinement was seen as a direct consequence of the Landau pole. However, the modern view based on lattice simulations is that the Landau pole is not an expression of confinement, but rather a sign that some non-perturbative effect, such as the Gribov ambiguity, invalidates the extension of the perturbative results to the IR regime. 
 \\
\\
The Gribov ambiguity or Gribov problem, first described by Gribov in \cite{gribov1978quantization}, demonstrates the non-uniqueness of the FP gauge-fixing beyond the perturbative level. To explain this, let us consider the Landau gauge, although an analogous argument can be cast in other gauges. If two gluon fields $A_{\mu}^a$ and $A_{\mu}^{'a}$ connected by a gauge transformation

\beq
A^{'a}_{\mu}=A^a_{\mu}-D_{\mu}^{ab}\alpha^b,
\eeq
they are said to be on the same $\textit{gauge orbit}$. Now if both fields are satisfying the same gauge fixing condition, i.e. $\partial_{\mu}A_{\mu}=\partial_{\mu}A'_{\mu}=0$, so that

\beq
-\partial_{\mu}D_{\mu}^{ab}\alpha^b=0,
\label{ooy}
\eeq
this means that the FP procedure failed to fully eliminate the multiple counting of physical states in the path integral due to gauge invariance. So, when eq. \eqref{ooy} has solutions for certain values of the gauge field, so-called $\textit{zero modes}$, this means the model still contains different gauge configurations, known as $\textit{Gribov copies}$. Notice that for small values of the coupling constant $g_c$, the LHS of eq. \eqref{ooy} reduces to $\partial^2 \alpha$, which has only positive eigenvalues. Therefore, the Gribov problem does not exist in the UV regime, and the FP procedure is valid there. Over the years, various attempts have been made to deal with this problem in the continuum functional approach, mainly by trying to evaluate the path integral in such a way that it contained no zero modes. The most notable attempts in this direction are the Gribov-Zwanziger (GZ) approach \cite{zwanziger1989local,zwanziger1993renormalizability,Zwanziger:2001kw} and the Refined Gribov-Zwanziger (RGZ) approach \cite{Dudal:2007cw,Dudal:2008sp,Dudal:2011gd,Capri:2015nzw}, recently formulated in a BRST invariant fashion \cite{Capri:2015ixa,Capri:2015nzw,Capri:2016aqq,Capri:2016gut,Capri:2017bfd}. For a nice overview of the Gribov problem and the (R)GZ approach, see \cite{Vandersickel:2012tz}.
\\
\\
The breakdown of the FP procedure and the existence of a Landau pole demonstrate the necessity of non-perturbative components that break with the standard FP gauge-fixed YM picture. This has lead to analytical QCD models that fundamentally differ from the model in eq. \eqref{pp1} in the IR, but preserves the standard FP predictions in the UV. Interestingly, the problem of the Landau pole in asymptotically free theories is in some cases circumvented by the introduction of an infrared gluon mass. Still, however, we want to recover the massless character
of the FP gauge fixing theory in the perturbative UV region. A model which implements an effective mass only in the IR region was first proposed in \cite{cornwall1982dynamical} based on the idea of a momentum-dependent or dynamical
gluon mass \cite{Parisi:1980jy, Bernard:1981pg}. For this, the Schwinger-Dyson (SD) equations are employed in order to get a suitable gap equation
that governs the evolution of the dynamical gluon mass $m(p)$, which vanishes for $p^2 \rightarrow \infty$. This setup preserves
both renormalizability and gauge invariance. 
\\
\\
The SD equations give relations between Green's functions that go beyond perturbation theory. In principle this would make them the most important analytical tool to get insight in the gluon propagator in the IR regime, and they are often employed as such \cite{papavassiliou2013effective,Aguilar:2014tka, Cyrol:2016tym,Huber:2018ned,Boucaud:2011ug}. However, in practice the SD equations are hard to work with because they entail an infinite set of coupled equations for the vertex functions, which somehow needs to be truncated. This requires involved techniques and calculations, with in some cases an important numerical part. Numerically, the IR regime of the gluon propagator is more rigorously described by lattice simulations. One important observation on the lattice is that the gluon propagator reaches a finite positive value in the deep IR for space-time
Euclidean dimensions $d > 2$, see e.g. \cite{ Cucchieri:2007md, Cucchieri:2007rg,Bogolubsky:2009dc, Maas:2008ri, Cucchieri:2009kk,Cucchieri:2009zt,Cucchieri:2010xr,Cucchieri:2011ig,Bornyakov:2009ug,Oliveira:2012eh,Bicudo:2015rma,Cucchieri:2016jwg,Duarte:2016iko,Dudal:2018cli,Boucaud:2018xup}. This saturation of the gluon propagator for small momenta $p$, see Figure \ref{pplr}, indicates massive behavior of the gluon in the IR regime. Massive-like behavior for the gluon propagator, known as the decoupling solution, has also emerged within other approaches, as the study of the Schwinger-Dyson equations, the Renornalization Group  and other techniques, see for instance \cite{Aguilar:2008xm,Aguilar:2015bud,Fischer:2008uz,Aguilar:2015nqa,Huber:2015ria,Fischer:2009tn,Weber:2011nw,Frasca:2007uz,Siringo:2015wtx} and references therein.

\subsection{Massive Yang-Mills model \label{massym}}

\begin{figure}[H]
	\center
	\includegraphics[width=8cm]{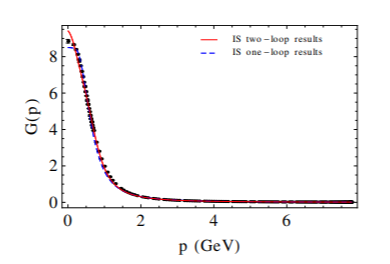}
	\caption{The gluon propagator in the IS scheme compared with the lattice result from \cite{Duarte:2016iko} for one- and two-loop corrections. Figure from \cite{Gracey:2019xom}.}
	\label{grac}
\end{figure}

The lattice results, as well as the fact that the Landau pole can be circumvented by an effective gluon mass, has stimulated research into effective massive models for the IR region of QCD. The aforementioned RGZ theory is an example of such a model. In this section we will discuss another example: the massive YM model from \cite{tissier2010infrared}, which is a particular case of the Curci-Ferrari (CF) model \cite{Curci:1976bt}. The action for this theory is  

\beq
S&=& \int d^4 x\left\{\frac{1}{4} G^a_{\mu \nu} G^a_{\mu \nu} + b^a \partial_{\mu}A^a_{\mu}+ \overline{c}^a \partial_{\mu}D_{\mu}^{ab}c^b+ \frac{1}{2}m^2 A^a_{\mu}A^a_{\mu}\right\},
\label{ppm}
\eeq
which is a Landau gauge FP Euclidean Lagrangian for pure gluodynamics, supplemented with a gluon mass term.
This term modifies the theory in the IR but preserves the FP perturbation theory for momenta $p \gg m$. It was argued that this model could be part of a complete gauge-fixing in the Landau gauge, since a CF gluon mass term may arise after the Gribov copies have been accounted for via an averaging procedure \cite{Serreau:2012cg}, see also \cite{Tissier:2017fqf} for a related discussion in a different gauge. The mass term breaks the BRST symmetry of the model, which means the unitarity of the model can be no longer proven \cite{de_Boer_1996}, although it is debatable whether some non-perturbative effect could restore unitarity. However, since the BRST breaking is soft, it does not spoil renormalizability. The Lagrangian \eqref{ppm} turns out to be still invariant under a modified BRST symmetry

\beq
s_m A_{\mu}^a&=&-D_{\mu}^{ab}c^b\nonumber\\
s_m c^a &=& \frac{1}{2}g f^{abc}c^b c^c\nonumber\\
s_m \overline{c}^a &=& b^a \nonumber \\
s_m b^a &=& i m^2 c^a,
\label{rapam}
\eeq
which is however not nilpotent since $s_m^2 \overline{c}^a\neq 0$.\\
\\
The massive YM model is not justified $a \,\, priori$ from first principles. Instead, its legitimacy is measured by how well it accounts for lattice results. In \cite{tissier2010infrared,tissier2011infrared,Gracey:2019xom},
it has been shown that, both at one and two-loop order, the model reproduces very well the lattice predictions for
the gluon and ghost propagator, see Figure \ref{grac}. It was also shown that with an adequate renormalization scheme, dubbed the Infrared Safe (IS) scheme, there is no Landau pole. \\
\\

The analysis of the gluon propagator in \cite{tissier2010infrared,tissier2011infrared,Gracey:2019xom} was done by perturbative loop calculations. How can the non-perturbative region be accessed with perturbative methods? In \cite{tissier2011infrared, Gracey:2019xom}, it is claimed that higher loop corrections for this model seem to be rather small, even for a significantly high coupling constant ($g=7.5$ in \cite{tissier2011infrared}). The
reason for this, as explaind in \cite{tissier2011infrared}, lies in the massive gluons. When momenta are much smaller than the gluon mass, all diagrams that include internal gluon lines are suppressed by inverse powers of the gluon mass. This means that higher order loop terms, which naturally possess more internal gluon lines, will be surpressed.
Thus, using an effective mass term makes perturbative loop calculations possible in an otherwise non-perturbative region.

\subsection{Positivity violation and complex poles \label{posval}}
The CF model is capable of reproducing, to high accuracy, the lattice results of the gluon propagator. This is despite the fact that the gluon propagator as derived from eq. \eqref{ppm} is not gauge-invariant and the model has no nilpotent BRST invariance. It should be emphasized here, however, that the goal of \cite{tissier2010infrared} and follow-up works was not to introduce a theory for massive gauge bosons, but to discuss a relatively simple and useful effective description of some non-perturbative aspects of QCD. Also, in this respect the CF model is not in a worse position than other 
models that try to go beyond standard perturbation theory, such as the GZ model, which also breaks BRST symmetry. One could even argue that unitarity of the gauge bosons sector, secured by a nilpotent BRST symmetry, is not so much an issue here as one expects the gauge bosons to be undetectable anyhow, due to confinement.
\\
\\
An interesting question is whether the massive YM model is also capable of reproducing other aspects of QCD observed on the lattice. One of the most intriguing observations of lattice QCD  in recent years is $\textit{positivity violation}$. Positivity violation means that in the KL spectral representation of the propagator 

\beq
G(p^2)=\int_0^{\infty} \frac{\rho(t)}{t+p^2} dt,
\eeq 
the spectral density function $\rho(t)$ is not positive everywhere. The spectral density function displays, for a certain two-point function such as the gluon propagator, the different states (1-particle states, bound states and multiparticle states) associated with different energy values. If the spectral density function violates positivity, the states it describes cannot be part of the physical state space. Positivity violation is therefore attributed to confinement \cite{cornwall2013positivity,Krein:1990sf,Roberts:1994dr,Lowdon:2017gpp}: the non-positivity of the spectral
function is seen as a reflection of the inability of the gluon to exist as a free physical particle.\\\\

\begin{figure}[H]
	\center
	\includegraphics[width=10cm]{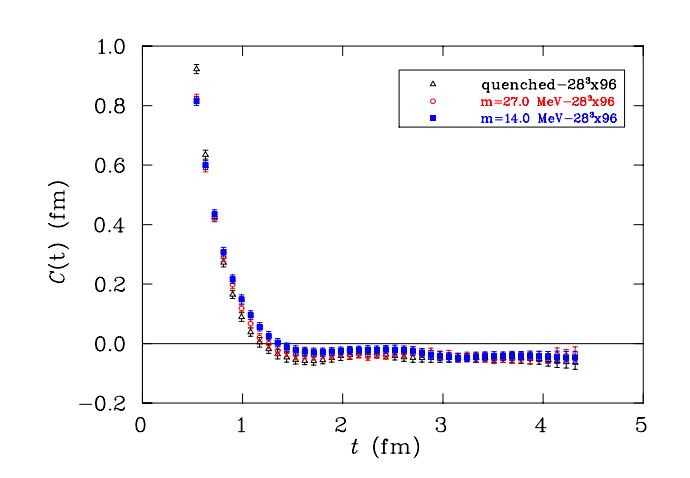}
	\caption{Lattice results for the real space propagator related to the gluon propagator, for the quenched case (no quarks) and light sea quark masses. The quenched approximation corresponds to the pure gauge theory for QCD. Figure from \cite{Bowman:2007du}.}
	\label{que2}
\end{figure}
\begin{figure}[H]
	\center
	\includegraphics[width=10cm]{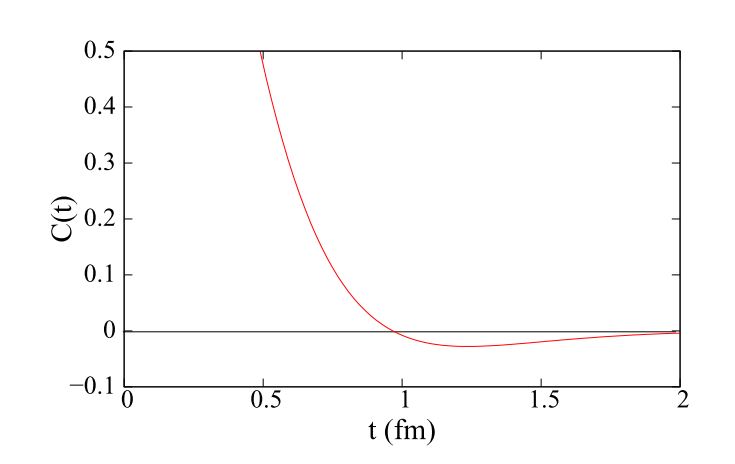}
	\caption{Real space propagator related to the one-loop corrected gluon propagator for the massive YM model. Figure from \cite{tissier2010infrared}.}
	\label{que}
\end{figure}

Positivity violation for the gluon propagator has been confirmed in both analytical studies of the SD equations \cite{PhysRevLett.79.3591, vonSmekal:1997ern, Alkofer_2004} and in lattice simulations \cite{Cucchieri:2004mf, Bowman:2007du}. In \cite{Alkofer:2000wg}, a complete overview of evidence for positivity violation is given. On the lattice, positivity violation is detected through the real space propagator $C(t)$ related to the gluon propagator, see Figure \ref{que2}. The real space propagator is defined by 

\beq
C(t)&=& \int_{-\infty}^{\infty} \frac{dp}{2\pi}e^{ipt}G(p),
\label{rs}
\eeq
that is, $C(t)$ is the Fourier transform of the gluon propagator $G(p)$. It can be shown, see for example \cite{Cucchieri:2004mf}, that positivity of the real space propagator implies positivity of the spectral density function. 
\\
\\
In \cite{tissier2010infrared}, the real space propagator for the gluon propagator in the CF model was obtained by inserting the one-loop gluon propagator derived from the action \eqref{ppm} into eq. \eqref{rs}, and it was found that the curve of $C(t)$ as observed on the lattice was reproduced, including the positivity violation, see Figure \ref{que}. On the one hand, it is remarkable that the CF model seems capable of reproducing the lattice results of the real space propagator. On the other hand, we should distinguish between the significance of positivity violation on the lattice and in the CF model. In lattice simulations, one starts out with a unitary (physical) YM model and subsequently observes positivity violation, i.e. non-physical behavior. In the CF model however, one starts out with a model without a nilpotent BRST symmetry. From the Kugo-Ojima
criterion \cite{kugo1979local}, it is known that nilpotency of the BRST symmetry is indispensable to formulate suitable conditions
for the construction of the states of the BRST invariant physical (Fock) sub-space, providing unitarity of the $S$-matrix. Indeed, in \cite{Ojima:1981fs,deBoer:1995dh} the existence of negative norm states in the $s_m$-invariant subspace (“the would-be physical
subspace”) was confirmed, see also section \ref{s5} for a detailed example. Therefore, to find non-physical behavior for a model without BRST symmetry is somewhat of a self fulfilling prophecy. \\
\\
In fact, the non-physical behaviour of the CF model can be detected in a step prior to the spectral density function. As was established in \cite{Kondo:2019rpa}, and recently in \cite{Fischer:2020xnb} in the context of SD equations, the use of
the Landau gauge in the massive YM model \eqref{ppm} leads to complex pole masses, which will obstruct the calculation
of the KL spectral function. Indeed, if at some order in perturbation theory (one-loop as in \cite{Kondo:2019rpa} for
example) a pair of Euclidean complex pole masses appear, at higher order these poles will generate branch points
in the complex $p^2$-plane at unwanted locations, i.e. away from the negative real axis, deep into the complex plane,
thereby invalidating a KL spectral representation. This can be appreciated by rewriting the Feynman integrals in terms of Schwinger or Feynman parameters, whose analytic properties can be studied through the Landau equations \cite{Eden:1966dnq}. Let us also refer to \cite{Baulieu:2009ha,Windisch:2012sz} for concrete examples. In chapter \ref{Amin} we will develop some methods to avoid complex poles in the unitary Higgs model.

\leavevmode
\\
Finally, it must be pointed out that also lattice simulations of the gluon propagator are not free of ``built-in'' non-physical features that could attribute to positivity violation. In all studies that show a violation of positivity, both in the context of SD equations and on the lattice, a gauge-dependent and therefore $\textit{a priori}$ non-physical Landau-gauge gluon propagator is used. Indeed, even while on the lattice it is in principle possible to work only with gauge-invariant quantities, in \cite{Cucchieri:2004mf, Bowman:2007du} a gauge-dependent environment was created, including a lattice gauge-fixing. This gauge-dependent environment also provided the opportunity to test the naturally gauge-dependent GZ proposal on the lattice, see for example \cite{Cucchieri:2010xr,cucchieri2013crossing, PhysRevD.81.074505,Cucchieri_2012,Cucchieri_2012b}. The use of the gauge-dependent propagator is justified in literature because for an unconfined field, the propagator presumably has a normal KL representation \cite{Cornwall:2009ud}, in line with the fact that predictions made from the perturbative gluon propagator of the gauge-fixed YM model agree with experimental observations \cite{Jegerlehner:2001fb,Jegerlehner:2002em,Martin:2015lxa,Martin:2015rea}. This motivated the hypothesis that gauge dependent propagators could give some piece of information about confinement. Nonetheless, the question can be asked whether the gauge field in its elementary, gauge-dependent form gives a complete representation of the gauge boson in all ranges of the energy scale. It can be hypothesized that these elementary fields are in fact part of a greater gauge-invariant configuration, which shares with the gauge-dependent field some essential, but not all, properties. In the next chapters, we will discuss two proposals to this effect: the non-local composite configuration $A^h$ in chapter \ref{Amin} and the local gauge-invariant composite operators proposed by 't Hooft \cite{tHooft:1980xss} in chapters \ref{VII} and \ref{VV}. The latter are unique to the Higgs model, but they are not less relevant to the above question since the $W,Z$ bosons are also expected to be confined in the low-energy regime. 
\newpage
\chapter{On a renormalizable class of gauge fixings for the gauge invariant operator $A^2_{\text{min}}$ \label{Amin}}

In this chapter we pursue the investigation \cite{Fiorentini:2016rwx} of the dimension two gauge invariant operator $A_{\min }^{2}$, obtained by minimizing the functional 
$\mathrm{Tr}\int d^{4}x\,A_{\mu }^{u}A_{\mu }^{u}$ along the gauge
orbit of $A_{\mu }$ \cite{Zwanziger:1990tn, DellAntonio:1989wae, DellAntonio:1991mms, vanBaal:1991zw}, namely
\begin{eqnarray}
A_{\min }^{2} &\equiv &\min_{\{u\}}\mathrm{Tr}\int d^{4}x\,A_{\mu
}^{u}A_{\mu }^{u}\;,
\nonumber \\
A_{\mu }^{u} &=&u^{\dagger }A_{\mu }u+\frac{i}{g}u^{\dagger }\partial _{\mu
}u\;.    \label{Aminn0}
\end{eqnarray}
As highlighted in \cite{Fiorentini:2016rwx}, the functional $A_{\min }^{2}$ enables us to introduce a non-local gauge invariant field configuration $A^h_\mu$ \cite{Lavelle:1995ty}  which turns out to be helpful to construct renormalizable BRST invariant YM theories which can be employed as effective massive theories to study non-perturbative
infrared aspects of confining YM theories in Euclidean space. As we will see, the massive YM model discussed in \ref{massym} is deeply related to $A_{\min }^{2}$, because $A^h_{\mu}$ and $A_{\mu}$ are equal in the Landau gauge, see eq. \eqref{land} below. See also \cite{Capri_2018b} for a supersymmetric extension of the composition $A^h_{\mu}(x)$. \\\\
In the present chapter we extend the analysis of the operator $A_{\min }^{2}$ to a general class of covariant gauges which share  great similarity with 't Hooft's $R_\zeta$-gauge discussed in section \ref{rxi}. In fact, as shown in \cite{Fiorentini:2016rwx}, the localization procedure for both $A_{\min }^{2}$ and $A^h_\mu$ requires the introduction of a dimensionless auxiliary Stueckelberg field $\xi$ which, as much as the Higgs field of 't Hooft's $R_\zeta$-gauge, will now enter explicitly the gauge condition through the appearance of a gauge massive paramater $\mu^2$.  This property will enable us to provide a fully BRST invariant mass for the auxiliary field $\xi$, a feature which might have helpful consequences in explicit loop calculations involving  $\xi$ in order to keep control of potential infrared divergences associated to its dimensionless nature.   Moreover, as in the case of $R_\zeta$-gauge, also the Faddeev-Popov ghosts will acquire a mass through the gauge-fixing. Of course, setting $\mu^2=0$, the linear covariant gauges discussed in \cite{Fiorentini:2016rwx} will be recovered. \\\\The chapter is organized as follows. In section \ref{2p1} we give a short presentation of the main properties of $A_{\min }^{2}$ and of the gauge invariant configuration $A^h_\mu$, reminding to Appendix \eqref{apb} for more specific details. Sections \ref{2p2},\ref{2p3},\ref{2p4},\ref{2p5} are devoted to the presentation of the local BRST invariant action for  $A_{\min }^{2}$ and $A^h_\mu$ as well as of the main properties of the aforementioned gauge-fixing. In section \ref{2p6} we establish the set of Ward identities fulfilled by the resulting action. These identities will be employed to characterize the most general allowed invariant counterterm through the procedure of the algebraic renormalization  \cite{Piguet:1995er}. In section \ref{2p7} a detailed analysis of the counterterm will be presented together with the renormalization factors needed to establish the all order renormalizability of the model. Section \ref{2p8} contains our conclusion. Finally, in Appendix \eqref{apc},  a second, equivalent, proof of the renormalizability of the model will be outlined by making use of a generalised gauge fixing and ensuing Ward identities.

\section{Brief review of the operator $A_{\min }^{2}$ and construction of a non-local gauge invariant and transverse gauge field $A^h_\mu$ \label{2p1}}
Here we will give a short overview of the operator $A_{\min }^{2}$, eq.\eqref{Aminn0}, reminding to the more complete Appendix \ref{apb} for details. \\\\In particular, looking at the the stationary condition for the functional \eqref{Aminn0},  one gets a non-local transverse
field configuration $A^h_\mu$, $\partial_\mu A^h_\mu=0$, which can be expressed as an infinite series in
the gauge field $A_\mu$, see  Appendix \ref{apb}, {\it i.e.}
\begin{eqnarray}
A_{\mu }^{h} &=&\left( \delta _{\mu \nu }-\frac{\partial _{\mu }\partial
	_{\nu }}{\partial ^{2}}\right) \phi _{\nu }\;,  \qquad  \partial_\mu A^h_\mu= 0 \;, \nonumber \\
\phi _{\nu } &=&A_{\nu }-ig\left[ \frac{1}{\partial ^{2}}\partial A,A_{\nu
}\right] +\frac{ig}{2}\left[ \frac{1}{\partial ^{2}}\partial A,\partial
_{\nu }\frac{1}{\partial ^{2}}\partial A\right] +O(A^{3})\;.  \label{min0}
\end{eqnarray}
Remarkably, as shown in Appendix \ref{apb}, the configuration $A_{\mu }^{h}$ turns out to be left invariant
by infinitesimal gauge transformations order by order in the gauge
coupling $g$ \cite{Lavelle:1995ty}:
\begin{eqnarray}
\delta A_{\mu }^{h} &=&0\;,  \nonumber \\
\delta A_{\mu } &=&-\partial _{\mu }\omega +ig\left[ A_{\mu },\omega \right]
\;.  \label{gio}
\end{eqnarray}
Making use of \eqref{min0}, the gauge invariant nature of expression \eqref{Aminn0} can be made manifest by rewriting it in terms of the  field
strength $F_{\mu \nu }$. In fact, as proven in
\cite{Zwanziger:1990tn}, it turns out that
\begin{eqnarray}
A_{\min }^{2} = \int d^4x A_{\mu }^{h} A_{\mu }^{h} &=&-\frac{1}{2}\mathrm{Tr}\int d^{4}x\left( F_{\mu
	\nu }\frac{1}{D^{2}}F_{\mu \nu }+2i\frac{1}{D^{2}}F_{\lambda \mu
}\left[ \frac{1}{D^{2}}D_{\kappa }F_{\kappa \lambda
},\frac{1}{D^{2}}D_{\nu }F_{\nu \mu }\right] \right.
\nonumber \\
&&-2i\left. \frac{1}{D^{2}}F_{\lambda \mu }\left[ \frac{1}{D^{2}}D_{\kappa
}F_{\kappa \nu },\frac{1}{D^{2}}D_{\nu }F_{\lambda \mu }\right] \right)
+O(F^{4})\;,  \label{zzw}
\end{eqnarray}
from which the gauge invariance becomes apparent. The operator $({D^{2}})^{-1}$ in expression 
\eqref{zzw} denotes the inverse of the Laplacian $D^2=D_\mu D_\mu$ with $D_\mu$ being the 
covariant derivative \cite{Zwanziger:1990tn}. Let us also underline that, in the Landau gauge 
$\partial_\mu A_\mu=0$, the operator $(A^h_\mu A^h_\mu)$ reduces to the operator $A^{2}$
\begin{equation}
(A^{h,a}_\mu A^{h,a}_\mu) \Big|_{\rm Landau} =  A^a_\mu A^a_\mu \;.      \label{land}
\end{equation}
This feature, combined with the gauge invariant nature of $(A^{h,a}_\mu A^{h,a}_\mu)$, implies that the 
anomalous dimension of $(A^{h,a}_\mu A^{h,a}_\mu)$ equals \cite{Fiorentini:2016rwx}, to all orders, that of the operator $A_{\mu }^{a}A_{\mu }^{a}$ of the Landau gauge, {\it i.e.} 
\begin{equation} 
\gamma_{(A^{h})^2} = \gamma_{A^{2}}\Big|_{\rm Landau} \;. \label{lld}
\end{equation} 
Moreover, as proven in \cite{Dudal:2002pq},  $\gamma_{A^{2}}\Big|_{\rm Landau}$ is not
an independent parameter, being given by 
\begin{equation}
\gamma_{A^{2}}\Big|_{\rm Landau} = \left(  \frac{\beta(a)}{a} + \gamma^{\rm Landau}_{A}(a)   \right)  \;, \qquad a = \frac{g^2}{16\pi^2}    \;,  \label{adl}
\end{equation}
where $(\beta(a), \gamma^{\rm Landau}_A(a))$ denote, respectively, the $\beta$-function and 
the anomalous dimension of the gauge field $A_\mu$ in the Landau gauge. This relation was conjectured and explicitly verified up to three-loop order in \cite{Gracey:2002yt}.

%%%%%%%%%%%%%%%%%%%%%%%%%%%%%%%%%%%%%%%%%%%%%%%%%%%%%
\section{A local action for   $A^h_\mu$\label{2p2}}
\label{local_framework}
%%%%%%%%%%%%%%%%%%%%%%%%%%%%%%%%%%%%%%%%%%%%%%%%%%%%%
Following \cite{Fiorentini:2016rwx},  a fully local framework for the gauge invariant operator $A_{\mu }^{h}$ can be achieved. To that end, we consider the local, BRST invariant, action
\begin{equation}
S_{inv} = \int d^{4}x\,\frac{1}{4}\,F^{a}_{\mu\nu}F^{a}_{\mu\nu} +
\int d^4x \left(\tau^{a}\,\partial_{\mu}A^{h,a}_{\mu}
+\frac{m^{2}}{2}\,A^{h,a}_{\mu}A^{h,a}_{\mu} + {\bar \eta}^a\partial_{\mu}D^{ab}_{\mu}(A^h)\eta^{b} \right)  \;,  \label{act1}
\end{equation}
where
\begin{equation}\label{local_Ah}
A^{h}_{\mu} \equiv A^{h,a}_{\mu}\,T^{a}=h^{\dagger}A_{\mu}h+\frac{i}{g}h^{\dagger}\partial_{\mu}h.
\end{equation}
with
\begin{equation}
h=e^{ig\xi}=e^{ig\xi^{a}T^{a}}.     \label{hxi}
\end{equation}
The matrices $\{T^a\}$ are the generators of the gauge group $SU(N)$ and $\xi^{a}$ is an auxiliary localizing Stueckelberg field. \\\\By expanding
(\ref{local_Ah}), one finds an infinite series whose first terms are
\begin{equation}\label{Ah_expansion}
(A^{h})^{a}_{\mu}=A^{a}_{\mu}- \partial_{\mu} \xi^a - gf^{abc} A^b_\mu \xi^c -\frac{g}{2}f^{abc}\xi^{b}\partial_\mu \xi^c
+ {\rm higher \; orders}\,.
\end{equation}
That the action $S_{inv}$ gives a local setup for the nonlocal operator $A^h_\mu$ of eq.\eqref{min0} follows by noticing that the Lagrange multiplier $\tau$ implements precisely the transversality condition
\begin{equation}
\partial_\mu A^h_\mu = 0  \;, \label{tr}
\end{equation}
which, when solved iteratively for the Stueckelberg field $\xi^a$, gives back  the expression \eqref{min0}, see  Appendix \ref{apb}. In addition, the extra ghosts $({\bar \eta}, \eta)$ account for the Jacobian arising from the functional integration over $\tau$ which gives a delta-function of the type  $\delta(\partial A^h)$. Finally, the term $\frac{m^{2}}{2}\,A^{h,a}_{\mu}A^{h,a}_{\mu}$  accounts for the inclusion of the gauge invariant operator  $A^{h,a}_{\mu}A^{h,a}_{\mu}$ through the mass parameter $m^2$ which, as mentioned before, can be used as an effective infrared parameter whose value can be estimated through comparison with the available lattice simulations on the two-point gluon correlation function, see \cite{Dudal:2007cw,Dudal:2008sp,Dudal:2011gd,tissier2010infrared,tissier2011infrared,Aguilar:2008xm,Aguilar:2015bud,Fischer:2008uz,Aguilar:2015nqa,Huber:2015ria,Fischer:2009tn,Weber:2011nw,Frasca:2007uz,Siringo:2015wtx,Cucchieri:2007md,Cucchieri:2007rg,Cucchieri:2011ig,Oliveira:2012eh,Cucchieri:2009kk,Cucchieri:2011aa,Bicudo:2015rma,Cucchieri:2016jwg} \\\\The local action $S_{inv}$, eq.\eqref{act1}, enjoys an exact BRST symmetry:
\begin{equation}
s S_{inv} = 0 \;, \label{brst_s}
\end{equation}
where the nilpotent BRST transformations are given by
\begin{eqnarray}
sA^{a}_{\mu}&=&-D^{ab}_{\mu}c^{b}\,,\nonumber \\
sc^{a}&=&\frac{g}{2}f^{abc}c^{b}c^{c}\,, \nonumber \\
s\bar{c}^{a}&=&ib^{a}\,,\nonumber \\
sb^{a}&=&0\,, \nonumber \\
s \tau^a & = & 0\,, \nonumber \\
s {\bar \eta}^a & = & s \eta^a = 0 \,, \nonumber \\
s^2 &=0 &\;.    \label{brst}
\end{eqnarray}
For the Stueckelberg field one has \cite{Dragon:1996tk},  with $i,j$ indices associated with a generic representation,
\begin{equation}
s h^{ij} = -ig c^a (T^a)^{ik} h^{kj}  \;, \qquad s (A^h)^a_\mu = 0  \;,  \label{brstst}
\end{equation}
from which the BRST transformation of the field $\xi^a$ can be evaluated iteratively, yielding
\begin{equation}
s \xi^a=  g^{ab}(\xi) c^b = - c^a + \frac{g}{2} f^{abc}c^b \xi^c - \frac{g^2}{12} f^{amr} f^{mpq} c^p \xi^q \xi^r + O(\xi^3)    \;.
\label{eqsxi}
\end{equation}

\section{Introducing the gauge fixing term $S_{gf}$ \label{2p3}}
As it stands, the action  \eqref{act1} needs to be equipped with the gauge fixing term, $S_{gf}$, which we choose as 
\begin{eqnarray}
S_{gf} &=&  \int d^4x \;s \left( {\bar c}^a (\partial_\mu A^a_\mu - \mu^2 \xi^a ) - {i} \frac{\alpha}{2} {\bar c^a} b^a \right)  \nonumber \\ 
&=&  \int d^4x \left( i b^a \partial_\mu A^a_\mu +  \frac{\alpha}{2} b^a b^a 
- i \mu^2 b^a \xi^a + {\bar c}^a \partial_\mu D^{ab}_\mu(A)c^b + \mu^2 {\bar c}^a g^{ab}(\xi) c^b  \right)    \;. \label{gf1}
\end{eqnarray}
Besides the traditional gauge parameter $\alpha$, we have now introduced a second gauge massive parameter $\mu^2$. As it will be clear in the next section, this massive parameter will provide a fully BRST invariant regularizing mass for the Stueckelberg field $\xi^a$, a feature which has helpful consequences when performing explicit loop calculations involving $\xi^a$. Setting $\mu^2=0$, the gauge fixing  \eqref{gf1} reduces to that of the usual linear covariant gauge \cite{Fiorentini:2016rwx}. Moreover, when $\mu^2=\alpha=0$, the Landau gauge, $\partial_\mu A^a_\mu=0$, is recovered. Nevertheless, it is worth underlining that both $\mu^2$ and $\alpha$ appear only in the gauge fixing term, which is an exact BRST variation. As such, $\mu^2$ and $\alpha$ are pure gauge parameters which will not affect the correlation functions of local BRST invariant operators. \\\\Though, before going any further, let us provide a few remarks related to the explicit presence  of the Stueckelberg field $\xi^a$ in eq. \eqref{gf1}. As it is easily realized, the field $\xi^a$ is a dimensionless field, a feature encoded in the fact that the invariant action $S_{inv}$ itself is an infinite series in powers of $\xi^a$. As in any local quantum field theory involving dimensionless fields, one has the freedom of performing arbitrary reparametrization of these fields as, for instance, in the case of  the two-dimensional non-linear sigma model \cite{Blasi:1988sh,Becchi:1988nh} and of  $N=1$ super YM in superspace  \cite{Piguet:1981fb,Piguet:1981hh}. In the present case, this means that we have the freedom of replacing $\xi^a$ by an arbitrary dimensionless function of $\xi^a$, namely
\begin{equation} 
\xi^{a} \rightarrow \omega^a(\xi) = \xi^a + a_1^{abc} \xi^b \xi^c + a_2^{abcd} \xi^b \xi^c \xi^d + a_3^{abcde} \xi^b \xi^c \xi^d \xi^e + ........   \label{rp} 
\end{equation}    
This freedom, inherent to the dimensionless nature of $\xi^a$, is clearly evidentiated at the quantum level by the fact that the Stueckelberg field renormalizes in a non-linear way  \cite{Fiorentini:2016rwx}, i.e. like eq. \eqref{rp}, expressing precisely the freedom one has in the choice of a reparametrization for $\xi^a$.  \\\\In our context, in eq.\eqref{gf1}, we could have been equally started  with a term like 
\begin{equation}
s \left( {\bar c}^a  \xi^a \right) \rightarrow      s \left( {\bar c}^a  \omega^a(\xi)  \right) =  s \left( {\bar c}^a  ( \xi^a + a_1^{abc} \xi^b \xi^c + a_2^{abcd} \xi^b \xi^c \xi^d +...)  \right)  \;. \label{rp1}
\end{equation}
Of course, as much as $\mu^2$ and $\alpha$, all coefficients $(a_1^{abc}, a_2^{abcd}, a_3^{abcde}, ...)$ are gauge parameters, not affecting the correlation functions of the gauge invariant quantities. Equation \eqref{rp1} expresses the freedom which one always has when dealing with a gauge fixing term which depends explicitly from a dimensionless field, as the term \eqref{gf1}. In particular, this freedom will persist through the renormalization analysis, meaning that the renormalization of the gauge fixing itself has to be determined modulo an exact BRST terms of the kind $s\left( {\bar c}^a  \omega^a(\xi)  \right)$. Alternatively, one could start directly with the generalized gauge-fixing 
\begin{eqnarray}
S_{gf}^{gen} &=&  \int d^4x \;s \left( {\bar c}^a (\partial_\mu A^a_\mu) - \mu^2 \omega^a(\xi) ) - {i} \frac{\alpha}{2} {\bar c^a} b^a \right)  \nonumber \\ 
&=&  \int d^4x \left( i b^a \partial_\mu A^a_\mu +  \frac{\alpha}{2} b^a b^a 
- i \mu^2 b^a \omega^a(\xi) + {\bar c}^a \partial_\mu D^{ab}_\mu(A)c^b + \mu^2 {\bar c}^a \frac{\partial \omega^{a}(\xi)}{\partial \xi^c} g^{cd}(\xi) c^d  \right)    \;, \nonumber \\
\label{gf12}
\end{eqnarray}
and take into account the  renormalization of the quantity $\omega^a(\xi)$, encoded in the infinte set of gauge parameters $(a_1^{abc}, a_2^{abcd}, a_3^{abcde}, ...)$. In the following, we shall make use of the gauge-fixing \eqref{gf1} and identify in the final counterterm the term which corresponds to the reparametrization \eqref{rp1}. Moreover, in the Appendix \ref{apc}, we shall provide a second proof of the renormalizability of the model by deriving the generalized Slavnov-Taylor identities corresponding to the gauge fixing term \eqref{gf12}. \\\\In summary, as starting point, we shall take  the local, BRST invariant action 
\begin{equation}
S = S_{inv} +  S_{gf} \;, \label{stact}
\end{equation}
with 
\begin{equation}
s S = 0\;, \label{sinv}
\end{equation}
where the BRST transformations are given by eqs.\eqref{brst},\eqref{brstst},\eqref{eqsxi}.  \\\\Let us proceed now by giving a look at the propagators of the elementary fields. 

\section{A look at the propagators of the elementary fields\label{2p4}} 

The propagators of the elementary fields are easily evaluated from the quadratic part of the action, eq.\eqref{stact}, {\it i.e.} 

\begin{eqnarray}
S_{quad.} & = & \int d^{4}x\; \left( \frac{1}{4} {(\partial_\mu A^a_\nu - \partial_\nu A^a_\mu)}^2 +ib^{a}\partial_{\mu}A_{\mu}^{a}+\frac{\alpha}{2}b^{a}b^{a}+\bar{c}^{a}\partial^{2}c^{a}-\mu^2 \bar{c}^{a}c^{a} \right.  \nonumber \\
&  & +\frac{m^{2}}{2}A_{\mu}^{a}A_{\mu}^{a}-m^{2}A_{\mu}^{a}\partial_{\mu}\xi^{a}+\frac{m^{2}}{2}\left(\partial_{\mu}\xi^{a}\right)\left(\partial_{\mu}\xi^{a}\right) \nonumber \\
&  & \Bigl.+\tau^{a}\partial_{\mu}A_{\mu}^{a}-\tau^{a}\partial^{2}\xi^{a}+\bar{\eta}^{a}\partial^{2}\eta^{a}-i\mu^2 b^{a}\xi^{a} \Bigr)  \nonumber \\
& = & \int d^{4}x\; \frac{1}{2}\left[\begin{array}{cccc}
A_{\mu}^{a} & b^{a} & \xi^{a} & \tau^{a}\end{array}\right] \times \nonumber \\
&  & \times  \left[\begin{array}{cccc}
\left(-\delta_{\mu\nu}\partial^{2}+\partial_{\mu}\partial_{\nu}+m^{2}\right) & -i\partial_{\mu} & -m^{2}\partial_{\mu} & -\partial_{\mu} \nonumber \\
i \partial_{\nu} & {\alpha} & - i\mu^{2} & 0 \nonumber \\
m^{2} \partial_{\nu} & -i\mu^{2} & -m^{2}\partial^{2} & -\partial^{2}\\
\partial_{\nu} & 0 & -\partial^{2} & 0
\end{array}\right] 
\left[\begin{array}{c}
A_{\mu}^{a}\\
b^{a}\\
\xi^{a}\\
\tau^{a}
\end{array}\right]  \nonumber \\
&&+\int d^4x \; \left(\bar{c}^{a}\partial^{2}c^{a}+\bar{\eta}^{a}\partial^{2}\eta^{a}-\mu^{2}\bar{c}^{a}c^{a} \right)  \;.   \label{qqp} 
\end{eqnarray}
Thus, for the propagators we get
\begin{eqnarray}
\left\langle A_{\mu}^{a}\left(p\right)A_{\nu}^{b}\left(-p\right)\right\rangle  & = & \delta^{ab}\left(\frac{P_{\mu\nu}}{p^{2}+m^{2}}+\frac{\alpha p^{2}L_{\mu\nu}}{\left(p^{2}+\mu^{2}\right)^{2}}\right) \nonumber \\
\left\langle A_{\mu}^{a}\left(p\right)b^{b}\left(-p\right)\right\rangle  & = & \delta^{ab}\left(\frac{p_{\mu}}{p^{2}+\mu^{2}}\right) \nonumber \\
\left\langle A_{\mu}^{a}\left(p\right)\xi^{b}\left(-p\right)\right\rangle  & = & \delta^{ab}\left(\frac{-i\alpha p_{\mu}}{\left(p^{2}+\mu^{2}\right)^{2}}\right) \nonumber \\
\left\langle A_{\mu}^{a}\left(p\right)\tau^{b}\left(-p\right)\right\rangle  & = & \delta^{ab}\left(\frac{i\mu^{2}p_{\mu}}{p^{2}\left(p^{2}+\mu^{2}\right)}\right) \nonumber \\
\left\langle b^{a}\left(p\right)b^{b}\left(-p\right)\right\rangle  & = & 0 \nonumber \\
\left\langle b^{a}\left(p\right)\xi^{b}\left(-p\right)\right\rangle  & = & \frac{\delta^{ab}i}{p^{2}+\mu^{2}}\nonumber \\
\left\langle b^{a}\left(p\right)\tau^{b}\left(-p\right)\right\rangle  & = & 0 \nonumber \\
\left\langle \xi^{a}\left(p\right)\xi^{b}\left(-p\right)\right\rangle  & = & \frac{\delta^{ab}\alpha}{\left(p^{2}+\mu^{2}\right)^{2}} \nonumber \\
\left\langle \xi^{a}\left(p\right)\tau^{b}\left(-p\right)\right\rangle  & = & \frac{\delta^{ab}}{p^{2}+\mu^{2}} \nonumber \\
\left\langle \tau^{a}\left(p\right)\tau^{b}\left(-p\right)\right\rangle  & = & -\frac{\delta^{ab}m^{2}}{p^{2}} \nonumber \\
\left\langle {\bar c}^{a}\left(p\right)c^{b}\left(-p\right)\right\rangle  & = & \frac{\delta^{ab}}{p^{2}+\mu^2} \nonumber \\
\left\langle {\bar \eta}^{a}\left(p\right)\eta^{b}\left(-p\right)\right\rangle  & = & \frac{\delta^{ab}}{p^{2}} \label{propqq}
\end{eqnarray}
where, $P_{\mu\nu}=\left(\delta_{\mu\nu}-\frac{p_{\mu}p_{\nu}}{p^{2}}\right)$
and $L_{\mu\nu}=\frac{p_{\mu}p_{\nu}}{p^{2}}$ are the transverse and longitudinal projectors. We see that all propagators have a nice ultraviolet behavior, fully compatible with the power-counting. Moreover, the role of the massive gauge parameter $\mu^2$ becomes now apparent: it gives a BRST invariant regularizing mass for the Stueckelberg field $\xi^a$. Observe in fact that, when $\mu^2=0$, the propagator of the Stueckelberg field is given by $\langle \xi(p)\xi(-p)\rangle_{\mu^2=0}   =  \frac{\alpha}{p^{4}}$ which might give rise to potential infrared divergences in some class of Feynman diagrams. Notice also that, as expected, the mass parameter $m^2$ appears in the transverse part of the gluon propagator, a feature which exhibits its physical meaning. In fact, being coupled to the gauge invariant operator $(A^{h,a}_{\mu}A^{h,a}_{\mu})$, the parameter $m^2$ will enter the correlation functions of physical operators, {\it i.e.} gauge invariant operators, allowing thus to parametrize in an effective way their infrared behavior.

\section{$A_{\min }^{2}$ versus the conventional Stueckelberg mass term \label{2p5}} \label{St} 
As done in \cite{Fiorentini:2016rwx}, before facing the analysis of the renormalizability of the action $S$, eq. \eqref{stact}, let us make a short comparison with the standard Stueckelberg mass term \cite{Ruegg:2003ps}, corresponding to the action 
\begin{equation} 
S_{Stueck} = \int d^{4}x\,\left(\frac{1}{4}\,F^{a}_{\mu\nu}F^{a}_{\mu\nu} +\frac{m^{2}}{2}\,A^{h,a}_{\mu}A^{h,a}_{\mu}  \right)  + S_{gf}  \;,  \label{stact1}
\end{equation}
where $S_{gf}$ is given by eq. \eqref{gf1}. One sees that the conventional Stueckelberg action corresponds to the addition of the gauge invariant operator $(A^{h,a}_{\mu}A^{h,a}_{\mu})$ without taking into account  the transversality constraint $\partial_\mu A^{h,a}_\mu=0$,  implemented in the action \eqref{stact} through the Lagrange multiplier $\tau^a$ and the corresponding ghosts $({\bar \eta}^a, \eta^a)$. The removal of the constraint $\partial_\mu A^{h,a}_\mu=0$ gives rise to the conventional Stueckelberg propagator, namely  
\begin{equation} 
\langle \xi^{a}\left(p\right)\xi^{b}\left(-p\right)\rangle_{Stueck}   =  \frac{\delta^{ab} p^2}{m^2 (p^2+\mu^2)^2} + \frac{\delta^{ab}\alpha}{\left(p^{2}+\mu^{2}\right)^{2}} \;. \label{stp}
\end{equation}
From this expression one easily understand the cause of the bad ultraviolet behavior of the Stueckelberg mass term, giving rise to its nonrenormalizability \cite{Ferrari:2004pd}.  We see in fact  that the mass parameter $m^2$ enters  the denominator of expression 
\eqref{stp}. As one easily figures out, this property jepardizes the renormalizability of the standard Stueckelberg formulation 
\cite{Ferrari:2004pd}. Due to the presence of the parameter $m^2$ in the denominator of expressions \eqref{stp}, non-renormalizable divergences in the inverse of the mass $m^2$ will show up, invalidating thus 
the perturbative loop expansion based on expression \eqref{stact1}. \\\\The role of the term  $\int d^4x\;\tau^{a}\,\partial_{\mu}A^{h,a}_{\mu}$, implementing the constraint $\partial_\mu A^{h,a}_\mu=0$, becomes now clear. It gives rise to a deep modification of the Stueckelberg propagator, removing precisely the first problematic term, $\frac{\delta^{ab} p^2}{m^2 (p^2+\mu^2)^2}$, from expression \eqref{stp}. We are left therefore only with the second piece, {\it i.e.}  $\frac{\delta^{ab}\alpha}{\left(p^{2}+\mu^{2}\right)^{2}}$, which does not pause any problem with the ultraviolet power-counting. It is this nice feature which will ensure the all order renormalizability of the action $S$, eq.\eqref{stact}, as we shall discuss in details in the next sections.

\section{Algebraic characterization  of the counterterm \label{2p6}}\label{alg1} 
We are now ready to start the analysis of the renormalizability of the action  $S$, eq. \eqref{stact}. Following the setup of the algebraic renormalization \cite{Piguet:1995er}, we proceed by establishing the set of Ward identities which will be employed for the study of the quantum corrections. To that end, we need to introduce a set of external BRST invariant sources $(\Omega^a_\mu, L^a, K^a)$ coupled to the non-linear BRST variations of the fields $(A^a_\mu, c^a, \xi^{a})$ as well as sources $(\mathcal{J}_{\mu}^{a}, \Xi_{\mu}^{a})$ coupled to the BRST invariant  composite operators $( A_{\mu}^{ha},  D_{\mu}^{ab}(A^{h}))$, 
\begin{equation}
s \Omega^a_\mu = s L^a = sK^a = s \mathcal{J}_{\mu}^{a} = s \Xi_{\mu}^{a} = 0 \;.   \label{invs}
\end{equation} 
We shall thus start with the BRST invariant complete action $\Sigma$ defined by  
\begin{eqnarray}
\Sigma & = & \int d^{4}x\left( \frac{1}{4}\left(F_{\mu\nu}^{a}\right)^{2}+ib^{a}\partial_{\mu}A_{\mu}^{a}+\bar{c}^{a}\partial_{\mu}D_{\mu}^{ab}c^{b}+\frac{\alpha}{2}\left(b^{a}\right)^{2}-iM^{ab}b^{a}\xi^{b}\right. \nonumber \\
&  & -N^{ab}\bar{c}^{a}\xi^{b}+M^{ab}\bar{c}^{a}g^{bc}\left(\xi\right)c^{c}+\bar{\eta}^{a}\partial_{\mu}D_{\mu}^{ab}\left(A^{h}\right)\eta^{b}+\frac{m^2}{2}A_{\mu}^{ha}A_{\mu}^{ha} \nonumber \\
&  & +\tau^{a}\partial_{\mu}A_{\mu}^{ha}-\Omega_{\mu}^{a}D_{\mu}^{ab}c^{b}+\frac{g}{2}f^{abc}L^{a}c^{b}c^{c}+K^{a}g\left(\xi\right)^{ab}c^{b}+\mathcal{J}_{\mu}^{a}A_{\mu}^{ha} \nonumber \\
&  & \Bigr.+\Xi_{\mu}^{a}D_{\mu}^{ab}\left(A^{h}\right)\eta^{b} \Bigl) \;, \label{cact}
\end{eqnarray}
where, for later convenience,  we have also introduced the BRST doublet of external sources $(M^{ab}, N^{ab})$ 
\begin{equation}
s M^{ab} = N^{ab} \;, \qquad s N^{ab} = 0  \;, 
\end{equation}
so that
\begin{equation}
s \Sigma = 0 \;. \label{Sg}
\end{equation} 
Notice that the invariant action $S$ of eq. \eqref{stact} is immediately recovered from the complete action $\Sigma$ upon setting the external sources  
$(\Omega_{\mu}^{a}=L^{a}=K^{a}=\mathcal{J}_{\mu}^{a}=\Xi_{\mu}^{a}=0)$ and $(M^{ab}=\delta^{ab}\mu^{2}, N^{ab}=0)$. \\\\It turns out that the complete action $\Sigma$ obeys the following Ward identities:  
\begin{itemize}
	\item the Slavnov-Taylor identity
	\begin{eqnarray}
	\mathcal{S}\left(\Sigma\right) & = & \int d^{4}x\left(\frac{\delta \Sigma}{\delta\Omega_{\mu}^{a}}\frac{\delta \Sigma}{\delta A_{\mu}^{a}}+\frac{\delta \Sigma}{\delta L^{a}}\frac{\delta \Sigma}{\delta c^{a}}+ib^{a}\frac{\delta \Sigma}{\delta\bar{c}^{a}}+\frac{\delta \Sigma}{\delta K^{a}}\frac{\delta \Sigma}{\delta\xi^{a}}+N^{ab}\frac{\delta \Sigma}{\delta M^{ab}}\right) = 0  \;,  \nonumber \\ \label{sti}
	\end{eqnarray}
	\item the equation of motion of the Lagrange multiplier  $b^{a}$ and of the antighost $\bar{c}^{a}$
	\begin{eqnarray}
	\frac{\delta \Sigma}{\delta b^{a}} & = & i\partial_{\mu}A_{\mu}^{a}+\alpha b^{a}-iM^{ab}\xi^{b} \;, \label{beq}
	\end{eqnarray}
	\begin{eqnarray}
	\frac{\delta \Sigma}{\delta\bar{c}^{a}}+\partial_{\mu}\frac{\delta \Sigma}{\delta\Omega_{\mu}^{a}}-M^{ab}\frac{\delta \Sigma}{\delta K^{b}} & = & N^{ab}\xi^{b} \;, \label{agh}
	\end{eqnarray}
	\item the ghost-number Ward identity 
	\begin{eqnarray}
	\int d^{4}x\left(c^{a}\frac{\delta \Sigma}{\delta c^{a}}-\bar{c}^{a}\frac{\delta \Sigma}{\delta\bar{c}^{a}}-\Omega_{\mu}^{a}\frac{\delta \Sigma}{\delta\Omega_{\mu}^{a}}-2L^{a}\frac{\delta \Sigma}{\delta L^{a}}-K^{a}\frac{\delta \Sigma}{\delta K^{a}}+N^{ab}\frac{\delta \Sigma}{\delta N^{ab}}\right) & = & 0  \label{ghn}
	\end{eqnarray}
	\item the equation of the Lagrange multiplier $\tau^{a}$
	\begin{eqnarray}
	\frac{\delta \Sigma }{\delta\tau^{a}}-\partial_{\mu}\frac{\delta \Sigma}{\delta\mathcal{J}_{\mu}^{a}} & = & 0 \;, \label{teq}
	\end{eqnarray}
	\item the  $\eta^{a}$ Ward identity
	\begin{eqnarray}
	\int d^{4}x\left(\frac{\delta \Sigma}{\delta\eta^{a}}+gf^{abc}\bar{\eta}^{b}\frac{\delta \Sigma}{\delta\tau^{c}}+gf^{abc}\Xi^{b}\frac{\delta \Sigma}{\delta\mathcal{J}_{\mu}^{c}}\right) & = & 
	0 \;, \label{etw}
	\end{eqnarray}
	\item the ${\bar \eta}^a$ antighost equation
	\begin{eqnarray}
	\frac{\delta \Sigma}{\delta\bar{\eta}^{a}}-\partial_{\mu}\frac{\delta \Sigma}{\delta\Xi_{\mu}^{a}} & = & 0 \;, \label{aetaw}
	\end{eqnarray}
	\item the $(\eta^a, \bar \eta^a)$ ghost number
	\begin{eqnarray}
	\int d^{4}x\left(\eta^{a}\frac{\delta \Sigma}{\delta\eta^{a}}-\bar{\eta}^{a}\frac{\delta \Sigma}{\delta\bar{\eta}^{a}}-\Xi^{a}\frac{\delta \Sigma}{\delta\Xi^{a}}\right) & = & 0 \;. \label{etew}
	\end{eqnarray}
\end{itemize} 

\newpage 

\begin{table} 
	\centering
	\begin{tabular}{|c|c|c|c|c|c|c|c|c|c|c|c|c|c||c|c}
		\hline 
		& $A_{\mu}^{a}$ & $b^{a}$ & $c^{a}$ & $\bar{c}^{a}$ & $\tau^{a}$ & $\eta^{a}$ & $\bar{\eta}^{a}$ & $\xi^{a}$  \tabularnewline
		\hline 
		\hline 
		dim. & 1 & 2 & 0 & 2 & 2 & 0 & 2 & 0 \tabularnewline
		\hline 
		c gh. number  & 0 & 0 & 1 & -1 & 0 & 0 & 0 & 0 \tabularnewline
		\hline 
		$\eta$ gh. number  & 0 & 0 & 0 & 0 & 0 & 1 & -1 & 0 \tabularnewline
		\hline    
	\end{tabular}
	\caption{The quantum numbers of the fields} 
	\label{tbb1}
\end{table} 
\begin{table}
	\centering
	\begin{tabular}{|c|c|c|c|c|c|c|c|c|c|c|c|c|c||c|c}
		\hline 
		& $\Omega_{\mu}^{a}$ & $L^{a}$ & $K^{a}$ & $\mathcal{J}_{\mu}^{a}$ & $\Xi_{\mu}^{a}$ & $M^{ab}$ & $N^{ab}$\tabularnewline
		\hline 
		\hline 
		dim. & 3 & 4 & 4 & 3 & 2 & 2 & 2\tabularnewline
		\hline 
		c gh. number   & -1 & -2 & -1 & 0 & 0 & 0 & 1\tabularnewline
		\hline 
		$\eta$ gh. number   & 0 & 0 & 0 & 0 & -1 & 0 & 0\tabularnewline
		\hline    
	\end{tabular}
	\caption{The quantum numbers of the sources}
	\label{tbb2} 
\end{table} 
\noindent All quantum numbers and dimensions of all fields and sources are displayed in  Tables \eqref{tbb1} and \eqref{tbb2}. \\\\In order to characterize the most general invariant counterterm  which can be freely added to all order
in perturbation theory, we follow the setup of the algebraic renormalization  \cite{Piguet:1995er} and 
perturb the classical action $\Sigma$, eq.\eqref{cact}, by adding an integrated local quantity in the fields and sources, 
$\Sigma^{ct}$, with dimension bounded by four and vanishing ghost number. We demand thus that the 
perturbed action, $(\Sigma +\varepsilon\Sigma^{ct})$, where $\varepsilon$ is an expansion parameter, 
fulfills, to the first order in $\varepsilon$, the same Ward identities obeyed by the classical action 
$\Sigma$, {\it i.e.} equations \eqref{sti}, \eqref{beq}, \eqref{ghn}, \eqref{teq}, \eqref{etw}  and \eqref{aetaw}. This amounts to impose the following 
constraints on $\Sigma$: 
\begin{equation}
{\cal B}_{\Sigma}  \Sigma^{ct} = 0 \;, \label{cc13} 
\end{equation} 
\begin{equation} 
\frac{\delta \Sigma^{ct} }{\delta b^{a}}  = 0   \;, \label{cc29}
\end{equation} 
\begin{equation}
\frac{\delta \Sigma^{ct}}{\delta\bar{c}^{a}}+\partial_{\mu}\frac{\delta \Sigma^{ct}}{\delta\Omega_{\mu}^{a}}-M^{ab}\frac{\delta \Sigma^{ct}}{\delta K^{b}}  = 0   \;, \label{cc31}
\end{equation}
\begin{equation}
\frac{\delta \Sigma^{ct} }{\delta\tau^{a}}-\partial_{\mu}\frac{\delta \Sigma^{ct}}{\delta\mathcal{J}_{\mu}^{a}}  =  0 \;,  \label{cc4}
\end{equation}
\begin{equation}
\int d^{4}x\left(\frac{\delta \Sigma^{ct}}{\delta\eta^{a}}+gf^{abc}\bar{\eta}^{b}\frac{\delta \Sigma^{ct}}{\delta\tau^{c}}+gf^{abc}\Xi^{b}\frac{\delta \Sigma^{ct}}{\delta\mathcal{J}_{\mu}^{c}}\right)  =  0 \;, \label{cc5}
\end{equation} 
\begin{equation}
\frac{\delta \Sigma^{ct}}{\delta\bar{\eta}^{a}}-\partial_{\mu}\frac{\delta \Sigma^{ct}}{\delta\Xi_{\mu}^{a}}  =  0 \;, \label{cc6}
\end{equation} 
where  ${\cal B}_{\Sigma} $ is the so-called nilpotent linearized Slavnov-Taylor operator  \cite{Piguet:1995er}, defined as 
\begin{eqnarray}
{\cal B}_{\Sigma}  &= &  \int d^{4}x\left(\frac{\delta \Sigma}{\delta\Omega_{\mu}^{a}}\frac{\delta}{\delta A_{\mu}^{a}}+\frac{\delta \Sigma}{\delta A_{\mu}^{a}}\frac{\delta}{\delta\Omega_{\mu}^{a}}+\frac{\delta \Sigma}{\delta L^{a}}\frac{\delta}{\delta c^{a}}+\frac{\delta \Sigma}{\delta c^{a}}\frac{\delta}{\delta L^{a}}+\frac{\delta \Sigma}{\delta K^{a}}\frac{\delta}{\delta\xi^{a}} \right) \nonumber \\ 
&+& \int d^4x \left( \frac{\delta \Sigma}{\delta\xi^{a}}\frac{\delta}{\delta K^{a}}+ib^{a}\frac{\delta}{\delta\bar{c}^{a}}+N^{ab}\frac{\delta}{\delta M^{ab}}\right)   \;, \label{lst} 
\end{eqnarray}
with
\begin{equation} 
{\cal B}_{\Sigma} {\cal B}_{\Sigma}  =  0 \;.      \label{nplst}
\end{equation}
The first condition, eq. \eqref{cc13}, tells us that the counterterm $\Sigma^{ct}$ belongs to the cohomology of the operator ${\cal B}_{\Sigma} $ in the space of the integrated local polynomials in the fields, sources and parameters, of dimension four and ghost number zero. Owing to the general results on the BRST cohomology of YM theories \cite{Piguet:1995er} and taking advantage of the analysis already done in \cite{Fiorentini:2016rwx}, the most general form for $\Sigma^{ct}$ can be written as  
\begin{eqnarray}
	\Sigma^{ct}  & = & \Delta_{cohom}+ {\cal B}_{\Sigma} \Delta^{\left(-1\right)} \;, \label{pct}
\end{eqnarray}
where  $\Delta_{cohom}$ identifies the cohomolgy of ${\cal B}_{\Sigma} $, {\it i.e.} the non-trivial solution of eq.\eqref{cc13}, and  $\Delta^{\left(-1\right)}$ stands for the exact part, {\it i.e.} for the trivial solution of  \eqref{cc13}. Notice that, according to the quantum numbers of the fields, $\Delta^{\left(-1\right)}$ is an integrated polynomial of dimension four,  c-ghost number -1 and $\eta$-number equal to zero. \\\\For $\Delta_{cohom}$,  we have 
\begin{eqnarray}
\Delta_{cohom} & = & \int d^{4}x\; \Bigl( \frac{a_{0}}{4}\left(F_{\mu\nu}^{a}\right)^{2}+a_{1}\left(\partial_{\mu}A_{\mu}^{ha}\right)\left(\partial_{\nu}A_{\nu}^{ha}\right)+a_{2}\left(\partial_{\mu}A_{\nu}^{ha}\right)\left(\partial_{\mu}A_{\nu}^{ha}\right) \Bigr.  \nonumber \\
&  & +a_{3}^{abcd}A_{\mu}^{ha}A_{\mu}^{hb}A_{\nu}^{hc}A_{\nu}^{hd}+\left(\partial_{\mu}\tau^{a}+\mathcal{J}_{\mu}^{a}\right)F_{\mu}^{a}\left(A,\xi\right)+a_{5}\left(\partial_{\mu}\bar{\eta}^{a}+\Xi_{\mu}^{a}\right)\left(\partial_{\mu}\eta^{a}\right) \nonumber \\
&  & \Bigl. +f^{abc}\left(\partial_{\mu}\bar{\eta}^{a}+\Xi_{\mu}^{a}\right)\eta^{b}G_{\mu}^{c}\left(A,\xi\right)+m^{2}I\left(A,\xi\right)   \Bigr)   \;, \label{e1} 
\end{eqnarray}
where $F_{\mu}^{a}\left(A,\xi\right)$, $G_{\mu}^{c}\left(A,\xi\right)$
and $I\left(A,\xi\right)$ are local functional of $A_{\mu}^{a}$
and $\xi^{a}$, with dimension 1, 1 and 2, respectively. To write  expression  \eqref{e1} we have taken into account the constraints \eqref{cc31}--\eqref{cc6}. Moreover,  from condition \eqref{cc13} one immediately gets  
\begin{equation}
{\cal B}_{\Sigma} F_{\mu}^{a}\left(A,\xi\right)= {\cal B}_{\Sigma}G_{\mu}^{c}\left(A,\xi\right)={\cal B}_{\Sigma} I\left(A,\xi\right)= 0 \;. \label{t1}
\end{equation} 
Proceeding as in \cite{Fiorentini:2016rwx}, equations \eqref{t1} are solved by 
\begin{equation}
F_{\mu}^{a}\left(A,\xi\right) = a_4 A_{\mu}^{ha} \;, \qquad  G_{\mu}^{c}\left(A,\xi\right)= a_6  A_{\mu}^{ha} \;, \qquad  I\left(A,\xi\right)= a_7  A_{\mu}^{ha}A_{\mu}^{ha} \;, \label{t1a} 
\end{equation} 
where $(a_4, a_6, a_7)$ are free coefficients. Therefore, 
\begin{eqnarray*}
	\Delta_{cohom} & = & \int d^{4}x\; \Bigl( \frac{a_{0}}{4}\left(F_{\mu\nu}^{a}\right)^{2}+a_{1}\left(\partial_{\mu}A_{\mu}^{ha}\right)\left(\partial_{\nu}A_{\nu}^{ha}\right)+a_{2}\left(\partial_{\mu}A_{\nu}^{ha}\right)\left(\partial_{\mu}A_{\nu}^{ha}\right) \Bigr. \nonumber \\
	&  & +a_{3}^{abcd}A_{\mu}^{ha}A_{\mu}^{hb}A_{\nu}^{hc}A_{\nu}^{hd}+a_{4}\left(\partial_{\mu}\tau^{a}+\mathcal{J}_{\mu}^{a}\right)A_{\mu}^{ha}+a_{5}\left(\partial_{\mu}\bar{\eta}^{a}+\Xi_{\mu}^{a}\right)\left(\partial_{\mu}\eta^{a}\right)\\
	&  & \Bigl. +a_{6}f^{abc}\left(\partial_{\mu}\bar{\eta}^{a}+\Xi_{\mu}^{a}\right)\eta^{b}A_{\mu}^{hc}+a_{7}m^{2}A_{\mu}^{ha}A_{\mu}^{ha} \Bigr) \;. \label{t1b}
\end{eqnarray*}
Let us discuss now the exact part of the cohomology of ${\cal B}_{\Sigma}$  which, taking into account the quantum numbers of the fields and sources,  can be parametrized as 
\begin{eqnarray*}
	\Delta^{\left(-1\right)} & = & \int d^{4}x \; \Bigl( f_{1}^{ab}\left(\xi,\alpha\right)\xi^{a}K^{b}+f_{2}^{ab}\left(\xi,\alpha\right)L^{a}c^{b}+f_{3}^{ab}\left(\xi,\alpha\right)\xi^{a}\left(\partial_{\mu}\Omega_{\mu}^{b}\right)+f_{4}^{ab}\left(\xi,\alpha\right)\left(\partial_{\mu}\xi^{a}\right)\Omega_{\mu}^{b}. \Bigr. \nonumber \\
	&  & +f_{5}^{ab}\left(\xi,\alpha\right)A_{\mu}^{a}\Omega_{\mu}^{b}+f_{6}^{ab}\left(\xi,\alpha\right)A_{\mu}^{a}\left(\partial_{\mu}\bar{c}^{b}\right)+f_{7}^{ab}\left(\xi,\alpha\right)\left(\partial_{\mu}A_{\mu}^{a}\right)\bar{c}^{b}\\
	&  & +f_{8}^{ab}\left(\xi,\alpha\right)\left(\partial_{\mu}\xi^{a}\right)\left(\partial_{\mu}\bar{c}^{b}\right)+f_{9}^{ab}\left(\xi,\alpha\right)\xi^{a}\left(\partial^{2}\bar{c}^{b}\right)+f_{10}^{ab}\left(\xi,\alpha\right)\bar{c}^{a}b^{b}\\
	&  & +f_{11}^{ab}\left(\xi,\alpha\right)\bar{c}^{a}\tau^{b}+f_{12}^{abc}\left(\xi,\alpha\right)\bar{\eta}^{a}\eta^{b}\bar{c}^{c}+f_{13}^{abc}\left(\xi,\alpha\right)\bar{c}^{a}\bar{c}^{b}c^{c}\\
	&  & \Bigl. +f_{14}^{abcd}\left(\xi,\alpha\right)M^{ab}\xi^{c}\bar{c}^{d} \Bigr) \;, \label{trv1}
\end{eqnarray*}
where $(f_{1}, ..., f_{14})$ are arbitrary coefficients. Imposing the constraint \eqref{cc29}, {\it i.e.}
\begin{equation}
\frac{\delta}{\delta b^{k}} {\cal B}_{\Sigma} \Delta^{\left(-1\right)} = 0 \;, \label{bct} 
\end{equation} 
and making use of the commutation relation 
\begin{equation}
\frac{\delta}{\delta b^{k}}{\cal B}_{\Sigma} ={\cal B}_{\Sigma} \frac{\delta}{\delta b^{k}}+i\left(\frac{\delta}{\delta\bar{c}^{k}}+\partial_{\mu}\frac{\delta}{\delta\Omega_{\mu}^{k}}-M^{kl}\frac{\delta}{\delta K^{l}}\right) \;, \label{cm}
\end{equation} 
one finds 
\begin{eqnarray*}
	\frac{\delta\Delta^{\left(-1\right)}}{\delta b^{k}}  =  f_{10}^{ak}\left(\xi,\alpha\right)\bar{c}^{a}  \qquad 
	\Rightarrow \qquad {\cal B}_{\Sigma}\frac{\delta\Delta^{\left(-1\right)}}{\delta b^{k}}  =  \frac{\delta \Sigma}{\delta K^{m}}\frac{\partial f_{10}^{ak}\left(\xi,\alpha\right)}{\partial\xi^{m}}\bar{c}^{a}+if_{10}^{ak}\left(\xi,\alpha\right)b^{a} \;. \label{cm1}
\end{eqnarray*}
Moreover, from 
\begin{eqnarray*}
	i\left(\frac{\delta\Delta^{\left(-1\right)}}{\delta\bar{c}^{k}}+\partial_{\mu}\frac{\delta\Delta^{\left(-1\right)}}{\delta\Omega_{\mu}^{k}}-M^{kl}\frac{\delta\Delta^{\left(-1\right)}}{\delta K^{l}}\right) & = & -i\partial_{\mu}\left(f_{6}^{ak}\left(\xi,\alpha\right)A_{\mu}^{a}\right)+if_{7}^{ak}\left(\xi,\alpha\right)\left(\partial_{\mu}A_{\mu}^{a}\right)\\
	&  & -i\partial_{\mu}\left(f_{8}^{ak}\left(\xi,\alpha\right)\left(\partial_{\mu}\xi^{a}\right)\right)+i\partial^{2}\left(f_{9}^{ak}\left(\xi,\alpha\right)\xi^{a}\right)\\
	&  & +if_{10}^{kb}\left(\xi,\alpha\right)b^{b}+if_{11}^{kb}\left(\xi,\alpha\right)\tau^{b}\\
	&  & +if_{12}^{abk}\left(\xi,\alpha\right)\bar{\eta}^{a}\eta^{b}+2if_{13}^{kbc}\left(\xi,\alpha\right)\bar{c}^{b}c^{c}\\
	&  & +if_{14}^{abck}\left(\xi,\alpha\right)M^{ab}\xi^{c}\\
	&  & -i\partial^{2}\left(f_{3}^{ak}\left(\xi,\alpha\right)\xi^{a}\right)+i\partial_{\mu}\left(f_{4}^{ak}\left(\xi,\alpha\right)\left(\partial_{\mu}\xi^{a}\right)\right) \\
	&  & +i\partial_{\mu}\left(f_{5}^{ak}\left(\xi,\alpha\right)A_{\mu}^{a}\right)-iM^{kl}f_{1}^{al}\left(\xi,\alpha\right)\xi^{a} \;, \label{cm2}
\end{eqnarray*}
it follows that 
\begin{eqnarray*}
	\frac{\delta}{\delta b^{k}} {\cal B}_{\Sigma} \Delta^{\left(-1\right)}  = 0  & = & \left[\frac{\partial f_{10}^{bk}\left(\xi,\alpha\right)}{\partial\xi^{m}}g^{mc}\left(\xi\right)-2if_{13}^{kbc}\left(\xi,\alpha\right)\right]c^{c}\bar{c}^{b}\\
	&  & +i\left[f_{10}^{ak}\left(\xi,\alpha\right)+f_{10}^{ka}\left(\xi,\alpha\right)\right]b^{a}\\
	&  & +i\left[-f_{6}^{ak}\left(\xi,\alpha\right)+f_{5}^{ak}\left(\xi,\alpha\right)+f_{7}^{ak}\left(\xi,\alpha\right)\right]\left(\partial_{\mu}A_{\mu}^{a}\right)\\
	&  & -i\left[\left(\partial_{\mu}f_{6}^{ak}\left(\xi,\alpha\right)\right)-\left(\partial_{\mu}f_{5}^{ak}\left(\xi,\alpha\right)\right)\right]A_{\mu}^{a}\\
	&  & +i\left[-\left(\partial_{\mu}f_{8}^{ak}\left(\xi,\alpha\right)\right)-\left(\partial_{\mu}f_{3}^{ak}\left(\xi,\alpha\right)\right)+\left(\partial_{\mu}f_{4}^{ak}\left(\xi,\alpha\right)\right)+\left(\partial_{\mu}f_{9}^{ak}\left(\xi,\alpha\right)\right)\right]\left(\partial_{\mu}\xi^{a}\right)\\
	&  & +i\left[-f_{8}^{ak}\left(\xi,\alpha\right)-f_{3}^{ak}\left(\xi,\alpha\right)+f_{4}^{ak}\left(\xi,\alpha\right)+f_{9}^{ak}\left(\xi,\alpha\right)\right]\left(\partial^{2}\xi^{a}\right)\\
	&  & +i\left[-\left(\partial^{2}f_{3}^{ak}\left(\xi,\alpha\right)\right)+\left(\partial^{2}f_{9}^{ak}\left(\xi,\alpha\right)\right)\right]\xi^{a}\\
	&  & +if_{11}^{kb}\left(\xi,\alpha\right)\tau^{b}+if_{12}^{abk}\left(\xi,\alpha\right)\bar{\eta}^{a}\eta^{b}
	+i\left[f_{14}^{abck}\left(\xi,\alpha\right)-\delta^{ka}f_{1}^{cb}\left(\xi,\alpha\right)\right]M^{ab}\xi^{c} \;, \label{cm3}
\end{eqnarray*}
form which we can derive relations among the coefficients $(f_{1}, ..., f_{14})$.  Let us start with 
\begin{equation} 
\left(\partial_{\mu}f_{6}^{ak}\left(\xi,\alpha\right)\right)-\left(\partial_{\mu}f_{5}^{ak}\left(\xi,\alpha\right)\right)=0 \qquad \Rightarrow  \qquad  f_{6}^{ab}  =  f_{5}^{ab}+\delta^{ab}a \;, \label{cm4}
\end{equation} 
where $a$ is a constant. Further
\begin{equation} 
-f_{6}^{ak}\left(\xi,\alpha\right)+f_{5}^{ak}\left(\xi,\alpha\right)+f_{7}^{ak}\left(\xi,\alpha\right)=0 \qquad \Rightarrow \qquad 
f_{7}^{ak}\left(\xi,\alpha\right)=\delta^{ab}a \;. \label{cm5}
\end{equation} 
Analogously 
\begin{equation} 
-\left(\partial^{2}f_{3}^{ak}\left(\xi,\alpha\right)\right)+\left(\partial^{2}f_{9}^{ak}\left(\xi,\alpha\right)\right)=0 \qquad \Rightarrow \qquad f_{9}^{ak}\left(\xi,\alpha\right)=f_{3}^{ak}\left(\xi,\alpha\right)+b\delta^{ak} \;, \label{cm6}
\end{equation} 
with $b$ a free constant. Next, from 
\begin{equation} 
\left[-\left(\partial_{\mu}f_{8}^{ak}\left(\xi,\alpha\right)\right)-\left(\partial_{\mu}f_{3}^{ak}\left(\xi,\alpha\right)\right)+\left(\partial_{\mu}f_{4}^{ak}\left(\xi,\alpha\right)\right)+\left(\partial_{\mu}f_{9}^{ak}\left(\xi,\alpha\right)\right)\right]   \;, \label{cm7}
\end{equation} 
we get 
\begin{equation}
f_{8}^{ak}\left(\xi,\alpha\right)=f_{4}^{ak}\left(\xi,\alpha\right)+c\delta^{ak} \;, 
\end{equation} 
with $c$ constant. Finally 
\begin{equation}
-f_{8}^{ak}\left(\xi,\alpha\right)-f_{3}^{ak}\left(\xi,\alpha\right)+f_{4}^{ak}\left(\xi,\alpha\right)+f_{9}^{ak}\left(\xi,\alpha\right)=0 \qquad \Rightarrow \qquad b=c  \;.  
\end{equation} 
Therefore,  $\Delta^{\left(-1\right)}$ becomes 
\begin{eqnarray*}
	\Delta^{\left(-1\right)} & = &
	\int d^{4}x \; \Bigl( f_{1}^{ab}\left(\xi,\alpha\right)\left(\xi^{a}K^{b}+M^{cb}\xi^{a}\bar{c}^{c}\right) \Bigr. \\
	&  & +f_{2}^{ab}\left(\xi,\alpha\right)L^{a}c^{b}+f_{3}^{ab}\left(\xi,\alpha\right)\xi^{a}\left(\left(\partial_{\mu}\Omega_{\mu}^{b}\right)+\left(\partial^{2}\bar{c}^{b}\right)\right)\\
	&  & +f_{4}^{ab}\left(\xi,\alpha\right)\left(\partial_{\mu}\xi^{a}\right)\left(\Omega_{\mu}^{b}+\left(\partial_{\mu}\bar{c}^{b}\right)\right)\\
	&  & +f_{5}^{ab}\left(\xi,\alpha\right)A_{\mu}^{a}\left(\Omega_{\mu}^{b}+\left(\partial_{\mu}\bar{c}^{b}\right)\right)\\
	&  & \Bigl. +f_{10}^{ab}\left(\xi,\alpha\right)\bar{c}^{a}b^{b}+\frac{1}{2i}\frac{\partial f_{10}^{ba}\left(\xi,\alpha\right)}{\partial\xi^{m}}g^{mc}\left(\xi\right)\bar{c}^{a}\bar{c}^{b}c^{c} \Bigr)  \;. \label{cm9} 
\end{eqnarray*}
We can now impose the constraint \eqref{cc5}  
\begin{eqnarray*}
	\int d^{4}x\left(\frac{\delta \Sigma^{ct}}{\delta\eta^{m}}+gf^{mnp}\bar{\eta}^{n}\frac{\delta \Sigma^{ct}}{\delta\tau^{p}}+gf^{mnp}\Xi^{n}\frac{\delta \Sigma^{ct}}{\delta\mathcal{J}_{\mu}^{p}}\right) & = & 0 \;, \label{cm10} 
\end{eqnarray*}

\begin{eqnarray*}
	\Rightarrow\int d^{4}x\left(a_{6}+a_{4}g\right)f^{mnp}\left(\partial_{\mu}\bar{\eta}^{n}+f^{mnp}\Xi^{n}\right)A_{\mu}^{hp} & = & 0 \;,  \label{cm11}
\end{eqnarray*}
from which we obtain $a_{6}=-a_{4}g$. \\\\As done in \cite{Fiorentini:2016rwx},  we can further reduce the number of parameters entering $\Sigma^{ct}$ by observing that, setting  
$K^{a}=M^{ab}=N^{ab}=\mathcal{J}_{\mu}^{a}=\Xi_{\mu}^{a}=m=0$, the complete action $\Sigma$, eq.\eqref{cact}, reduces to that or ordinary YM theory in the linear covarinat gauges, as  integration over $\tau^{a}$, $\eta^{a}$ and $\overline{\eta}^{a}$
gives a unity. As a consequence, making use of the well known renormalization of standard YM theory in the linear covariant gauges \cite{Piguet:1995er}, we get  $a_{1}=a_{2}=a_{3}^{abcd}=0$, $a_{5}=a_{4}$, as well as 

\begin{equation}
f_{2}^{ab}\left(\xi,\alpha\right)  =  \delta^{ab}d_{1}\left(\alpha\right) \;, \qquad  f_{5}^{ ab}\left(\xi,\alpha\right)  =  \delta^{ab}d_{2}\left(\alpha\right) \;, \label{cm12} 
\end{equation}
with $(d_{1}, d_{2})$ free parameters. In addition, we also have 
\begin{eqnarray*}
	f_{3}^{ab}\left(\xi,\alpha\right)=f_{4}^{ab}\left(\xi,\alpha\right)=f_{10}^{ab}\left(\xi,\alpha\right) & = & 0 \;. \label{cm13} 
\end{eqnarray*}
Hence
\begin{eqnarray*}
	\Delta_{cohom}   & = & \int d^{4}x\; \Bigl( \frac{a_{0}}{4}\left(F_{\mu\nu}^{a}\right)^{2}+a_{4}\left(\left(\partial_{\mu}\tau^{a}+\mathcal{J}_{\mu}^{a}\right)A_{\mu}^{ha}+\left(\partial_{\mu}\bar{\eta}^{a}+\Xi_{\mu}^{a}\right)D^{ab}\left(A^{h}\right)\eta^{b}\right)  \Bigr. \\
	&  & \Bigl. \; \; \; \; \; +a_{7}m^{2}A_{\mu}^{ha}A_{\mu}^{ha} \Bigr) \;, \label{cm14} 
\end{eqnarray*}
and

\begin{eqnarray*}
	\Delta^{\left(-1\right)}  =  \int d^{4}x \Bigl( f_{1}^{ab}\left(\xi,\alpha\right)\left(\xi^{a}K^{b}+M^{cb}\xi^{a}\bar{c}^{c}\right) 
	+d_{1}\left(\alpha\right)L^{a}c^{a}+d_{2}\left(\alpha\right)A_{\mu}^{a}\left(\Omega_{\mu}^{a}+\left(\partial_{\mu}\bar{c}^{a}\right)\right) \Bigr) \;.  \label{cm15}
\end{eqnarray*}
Let us end this section by  rewriting the final expression of the most general invariant counterterm $\Sigma^{ct}$ in its parametric form \cite{Piguet:1995er}, a task that will simplify the analysis of the renormalziation factors, namely  

\begin{eqnarray}
\Sigma^{ct}  & = & -a_{0}g\frac{\partial \Sigma}{\partial g}+d_{2}\left(\alpha\right)2\alpha\frac{\partial \Sigma}{\partial\alpha}+a_{7}m^{2}\frac{\partial \Sigma}{\partial m^{2}} \nonumber \\
&  & +\int d^{4}x\; \Bigl( a_{4}\left(-\tau^{a}\frac{\delta \Sigma}{\delta\tau^{a}}+\mathcal{J}_{\mu}^{a}\frac{\delta \Sigma}{\delta\mathcal{J}_{\mu}^{a}}-\bar{\eta}^{a}\frac{\delta \Sigma}{\delta\bar{\eta}^{a}}+\Xi_{\mu}^{a}\frac{\delta \Sigma}{\delta\Xi_{\mu}^{a}}\right) \Bigr.  \nonumber \\
&  & -\left(f_{1}^{ab}\left(\xi,\alpha\right)+\frac{\partial f_{1}^{kb}\left(\xi,\alpha\right)}{\partial\xi^{a}}\xi^{k}\right)K^{b}\frac{\delta \Sigma}{\delta K^{a}}+f_{1}^{ab}\left(\xi,\alpha\right)\xi^{a}\frac{\delta \Sigma}{\delta\xi^{b}} \nonumber \\
&  & +d_{2}\left(\alpha\right)A_{\mu}^{a}\frac{\delta \Sigma}{\delta A_{\mu}^{a}}-d_{2}\left(\alpha\right)b^{a}\frac{\delta \Sigma}{\delta b^{a}}
-d_{2}\left(\alpha\right)\Omega_{\mu}^{a}\frac{\delta \Sigma}{\delta\Omega_{\mu}^{a}}-d_{2}\left(\alpha\right)\bar{c}^{a}\frac{\delta \Sigma}{\delta\overline{c}^{a}} \nonumber \\
&  & -d_{1}\left(\alpha\right)c^{a}\frac{\delta \Sigma}{\delta c^{a}}+d_{1}\left(\alpha\right)L^{a}\frac{\delta \Sigma}{\delta L^{a}} +\left(-f_{1}^{cb}\left(\xi,\alpha\right)+d_{2}\left(\alpha\right)\delta^{cb}\right)N^{ab}\frac{\delta \Sigma}{\delta N^{ac}} \nonumber \\
&  & \Bigl.+\left(d_{2}\left(\alpha\right)\delta^{ab}-f_{1}^{ab}\left(\xi,\alpha\right)\right)M^{cb}\frac{\delta \Sigma}{\delta M^{ca}} +\frac{\partial f_{1}^{ab}\left(\xi,\alpha\right)}{\partial\xi^{k}}M^{cb}\xi^{a}g\left(\xi\right)^{kd}c^{d}\bar{c}^{c}  \Bigr) \;. \label{ctf}
\end{eqnarray}

\section{Analysis of the counterterm and renormalization factors \label{2p7}}\label{alg2} 
Having determined the most general form of the local invariant counterterm, eq.\eqref{ctf}, let us turn to its physical meaning. 
As already mentioned before, in order to determine the renormalization of the fields, sources and parameters, we have to 
pay attention to the fact that, due to the explicit dependence of the gauge fixing from the Stueckelberg field $\xi^a$,  the renormalization of the gauge fixing itself is determined up to an ambiguity of the type of eq.\eqref{rp1}, which would correspond to the renormalization of the quantity $\omega^a(\xi)$, {\it i.e.} of the gauge parameters $(a_1^{abc}, a_2^{abcd}, a_3^{abcde}, ...)$. To that end, it will be sufficient to analyse the last two terms of the expression for $\Sigma^{ct}$, eq.\eqref{ctf}, which, upon setting the sources $(M^{ab},N^{ab})$ to their physical values, namely $(M^{ab}= \delta^{ab} \mu^2, N^{ab}=0)$, becomes 
\begin{equation}
\left( d_2(\alpha) - f_1(0, \alpha) \right) \mu^2 \frac{\partial \Sigma}{\partial \mu^2}  + \mu^2 \int d^4x \left( {\tilde f}_1^{ab}(\xi, \alpha) (i b^b \xi^a - {\bar c}^b g^{ak}(\xi)c^k) + 
\frac{\partial {\tilde f}_1^{ab}(\xi, \alpha) }{\partial \xi^k} {\bar c}^b \xi^a g^{kd}(\xi) c^d \right)   \label{ct1}
\end{equation}
where we have set 
\begin{equation} 
f_1^{ab}(\xi, \alpha) = f_1^{ab}(0, \alpha) + {\tilde f}_1^{ab}(\xi, \alpha)  \;, \label{fd}  
\end{equation}
with $f_1^{ab}(0, \alpha) = \delta^{ab} f_1(0,\alpha)$ being the first, $\xi^a$-independent,  term of the Taylor expansion of $f_1^{ab}(\xi, \alpha)$ in powers of $\xi^a$ and $ {\tilde f}_1^{ab}(\xi, \alpha)$ denoting the $\xi$-dependent remaining terms. Of course, $f_1^{ab}(0, \alpha) = \delta^{ab} f_1(0,\alpha)$ is just a constant. \\\\Furthermore, we observe that expression \eqref{ct1} can be rewritten as 
\begin{equation}
\left( d_2(\alpha) - f_1(0, \alpha) \right) \mu^2 \frac{\partial \Sigma}{\partial \mu^2}  + \mu^2 \int d^4x \;s \left( {\tilde f}_1^{ab}(\xi,\alpha) {\bar c}^b \xi^a \right)  \;, \label{ct2} 
\end{equation}
or, equivalently 
\begin{equation}
\left( d_2(\alpha) - f_1(0, \alpha) \right) \mu^2 \frac{\partial \Sigma}{\partial \mu^2}  + \mu^2 \int d^4x \;s \left(  {\bar c}^b {\tilde \omega}^b(\xi, \alpha) \right) \;, \label{ct3}
\end{equation}
with ${\tilde \omega}^b(\xi, \alpha) = {\tilde f}_1^{ab}(\xi,\alpha) \xi^a$. \\\\We are now able to unravel the meaning of this term. First, the term $\left( d_2(\alpha) - f_1^{aa}(0, \alpha) \right)$ corresponds to a multiplicative renormalization of the gauge massive parameter $\mu^2$. This follows by observing that, being $\mu^2$ a space-time independent parameter, its renormalization must be given by a field independent space-time constant factor, {\it i.e.} precisely by $\left( d_2(\alpha) - f_1^{aa}(0, \alpha) \right)$. On the other hand, the term $\int d^4x \;s \left(  {\bar c}^b {\tilde \omega}^b(\xi, \alpha) \right)$ is of the type of eq.\eqref{rp1}, thus corresponding to the ambiguity inherent to the gauge fixing discussed before. As already mentioned, this term can be handled by starting with the generalised gauge fixing  \eqref{gf12}, whose algebraic renormalization can be faced by employing the Ward identities displayed in Appendix \eqref{apc}. Doing so, the term $\int d^4x \;s \left(  {\bar c}^b {\tilde \omega}^b(\xi, \alpha) \right)$ will correspond to a renormalization of the gauge fixing function $\omega^a(\xi)$, {\it i.e.} of the gauge parameters $(a_1^{abc}, a_2^{abcd}, a_3^{abcde}, ...)$. \\\\We can now read off the renormalization factors, {\it i.e.} 
\begin{equation} 
\Sigma(\Phi) +  \varepsilon  \Sigma^{ct}(\Phi)   =   \Sigma(\Phi_0) + O(\varepsilon^2) \;, \label{ra1}
\end{equation}
with
\begin{eqnarray}
\Phi_0= Z_{\Phi} \Phi  + O(\varepsilon^2) \;, \label{ra2}
\end{eqnarray}
where $\Phi$ stands for a short-hand notation for all fields, sources and parameters. Specifically, for the renormalization factors one finds:

\begin{eqnarray}
A_0&=&Z_A^{1/2} A_{\mu}\;,  \,\,\,  b_0=Z_b^{1/2}b\;,  \,\,\,  c_0=Z_c^{1/2}c \;, \,\,\, \bar{c}_0=Z_{\bar{c}}^{1/2}\bar{c} \;,  \\ 
\xi_0^a &= &Z^{ab}_{\xi}(\xi)\xi^b \;  \,\,\, \tau_0=Z_{\tau}^{1/2} \tau \;,  \Omega_0 =Z_{\Omega} \Omega \;, \,\, \,L_0=Z_L L \; \,\,\,  \\
K_0^a&=& Z_K^{ab} (\xi) K^b\;, \,\,\, m_0^2=Z_{m^2} m^2\;, \,\,\,\mathcal{J}_0=Z_{\mathcal{J}}\mathcal{J}\;,\\
g_0&=&Z_g g \;, \,\,\, \alpha_0=Z_{\alpha} \alpha \;, \,\,\,\bar{\eta}_0=Z_{\bar{\eta}}^{1/2}\bar{\eta} \;, \,\,\, \eta_0=Z_{\eta}^{1/2} \eta \;,  \\
\Xi_0&=&Z_{\Xi} \Xi\;, \,\,\, \mu^2_0=Z_{\mu^2} \mu^2 \;, \label{ra3}
\end{eqnarray}
where
\begin{eqnarray}
Z_g&&=1-\varepsilon \frac{a_0}{2} \nonumber \\
Z^{1/2}_A&&=Z^{-1}_{\Omega}=Z^{-1/2}_{\bar{c}}=Z^{-1/2}_b=Z^{1/2}_{\alpha}=1+\varepsilon d_2(\alpha) \nonumber \\
Z_\xi^{ab}&&=\delta^{ab}+\varepsilon f_{1}^{ab}(\xi,\alpha) \nonumber \\
Z_L&&=Z^{-1/2}_c=1+\varepsilon d_1(\alpha) \nonumber \\
Z_{\bar{\eta}}&&=Z_{\eta}=Z^2_{\Xi}=Z^{1/2}_{\tau}=Z_{\mathcal{J}}=1+\varepsilon a_4 \nonumber \\
Z_{m^2}&&=1+\varepsilon a_7 \nonumber \\
Z_{\mu^2}&&=1+\varepsilon (d_2-f_2(0,\alpha)) \nonumber \\
Z_{K}^{ab}&&=\delta^{ab}-\varepsilon\left(f_{1}^{ab}(\xi,\alpha)+\frac{\partial f_{1}^{kb}(\xi,\alpha)}{\partial \xi^a}\xi^{k}\right) \;. \label{ra4}
\end{eqnarray}
Notice that, as expected, the dimensionless field $\xi^a$ renormalizes in a non-linear way through the quantity $f_{1}^{ab}(\xi,\alpha)$ which is a power series in $\xi^a$.  Equations \eqref{ra1} and \eqref{ra4} establish the renormalizability of the complete action $\Sigma$, eq.\eqref{cact}, and thus of the invariant action $S$ of expression \eqref{stact}, up to a BRST exact unphysical ambiguity of the type of eq.\eqref{rp1}. As already mentioned, the explicit inclusion of such an ambiguity will be provided in Appendix \eqref{apc}.

\section{Conclusion \label{2p8}}

In this chapter the gauge invariant operator  $A_{\min }^{2}$, eq.\eqref{Aminn0}, and corresponding gauge invariant transverse field configuration $A^{ah}_\mu$, eq.\eqref{min0},  have been investigated in a general class of gauge fixings, eq.\eqref{gf1} and eq.\eqref{gf12},  which share similarities with 't Hooft's $R_\zeta$-gauge used in the analysis of YM theory with spontaneous symmetry breaking.  As shown in \cite{Fiorentini:2016rwx}, a local setup can be constructed for both $A_{\min }^{2}$ and $A^{ah}_\mu$, being summarised by the local and BRST invariant action \eqref{act1}. The localization procedure makes use of an auxiliary dimensionless Stueckelberg field $\xi^a$. However, despite the presence of the field $\xi^a$ and unlike the conventional non-renormalizable Stueckelberg mass term, the present construction gives rise to a perfectly well behaved model in the ultraviolet which turns out to be renormalizable to all orders, as discussed in details in sections \eqref{alg1} and \eqref{alg2} as well as in Appendix \eqref{apc}. In particular, the pivotal role of the transversality constraint $\partial_\mu A^{ah}_\mu =0$ has been underlined throughout the paper. It is precisely the direct implementation of this constraint in the local action \eqref{act1} which makes a substantial difference with respect to the conventional Stueckelberg theory. In fact, as pointed out in section \eqref{St}, it removes exactly the component of the Stueckelberg propagator which gives rise to non-renormalizable ultraviolet divergences, see eq.\eqref{stp} versus eqs.\eqref{propqq}. In particular, form eqs.\eqref{propqq}, one sees that, similar to what happens in the case of  
't Hooft's $R_\zeta$-gauge, the use of the general class of gauge fixings \eqref{gf1} and \eqref{gf12} provide a mass $\mu^2$ for the dimensionless Stueckelberg field $\xi^a$. This a welcome feature which can be effectively employed as a fully BRST invariant infrared regularization for  $\xi^a$  in explicit higher loop calculations.

\newpage	

\chapter{Some remarks on the spectral functions of the Abelian Higgs Model \label{VII}}

%%%%%%%%%%%%%%%%%%%%%%%%%%%%%%%%%%%%%%%%%%%%%%%%%%%%%%%%%%%

In section \ref{posval}, we have discussed the two main observations in lattice QCD in recent years: massive behavior of the gluon propagator, and positivity violation of its spectral density function. 
In this context, it is worthwhile to investigate the spectral properties
of massive gauge models to try and shed some light on the infrared behavior of their fundamental fields in an analytical way. The direct comparison between a massive model that violates BRST, such as the massive YM model from section \ref{massym}, and a model that preserves the original nilpotent BRST symmetry, such as the Higgs model, can be particularly enlightening. In any case, the explicit determination of the spectral properties of
Higgs theories and the study of the role played by gauge symmetry there is an interesting pursue on its own.\\
\\
Most articles on massive YM models employ the renormalizable Landau gauge, although it was noticed that this gauge might not be the preferred gauge in non-perturbative calculations \cite{Oehme:1979ai}. For the Higgs model, one can fix the gauge by means of 't Hooft $R_{\xi}$-gauge, see section \ref{rxi}. Understanding the different gauges and their influence on the spectral properties is a delicate subject. This gave us further reason to undertake a systematic study of  the spectral properties of Higgs models. In this chapter, we present the results for the simplest case: that of the $U(1)$ Abelian Higgs model. In fact, it turned out that this model is already very illuminating on aspects like positivity of the spectral function, gauge-parameter independence of physical quantities and unitarity. Of course, these properties are not unknown in the Abelian case. This chapter should therefore not be seen as giving any new information on the physical properties of the Abelian model. Rather, exactly because these properties are so well-known, we are in a better position to understand the problems that we face when calculating  the analytic structure behind some of them within a gauge-fixed setup. This chapter is therefore a first attempt to understand  analytically  the spectral properties of a Higgs-gauge model  in contrast to those of a non-unitary massive model. As such, it is laying the groundwork  for the next chapters.\\
\\
The $U(1)$ Higgs model is known to be unitary \cite{gieres1997symmetries, t1981recent} and renormalizable \cite{Becchi:1974md}. In this work, we consider two propagators: that of the photon, and that of the Higgs scalar field. They are obtained through the calculation of the one-loop corrections to the corresponding $1PI$ two-point functions. After adopting the $R_{\xi}$-gauge, we are left with an exact BRST nilpotent  symmetry.   Of course, the correlation function of BRST invariant quantities should be independent of the gauge parameter. Since the transverse component of the photon propagator is gauge invariant, we should find that the one-loop corrected transverse propagator does not depend on the gauge parameter. As a consequence, the photon pole mass will neither. This property has been proven before by the use of the Nielsen identities, \cite{Haussling:1996rq}, see also \cite{Nielsen:1975fs,Piguet:1984js,gambino2000nielsen}, but never in a direct calculation. The same goes for the Higgs particle propagator: the gauge independence of its pole mass was proven in \cite{Haussling:1996rq}, but never in a direct loop calculation to our knowledge.  We underline here the importance of properly taking into account the tadpole contributions  \cite{Martin:2015lxa,Martin:2015rea}  or, equivalently, the effect on the propagators of quantum corrections of the Higgs vacuum expectation value. Armed with the one-loop results, we are able to calculate the spectral properties of the respective propagators for different values of the gauge parameter. Finally, we compare our results with those of a non-unitary massive Abelian model, to clearly pinpoint at the level of spectral functions the differences (and issues) of both unitary and non-unitary massive vector boson models.\\
\\
This chapter is organized as follows. In section \ref{s2}, we review the $U(1)$ Higgs model and its gauge fixing, as well as the tree-level field propagators and vertices. In section \ref{jaaa}, we calculate the one-loop propagator of both the photon field and the Higgs field, showing the gauge-parameter independence of the transverse photon propagator and of the Higgs pole mass up to one-loop order. In section \ref{s4}, we calculate the spectral function of both propagators. In section \ref{ont} we discuss some subtleties of the Higgs spectral function and in section \ref{s5} we compare our results with those of a non-unitary massive Abelian model. We also  address  the residue computation. Section \ref{s6} collects our conclusions and outlook.

\section{Abelian Higgs model: some essentials \label{s2}}
We start from the Abelian Higgs classical action with a manifest global $U(1)$ symmetry
\beq
S = \int d^4x \left\{\frac{1}{4} F_{\m\n}F_{\m\n} + (D_{\m}\vf)^{\dagger} D_{\m}\vf +\frac{\l}{2}\left(\vf^{\dagger}\vf-\frac{v^2}{2}\right)^2\right\}\label{higgsqed},
\eeq
where
\beq
F_{\m\n}=\pa_{\m}A_{\n}-\pa_{\n}A_{\m},\nonumber\\
D_{\m}\vf = \pa_{\m}\vf+ieA_{\m}\vf
\eeq
and the parameter $v$ gives the minimizing value of the scalar field  to first order in $\hbar$, $\vf_0 =v$.
The spontaneous symmetry breaking is implemented by expressing  the scalar field as an expansion around its minimizing value,  namely
\beq
\vf=\frac{1}{\sqrt{2}}((v+h)+i\r)\label{higgs},
\eeq
where the real part $h$ is identified as  the Higgs field and $\rho$ is the (unphysical) Goldstone boson, with $\langle \rho \rangle=0$.  Here we choose to expand around the classical value of the minimizing value, so that $\langle h \rangle$ is zero at the classical level, but receives loop corrections\footnote{ There is of course an equivalent procedure of fixing $\langle h \rangle$ to zero at all orders, by expanding $\vf$ around the full minimizing value:
	$\vf=\frac{1}{\sqrt{2}}((\langle \vf \rangle+h)+i\r)$. In the Appendix \ref{v} we explicitly show that---as expected---both procedures give the same final results up to a given order.} . The action \eqref{higgsqed} now becomes
\beq
S&=&\int d^4 x \,\left\{\frac{1}{4} F_{\m\n}F_{\m\n}+\ha\pa_{\m}h\pa_{\m}h+\ha\pa_{\m}\r\pa_{\m}\r - e\,\r\,\pa_{\m}h\, A_{\m}+e\,(h+v)A_{\m}\pa_{\m}\r\right. \nonumber\\
&+&\left.\frac{1}{2} e^2 A_{\m}[(h+v)^2 + \r^2]A_{\m}+\frac{1}{8}\l(h^2+2h v +\r^2)^2\right\}\label{fullaction2}
\eeq
and we notice that both the gauge field and the Higgs field have acquired the following masses
\beq
m^2 = e^2 v^2,\,\, m_{h}^2 = \l v^2.
\eeq
With this parametrization, the Higgs coupling $\lambda$ and  the parameter $v$ can be fixed in terms of $m$, $m_h$ and $e$, whose values  will be suitably chosen later on in the text.\\
\\
Even in the broken phase, the action \eqref{fullaction2} is left invariant by the following gauge transformations
\beq
\d A_{\m}&=&-\pa_{\m}\w,\,\,\d \vf = ie\w\vf,\,\,\d\vf^{\dagger}=-ie\w\vf^{\dagger},\nonumber\\
\d h &=&-e\w\r,\,\, \d \r =e\w(v+h).
\label{hh5}
\eeq
where $\omega$ is the gauge parameter.
\subsection{Gauge fixing}
Quantization of the theory \eqref{fullaction2} requires a proper gauge fixing. We shall employ the gauge fixing term
\beq
S_{gf}=\int d^4x \left\{\frac{1}{2\xi}\left(\partial_{\m}A_{\m}+\xi m \rho\right)^2\right\},
\label{34}
\eeq
known as the 't Hooft or $R_{\xi}$-gauge, which has the pleasant property of cancelling the mixed term $\int d^4x (ev\;A_{\m}\pa_{\m}\r)$ in the expression \eqref{fullaction2}. Of course, \eqref{34} breaks the gauge invariance of the action. As is well known, the latter is replaced by the BRST invariance. In fact, introducing the FP ghost fields $\bar{c},c$ as well as the  auxiliary field $b$, for the BRST transformations we have
\beq
sA_{\m}&=&-\pa_{\m}c,\nonumber\\
s c&=& 0,\nonumber\\
s \vf &=& iec\vf ,\nonumber\\
s\vf^{\dagger} &=& -ie c \vf ^{\dagger},\nonumber\\
s h &=& -e c \r,\nonumber\\
s\r &=& e c(v+ h),\nonumber\\
s\bar{c}&=&ib,\nonumber\\
s b&=&0.
\label{brst2}
\eeq
Importantly, the operator $s$ is nilpotent, i.e.~$s^2=0$, allowing to work with the so-called BRST  cohomology, a useful concept to prove unitarity and renormalizability of the Abelian Higgs model \cite{Becchi:1974md,Becchi:1974xu,kugo1979local}.\\
\\
We can now introduce the gauge fixing in a BRST invariant way via
\beq
\mathcal{S}_{gf}&=&s\int d^d x \left\{-i\frac{\e}{2}\bar{c}b+\bar{c}(\pa_{\m}A_{\m} +\e m \r)\right\},\\
&=&\int d^d x \left\{\frac{\xi}{2}b^2+ib\pa_{\mu}A_{\m}+i b\xi m \r+\bar{c}\pa^2c-\xi m^2 \bar{c}c- \xi me\bar{c}hc\right\}.
\eeq
Notice that the ghosts $(\bar c,c)$ get a gauge parameter dependent mass, while interacting directly  with the Higgs field.\\
\\
The total gauge fixed BRST invariant action then becomes
\beq
S&=&\int d^4 x \,\Bigg\{\frac{1}{4} F_{\m\n}F_{\m\n} +\ha\pa_{\m}h\pa_{\m}h+\ha\pa_{\m}\r\pa_{\m}\r - e\,\r\,\pa_{\m}h\, A_{\m}+ e\,h A_{\m}\pa_{\m}\r+ \frac{1}{2} m^2 A_{\mu}A_{\mu} \nonumber\\
&+& \frac{1}{2} e^2 A_{\m}[h^2 +2vh+ \r^2]A_{\m}+\frac{1}{8}\l(h^2 +\r^2)(h^2 +\r^2+4h v)+\ha m_h^2 h^2+m A_{\mu}\pa_{\mu}\rho+\frac{\xi}{2}b^2+ib \pa_{\m}A_{\m} \nonumber\\
&+&ib \xi m \rho +\bar{c}(\pa^2)c - m^2 \e c\bar{c}-m\e e \bar{c}c h\Bigg\}, \label{fullaction}
\eeq
with
\beq
s S = 0\;. \label{brstinvact}
\eeq
In Appendix \ref{FR} we collect the propagators and vertices corresponding to the action \eqref{fullaction} of the Abelian Higgs model in the $R_{\xi}$ gauge.

\section{Photon and Higgs propagators at one-loop \label{jaaa}}
In this section we obtain the one-loop corrections to the photon propagator, as well as to the propagator of the Higgs boson.  This requires the calculation\footnote{We have used the techniques of modifying integrals into ``master integrals'' with momentum-independent numerators from \cite{passarino1979one}.}, in section \ref{AA} and \ref{hh8}, of the Feynman diagrams as shown in Figure \ref{les8} and Figure \ref{more8}. Notice that the last four diagrams in Figure \ref{les8} and Figure \ref{more8} vanish  for $\langle h \rangle=0$. Since we have chosen to expand the $\vf$ field around its classical minimizing value $v$ (cf. \eqref{higgs}), $\langle h \rangle$ has loop contributions which are nonzero and the resulting tadpole diagrams have to be included in the quantum corrections for the propagators\footnote{The diagrams with tadpole balloons are not part of the standard definition of one-particle irreducible diagrams that contribute to the self-energies. However, since the momentum flowing in the vertical $h$-field (dashed) line is zero, they can be effectively included as a momentum-independent term in the self-energies.}. Of course the final result for the propagators would be the same had we chosen to expand the $\vf$ field around its full minimizing value and required $\langle h\rangle=0$.
In fact, including the tadpole diagrams in our formulation has the same effect as shifting the masses of the fields to include the one-loop corrections to the Higgs minimizing value $\langle \vf \rangle$, calculated by imposing $\langle h \rangle=0$ (see Appendix \ref{v} for the technical details). These diagrams can actually be seen as a correction to the tree-level mass term: in the spontaneously broken phase the gauge boson mass is given by $m=e\langle \varphi \rangle$, depending thus on $\langle \varphi \rangle$ that receives quantum corrections order by order. Therefore, the full inverse photon propagator can be written as
\beq
G_{AA}^{-1}(p^2)&=& p^2 +e^2v^2 + \left(\textrm{1PI diagrams}\right) + \left(\textrm{diagrams with tadpoles}\right)\nonumber\\
&=& p^2 + e^2\langle \varphi \rangle^2+ \left(\textrm{1PI diagrams}\right)
\,,
\eeq
where the equalities are to be understood up to a given order in perturbation theory and a similar reasoning can be drawn for the Higgs propagator. In what follows, we shall proceed with the expansion adopted in eq. \eqref{higgs} and include the tadpole diagrams explicitly in our self-energy results. The calculations are done for arbitrary dimension $d$. In section \ref{d} we will analyze the results for $d=4-\epsilon$, making use of the techniques of dimensional regularization in  the $\overline{\text{MS}}$ scheme.
\subsection{Corrections to the photon self-energy\label{AA}}
\begin{figure}[t]
	\includegraphics[width=\textwidth]{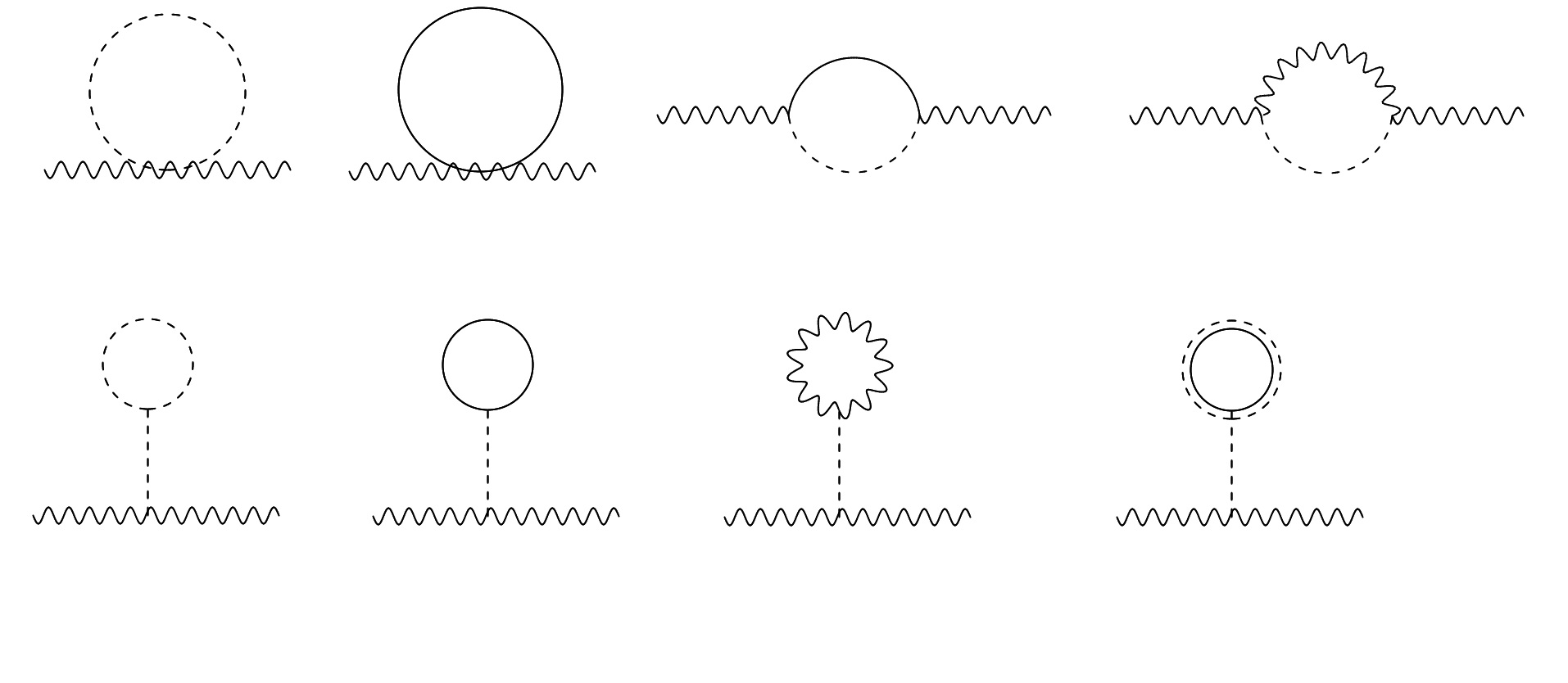}
	\caption{Contributions to one-loop photon self-energy  in the Abelian Higgs Model, including tadpole contributions in the second line. Wavy lines represent the photon field, dashed lines the Higgs field, solid lines the Goldstone boson and double lines the ghost field.}
	\label{les8}
\end{figure}
The first diagram contributing to the photon self-energy is the Higgs boson snail (first diagram in the first line of Figure \ref{les8}) and gives a contribution
\beq
\Gamma_{A_{\m}A_{\n},1}(p^2)&=&\frac{-4e^2}{(4\pi)^{d/2}}\frac{\Gamma(2-d/2)}{2-d}\frac{m_{h}^{d-2}}{2}\delta_{\m\n}.
\label{besu4}
\eeq
The second diagram is the Goldstone boson snail (second diagram in the first line of Figure \ref{les8})
\beq
\Gamma_{A_{\m}A_{\n},2}(p^2)	&=&\frac{-4e^2}{(4\pi)^{d/2}}\frac{\Gamma(2-d/2)}{2-d}\frac{(\e m^2)^{d/2-1}}{2}\delta_{\m\n}.
\eeq
Being momentum-independent, the only effect of these first two diagrams is to renormalize the mass parameters $(m^2_h,m^2)$.
\\
The third term contributing to the photon propagator is the Higgs-Goldstone sunset (third diagram in first line of Figure \ref{les8})
\beq
\Gamma_{A_{\m}A_{\n},3}(p^2) &=&\frac{4 e^2}{(4\pi)^{d/2}} \frac{\Gamma(2-d/2)}{2-d}\int_0^1 dx \Bigg[ K_{d/2-1}(m_h^2,\xi m^2){\mc P}_{\m\n}+\Bigg(K_{d/2 -1}(m_h^2,\xi m^2)\nonumber\\
&+&\frac{(2-d)}{4}(1-4x(1-x))p^2 K_{d/2 -2}(m_h^2,\xi m^2)\Bigg)\mathcal{L}_{\m\n} \Bigg],
\eeq
where we used the definitions
\beq
K_{\a}(m_1^2,m_2^2)\equiv \Big(p^2x(1-x)+xm_1^2+(1-x)m_2^2\Big)^{\a},
\eeq
and 
\beq
\mathcal{P}_{\m\n}&=&\delta_{\m\n}-\frac{p_{\m}p_{\n}}{p^2},\\
\mathcal{L}_{\m\n}&=&\frac{p_{\m}p_{\n}}{p^2},
\eeq
which are the transversal and longitudinal projectors, respectively.
The fourth term contributing to the photon propagator is the Higgs-photon sunset (fourth diagram in first line of Figure \ref{les8})
\beq
\Gamma_{A_{\m}A_{\n},4}(p^2)
&=&\frac{4 e^2}{(4\pi)^{d/2}}\frac{\Gamma(2-d/2)}{2-d}\int_{0}^{1} dx \Bigg[\Big((2-d)m^2 K_{d/2 -2}(m_h^2,m^2)+K_{d/2-1}(m_h^2,m^2)\nonumber\\
&-&K_{d/2 -1}(m_h^2,\xi m^2)\Big){\cal P}_{\m\n}+\Big((2-d)m^2 K_{d/2 -2}(m_h^2,m^2)\nonumber\\
&+&K_{d/2 -1}(m_h^2,m^2)-K_{d/2 -1}(m_h^2,\xi m^2)\nonumber\\
&&+(2-d)p^2 x^2 (K_{d/2-2}(m_h^2,m^2)-K_{d/2-1}(m_h^2,\xi m^2))\Big)\mathcal{L}_{\m\n}\Bigg].
\eeq
Finally, we have four tadpole (balloon) diagrams. The Higgs boson balloon (first diagram of the last line in Figure \ref{les8})
\beq
\Gamma_{A_{\m}A_{\n},5}(p^2)
&=& \frac{4e^2}{(4\pi)^{d/2}}\frac{\Gamma(2-d/2)}{(2-d)}\frac{3}{2}m_h^{d/2-1}\delta_{\m\n},
\eeq
the Goldstone boson balloon (second diagram of the last line in Figure \ref{les8})
\beq
\Gamma_{A_{\m}A_{\n},6}(p^2)
&=& \frac{4e^2}{(4\pi)^{d/2}}\frac{\Gamma(2-d/2)}{(2-d)}\frac{1}{2}(\xi m)^{d/2-1}\delta_{\m\n},
\eeq
the photon balloon (third diagram of the last line in Figure \ref{les8})
\beq
\Gamma_{A_{\m}A_{\n},7}(p^2)&=&2e^2 \frac{m^2}{m_h^2}\int \frac{d^dk}{(2\pi)^d}\left( \frac{1}{k^2+m^2}(d-1)+\frac{\xi}{k^2+\xi m^2}\right)\delta_{\m\n},
\eeq
and finally, the ghost balloon (fourth diagram of the last line in Figure \ref{les8})
\beq
\Gamma_{A_{\m}A_{\n},8}(p^2)&=&-2e^2 \frac{m^2}{m_h^2}\int \frac{d^dk}{(2\pi)^d}\frac{\xi}{k^2+\xi m^2}\delta_{\m\n}.
\label{absu4}
\eeq
Combining all these contributions \eqref{besu4}-\eqref{absu4}, we find
\beq \label{fotonzelf}
\Gamma_{A_{\m}A{\n}}(p^2)&=&\frac{4 e^2}{(4\pi)^{d/2}}\frac{\Gamma(2-d/2)}{2-d}\int_{0}^{1} dx\Big((2-d)m^2 K_{d/2-2}(m^2,m_h^2)+K_{d/2-1}(m^2,m_h^2)\nonumber\\
&+&m_{h}^{d-2}+\frac{m^d}{m_h^2}(d-1)\Big)\mathcal{P}_{\m\n}\nonumber\\
&+&\frac{4 e^2}{(4\pi)^{d/2}}\frac{\Gamma(2-d/2)}{2-d}\int_{0}^{1} dx\Bigg(\frac{2-d}{4}(1-4x)p^2 K_{d/2 -2}(m_h^2,\xi m^2)\nonumber\\
&+&(2-d)(m^2+p^2 x^2)K_{d/2 -2}(m_h^2,m^2)\nonumber\\
&+&K_{d/2 -1}(m_h^2, m^2)+m_{h}^{d-2}+\frac{m^d}{m_h^2}(d-1)\Bigg)\mathcal{L}_{\m\n}.
\eeq
Defining
\beq
\Gamma_{A_\m A_\n}=\Pi^{\perp}_{AA}(p^2)\mathcal{P}_{\m\n}+\Pi^{\parallel}_{AA}(p^2) \mathcal{L}_{\m\n},
\eeq
it follows that
\beq
\partial_{\xi}\Pi^{\perp}_{AA}=0. \label{xiperpind}
\eeq
As expected, eq.\eqref{xiperpind}  expresses the gauge parameter independence of the gauge invariant transverse component of the photon propagator \cite{Haussling:1996rq}.

\subsection{Corrections to the  Higgs self-energy  \label{hh8}}

\begin{figure}[t]
	\includegraphics[width=\textwidth]{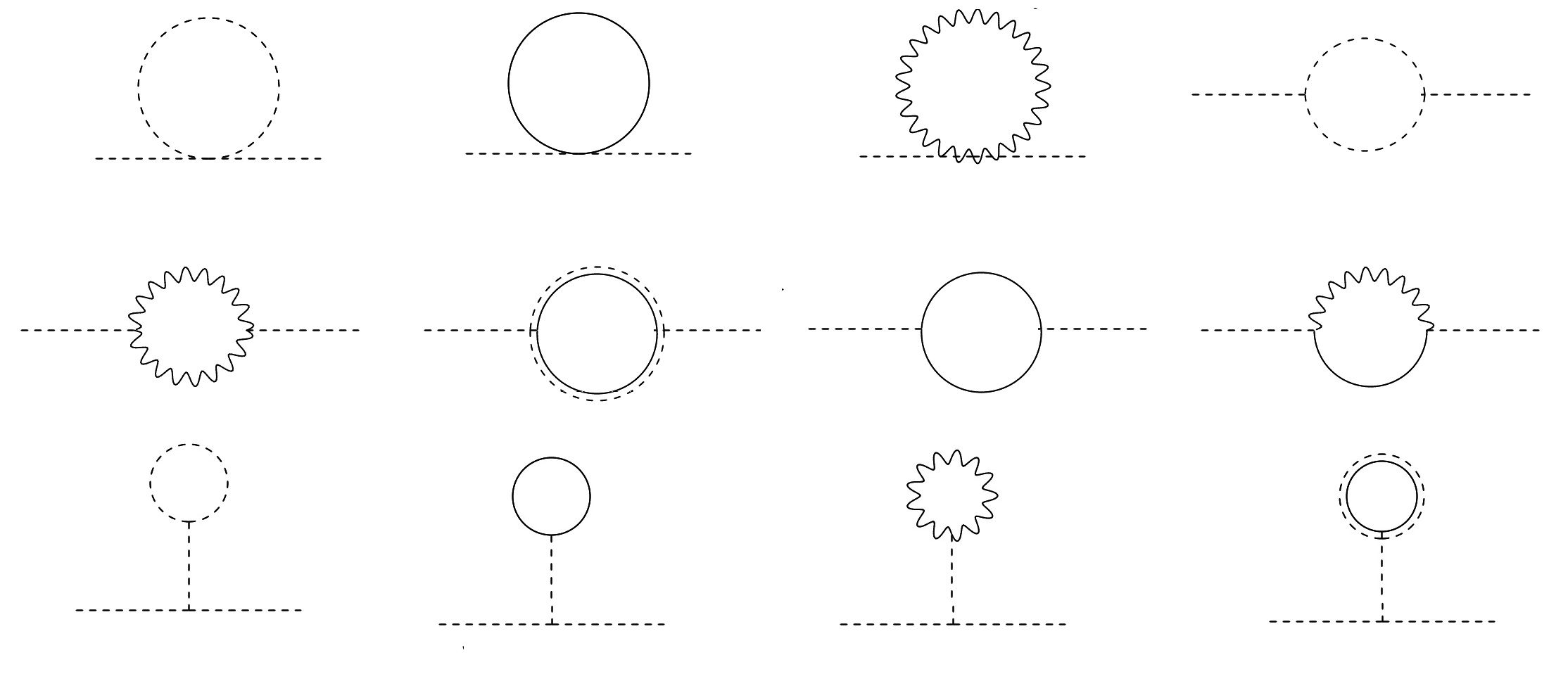}
	\caption{Contributions to the one-loop Higgs self-energy. Line representations are as in Figure 1. }
	\label{more8}
\end{figure}
The first diagrams contributing to the Higgs self-energy are of the snail type, renormalizing the masses of the internal fields.
\\
\\
The Higgs boson snail (first diagram in the first line of Figure \ref{more8})
\beq
\Gamma_{hh,1}(p^2)&=&-3\frac{\lambda}{(4\pi)^{d/2}} \frac{\Gamma(2-d/2)}{(2-d)}m_h^{d-2},
\label{wat8}
\eeq
the Goldstone boson snail  (second diagram in the first line of Figure \ref{more8})
\beq
\Gamma_{hh,2}(p^2) &=& -\frac{\lambda}{(4\pi)^{d/2}} \frac{\Gamma(2-d/2)}{(2-d)}(\e m^2)^{d/2-1}
\label{ii}
\eeq
and the photon snail  (third diagram in the first line of Figure \ref{more8})
\beq
\Gamma_{hh,3}(p^2) &=& -2\frac{e^2}{(4\pi)^{d/2}} \frac{\Gamma(2-d/2)}{(2-d)}\Big( (d-1)m^{d-2}+\e(\e m^2)^{d/2-1}\Big).
\eeq
Next, we meet a couple of sunset diagrams. The Higgs boson sunset (fourth diagram in the first line of Figure \ref{more8}):
\beq
\Gamma_{hh,4}(p^2)
&=& \frac{9}{2}\frac{\lambda}{(4\pi)^{d/2}} \frac{\Gamma(2-d/2)}{(2-d)}(2-d)m_h^2 \int_0^1 dx  K_{d/2-2}(m_h^2,m_h^2),
\eeq
the photon sunset (first diagram in the second line of Figure \ref{more8}):
\beq
\Gamma_{hh,5}(p^2)	&=&  e^2 \frac{\Gamma(2-d/2)}{2-d}\frac{1}{(4\pi)^{d/2}}\int_0^1 dx\Bigg[(2-d)\Big(2m^2(d-1)+2p^2+\frac{p^4}{2m^2}\Big)K_{d/2-2}(m^2,m^2)\nonumber\\
&-&(2-d)\Big(2p^2+\frac{p^4}{m^2}+\xi^2 m^2+2p^2\xi-2\xi m^2+m^2\Big) K_{d/2-2}(m^2,\xi m^2)\nonumber\\
&+&(2-d)\Big( 2 \xi p^2+ 2 \xi^2 m^2+ \frac{p^4}{2 m^2}\Big) K_{d/2-2}(\xi m^2, \xi m^2)\nonumber\\
&+&2(\xi -1) (m^2)^{d/2-1}+2(1-\xi)(\xi m^2)^{d/2-1}\Big],
\eeq
the ghost sunset (second diagram in the second line of Figure \ref{more8}):
\beq
\Gamma_{hh,6}(p^2)
&=&-\frac{e^2}{(4\pi)^{d/2}} \frac{\Gamma(2-d/2)}{(2-d)} (2-d) m^2 \e^2 \int_0^1 dxK_{d/2-2}(\e m^2, \e m^2),
\eeq
the Goldstone boson sunset (third diagram in the second line of Figure \ref{more8}):
\beq
\Gamma_{hh,7}(p^2)&=&\frac{1}{2}\frac{\lambda}{(4\pi)^{d/2}} \frac{\Gamma(2-d/2)}{(2-d)}(2-d) m_h^2 \int_0^1 dx K_{d/2-2}(\e m^2, \e m^2)^{d/2-2}
\eeq
and a mixed Goldstone-photon sunset (fourth diagram in the second line of Figure \ref{more8}):
\beq
\Gamma_{hh,8}(p^2)
&=& e^2 \frac{\Gamma(2-d/2)}{2-d}\frac{1}{(4\pi)^{d/2}}\int_0^1 dx\Bigg[(2-d)\Big(2p^2+\frac{p^4}{m^2}+\xi^2 m^2\\
&+&2p^2\xi-2\xi m^2+m^2\Big) K_{d/2-2}(m^2,\xi m^2)\nonumber\\
&-&(2-d)\Big(\xi^2 m^2+\frac{p^4}{m^2}+2 p^2 \xi \Big)  K_{d/2-2}(\xi m^2,\xi m^2)\nonumber\\
&+& 2 \Big(1-\xi -\frac{p^2}{m^2}\Big)  (m^2)^{d/2-1}\nonumber\\
&+& 2\Big( 2\xi - 1+\frac{p^2}{m^2}\Big)(\xi m^2)^{d/2-1}\Bigg].
\eeq
Finally, we have the tadpole diagrams. The Higgs balloon (first diagram on the third line of Figure \ref{more8}):
\beq
\Gamma_{hh,9}(p^2) &=&9 \frac{\lambda}{(4\pi)^{d/2}} \frac{\Gamma(2-d/2)}{(2-d)}m_h^{d-2},
\eeq
the photon balloon (second diagram on the third line of Figure \ref{more8}):
\beq
\Gamma_{hh,10}(p^2) &=& 6\frac{e^2}{(4\pi)^{d/2}} \frac{\Gamma(2-d/2)}{(2-d)}\Big( (d-1)m^{d-2}+\e(\e m^2)^{d/2-1}\Big),
\eeq
the Goldstone boson balloon (third diagram on the third line of Figure \ref{more8}):
\beq
\Gamma_{hh,11}(p^2) &=& 3 \frac{\lambda}{(4\pi)^{d/2}} \frac{\Gamma(2-d/2)}{(2-d)}(\e m^2)^{d/2-1}
\label{uu}
\eeq
the ghost balloon (fourth diagram on the third line of Figure \ref{more8}):	
\beq
\Gamma_{hh,12}(p^2)&=&-6 \frac{e^2 \e}{(4\pi)^{d/2}} \frac{\Gamma(2-d/2)}{(2-d)} (\e m^2)^{d/2-1}.
\label{poi8}
\eeq
Putting together eqs.~\eqref{wat8} to \eqref{poi8} we find the total one-loop correction to the Higgs boson self-energy,
\beq
\Pi_{hh}(p^2)\equiv\Gamma_{hh}(p^2)&=&\frac{\Gamma(2-d/2)}{2-d}\frac{1}{(4\pi)^{d/2}}\int_0^1 dx\Bigg[
(2-d)e^2\Big(2m^2(d-1)+2p^2+\frac{p^4}{2m^2}\Big)K_{d/2-2}(m^2,m^2)\nonumber\\
&+& \frac{9}{2}\lambda(2-d)m_h^2 K_{d/2-2}(m_h^2,m_h^2)\nonumber\\
&+&e^2\left(-2 \frac{p^2}{m^2}+4(d-1)\right) (m^2)^{d/2-1}\nonumber\\
&+& 6 \lambda (m_h^2)^{d/2-1}\nonumber\\
&+&(2-d)\Big(-\frac{p^4}{2m^2}e^2+\frac{\lambda}{2}m_h^2\Big) K_{d/2-2}(\xi m^2, \xi m^2)\nonumber\\
&+& 2(\frac{p^2}{m^2}e^2+\lambda) (\xi m^2)^{d/2-1}\Big].
\label{pop}
\eeq

\subsection{Results for $d=4-\epsilon$ \label{d}}
For $d=4$, the 2-point functions  are divergent. We therefore follow the standard procedure of dimensional regularization, as we have no chiral fermions present. Thus, we choose $d=4-\epsilon$ with $\epsilon$ an infinitesimal parameter, and analyze the solution in the limit $\epsilon \rightarrow 0$.
\\
\\
Let us start with the photon 2-point function, given for arbitrary dimension $d$ by \eqref{fotonzelf}. The mass dimension of the coupling constant $e$ is $[e]=2-d/2=\epsilon/2$, and redefining $e\rightarrow e\tilde{\mu}^{\epsilon/2}=e\tilde{\mu}^{2-d/2}$ we put the dimension on $\tilde{\mu}$, while $e$ is dimensionless. Using
\beq
\frac{4 e^2}{(4\pi)^{d/2}}\frac{\Gamma(2-d/2)}{2-d}&\overset{d\rightarrow 4-\epsilon}{=}&
-2\frac{e^2}{(4\pi)^2} \left(\frac{2}{\epsilon}+1+\text{ln}(\mu^2)\right),
\eeq
where we defined
\beq
\mu^2=\frac{4\pi \tilde{\mu}^2}{e^{\gamma_E}},
\eeq we find for the divergent part of the transverse photon 2-point function:
\beq
\Pi^\perp_{AA,div}(p^2)
&=&\frac{2}{\epsilon}\frac{e^2}{(4\pi)^2}\left(\frac{p^2}{3}+6( \frac{g^2}{\lambda}-\frac{1}{2})m^2+3m_h^2\right)
\eeq
and these infinities are, following the $\overline{\text{MS}}$-scheme, cancelled by the corresponding counterterms. The renormalized correlation function is then finite in the limit $d \rightarrow 4$ and we find the one-loop correction
\beq
\Pi_{AA}(p^2)&=& 2\frac{e^2}{(4\pi)^2}\int_{0}^{1} dx \,\,\Bigg\{p^2 x(1-x)+m^2x \nonumber\\
&+&m_h^2(1-x)(1-\ln\frac{p^2 x(1-x)+m^2x+m_h^2(1-x)}{\m^2})+m_h^2(1-\ln\frac{m_h^2}{\m^2})\nonumber\\
&+&\frac{m^4}{m_h^2}(1-3 \ln \frac{m^2}{\m^2})+2m^2 \ln \frac{p^2 x(1-x)+m^2x+m_h^2(1-x)}{\m^2}\Bigg\}.
\label{AAF}
\eeq
In the same way, we find the divergent part of the Higgs boson 2-point function:
\beq
\Pi_{hh,div}(p^2)&=& -\frac{1}{2\epsilon}\frac{1}{(4\pi)^2}\Big(e^2(12p^2-4\xi p^2)+\lambda(8m_h^2-4\xi m^2)\Big),
\eeq
which is canceled by the  corresponding counterterm.  Therefore,  the one-loop correction to the Higgs boson propagator reads
\beq
\Pi_{hh}(p^2)&=& \frac{1}{(4\pi)^2}\int_{0}^{1} dx
\Bigg\{
e^2\Bigg[p^2(1-\ln\frac{m^2}{\m^2}-2\ln \frac{p^2 x(1-x)+m^2}{\m^2}) \nonumber\\
&-&\frac{p^4}{2m^2} \ln \frac{p^2 x(1-x)+m^2}{\m^2}-6m^2(1-\ln \frac{m^2}{\m^2}+\ln \frac{p^2 x(1-x)+m^2}{\m^2})\Bigg]\nonumber\\
&+&\lambda \Big[\frac{1}{2}m_h^2(-6+6\ln \frac{m_h^2}{\m^2}-9\ln \frac{p^2 x(1-x)+m_h^2}{\m^2})\Big]\nonumber\\
&-&\Bigg[\xi (e^2p^2+\l m^2)(1- \ln \frac{\xi m^2}{\m^2})-(e^ 2\frac{p^4}{2m^2}-\lambda\frac{m_h^2}{2})\ln \frac{p^2 x(1-x)+\xi m^2}{\m^2}\Bigg]
\Bigg\}.
\label{dk2}
\eeq

\subsection{One-loop propagators for the  elementary fields \label{elm}}
\begin{figure}[H]
	\center
	\includegraphics[width=12cm]{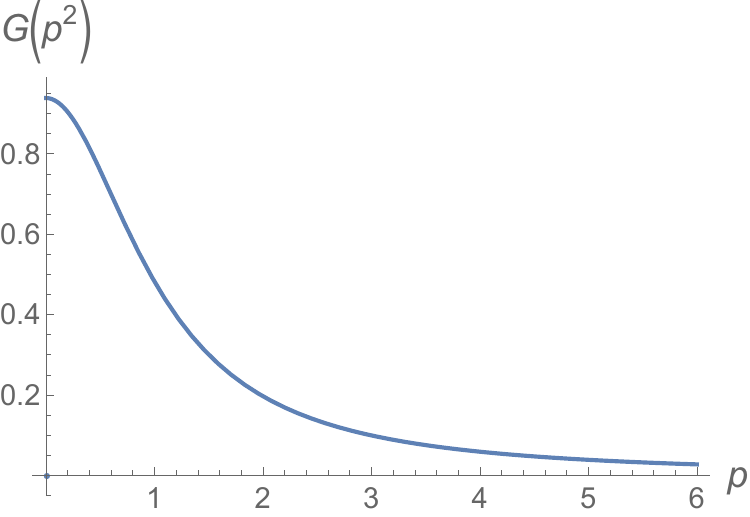}
	\caption{Resummed form factor for the photon operator, with $p$ given in units of the energy scale $\mu$, for the parameter values $e=1$, $v=1 \, \mu$, $\lambda=\frac{1}{5}$.}
	\label{propAAA}
\end{figure}

For the photon field, the transverse part of the propagator $G_{\m\n}^{AA}(p^2)$ up to order $\hbar$ is given in  momentum space by

\beq
\braket{A_{\mu}(p) A_{\n}(-p)}&=&\frac{1}{p^2+m^2}+\frac{1}{(p^2+m^2)^2}\Pi_{AA}(p^2)+\mathcal{O}(\hbar^2)
\eeq
and we can approach the all-order form factor with the resummed approximation

\beq
G^{T}_{AA} (p^2) &=& \frac{1}{ p^2+m^2-\Pi(p^2)},
\label{dk}
\eeq
shown in Figure \ref{propAAA}.

\begin{figure}[H]
	\center
	\includegraphics[width=12cm]{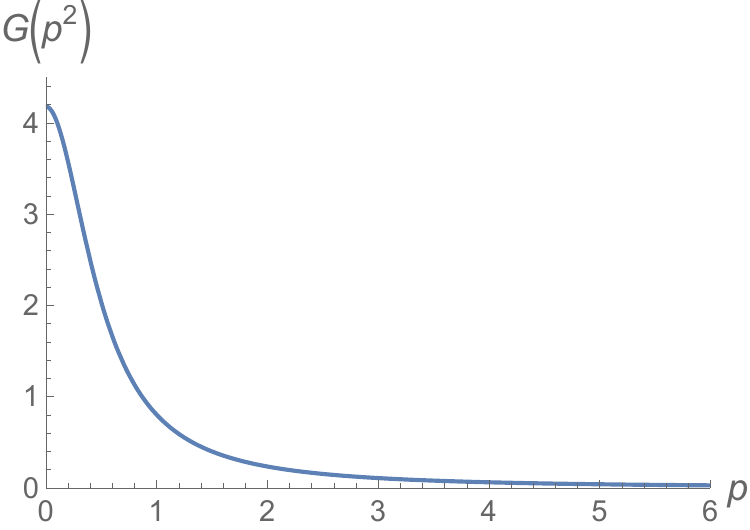}
	\caption{Resummed form factor for the Higgs operator, with $p$ given in units of the energy scale $\mu$, for the parameter values $e=1$, $v=1 \, \mu$, $\lambda=\frac{1}{5}$.}
	\label{prophhh}
\end{figure}

For the Higgs field, we find
\beq
\braket{h(p) h(-p)}&=&\frac{1}{p^2+m_h^2}+ \frac{1}{(p^2+m_h^2)^2} \Pi_{hh}(p^2)+\mathcal{O}(\hbar^2).
\label{kik}
\eeq
Before making the resummed approximation, we notice that the resummation is based on the assumption that the second term in \eqref{kik} is much smaller than the first term. Then, we see that \eqref{kik} contains terms of the order of $ \frac{p^4}{(p^2+m_h^2)^2} \ln \frac{p^2x(1-x)+m_h^2}{\mu^2}$, which cannot be resummed for big values of $p$. We therefore use the identity

\beq
p^4 &=& (p^2+m_h^2)^2-m_h^4-2p^2m_h^2,
\label{u1}
\eeq
to rewrite 

\beq
\frac{p^4}{(p^2+m_h^2)^2} \ln \frac{p^2x(1-x)+m_h^2}{\mu^2} &=& \ln \frac{p^2x(1-x)+m_h^2}{\mu^2} -\frac{(m^4+2p^2m^2)}{(p^2+m^2)^2} \ln \frac{p^2x(1-x)+m_h^2}{\mu^2}.
\label{jju5}
\eeq
The last two terms in \eqref{jju5} can be safely resummed. We rewrite

\beq
\frac{{\Pi}_{hh}(p^2)}{(p^2+m_h^2)^2}&=& \frac{\hat{\Pi}_{hh}(p^2)}{(p^2+m_h^2)^2}+C_{hh}(p^2),
\label{jju2}
\eeq
with
\beq
\hat{\Pi}_{hh}(p^2)&=& \frac{1}{(4\pi)^2}\int_{0}^{1} dx
\Bigg\{
e^2\Bigg[p^2(1-\ln\frac{m^2}{\m^2}-2\ln \frac{p^2 x(1-x)+m^2}{\m^2}) \nonumber\\
&+&\frac{(m_h^4+2p^2m_h^2)}{2m^2} \ln \frac{p^2 x(1-x)+m_h^2}{\m^2}-6m^2(1-\ln \frac{m^2}{\m^2}+\ln \frac{p^2 x(1-x)+m^2}{\m^2})\Bigg]\nonumber\\
&+&\lambda \Big[\frac{1}{2}m_h^2(-6+6\ln \frac{m_h^2}{\m^2}-9\ln \frac{p^2 x(1-x)+m_h^2}{\m^2})\Big]-\Bigg[\xi (e^2p^2+\l m^2)(1- \ln \frac{\xi m^2}{\m^2})\nonumber\\
&+&(e^ 2\frac{(m_h^4+2p^2m_h^2)}{2m^2}+\lambda\frac{m_h^2}{2})\ln \frac{p^2 x(1-x)+\xi m^2}{\m^2}\Bigg]
\Bigg\}.
\label{dk6}
\eeq
and
\beq
C_{hh}(p^2)&=& - \frac{e^2}{2m^2(4\pi)^2}\int_{0}^{1} dx \Big\{ \ln\left( \frac{p^2 x(1-x)+m_h^2}{\m^2}\right)- \ln\left( \frac{p^2 x(1-x)+\xi m^2}{\m^2}\right)\Big\}
\eeq
and the resummed approximation becomes

\beq
G_{hh}(p^2)&=& \frac{1}{p^2+m^2-\hat{\Pi}(p^2)}+C_{hh}(p^2),
\label{ghghgh}
\eeq
which is shown in Figure \ref{prophhh}.
Notice that the dependence on the Feynman parameter $x$ for all integrals is restricted to functions of the type $\int_0^1 dx \ln \frac{p^2 x(1-x)+x m_1^2+(1-x)m_2^2}{\mu^2}$. These functions have an analytical solution, depicted in Appendix \ref{apfeyn}. Since the transverse component $A^T_\mu$ of the Abelian gauge field is gauge invariant, it turns out that the transverse  photon propagator is independent from the gauge parameter $\xi$,  while the Higgs propagator does depend on $\xi$, in agreement with the Nielsen identities  analyzed in \cite{Haussling:1996rq}.

\section{Spectral properties of the propagators \label{s4}}
In this section we will  investigate  the spectral properties corresponding to the connected propagators of the last section. Strictly speaking, the calculation of the spectral properties should only be done to  first order\footnote{This would correspond to first order in the gauge coupling $e^2$ and in the Higgs coupling $\lambda$ neglecting the implicit coupling dependence in the masses.} in
$\hbar$,  since the one-loop corrections to the propagators have been evaluated up to this order.  In practice, however, for small values of the coupling   constants  the higher-order contributions become negligible, and one could treat the one-loop solution as the all-order solution without a significant numerical difference. Even so, when looking for analytical rather than numerical results---for example a gauge parameter dependence---we should restrict ourselves to the first-order results. We shall see the crucial difference between both approaches.
\\
\\
To plot the spectral properties of our model we choose some specific values of the parameters $\{m,m_h,\mu,e\}$. We want to restrict ourselves to the case where the Higgs particle is a stable particle, so we need $m_h^2 < 4m^2$. Furthermore, given the Abelian nature of the model, and thus a weak coupling regime in the infrared, we can choose an energy scale $\mu$ that is sufficiently small w.r.t.~the elusive Landau pole (that is exponentially large) and a corresponding small value for the coupling constant $e$. The particular values chosen per graph are denoted in the figure captions. Notice that by choosing $\mu$ and $e$, we are implicitly fixing the Landau pole $\Lambda$, with $\mu\ll \Lambda$, see \cite{irges2017renormalization} for more details.  We have checked that results are as good as independent from the choice of $\mu$ over a very wide range of $\mu$-values.
\\
\\
We start by calculating the pole mass in section \ref{31}. The pole mass is the actual physical mass of a particle that enters the energy-momentum dispersion relation. It is an observable for both the photon and the Higgs boson and should therefore not depend on the gauge parameter $\xi$. We will also discuss the residue to first order and compare these with the output from the Nielsen identities \cite{Haussling:1996rq}. In section \ref{32} we show how to obtain the spectral function to first order from the propagator. In section \ref{ont} we will discuss some more details about the Higgs spectral function.

\subsection{Pole mass, residue and Nielsen identities \label{31}}
The pole mass for any massless or massive field excitation is obtained by calculating the pole of the resummed connected propagator
\beq
G(p^2)= \frac{1}{p^2+m^2-\Pi(p^2)},
\label{245}
\eeq
where $\Pi(p^2)$ is the self-energy correction. The pole of the propagator is thus equivalently defined by the equation
\beq
p^2+m^2-\Pi(p^2)=0 \, \label{ppp}
\eeq
and its solution defines the pole mass $p^2=-m_{pole}^2$. As consistency requires us to work up to a fixed order in perturbation theory, we should solve eq.\eqref{ppp} for the pole mass in an iterative fashion. To first order in  $\hbar$ , we find
\beq
m_{pole}^2=m^2-\Pi^{1-loop}(-m^2)+\mathcal{O}(\hbar^2),
\label{pol}
\eeq
where $\Pi^{1-loop}$ is the first order, or one-loop, correction to the propagator.

Next, we also want to compute the residue $Z$, again up to order   $\hbar$. In principle, the residue is given by
\beq
Z= \lim_{p^2 \rightarrow -m_{pole}^2} (p^2+m_{pole}^2)G(p^2).
\eeq
We write \eqref{245} in a slightly different way
\beq
G(p^2)&=& \frac{1}{p^2+m^2-\Pi(p^2)}\nonumber\\
&=&\frac{1}{p^2+m^2-\Pi^{1-loop}(-m^2)-(\Pi(p^2)-\Pi^{1-loop}(-m^2))}\nonumber\\
&=&\frac{1}{p^2+m_{pole}^2-\widetilde{\Pi}(p^2)},
\label{op}
\eeq
where we defined $\widetilde{\Pi}(p^2)=\Pi(p^2)-\Pi^{1-loop}(-m^2)$. At one-loop, expanding  $\widetilde{\Pi}(p^2)$ around $p^2=-m^2_{pole}=-m^2+\mathcal{O}(\hbar)$  gives the residue
\beq
Z&=&\frac{1}{1-\partial_{p^2} \Pi(p^2)\vert_{p^2=-m^2}}=1+\partial_{p^2} \Pi(p^2)\vert_{p^2=-m^2}+\mathcal{O}(\hbar^2).
\label{kak}
\eeq
In \cite{Haussling:1996rq}, for the Abelian Higgs model, the Nielsen identities were obtained for both the photon and the Higgs boson. It was found that for the photon propagator, the transverse part is explicitly independent of $\xi$ to all orders of perturbation theory, giving the Nielsen identity:
\beq
\partial_{\xi}(G^{\perp}_{AA})^{-1}(p^2)=0
\label{ppo}
\eeq
and consequently
\beq
\partial_{\xi}\partial_{p^2}(G^{\perp}_{AA})^{-1}(p^2) \vert_{p^2=-m_{pole}^2}&=&0,\nonumber\\
\partial_{\xi}(G^{\perp}_{AA})^{-1}(-m_{pole}^2)&=&0,
\eeq
confirming the gauge independence of the residue and the pole mass. Of course, this is not unexpected since the transverse part of an Abelian gauge field propagator can be written as
\begin{equation}
{\cal P}_{\mu\nu} \braket{A_\mu A_\nu}_{conn} \propto \braket{ A_\mu^T A_\mu^T}\,,\qquad A_\mu^T= {\cal P}_{\mu\nu} A_\nu
\end{equation}
and the transverse component $A_\mu^T$  is gauge invariant under Abelian gauge transformations.\\
\\
We can now compare the outcome of the Nielsen identities with  our one-loop calculation \eqref{dk}. Indeed, to the first order,
eq.\eqref{dk}  is an explicit demonstration of the identity \eqref{ppo}. For the Higgs boson, the Nielsen identity is a bit more complicated, and is given by
\beq
\partial_{\xi}G^{-1}_{hh}(p^2)=-\partial_{\chi}G^{-1}_{Y_1h}(p^2)G_{hh}^{-1}(p^2),
\label{sk}
\eeq
where  $G^{-1}_{Y_1h}(p^2)$ stands for a non-vanishing $1PI$ Green function which can be obtained from the extended  BRST symmetry which also acts on the gauge parameter \cite{Piguet:1984js}. To be more precise, $Y_1$ is a local source coupled to the BRST variation of the Higgs field (see \eqref{brst2}), while $\chi$ is coupled to the integrated  composite operator $\int d^4x\left(-\frac{i}{2}\bar c b+mc\rho\right)$. Acting with $\partial_\chi$ inserts the latter  composite operator with zero momentum flow into the $1PI$ Green function $\braket{(sh) h }$, see \cite{Haussling:1996rq} for the explicit expression of $G^{-1}_{Y_1h}(p^2)$ in terms of Feynman diagrams. As a consequence, the Higgs propagator $G_{hh}(p^2)$ is not gauge independent, in agreement  with our results \eqref{dk2}. From \eqref{sk} we further find
\beq
\partial_{\xi}\partial_{p^2}G^{-1}_{hh}(p^2) \vert_{p^2=-m_{pole}^2}&=&-\partial_{\chi}G^{-1}_{Y_1h}(-m_{pole}^2)\partial_{p^2}G_{hh}^{-1}(p^2)\vert_{p^2=-m_{pole}^2},
\eeq
which means that the residue is not gauge independent, as $G^{-1}_{Y_1h}(p^2)$ does not necessarily vanish at the pole. We can confirm this for the one-loop calculation, see Figure \ref{Z}.
\\
\\
Furthermore, we do have
\beq
\partial_{\xi}G^{-1}_{hh}(-m_{pole}^2)&=&0,
\eeq
so that the Higgs pole mass is indeed gauge independent, the expected result for the physical (observable) Higgs mass. This can be confirmed to one-loop order by using eq. \eqref{pol}, see also Figure \ref{realpole}. Explicitly, in eq. \eqref{ghghgh} for $G^{-1}_{hh}(-m_h^2)$ all the gauge parameter dependence drops out, which means that the Higgs  pole mass is gauge independent to first order in  $\hbar$.
\begin{figure}[t]
	\center
	\includegraphics[width=10cm]{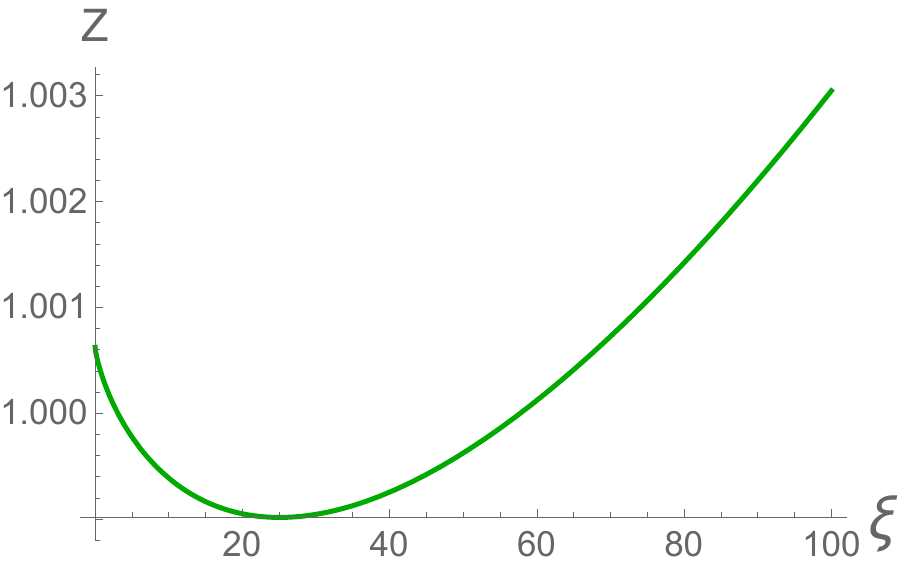}
	\caption{Gauge dependence of the residue of the pole for the Higgs field, for the parameter values $m=2$  \text{GeV}, $m_h=\frac{1}{2}$ GeV, $\mu=10$ GeV, $e=\frac{1}{10}$.}
	\label{Z}
\end{figure}
	\subsection{Obtaining the spectral function \label{32}}
We can try to determine the spectral function itself to first order. To do so, we compare the K\"all\'{e}n-Lehmann spectral representation for the propagator
\beq
G(p^2)=\int_0^{\infty} dt \frac{\rho (t)}{t+p^2},
\label{ltt5}
\eeq
where $\rho(t)$ is the spectral density function, with the propagator \eqref{op} to first order, written as
\beq
G(p^2)&=&\frac{Z}{(p^2+m_{pole}^2-\widetilde{\Pi}(p^2))Z}\nonumber\\
&=&\frac{Z}{p^2+m_{pole}^2-\widetilde{\Pi}(p^2)+(p^2+m_{pole}^2)\frac{\partial \widetilde{\Pi}(p^2)}{\partial p^2}\vert_{p^2=-m^2}}\nonumber\\
&=&\frac{Z}{p^2+m_{pole}^2}+Z\left(\frac{\widetilde{\Pi}(p^2)-(p^2+m_{pole}^2)\frac{\partial \widetilde{\Pi}(p^2)}{\partial p^2}\vert_{p^2=-m^2}}{(p^2+m_{pole}^2)^2}\right),
\label{12}
\eeq
where in the last line we used a first-order Taylor expansion so that the propagator has an isolated pole at $p^2=-m_{pole}^2$. In  \eqref{ltt5} we can isolate this pole in the same way, by defining the spectral density function as  $\rho(t)=Z \delta(t-m_{pole}^2)+\widetilde{\rho}(t)$, giving
\beq
G(p^2)=\frac{Z}{p^2+m_{pole}^2}+ \int_0^{\infty} dt\frac{\widetilde{\rho}(t)}{t+p^2}
\label{22}
\eeq
and we identify the second term in each of the representations \eqref{12} and \eqref{22} as the $\textit{reduced propagator}$
\beq
\widetilde{G}(p^2)&\equiv& G(p^2)-\frac{Z}{p^2+m_{pole}^2},
\eeq
so that
\beq
\widetilde{G}(p^2)= \int_0^{\infty}dt \frac{\widetilde{\rho}(t)}{t+p^2} &=&Z\left(\frac{\widetilde{\Pi}(p^2)-(p^2+m_{pole}^2)\frac{\partial \widetilde{\Pi}(p^2)}{\partial p^2}\vert_{p^2=-m^2}}{(p^2+m_{pole}^2)^2}\right).
\label{pp}
\eeq
Finally, using Cauchy's integral theorem in complex analysis, we can find $\widetilde{\rho}(t)$ as a function of $\widetilde{G}(p^2)$, giving
\beq
\widetilde{\rho}(t)=\frac{1}{2\pi i}\lim_{\epsilon\to 0^+}\left(\widetilde{G}(-t-i\epsilon)-\widetilde{G}(-t+i\epsilon)\right).
\label{key}
\eeq
\subsection{Spectral density functions \label{TT}}

\begin{figure}[H]
	\center
	\includegraphics[width=12cm]{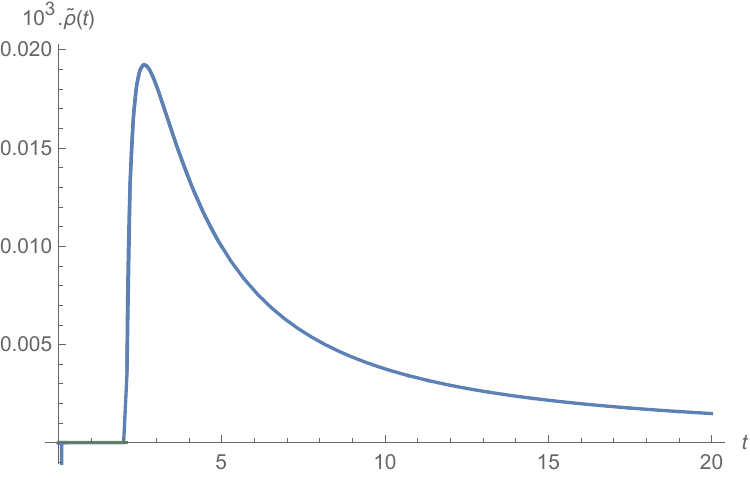}
	\caption{Spectral function for the transverse part of the reduced photon propagator $\langle A(p)A(-p) \rangle^T$, with $t$ given in units of $\mu^2$, for the parameter values $e=1$, $v=1 \, \mu$, $\lambda=\frac{1}{5}$.}
	\label{Y4}
\end{figure}

\begin{figure}[H]
	\center
	\includegraphics[width=12cm]{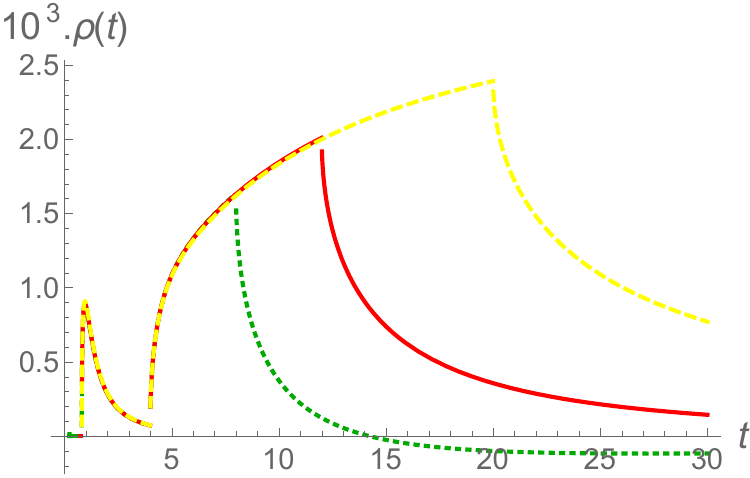}
	\caption{Spectral function for the reduced Higgs propagator $\langle h(p)h(-p) \rangle$, for gauge parameters $\xi = 2$ (Green, dotted), $\xi = 3$ (Yellow, dashed),
		$\xi= 5$ (Red, Solid),  with $t$ given in units of $\mu^2$, for the parameter values $e=1$, $v=1 \, \mu$, $\lambda=\frac{1}{5}$.}
	\label{Y38}
\end{figure}

We are now ready to plot the spectral density function for the photon propagator and the Higgs boson propagator. For this section we shall write all quantities as a function of the renormalization scale $\mu$ and choose the parameters $e=1$, $v=1 \, \mu$, $\lambda=\frac{1}{5}$, so that $m=1\, \mu$ and $m_h=\frac{1}{\sqrt{5}}\, \mu$. For this choice of parameters, all one-loop corrections computed are within $20\%$ of the tree-level results, indicating that our perturbative approximation is under control.
\\\\
Since in the Abelian case the transverse component of the gauge field $A^T_{\m}(x)$ is gauge invariant, it turns out that the corresponding propagator \eqref{dk} is independent from the gauge parameter $\xi$, and so are its pole mass, residue and spectral function.  Following the steps  from section \ref{31}, we find the first-order pole mass of the transverse photon to be
\beq
m_{pole}^2 &=& 1.05417\, \mu^2
\label{polem3}
\eeq
and the first-order residue
\beq
Z= 0.984983.
\eeq
These values are small corrections of the tree-level ones, $m^2=\mu^2$ and $Z_{\text{tree}}=1$, so that the one-loop approximation appears to be consistent.
\\
\\
The spectral function is given in Figure \ref{Y4}. We can distinguish a two-particle state threshold at $t=(m+m_h)^2=2.09 \, \mu^2$, and the spectral density function is positive, adequately describing the physical photon excitation.
For the Higgs fields, following the steps from section \ref{31}, we find the pole mass to first order in $\hbar$ to be

\beq
m_{h,pole}^2 &=& 0.237987 \, \mu^2=1.1899\, m_h^2,
\label{higgsmass}
\eeq
for all values of the parameter $\xi$. This means that while the Higgs propagator \eqref{dk2} is gauge-dependent, the pole mass is gauge-independent. This is in full agreement  with the Nielsen identities of the Abelian $U(1)$ Higgs model studied in \cite{Haussling:1996rq}. For the residue, we distinguish three regions:
\\
\begin{itemize}
	\item $\xi< \frac{1}{20}=\frac{\lambda}{4e^2}$: for these values $m_h > 2 \sqrt{\xi} m $, which means the Higgs particle is unstable and can decay into two Goldstone fields $\r(x)$. Of course, this process is physically impossible because the Goldstone boson itself is not physical. It therefore clearly demonstrates the unphysical nature of the propagator $\braket{h(x)h(y)}$. For these values of $\xi$, the pole mass is a real value inside the branch cut created by the two-particle state of Goldstone excitations. This  means that we cannot properly define the derivative of the one-loop correction to obtain the corresponding residue through eq. \eqref{kak}.
	\item $\xi \leq 3$: for these values we find $Z > 1$.
	\item $\xi >3$: for these values we find $Z<1$.
\end{itemize}
In Figure \ref{Y38}, we find the spectral density functions for three values of $\xi:2, 3, 5$. For small $t$, their behaviour is the same, with a two-particle state for the Higgs field at $t=(m_h+m_h)^2= 0.8\, \mu^2$, and a two-particle state for the photon field, starting at $t=(m+m)^2=4 \mu^2$. Then, we see that there is a negative contribution, different for each diagram, at $t=(\sqrt{\xi} m+\sqrt{\xi} m)^2$. This corresponds to the threshold for creation of two (unphysical) Goldstone bosons. This negative contribution eventually overcomes the other ones, leading to a negative regime in the spectral function, independently of the value of $\xi$. This feature is consistent with the large-momentum behaviour of the Higgs propagator \eqref{dk2}, for a detailed discussion see Appendix \ref{lot}. 		As one lowers the value of the gauge parameter $\xi$, this unphysical threshold is shifted towards lower $t$'s
and may occur for momentum values lower than the physical two-particle states of two Higgs or two photons. As discussed above, for $\xi< \frac{\lambda}{4e^2}$ even the one-particle delta peak becomes located within the unphysical Goldstone production region and the standard interpretation of the spectral properties is completely lost. It is therefore clear that this correlation function does not display the desired spectral properties to describe the Higgs mode in this theory, indicating the necessity of resorting to another operator as we shall do in what follows.

\section{Some subtleties of the Higgs spectral function \label{ont}}
In this section we  discuss some subtleties that arose during the analysis of the spectral function of the Higgs boson.

\subsection{A slightly less correct approximation for the pole mass}	
\begin{figure}[H]
	\includegraphics[width=16cm]{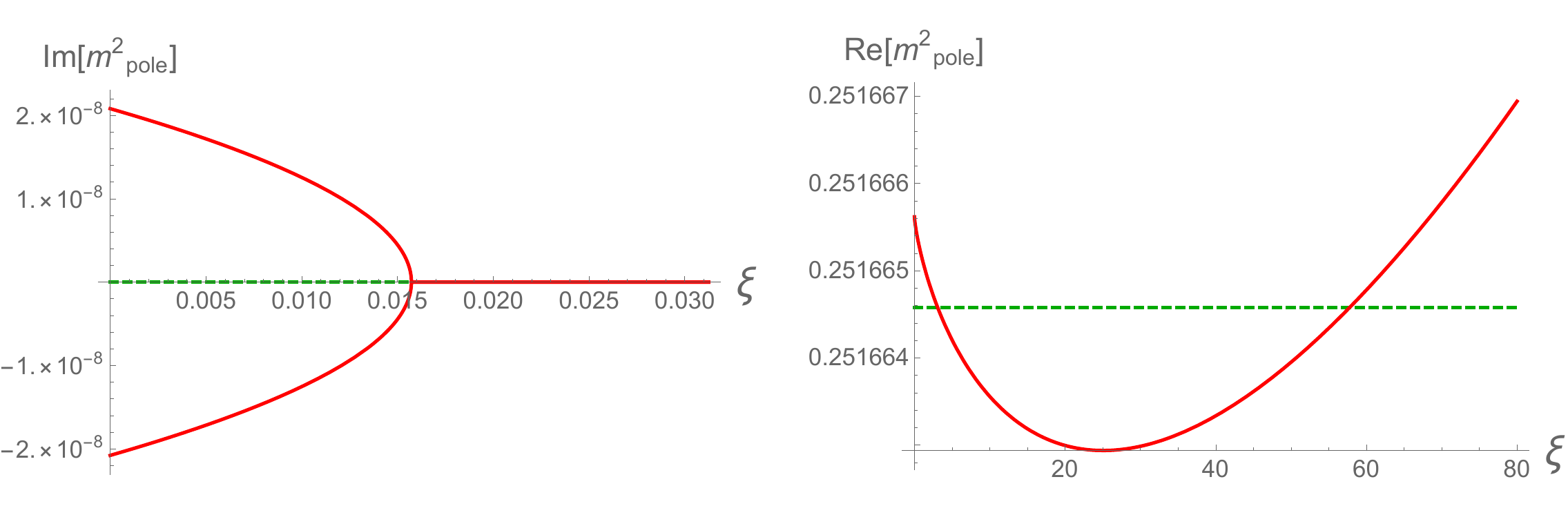}
	\caption{Gauge dependence of the Higgs pole mass obtained iteratively to first order (Green) and the approximated pole mass (Red), for the parameter values $m=2$  \text{GeV}, $m_h=\frac{1}{2}$ GeV, $\mu=10$ GeV, $e=\frac{1}{10}$. Up: real part, down: imaginary part. }
	\label{realpole}
\end{figure}
In the previous two sections we have obtained strictly first-order expressions. In practice, for small values of the coupling parameter $e^2$, we could think about making the approximation
\beq
G(p^2)&=& \frac{1}{p^2+m^2-\Pi(p^2)}\approx  \frac{1}{p^2+m^2-\Pi^{1-loop}(p^2)},
\eeq
in which case one can fix the pole mass by locating the root of
\beq
p^2+m^2-\Pi^{1-loop}(p^2)=0.
\label{ixi}
\eeq
The difference between the pole masses obtained by the iterative method \eqref{pol} and the approximation \eqref{ixi} is very small, of the order $10^{-6}$ GeV for our set of parameters. However, it is rather interesting to notice that the pole mass of the Higgs boson becomes gauge dependent in the approximation \eqref{ixi}. This is no surprise, as the validity of the Nielsen identities is understood either in an exact way, or in a consistent order per order approximation. The previous approximation is neither.
\\
\\
In Figure \ref{realpole} one can see the gauge dependence of the approximated pole mass of the Higgs, in contrast with the first order pole mass. Even worse, for very small values of $\xi$, the approximated pole mass gets complex (conjugate) values. This is due to the fact that the threshold of the branch cut, the branch point, for \eqref{pop} is $\xi$-dependent, as we will see in the next section.

\subsection{Something more on the branch points}
\begin{figure}[H]
	\includegraphics[width=10cm]{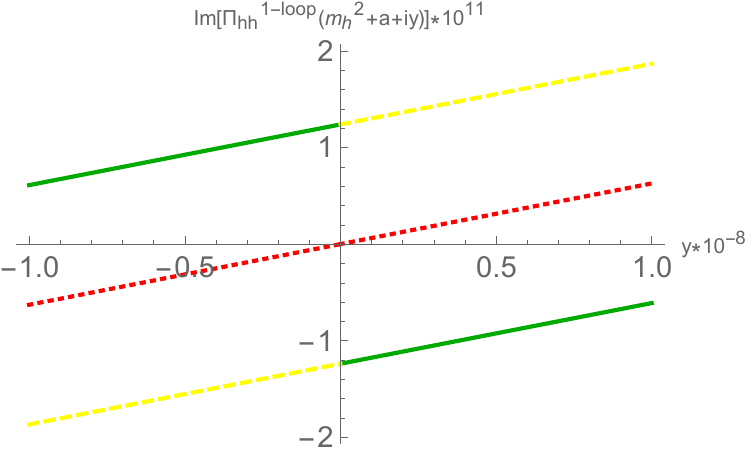}
	\centering
	\caption{Behaviour of the one-loop correction of the Higgs propagator $\Pi_{hh}(p^2)$ around the pole mass, for the values $a=-10^{-6}$ (Yellow, dashed), $a=0$ (Red, dotted), $a=10^{-6}$ (Green, solid). The value $x$ is a small imaginary variation of the argument in $\Pi_{hh}(p^2)$.  Only for $a=0$ we find a continuous function at $x=0$, meaning that for any other value, we are on the branch cut.}
	\label{less}
\end{figure}

The existence of a diagram with two internal Goldstone lines (see Figure \ref{more}) leads to a term proportional to $\int_0^1 dx \ln (p^2 x(1-x)+\xi m^2)$ in the Higgs propagator \eqref{pop}. This means that for small values of $\xi$, the threshold for the branch cut of the propagator will be $\xi$-dependent too. Let us look at the Landau gauge $\xi=0$. In this gauge, the above $\ln$-term is proportional to $\ln(p^2)$, due to the now massless Goldstone bosons. This logarithm has a branch point at $p^2=0$, meaning that the pole mass will be lying on the branch cut. Since the first order pole mass is real and gauge independent, this means that
$\Pi_{hh}^{1-loop}(-m_{h}^2)$ is a singular real point on the branch cut. In the slightly less correct approximation of the last section, we will find complex conjugate poles as in Figure \ref{realpole}. This is explained by the fact that for every real value different from $p^2=-m_h^2$, we are on the branch cut, see Figure \ref{less} .  \\

Another consequence of the fact that, for small $\xi$, the pole mass is a real point inside the branch cut is that $\Pi_{hh}(p^2)$ is non-differentiable at $p^2=-m_h^2$ and we cannot extract a residue for this pole. In order to avoid such a problem, we should move away from the Landau gauge and take a larger value for $\xi$, so that the threshold for the branch cut will be smaller than $-m_h^2$. For this we need that $4 \xi m^2 > m_h^2$, which in the case of our parameters set means to require that $\xi > \frac{1}{64}$, in accordance with Figure \ref{realpole}.

\subsection{Asymptotics of the spectral function}
Away from the Landau gauge, we see on Figure \ref{Y38} that for e.g.~$\xi=2$ the Higgs spectral function is not non-positive everywhere, while for e.g.~$\xi=4$ it is positive definite, with a turning point at $\xi=3$. How can we explain this difference? The answer can be related to the UV behaviour of the propagator. For $p^2\rightarrow \infty$, the Higgs boson propagator at one-loop behaves as
\beq
G_{hh}(p^2)=\frac{\mathcal{Z}}{p^2 \ln \frac{p^2}{\m^2}},
\eeq
with $\mathcal{Z}$ depending on the gauge parameter $\xi$. Now, one can show (see Appendix \ref{lot}) that for $\mathcal{Z}>0$, $\rho(t)$ becomes negative for a large value of $t$. For our parameter set, we find that for large momenta
\beq
G_{hh}^{-1}(p^2)\to (3-\xi)\frac{p^2\ln(p^2)}{1600 \pi^2},\quad\text{for}\quad p^2      \rightarrow \infty,
\eeq
so that for $\xi < 3$, we indeed find $\mathcal{Z}>0$. This indicates that the large momentum behaviour of the propagator makes a difference around $\xi=3$, and determines the positivity of the spectral function, a known fact \cite{Oehme:1979ai,oehme1990superconvergence,Alkofer:2000wg}. This being said, at the same time we cannot trust the propagator values for $p^2 \to\infty$ without taking into account the renormalization group (RG) effects and in particular the running of the coupling, which is problematic for non-asymptotically free gauge theories as the Abelian Higgs model.

\section{A non-unitary $U(1)$ model \label{s5}}
In this section, we will discuss an  Abelian model of the Curci-Ferrari (CF) type  \cite{curci1976slavnov}, in order to compare it  with the Higgs model \eqref{fullaction2}. Both models are massive $U(1)$ models with a BRST symmetry. However, while the BRST operator $s$ of the Higgs model is nilpotent, this is not true for the CF-like model. We know the that the Higgs model is unitary but, by the criterion of  \cite{kugo1979local}, the CF model is most probably not.\\
\\
In section \ref{iiu}  we discuss some essentials for the CF-like model: the action with the modified BRST symmetry, tree-level propagators and vertices. In section \ref{ij} we discuss the one-loop propagators for the photon and scalar field and extract the spectral function. In section \ref{ik} we introduce a local composite field operator that is left invariant by the modified BRST symmetry ot the CF model. The spectral properties of this composite state's propagator will tell us something about the (non-)unitarity of the model, since for unitary models, we expect the propagator of a BRST invariant composite operator to be gauge parameter independent, and the spectral function to be positive definite.

\subsection{CF-like U(1) model: some essentials \label{iiu}}
We start with the action of the CF-like $U(1)$ model
\beq
S_{CF}&=& \int d^d x \left\{\frac{1}{4}F_{\m\n}F_{\m\n}+\frac{m^2}{2}A_{\m}A_{\m}+  (D_{\m}\vf)^{\dagger} D_{\m}\vf +m_{\vf}^2 \vf^{\dagger}\vf+\lambda (\vf \vf^{\dagger})^2\right.\nonumber\\
&-& \left.\a \frac{b^2}{2}+ b\pa_{\mu}A_{\m}+\bar{c} \pa^2 c- \a m^2 \bar{c} c\right\},
\label{loll}
\eeq
where the mass term $\frac{m^2}{2}A_{\m}A_{\m}$ is put in by hand rather than coming from a spontaneous symmetry breaking, and we have fixed the gauge in the linear covariant gauge with gauge parameter $\a$. The mass term breaks the BRST symmetry \eqref{brst2} in a soft way. This Abelian CF action is however invariant under the modified BRST symmetry, $s_mS_{CF}=0$, with\footnote{This is the Abelian version of the variation \eqref{brst2}. For computational purposes, we have also made a rescaling $ib \rightarrow b $. Notice that higher order $\alpha$-dependent terms present in the CF model are absent in the Abelian limit \cite{curci1976class,Delduc:1989uc}.}
\beq
s_m A_{\m}&=&-\pa_{\m}c,\nonumber\\
s_m c&=& 0,\nonumber\\
s_m \vf &=& iec\vf ,\nonumber\\
s_m \vf^{\dagger} &=& -ie c \vf ^{\dagger},\nonumber\\
s_m \bar{c}&=&b,\nonumber\\
s_m b&=&-m^2 c.
\label{hallo}
\eeq
As noticed before in our Introduction, this modified BRST symmetry is not nilpotent since $s_m^2 \bar{c} \neq 0$.

From the quadratic part of \eqref{loll} we find the following propagators at tree-level
\beq
\langle A_{\m}(p)A_{\n}(-p)\rangle &=& \frac{1}{p^2+ m^2}{\mc P}_{\m\n}+\frac{\a}{p^2+\a m^2}\mathcal{L}_{\m\n},\nonumber\\
\langle A_{\m}(p)b(-p)\rangle &=& i \frac{p_{\m}}{p^2+\a m^2},\nonumber\\
\langle b(p)b(-p) \rangle &=& -\frac{m^2}{p^2+\a m^2},\nonumber\\
\langle \vf^{\dagger}(p)\vf(-p) \rangle &=& \frac{1}{p^2+m_{\vf}^2},\nonumber\\
\braket{\bar{c}(p)c(-p)}&=&-\frac{1}{p^2+\a m^2},
\eeq
while from the interaction terms we find the vertices
\beq
\Gamma_{A_{\m}\vf^{\dagger}\vf }(-p_1,-p_2,-p_3)&=&e( p_{3,\m}- p_{2,\m})\delta(p_1+p_2+p_3),\nonumber\\
\Gamma_{A_{\m}A_{\n}\vf^{\dagger}\vf }(-p_1,-p_2,-p_3,-p_4)&=&-2e^2\delta_{\m\n}\delta(p_1+p_2+p_3+p_4),\nonumber\\
\Gamma_{\vf^{\dagger} \vf\vf^{\dagger} \vf}(-p_1,-p_2,-p_3,-p_4)&=& -4\lambda\delta(p_1+p_2+p_3+p_4).
\eeq

\subsection{Propagators and spectral functions \label{ij}}
\begin{figure}[t]
	\includegraphics[width=10cm]{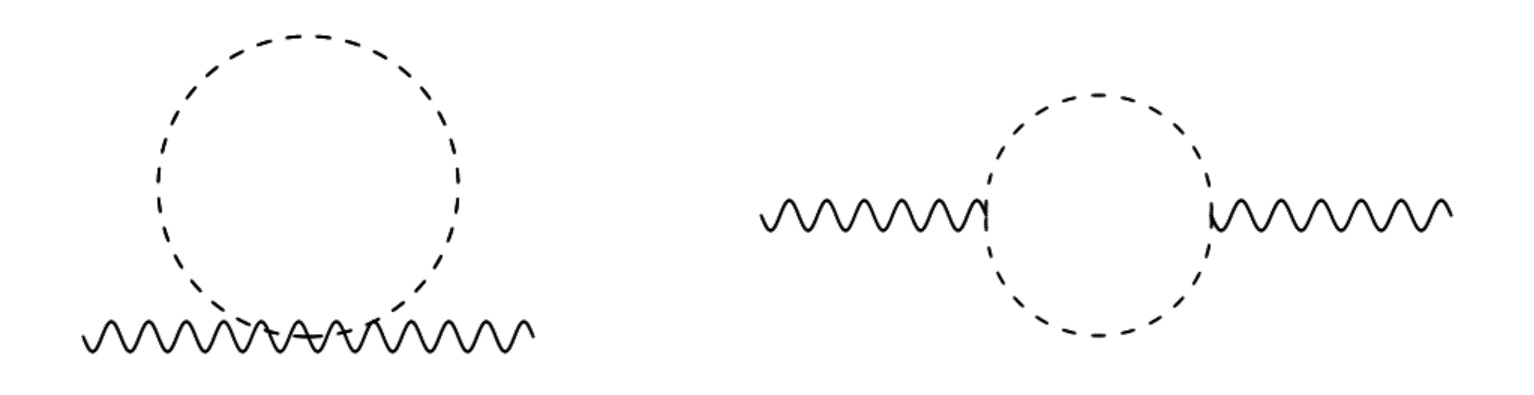}
	\caption{Contributions to one-loop CF photon self-energy. Wavy lines represent the photon field and dashed lines represent the scalar field.}
	\label{fruit}
\end{figure}
\begin{figure}[t]
	\includegraphics[width=10cm]{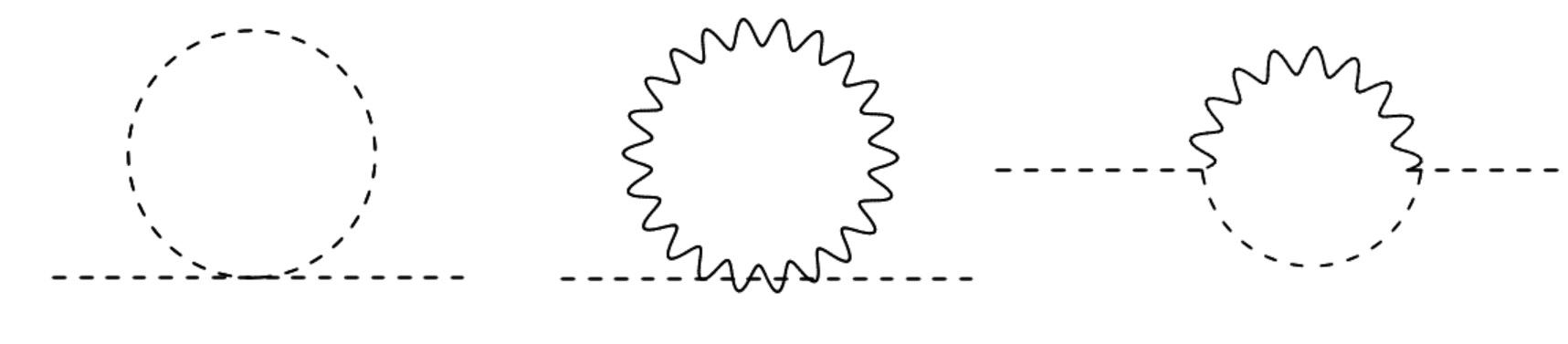}
	\caption{Contributions to the one-loop CF scalar self-energy. Line representation as in Figure \ref{fruit}.}
	\label{7}
\end{figure}
\begin{figure}[t]
	\center
	\includegraphics[width=10cm]{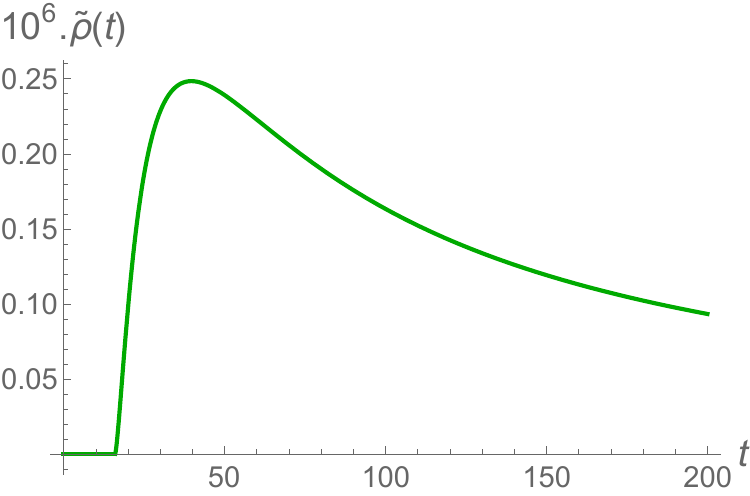}
	\caption{The reduced spectral function of the photon field in the Abelian CF model, with $t$ given in $\text{GeV}^2$, for the parameter values $m=2$  \text{GeV}, $m_h=\frac{1}{2}$ GeV, $\mu=10$ GeV, $e=\frac{1}{10}$. The first-order pole mass lies at $t=0.25151 \, \text{GeV}^2$. The photon two-particle state starts at $t^*=4m_{\vf}^2=16 \,\text{GeV}^2$. }
	\label{bbppp}
\end{figure}

The one-loop corrections to the photon and scalar  self-energies  are given in Figure \ref{fruit} and \ref{7}. Without going through the calculational details, we will directly give here the propagators in $d=4$ and discuss some curiosities. The inverse connected photon propagator,
\beq
(G^{\perp}_{AA})^{-1}(p^2)&=&p^2+m^2+\frac{e^2}{(4\pi)^2}\int_0^1 dx K(m_{\varphi}^2,m_{\varphi}^2)(1-\ln \frac{K(m_{\varphi}^2,m_{\varphi}^2)}{\m^2})-m_{\vf}^2(1-\ln \frac{m_{\vf}^2}{\m^2}),
\label{opop}
\eeq
is independent of the gauge parameter. The threshold of the branch cut is given by $t^*=-4 m_{\varphi}^2$, and to avoid a pole mass lying on the branch cut, we need to choose here $m^2 < 4m_{\varphi}^2$. Choosing $m=\frac{1}{2}$  \text{GeV}, $m_{\vf}=2$ GeV, $\mu=10$ GeV, $e=\frac{1}{10}$, we find a positive spectral function, see Figure \ref{bbppp}.

\begin{figure}[H]
	\includegraphics[width=16cm]{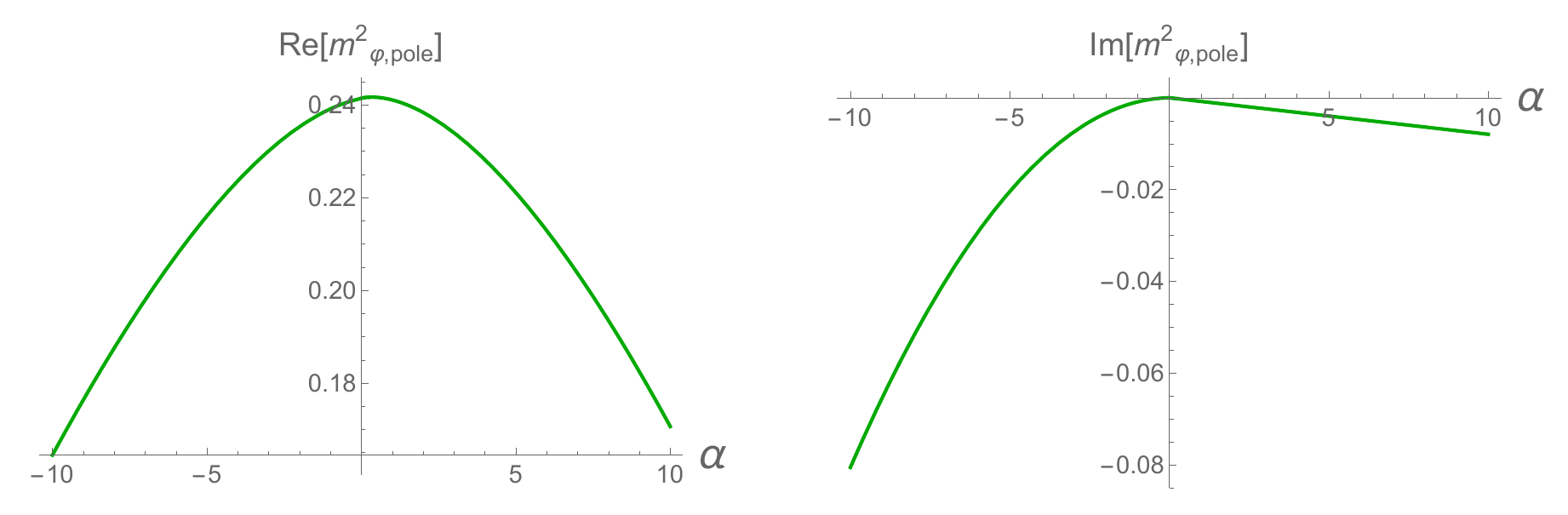}
	\caption{Gauge dependence of the first order pole mass for the scalar field. Left: Real part, Right: Imaginary part. The chosen parameter values are $m=2$  \text{GeV}, $m_{\vf}=\frac{1}{2}$ GeV, $\mu=10$ GeV, $e=\frac{1}{10}$.}
	\label{235}
\end{figure}

More interestingly, the scalar propagator
\beq
G_{\vf \vf}^{-1}(p)&=&p^2+m_{\vf}^2-\frac{e^2}{(4\pi)^2} \int_0^1 dx \Big(m^2-\a^2 m^2-\a K(m_{\vf}^2,m^2)(1-2 \ln \frac{K(m_{\vf}^2,\a m^2)}{\m^2})\\
&+&4K(m_{\vf}^2,0) \frac{p ^2}{m^2}(1-\ln \frac{K(m_{\vf}^2,0)}{\m^2})-2K(m_{\vf}^2,m^2) \frac{p^2}{m^2}(1-\ln \frac{K(m_{\vf}^2,m^2)}{\m^2})\nonumber \\
&-& 2K(m_{\vf}^2,\a m^2) \frac{p^2}{m^2}(1-\ln \frac{K(m_{\vf}^2,\a m^2)}{\m^2})+8 \frac{p^4}{m^2}x^2 \ln \frac{K(m_{\vf}^2,0)}{\m^2}\nonumber \\
&-& 4\frac{p^4}{m^2}x^2 \ln \frac{K(m_{\vf}^2,m^2)}{\m^2}-4 p^2 \ln \frac{K(m_{\vf}^2,m^2)}{\m^2}-(4\a p^2 x - \a p^2 x^2 +4 \frac{p^2}{m^2}x^2)\ln \frac{K(m_{\vf}^2,\a m^2)}{\m^2}\nonumber \\
&-& 3m^2 \ln \frac{m^2}{\m^2}+ \a m^2 \ln \frac{\a m^2}{\m^2}\Big)+\frac{\l}{(4\pi)^2} m_{\vf}^2(1-\ln \frac{m_{\vf}^2}{\m^2}),
\eeq
is $\a$-dependent, and so is the iterative first-order pole mass $m_{\vf, pole}^2=m_{\vf}^2-\Pi^{1-loop}(-m_{\vf}^2)$. This field can thus not represent a physical particle. For any value other than the Landau gauge $\a=0$ we furthermore get complex poles, see Figure \ref{235}.
\\
\\
From the fact that we find gauge dependent (complex) pole masses for the scalar field, we can already draw the conclusion that the CF model does not describe a physical scalar field. In the next section we will explicitly verify the non-unitary of this model in yet another way. Essentially, our findings so far mean that in the CF setting, the unphysical gauge parameter $\alpha$ plays here a quite important role,  just like the coupling: different values of the gauge parameter label dfferent theories. This can also be seen from another example: the one-loop vacuum energy of the model will now not only depend on $m$, but also on $\alpha$.

\subsection{Gauge invariant operator \label{ik}}
The Abelian CF model allows us to  construct  a  BRST invariant composite operator $\left(\frac{b^2}{2}+m^2 \bar{c} c \right)$, with
\beq
s_m \left(\frac{b^2}{2}+m^2 \bar{c} c\right)=0.    \label{cop}
\eeq
Although $s_m^2\neq0$  and we can therefore   no longer introduce the BRST cohomology classes, we can still use the fact that $s_m$ is a symmetry generator, thereby defining a would-be physical subspace as the one being annihilated by $s_m$.  A Fock space analogue of this operator was introduced in \cite{Ojima:1981fs},  where  it was established that it has negative norm. As a consequence, it was shown that the physical subspace relating to the symmetry generator $s_m$ was not well-defined, as it contains ghost states. Several more such states were identified later on in \cite{deBoer:1995dh}.
Up to leading order, the connected propagator of the composite operator in eq.\eqref{cop} reads
\beq
G_{\frac{b^2}{2}+m^2 \bar{c} c}(p^2)=\left\langle\left( \frac{b^2}{2}+m^2 \bar{c} c\right) , \left(\frac{b^2}{2}+m^2 \bar{c}c\right) \right\rangle=\frac{1}{4}\langle b^2,b^2\rangle + m^4\langle \bar{c}c, \bar{c}c\rangle,
\label{566}
\eeq

\begin{figure}[t]
	\includegraphics[width=10cm]{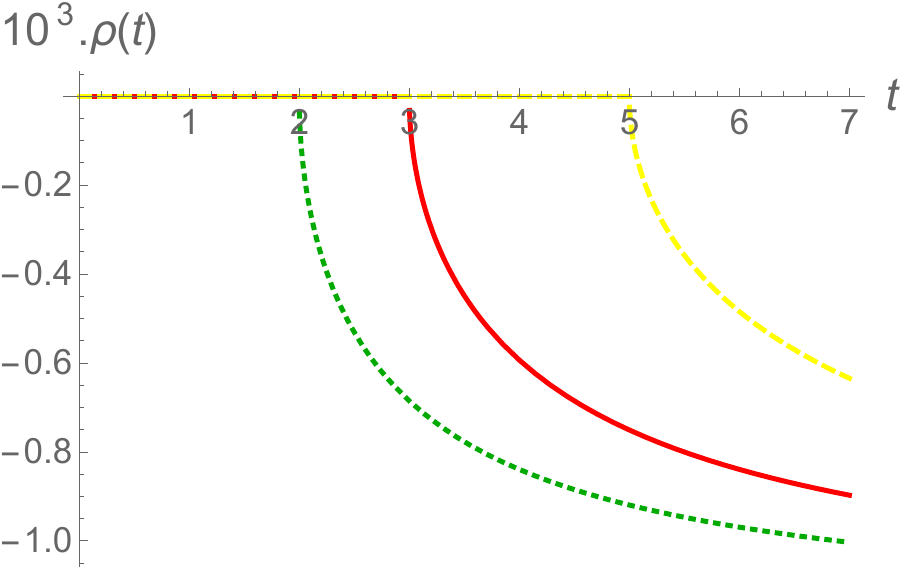}
	\caption{The spectral function of the composite operator $\frac{b^2}{2}+m^2 \bar{c} c$, for $\a=2$ (Green, dotted), $\a=3$ (Red, solid), $\a=5$ (Yellow, dashed). The chosen parameter values are $m=\frac{1}{2}$  \text{GeV}, $\mu=10$ GeV. The treshold for the branch cut of the propagator is given by $t^*=(\sqrt{\a}m+\sqrt{\a}m)^2$, being $t^*=(2 \, \text{GeV}^2,3\, \text{GeV}^2,5\, \text{GeV}^2)$ for $\xi=(2,3,5)$.}
	\label{bbpp}
\end{figure}

We thus find the propagator \eqref{566} to be
\beq
G_{\frac{b^2}{2}+m^2 \bar{c} c}(p^2)&=&-\frac{3}{4}m^2 \int \frac{d^d k}{(2\pi)^d} \frac{1}{k^2+\a m^2}\frac{1}{(k-p)^2+\a m^2} \nonumber\\
&=&-\frac{3}{4}m^2 \frac{1}{(4\pi)^{d/2}}\Gamma(2-d/2) \int^1_0 dx K_{d/2-2}(\a m^2, \a m^2)
\label{hhu}
\eeq
and this gives for $d=4$, using the $\overline{\text{MS}}$-scheme
\beq
G_{\frac{b^2}{2}+m^2 \bar{c} c}(p^2)&=&\frac{3}{4} \frac{m^2}{(4\pi)^2}\int_0^1 dx \ln \left(\frac{K(\a m^2, \a m^2)}{\m^2}\right).
\eeq
Clearly, the propagator is depending on the gauge parameter $\a$, a not so welcome feature for a presumably physical object.\\
\\
We can also find the spectral function immediately from the propagator by again relying on \eqref{key}. In Figure \ref{bbpp}, one sees that the spectral function is negative for different values of $\a$. Both the $\a$-dependence and the negative-definiteness of the spectral functions demonstrate the non-unitarity of the Abelian CF model. To our knowledge, this is the first time that  ghost-dependent invariant operators  in the physical subspace of a CF model have been constructed from the functional viewpoint\footnote{A similar result can be checked to hold for the original non-Abelian CF model, by adding a few higher order terms to the here introduced Abelian operator. This means that  a non-Abelian version of the operator \eqref{cop}, invariant under the BRST transformation \eqref{brst2}, can be written down.}, complementing the (asymptotic) Fock space analyses of \cite{Ojima:1981fs,deBoer:1995dh}.

\section{Conclusion \label{s6}  }
In this chapter we have studied the K\"all\'{e}n-Lehmann spectral  properties  of the $U(1)$ Abelian Higgs model in the $R_{\xi}$ gauge, and that of a $U(1)$ Curci-Ferrari
(CF) like model. Our main aim was to disentangle in this analytical, gauge-fixed setup what  is  physical  and  what  is  not  at  the  level  of  the elementary particle propagators, in conjunction with the Nielsen identities. Special attention was given to the role played by gauge (in)dependence of different quantities and by the correct implementation of the results up to a given order in perturbation theory.
In particular, calculating the spectral function for the Higgs propagator in the $U(1)$ model, it became apparent that an unphysical occurrence of complex poles, as well as a gauge-dependent pole mass, are caused by the use of the resummed (approximate) propagator as being exact. Indeed, for small coupling constants, the one-loop correction gives a good approximation of the all-order loop correction, and this is a much used method to find numerical results \cite{tissier2010infrared, tissier2011infrared,hayashi2018complex}. However, for analytical purposes, one should stick to the order at which one has calculated the propagator. As we have illustrated, at least in the $U(1)$ Abelian Higgs model case, one will then find a real and gauge independent pole mass for the Higgs boson, in accordance with what the Nielsen identities dictate \cite{Haussling:1996rq}.
\\
\\
Another issue faced here was the fact that the branch point for the Higgs propagator is $\xi$-dependent, being located at $p^2=0$ for the Landau gauge $\xi=0$. For small values of $\xi$, the pole mass has a real value.  However, its value is located   on the branch cut, making it impossible to define a residue at this point, and therefore a spectral function. This means that in order to formulate a spectral function, we should move away from the Landau gauge.
These issues with unphysical (gauge-variant) thresholds are nothing new, see for example \cite{Binosi:2009qm}. They reinforce  in a natural way the need to work with gauge-invariant field operators to correctly describe the observable excitations of a gauge theory.
\\
\\
For the photon, the (transverse) propagator is gauge independent (even BRST invariant), and consequently so are the pole mass, residue and spectral function. For the Higgs boson, the propagator, residue and spectral function are gauge dependent, while the pole mass is gauge independent, in line with the latter being an observable quantity. Notice that the residue of the two-point function does not need to be gauge independent,
since this does not follow from the Nielsen identities, as we discussed in our main text. Rather, the Nielsen identities can be used to show that  the residues of the pole masses in $S$-matrix elements are gauge independent, that is the residues of the singularities in observable scattering amplitudes, see \cite{Grassi:2000dz,Grassi:2001bz}. These residues can evidently be different per scattering process (and per different mass pole). The fact that the Higgs propagator is gauge dependent is not surprising, given that the Higgs field is not invariant under the Abelian gauge transformation.\\
\\
Concluding, in this chapter several tools have been worked out to determine spectral properties in perturbation theory. We worked up to first order in
$\hbar$
, but everything can be consistently extended to higher orders.  We paid attention how to avoid problems with complex poles and to the pivotal important role of the Nielsen identities, which are intimately related to the exact nilpotent BRST  invariance of the model. These tools will turn out to be quite useful for forthcoming work on the spectral properties of HYM theories. For these theories, the Nielsen identities are well established \cite{gambino2000nielsen}, with supporting lattice data \cite{maas2014two}, providing thus  a solid foundation to compare any results with.\\
\\
In the next chapters we will consider gauge-invariant operators and study their spectral properties using the same techniques of this paper. If the elementary fields are not gauge invariant (like the Higgs field, but also the gluon field in QCD), these
aforementioned gauge-invariant operators will turn out to be composite in nature. Such an approach has recently been  addressed in \cite{Maas:2017xzh,Maas:2017wzi}, based on the seminal observations of Fr\"ohlich-Morchio-Strocchi \cite{Frohlich:1980gj,Frohlich:1981yi}, in which composite operators with the same global quantum numbers (parity, spin, \ldots) as the elementary particles are constructed.\footnote{ For a recent discussion of the renormalization
	properties of higher dimensional gauge invariant operators in
	HYM models see the recent results by \cite{Binosi:2019ecz}.} These composite states will enable us to access directly  the physical spectrum of the theory.   Moreover, we notice that the spectral properties
and the behaviour in the complex momentum  plane of a (gauge-invariant) composite operator will nontrivially depend on the spectral properties of its gauge-variant constituents. This gives another motivation why it is meaningful to study spectral properties of gauge-variant propagators.  Another nice illustrative example of this interplay is the Bethe-Salpeter study of glueballs in pure gauge theories \cite{Sanchis-Alepuz:2015hma}, based on spectral properties of constituent gluons and ghosts \cite{Strauss:2012dg}.

\newpage
	
\chapter{Gauge-invariant spectral description of the $U(1)$ Higgs model from local composite operators \label{VV}}

An essential aspect of gauge theories is that all physical observable quantities have to be gauge invariant \cite{Peskin:1995ev,tHooft:1980xss}. Therefore, a formulation of the properties of the elementary excitations in terms of gauge-invariant variables is very welcome. Such an endeavour has been addressed by several authors\footnote{See \cite{Maas:2017wzi,maas2019observable}   for a general review on this matter.}   \cite{hooft1980we,hooft2012nonperturbative,frohlich1980higgs, frohlich1981higgs}, who have been able to construct, out of the elementary fields, a set of local gauge invariant composite operators which can  effectively implement a gauge invariant framework by using the tools of quantum gauge field theories: renormalizability, locality, Lorentz covariance and BRST exact symmetry. \\\\
The aim of this chapter is that of discussing the features of two local gauge invariant operators within the framework of the $U(1)$ Abelian Higgs model discussed in chapter \ref{VII}. Following \cite{hooft1980we,hooft2012nonperturbative},  we shall consider the two local  composite operators $O(x)$ and $V_\mu(x)$ invariant under \eqref{hh5}, given by 
\beq
O(x) &=& 1/2 (h^2(x) +2 v h(x) + \rho^2(x))=\vf^{\dagger}(x) \vf(x)-\frac{v^2}{2} \;,\nonumber\\
V_{\mu}(x) &=&  -i \vf^{\dagger}(x) (D_\mu \vf)(x) \;.  \label{op12}
\eeq
The relevance of these operators can be understood by using the expansion \eqref{higgs} and retaining the first order terms. For the two-point correlator of the scalar operator one finds (cf. eq. \eqref{exp14} for the full expression):
\beq
\braket{{O}(x) {O}(y)}  &\sim&  v^2\langle h(x) h(y) \rangle_{(\text{tree-level)}} + O(\hbar) + \braket{O\left(h^3;h\rho^2;\rho^4\right)} \;,  \label{mean}
\eeq
while the contributions to the vector operator at lowest order in the fields read
\beq	
V_{\mu}(x) &\sim & \frac{e v^2}{2} A_\mu(x)  + {\text{total derivative}}\; +\; { \text {higher orders}} \;.  \label{mean2}
\eeq
We see therefore that the gauge-invariant operator ${O}(x)$ is related to the Higgs excitation, while $ V_{\mu}(x) $ is associated with the photon. \\
\\
In this chapter, we shall compute the BRST-invariant two-point correlation functions
\beq
\langle O(x) O(y) \rangle\;,   \; \; \; \; \; \; \langle V_{\mu}(x) V_{\nu}(y) \rangle \;, \label{cf}
\eeq
at one-loop order in the 't Hooft  $R_{\xi}$-gauge and discuss the differences with respect to the corresponding one-loop elementary propagators $\langle h(x) h(y) \rangle$ and $\langle A_\mu(x) A_\nu(y) \rangle$ already evaluated in the last chapter. Let us point out that the two local operators $(V_{\mu}(x), O(x))$ belong to the cohomology of the BRST operator \cite{Piguet:1995er}, i.e.
\beq
s V_{\mu}(x) & = & 0 \;, \qquad \qquad V_{\mu}(x) \neq s\Delta_\mu(x) \nonumber \\
s O(x) & = & 0 \;, \qquad \qquad O(x) \neq s \Delta(x) \label{coh} \;,
\eeq
	for any local quantities $(\Delta_\mu(x), \Delta(x))$.
\\\\As expected, both correlation functions of eq. \eqref{cf}
turn out to be independent from the gauge parameter $\xi$. Moreover, we shall show that the one-loop pole masses of $\langle V_{\mu}(x) V_{\nu}(y) \rangle_T $ and $\langle O(x) O(y) \rangle$ are exactly the same as those of the elementary propagators $\langle A_\mu(x) A_\nu(y) \rangle_{T}$ and
$\langle h(x) h(y) \rangle$, where $\langle A_\mu(x) A_\nu(y) \rangle_{T}$ stands for the transverse component of $\langle A_\mu(x) A_\nu(y) \rangle$, i.e.
\beq
\langle A_\mu(x) A_\nu(y) \rangle_{T} = \left(\delta_{\mu \rho} - \frac{\partial_\mu \partial_\rho}{\partial^2}\right)\langle A_\rho(x) A_\nu(y) \rangle\, .
\eeq
This important feature makes apparent the fact that the operators $V_{\mu}(x)$ and  $ O(x) $ give a gauge invariant picture for the photon and Higgs modes. In addition, the correlation functions $\langle V_{\mu}(x) V_{\nu}(y) \rangle_T $ and $\langle O(x) O(y) \rangle$ exhibit a spectral KL representation with positive spectral densities, allowing for a physical interpretation in terms of particles. This property is in sharp contrast with the one-loop spectral density of the elementary non gauge invariant Higgs propagator $\langle h(x) h(y) \rangle$, which displays an explicit dependence on the gauge parameter $\xi$, as established in the last chapter. Moreover, the longitudinal part of the correlator $\langle V_{\mu}(x) V_{\nu}(y) \rangle$ -- which is independently gauge invariant -- is shown to exhibit the pole mass of the Higgs excitation. This last feature reinforces the consistency of the present description of the physical degrees of freedom of the theory, since the only physically expected elementary excitations are indeed the Higgs and the photon ones.
Let us also underline that, to our knowledge, this is the first explicit one-loop calculation of the gauge-invariant correlators \eqref{cf} and of their analytical properties.
 \\\\This chapter is organized as follows. In section \ref{fmsp} we compute at one-loop order the two-point functions for the composite operators in. In section \ref{IIII}, we provide the detailed analysis of the spectral properties of the composite operators, and compare them with the spectral properties of the elementary fields. The unitary limit, in which the gauge parameter $\xi$ tends to infinity, is investigated in section \ref{unitarylimit}. Section \ref{VI} collects our conclusion and outlook.  The final Appendices contain the derivation of the Feynman rules and of the diagrams contributing to \eqref{cf}.
\\\\
\section{The correlation functions  $\langle O(x) O(y) \rangle$ and $ \langle V_{\mu}(x) V_{\nu}(y) \rangle$  at one loop order \label{fmsp}}
\begin{figure}[H]
	\center
	\includegraphics[width=12cm]{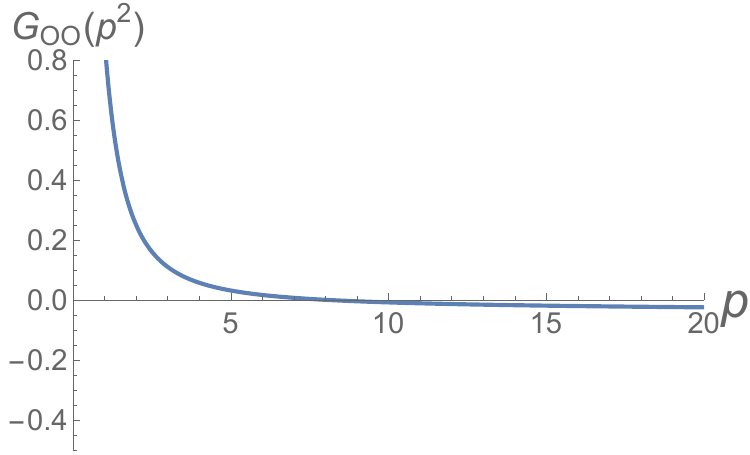}
	\caption{Resummed form factor for the scalar composite operator.
		The momentum $p$ is given in units of the energy scale $\mu$, for the parameter values $e=1$, $v=1 \, \mu$, $\lambda=\frac{1}{5}$.}
	\label{prophh1}
\end{figure}

We study the two-point correlation functions of the local gauge invariant operators $(V_{\mu}(x), O(x))$. For the correlator of the scalar composite operator we get:	
\beq
\braket{{O}(x) {O}(y)} &=& v ^2 \braket{h(x) h(y)}+ v  \braket{h(x) \rho (y)^2}+ v  \braket{h(x), h(y)^2}+ \nonumber \\
&&+\frac{1}{4} \Big( \braket{h(x)^2 \rho (y)^2}+\braket{h(x)^2 h(y)^2}+ \braket{\rho (x)^2 \rho (y)^2}\Big). \label{exp14}
\eeq
Individually, the terms in the expansion \eqref{exp14} are not gauge invariant, but their sum is. We can now analyze the connected diagrams for each term, up to one-loop order, through the action \eqref{fullaction2}. We calculated the one-loop diagrams in Appendix \ref{A}. Looking at the diagrams in Figure \ref{Yw}, we can see that the correlation function $\braket{O(p)O(-p)}$ will have the following structure
\beq
\braket{O(p)O(-p)}^{1-loop} &=&  \frac{A_{fin}(p^2)+\delta A_{div}(p^2) }{(p^2+m_h^2)^2} + \frac{B_{fin}(p^2)+\delta B_{div}(p^2)}{(p^2+m_h^2)}\nonumber\\
&+& C_{fin}(p^2)+\delta C_{div}(p^2),
\label{finn}
\eeq
where $(A_{fin}, B_{fin}, C_{fin})$ stand for the finite parts and $(\delta A_{div}, \delta B_{div}, \delta C_{div})$ for the purely divergent terms, i.e. the pole terms in $\frac{1}{\epsilon}$ obtained by means of the  dimensional regularization, namely

\beq
\delta A_{div}(p^2)&\overset{\epsilon \rightarrow 0}{=}&\frac{ v^2}{8\pi^2 \epsilon} \Big(2v^2 \lambda^2-e^2(p^2(-3+\xi)+v^2\lambda \xi)\Big)\nonumber \\
\delta B_{div}(p^2)&\overset{\epsilon \rightarrow 0}{=}&\frac{v^2 (6 e^4-\lambda^2+e^2 \lambda \xi)}{8 \pi^2 \epsilon \lambda}\nonumber \\
\delta C_{div}(p^2) &\overset{\epsilon \rightarrow 0}{=}& \frac{1}{8 \pi^2 \epsilon}
\eeq
while 
\beq
A_{fin} (p^2)&=& \frac{v^2}{(4\pi)^2}\int_{0}^{1} dx
\Bigg\{
e^2\Bigg[p^2(1-\ln\frac{m^2}{\m^2}-2\ln \frac{p^2 x(1-x)+m^2}{\m^2}) \nonumber\\
&-&\frac{p^4}{2m^2} \ln \frac{p^2 x(1-x)+m^2}{\m^2}-6m^2(1-\ln \frac{m^2}{\m^2}+\ln \frac{p^2 x(1-x)+m^2}{\m^2})\Bigg]\nonumber\\
&+&\lambda \Big[\frac{1}{2}m_h^2(-6+6\ln \frac{m_h^2}{\m^2}-9\ln \frac{p^2 x(1-x)+m_h^2}{\m^2})\Big]\nonumber\\
&-&\Bigg[\xi (e^2p^2+\l m^2)(1- \ln \frac{\xi m^2}{\m^2})-(e^ 2\frac{p^4}{2m^2}-\lambda\frac{m_h^2}{2})\ln \frac{p^2 x(1-x)+\xi m^2}{\m^2}\Bigg]\Bigg\}\nonumber \\
B_{fin}(p^2)&=&  \frac{1}{(4\pi)^2m_h^2}\int_{0}^{1} dx \Bigg\{-m^2 \xi  m_h^2 \ln \left(\frac{m^2 \xi }{\mu ^2}\right)+m^2 \xi  m_h^2+m_h^4 \Big(3 \ln \left(\frac{ m_h^2+p^2 (1-x) x}{\mu ^2}\right)\nonumber \\
&+&\ln \left(\frac{m^2 \xi+p^2 (1-x) x}{\mu ^2}\right)\Big)-3 m_h^4 \ln \left(\frac{m_h^2}{\mu ^2}\right)+3 m_h^4+2 m^4-6 m^4 \ln \left(\frac{m^2}{\mu ^2}\right)\Bigg\}\nonumber \\
C_{fin}(p^2) &=& -\frac{1}{2(4\pi)^2}\int_{0}^{1} dx \Bigg\{\ln \left(\frac{m_h^2+p^2 (1-x) x}{\mu ^2}\right)+\ln \left(\frac{m^2 \xi +p^2 (1-x) x}{\mu ^2}\right)\Bigg\}.
\eeq
The divergent terms $(\delta A_{div}, \delta B_{div}, \delta C_{div})$ can be eliminated by means of the Lagrangian counterterms as well as by suitable counterterms in the external part of the action $S_J$ accounting for the introduction of the composite operator $O(x)$, i.e. 

\beq
S_J&=& S +
\int d^4 x \Bigg[ ( 1+\delta Z_{div}^0) J(x) O(x) + ( 1+\delta Z_{div}) \frac{(J(x))^2}{2}\Bigg],
\label{1o}
\eeq
where $J(x)$ is a BRST invariant dimension two source needed to define the generator $Z^c(J)$ of the connected Green function $\braket{O(x)O(y)}$:

\beq
\braket{O(x)O(y)}=\frac{\delta^2 Z^c(J)}{\delta J(x) \delta J(y)}\vert_{J=0}.
\eeq
It is worth emphasizing here that we have the freedom of introducing a pure contact BRST invariant term in the external source $J(x)$:

\beq
\int d^4 x \,\frac{\alpha}{2}\,J^2(x),
\label{3d}
\eeq
which can be arbitrarily added to the action \eqref{1o}. Including such a term in \eqref{1o} will have the effect of adding a dimensionless constant to $G_{OO}=\braket{O(p)O(-p)}$, i.e.

\beq
G_{OO}(p^2)\rightarrow G_{OO}(p^2)+\alpha
\label{3e}.
\eeq
In particular, $\alpha$ can be chosen to be equal to $-G_{OO}(0)$, implying then that the modified Green's function

\beq
G_{OO}(p^2)-G_{OO}(0)
\label{dv1}
\eeq
will obey a one substracted KL representaion
.
\\
\\
Inserting the unity 

\beq
1= (p^2+m_h^2)/(p^2+m_h^2) = ((p^2+m_h^2)/(p^2+m_h^2))^2,
\label{2o}
\eeq
for the finite part of $\braket{O(p)O(-p)}$, we write 
\beq
\langle {O}(p) {O}(-p) \rangle_{fin} &=& \frac{v^2}{p^2+m_h^2}+\frac{\hbar v^2}{(p^2+m_h^2)^2}\Pi (p^2)+\mathcal{O}(\hbar^2)
\eeq
where
\beq
\Pi_{OO}(p^2) &=&  \frac{1}{v^2}\Big((A_{fin}(p^2)) + (p^2+m_h^2)(B_{fin}(p^2))+ (C_{fin}(p^2)) (p^2+m_h^2)^2\Big), \nonumber
\\
&=&\frac{1}{32 \pi ^2  v^2 m_h^2}
\int_0^1 dx \Bigg\{-8 m_h^2 m^4-2 m^2 p^2 (m_h^2+6 m^2) \ln \left(\frac{m^2}{\mu ^2}\right)+
\nonumber\\
&& + m_h^2 \Big[-(p^2-2 m_h^2)^2 \ln \left(\frac{m_h^2+p^2(1-x) x}{\mu ^2}\right) 
%\nonumber \\
%&&
-(12 m^4+4 m^2 p^2+p^4) \ln \left(\frac{m^2+p^2(1-x) x}{\mu ^2}\right)\Big]+
\nonumber \\
&&+2 p^2 (3 m_h^4+m_h^2 m^2+2 m^4)-6 m_h^4 p^2 \ln \left(\frac{m_h^2}{\mu ^2}\right)\Bigg\},
\label{ark3}
\eeq
and since \eqref{ark3} contains terms of the order of $\frac{p^4}{p^2+m^2}\ln (p^2)$,
we follow the steps \eqref{u1}-\eqref{jju2} to find the resummed form factor in the one-loop approximation
\beq
G_{{O}{O}}(p^2)
&=&\frac{v^2}{p^2+m_h^2-\hat{\Pi}_{OO}(p^2)}+C_{OO}(p^2)
\eeq
with

\beq
\hat{\Pi}_{OO}(p^2)&=&\frac{1}{32 \pi ^2  v^2 m_h^2}
\int_0^1 dx \Bigg\{-8 m_h^2 m^4-2 m^2 p^2 (m_h^2+6 m^2) \ln \left(\frac{m^2}{\mu ^2}\right)+
\nonumber\\
&& + m_h^2 \Big[3(m_h^4+2m_h^2p^2) \ln \left(\frac{m_h^2+p^2(1-x) x}{\mu ^2}\right) 
\nonumber \\
&&
-(12 m^4+4 m^2 p^2-m_h^4-2p^2m_h^2) \ln \left(\frac{m^2+p^2(1-x) x}{\mu ^2}\right)\Big]+
\nonumber \\
&&+2 p^2 (3 m_h^4+m_h^2 m^2+2 m^4)-6 m_h^4 p^2 \ln \left(\frac{m_h^2}{\mu ^2}\right)\Bigg\},
\label{dk34}
\eeq
and 

\beq
C_{OO}(p^2)&=&-\frac{1}{32 \pi ^2  }
\int_0^1 dx \Bigg\{\ln \left(\frac{m_h^2+p^2x(1-x)}{\mu^2}\right)+\ln \left( \frac{ m^2 + p^2 x (1-x) }{\mu^2}\right)\Bigg\}.
\label{grgr4}
\eeq
The form factor is depicted in Figure \ref{prophh1}. Notice that the Green function $G_{OO}(p^2)$ becomes negative for large enough values of the momentum $p$. As one realizes from expression \eqref{dk34}, this feature is due to the growing in the UV region of the logarithms contained in the term $C_{OO}(p^2)$, see eq.~\eqref{grgr4}. It is worth mentioning that this behaviour is also present when the parameter $v$ is completely removed from the theory. In fact, setting $v=0$, the action $S_0$ in eq.~\eqref{higgsqed}, reduces to that of massless scalar QED, namely

\beq
\label{dd2}
\left.S_0\right|_{v=0}=\int d^4x \left(\frac{F_{\mu\nu}^2}{4}+ (D_\mu\varphi)^\dagger(D_\mu\varphi) + \frac{\lambda}{2}(\varphi^\dagger\varphi)^2\right)\,,
\eeq

with

\beq
\label{dd3}
\left.\varphi\right|_{v=0}=\frac{1}{\sqrt 2}(h+i\rho).
\eeq
Of course, when $v=0$, the operators $O=\varphi^\dagger \varphi$ and $V_\mu=-i \varphi^\dagger D_\mu \varphi$ are still gauge invariant. Though, from eqs.~\eqref{dk34}-\eqref{grgr4}, computing $\braket{O(p)O(-p)}_{v=0}$, one immediately gets

\begin{equation}\label{dd1}
\braket{O(p)O(-p)}_{v=0}=\left.C_{OO}\right|_{v=0}=-\frac{1}{16\pi^2} \int_0^1 dx \ln \frac{p^2x(1-x)}{\mu^2}.
\end{equation}

This equation precisely shows that the term $C_{OO0}$, and thus the negative behaviour for large enough values of $p$, is what one usually obtains in a theory for which $v=0$, making evident that the presence of $C_{OO}$ is not peculiarity of the $U(1)$ Higgs model in the $U(1)$ Higgs model, on the contrary. However, in addition to the term $C_{OO}$ and unlike massless scalar QED, the correlation function $\braket{O(p)O(-p)}$ of the $U(1)$ Higgs model exhibits the term $\frac{v^2}{p^2+m_h^2-\tilde\Pi_{OO}}$, which will play a pivotal role. Indeed, as we shall see later on, this term, originating from the expansion of $\varphi$ around the minimum of the Higgs potential, $\phi=\frac{1}{\sqrt 2}(v+h+i\rho)$, will enable us to devise a gauge invariant description of the elementary excitations of the model.

\begin{figure}[H]
	\center
	\includegraphics[width=12cm]{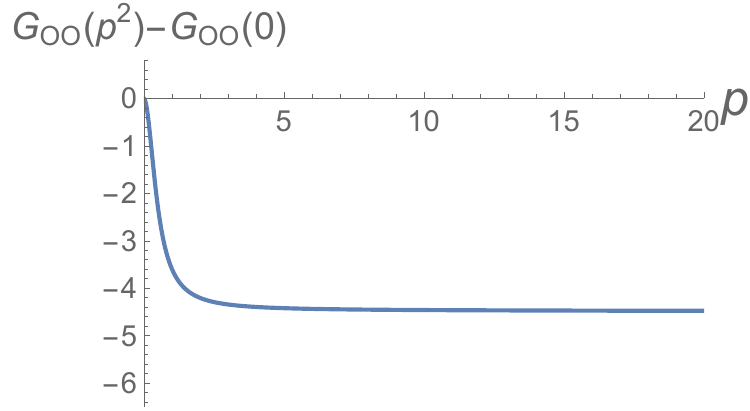}
	\caption{The resummed form factor with one subtraction.
		The momentum $p$ is given in units of the energy scale $\mu$, for the parameter values $e=1$, $v=1 \, \mu$, $\lambda=\frac{1}{5}$.}
	\label{prophh2}
\end{figure}

\begin{figure}[H]
	\center
	\includegraphics[width=12cm]{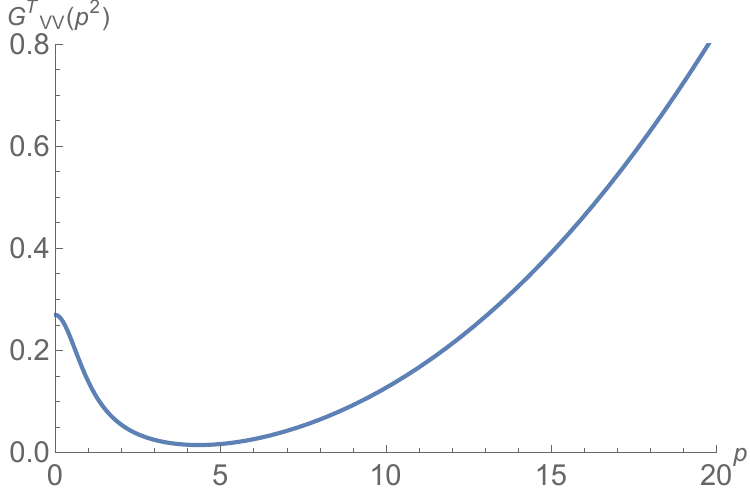}
	\caption{Resummed form factor for the vector composite operator (left), the resummed form factor with one subtraction (middle), and the first derivative of the form factor (left).
		The momentum $p$ is given in units of the energy scale $\mu$, for the parameter values $e=1$, $v=1 \, \mu$, $\lambda=\frac{1}{5}$.}
	\label{propAAT1}
\end{figure}

\begin{figure}[H]
	\center
	\includegraphics[width=12cm]{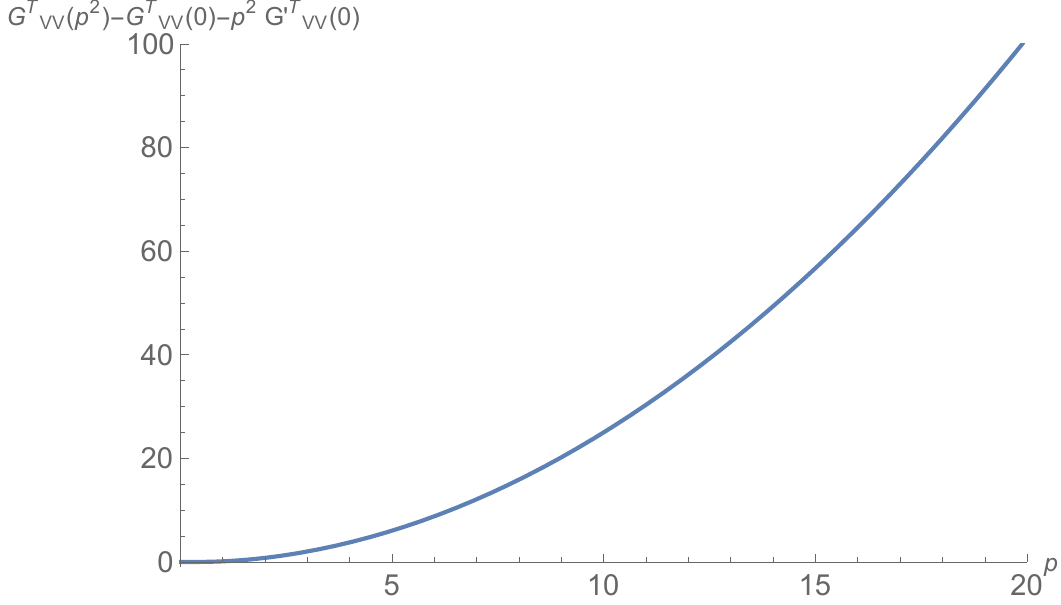}
	\caption{Resummed form factor for the vector composite operator.
		The momentum $p$ is given in units of the energy scale $\mu$, for the parameter values $e=1$, $v=1 \, \mu$, $\lambda=\frac{1}{5}$.}
	\label{propAAT13}
\end{figure}

Let us end the analysis of the correlation function $G_{OO}(p^2)$ by displaying the behaviour of its first derivative, $\frac{\partial G_{OO}(p^2)}{\partial p^2}$, as well as of the one subtracted correlator $G_{OO}(p^2) - G_{OO}(0)$,
see Figure \ref{prophh2}. The first derivative, as expected, is negative while, unlike $G_{OO}(p^2)$, decays to zero for $p^2\to\infty$. The quantity $\frac{\partial G_{OO}(p^2)}{\partial p^2}$ will be helpful when discussing the spectral representation corresponding to $\braket{O(p)O(-p)}$.
\begin{figure}[H]
	\center
	\includegraphics[width=12cm]{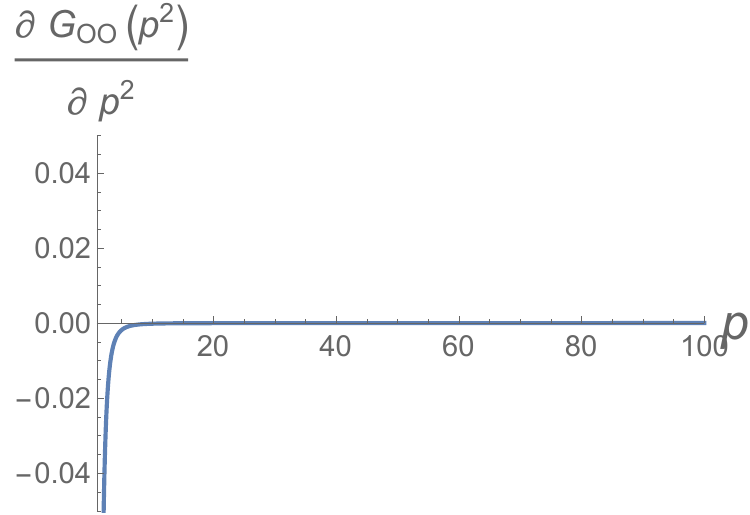}
	\caption{The first derivative of the form factor (left).
		The momentum $p$ is given in units of the energy scale $\mu$, for the parameter values $e=1$, $v=1 \, \mu$, $\lambda=\frac{1}{5}$.}
	\label{prophh3}
\end{figure}
Then, for the vectorial composite operator ${V}_{\mu}(x)$, we first observe that

\beq
V_{\mu}(x)&=& -i \varphi^{\dagger}(x) (D_{\mu}\varphi)(x)\nonumber \\
&=& e \vf^{\dagger}(x) A_{\mu}(x) \vf(x) -\frac{1}{2} i\varphi^{\dagger}(x)\partial_{\mu} \vf(x) +\frac{1}{2}i \varphi(x)\partial_{\mu} \vf^{\dagger}(x)- i \partial_{\mu} O(x),  
\eeq
and since we know that the last term is gauge-invariant, the first three terms together are also gauge-invariant. We thus define a new gauge-invariant operator 

\beq
V'_{\mu}(x)
&=& e \vf^{\dagger}(x) A_{\mu}(x) \vf(x) -\frac{1}{2} i\varphi^{\dagger}(x)\partial_{\mu} \vf(x) +\frac{1}{2}i \varphi(x)\partial_{\mu} \vf^{\dagger}(x),  
\eeq
expanding the scalar field $\vf(x)$ we find

\beq
V'_{\mu}(x)
&=& \frac{1}{2} e (v+h(x))^2 A_{\mu}(x) + \frac{1}{2}e \rho^2(x) A_{\mu}(x)+(v+h(x))\partial_{\mu} \rho (x)-\rho(x) \partial_{\mu} h(x)
\eeq
so that 
\beq
\braket{{V'}_{\m}(x){V'}_{\n}(y)} &\overset{\varphi \rightarrow \frac{1}{\sqrt{2}}(v + h +i \rho)}{=}&-\frac{1}{4}\Bigg\{-e^2 v^4  \langle A_{\mu }(x) A_{\nu}(y) \rangle-4 e^2 v^3  \langle h(x)  A_{\mu }(x)A_{\nu}(y) \rangle \nonumber \\
&&-2 e^2 v^2  \langle h(x)^2A_{\mu }(x)  A_{\nu}(y) \rangle-4 e^2 v^2  \langle h(x)  A_{\mu }(x) h(y) A_{\nu}(y)\rangle \nonumber \\
&&-2 e^2 v^2 \langle \rho (x)^2A_{\mu }(x) A_{\nu}(y)  \rangle-2 e v^2 \partial ^x{}_{\mu } \langle h(x) \rho (x) A_v(y) \rangle \nonumber \\
&&+4 e v^2  \langle \partial ^x{}_{\mu } h(x)\rho (x)A_{\nu}(y) \rangle -2 e v^3\partial ^x{}_{\mu } \langle \rho (x) A_{\nu}(y) \rangle\nonumber \\
&&-4 e v^2 \partial ^x{}_{\mu } \langle  \rho (x) h(y) A_v(y) \rangle-2 v \partial ^x{}_{\mu }\partial ^y{}_{\nu } \langle h(x) \rho (x) \rho (y) \rangle \nonumber \\
&&+4 v \partial ^y{}_{\nu } \langle \partial ^x{}_{\mu }h(x) \rho (x)\rho (y) \rangle-\partial ^x{}_{\mu }\partial ^y{}_{\nu } \langle h(x) \rho (x) h(y) \rho (y) \rangle \nonumber \\
&&+4 \langle\partial ^x{}_{\mu }h(x)\rho (x) h(y) \partial ^y{}_{\nu }\rho (y) \rangle-v^2\partial ^x{}_{\mu }\partial ^y{}_{\nu }  \langle\rho (x) \rho (y) \nonumber \rangle \Bigg\}\nonumber \\
&&+O(\hbar^2),
\label{exp2}
\eeq
where we have discarded the terms that do not have one-loop contributions.
In momentum space, we can split the two-point function into transverse and longitudinal parts in the usual way:
\beq
\langle {V'}_{\mu}(p){V'}_{\nu}(-p)\rangle =\langle {V'}(p) {V'}(-p)\rangle^T \mathcal{P}_{\mu\nu}+\langle V'(p)V'(-p)\rangle^L \mathcal{L}_{\mu\nu} \,,
\eeq
where we have introduced the transverse and longitudinal projectors, given respectively by
\beq
\mathcal{P}_{\mu\nu}(p)&=& \delta_{\mu\nu}-\frac{p_{\mu}p_{\nu}}{p^2}\,,
\nonumber\\
\mathcal{L}_{\mu\nu}(p)&=& \frac{p_{\mu}p_{\nu}}{p^2}\,.
\eeq
At tree-level, we find in momentum space
\beq
\langle V'_{\mu}(p)V'_{\nu}(-p)\rangle_{\text{\text{tree}}}&=&-\frac{1}{4}\left(-e^2 v^4  \langle  A_{\mu }(p)A_{\nu}(-p) \rangle-v^2p_{\mu }p_{\nu }  \langle\rho (p) \rho (-p) \rangle\right) \nonumber \\
&=&\frac{1}{4}\left(e^2 v^4\frac{1}{p^2+m^2} \mathcal{P}_{\mu\nu}+e^2 v^4\frac{\xi}{p^2+\xi m^2}\mathcal{L}_{\mu \nu}+ v^2 \frac{p^2}{p^2+\xi m^2}\mathcal{L}_{\mu \nu}\right) \nonumber \\
&=&\frac{e^2 v^4}{4}\frac{1}{p^2+m^2} \mathcal{P}_{\mu\nu}+v^2\mathcal{L}_{\mu \nu}.
\eeq
%which means that the transverse part has a \text{tree}-level mass equal to the photon mass, while the longitudinal part has the same tree-level mass as the Higgs field. This nice feature was also observed in \cite{maas2019observable}.%
We can now analyze the connected diagrams for each term, up to one-loop order, through the action \eqref{fullaction2}. We calculated the one-loop diagrams in Appendix \ref{ah}. Let us start with the transverse part. Looking at the diagrams in Figure \ref{Y22}, we can see that the one-loop correlation function will have the following structure
\beq
\braket{V'(p)V'(-p)}^{T,1-loop} &=&  \frac{A^V_{fin}(p^2)+\delta A^V_{div}(p^2)}{(p^2+m^2)^2} + \frac{B^V_{fin}(p^2)+\delta B^V_{div}(p^2)}{(p^2+m^2)}\nonumber \\
&+& C^V_{fin}(p^2)+\delta C^V_{div}(p^2) 
\label{finn}
\eeq
where $(A^V_{fin}, B^V_{fin}, C^V_{fin})$ stand for the finite parts and $(\delta A^V_{div}, \delta B^V_{div}, \delta C^V_{div})$ for the purely divergent terms, i.e. the pole terms in $\frac{1}{\epsilon}$ obtained by means of the  dimensional regularization, namely

\beq
\delta A^V_{div}&\overset{\epsilon \rightarrow 0}{=}&\frac{e^4 v^4}{2 (4\pi)^2 \epsilon} \Big(\frac{1}{3}p^2-6(\frac{e^2}{\lambda}-\frac{1}{2})e^2 v^2+3\lambda v^2\Big)\nonumber \\
\delta B^V_{div}&\overset{\epsilon \rightarrow 0}{=}&\frac{v^2 }{(4 \pi)^2 \epsilon}(6 \frac{e^6v^2}{\lambda}-3 e^4 v^2-\frac{e^2p^2}{3}+3e^2 \lambda v^2)\nonumber \\
\delta C^V_{div} &\overset{\epsilon \rightarrow 0}{=}& \frac{1}{6 (4\pi)^2 \epsilon}(9e^2v^2-p^2-3\lambda v^2)
\eeq
and 
\beq
A^V_{fin} &=&\frac{e^4v^4}{2(4\pi)^2}\int_{0}^{1} dx \,\,\Bigg\{p^2 x(1-x)+m^2x \nonumber\\
&+&m_h^2(1-x)(1-\ln\frac{p^2 x(1-x)+m^2x+m_h^2(1-x)}{\m^2})+m_h^2(1-\ln\frac{m_h^2}{\m^2})\nonumber\\
&+&\frac{m^4}{m_h^2}(1-3 \ln \frac{m^2}{\m^2})+2m^2 \ln \frac{p^2 x(1-x)+m^2x+m_h^2(1-x)}{\m^2}\Bigg\}\nonumber \\
B^V_{fin}&=& \frac{m^2}{18 m_h^2 p^2 (4\pi)^2} \int_0^1 dx   \Bigg\{3 m_h^4 \left(m_h^2-m^2-7 p^2\right) \ln \left(\frac{m_h^2}{\mu ^2}\right)\nonumber\\
&-&3 m_h^2 \left(2 p^2 \left(m_h^2-5 m^2\right)+\left(m_h^2-m^2\right){}^2+p^4\right) \ln \left(\frac{x m_h^2+m^2 (1-x)+p^2 (1-x) x}{\mu ^2}\right)\nonumber\\
&-&3 \left(m_h^3-m^2 m_h\right){}^2+9 p^2 \left(m^2 m_h^2+3 m_h^4+2 m^4\right)+2 p^4 m_h^2\nonumber\\
&-&3 m^2 \left(m_h^2 \left(p^2-m^2\right)+m_h^4+18 m^2 p^2\right) \ln \left(\frac{m^2}{\mu ^2}\right)\Bigg\}\nonumber \\ 
C^V_{fin} &=& \frac{1}{36 (4 \pi )^2 p^2}\int_0^1 dx \Bigg\{3 m^2 \left(m_h^2-m^2+p^2\right) \ln \left(\frac{m^2}{\mu ^2}\right)\nonumber \\
&+&3 m_h^2 \left(-m_h^2+m^2+p^2\right) \ln \left(\frac{m_h^2}{\mu ^2}\right)+6 m_h^2 \left(p^2-m^2\right)-5 p^2 \left(3 m_h^2-9 m^2+p^2\right)\nonumber \\
&+&3 \left(2 m_h^2 \left(p^2-m^2\right)+m_h^4+m^4-10 m^2 p^2+p^4\right) \ln \left(\frac{ p^2 x(1-x)+xm^2+(1-x) m_h^2}{\mu ^2}\right)\nonumber \\
&+& 3 m_h^4+3 m^4-54 m^2 p^2+3 p^4\Bigg\}.
\eeq
The divergent terms $(\delta A^V_{div}, \delta B^V_{div}, \delta C^V_{div})$ can be eliminated by means of the Lagrangian counterterms as well as by suitable counterterms in the external part of the action $S^V_J$ accounting for the introduction of the composite operator $V'_{\m}(x)$, i.e. 

\beq
S^V_J&=& S +
\int d^4 x \Bigg[ ( 1+\delta Z_{div}^{V,0}) J_{\m}(x) V_{\m}(x) + ( 1+\delta Z^V_{div}) \frac{J_{\m}(x)J_{\m}(x)}{2}\Bigg],
\label{1o1}
\eeq
where $J_{\m}(x)$ is a BRST invariant dimension one source needed to define the generator $Z^c(J)$ of the connected Green function $\braket{V'_{\m}(x)V'_{\n}(y)}$:

\beq
\braket{V'_{\m}(x)V'_{\n}(y)}=\frac{\delta^2 Z^c(J)}{\delta J_{\m}(x) \delta J_{\n}(y)}\vert_{J=0},
\eeq
and like in the scalar case, we have the freedom of introducing a pure contact BRST invariant term in the external source $J_{\m}(x)$:

\beq
\int d^4 x \,\frac{1}{2}(\b v^2 \,J_{\m}(x)J_{\m}(x)+\gamma J_{\m}(x)\partial^2 J_{\m}(x)+\sigma (\partial_{\m}J_{\m}(x))^2),
\label{3dd}
\eeq
which can be arbitrarily added to the action \eqref{1o1}. Including such a term in \eqref{1o1} will have the effect of adding a dimensionless constant to $G^T_{VV}(p^2)=\braket{V'(p)V'(-p)^T}$, i.e.

\beq
G^T_{VV}(p^2)\rightarrow G^T_{VV}(p^2)+\beta v^2+\gamma p^2 
\label{3e},
\eeq
where we notice that the last term in \eqref{3dd} does not contribute to the transversal part of the propagator.
In particular, $\beta$ and $\gamma$ can be choosen so that \eqref{3e} becomes

\beq
G^T_{VV}(p^2)-G^T_{VV}(0)-p^2 \frac{\partial G^T_{VV}(p^2)}{\partial p^2}\vert_{p=0},
\label{dv2}
\eeq
see Figure \ref{propAAT13}. It is a Green's function that obeys a two substracted KL representation, see section \ref{IIII}.
Following the steps in the same way as for the scalar composite field, \eqref{1o}-\eqref{2o}, we find 

\beq
\langle V'(p)V'(-p)\rangle^T&=&\frac{e^2 v^4}{4}\frac{1}{p^2+m^2}+ \frac{\hbar e^2 v^4}{4}\frac{\Pi^T_{VV}(p^2)}{(p^2+m^2)^2}+ \mathcal{O}(\hbar^2),
\eeq
with 
\beq
\Pi^T_{VV}(p^2)
&=&-\frac{1}{9 (4\pi)^2 e^2 v^4 m_h^2}\int_0^1 dx\Bigg\{-18 m^4 (m_h^4+m^4)+9 m_h^2 p^4 (m_h^2+m^2) \nonumber \\
&-&3 m_h^2 p^2 \big[2 p^2 (m_h^2-5 m^2)+(m_h^2-m^2)^2+p^4\big] \ln \left(\frac{m_h^2 (1-x)+m^2 x+p^2 (1-x) x}{\mu ^2}\right)+2 m_h^2 p^6 \nonumber \\
&+&3 m_h^4 \ln \left(\frac{m_h^2}{\mu ^2}\right) \big[p^2 (m_h^2+11 m^2)+6 m^4-p^4\big] \nonumber \\
&+&3 m^2 \big[-m_h^2 p^4+p^2 (-m_h^4+m_h^2 m^2+36 m^4)+18 m^6\big] \ln \left(\frac{m^2}{\mu ^2}\right) \nonumber \\
&-&3 p^2 (m_h^6+10 m_h^4 m^2+m_h^2 m^4+12 m^6)\Bigg\},
\eeq
and following the steps \eqref{u1}-\eqref{jju2}, we find 
\beq
G_{VV}^T&=& \frac{e^2 v^4}{4}\Big(\frac{1}{p^2+m^2-\Pi^T_{VV}(p^2)}\Big)+ C_{VV}(p^2)
\label{GAAA}
\eeq
with 

\beq
\hat{\Pi}^{T}_{VV}(p^2)
&=&-\frac{1}{9 (4\pi)^2 e^2 v^4 m_h^2}\int_0^1 dx\Bigg\{-18 m^4 (m_h^4+m^4)+9 m_h^2 p^4 (m_h^2+m^2) \nonumber \\
&-&3 m_h^2 \big[2 (-2m^2p^2-m^4) (m_h^2-5 m^2)\\
&+&p^2(m_h^2-m^2)^2-2m^2p^2-m^4 \big] \ln \left(\frac{m_h^2 (1-x)+m^2 x+p^2 (1-x) x}{\mu ^2}\right) \nonumber \\
&+&2 m_h^2 p^6+3 m_h^4 \ln \left(\frac{m_h^2}{\mu ^2}\right) \big[p^2 (m_h^2+11 m^2)+6 m^4-p^4\big] \nonumber \\
&+&3 m^2 \big[-m_h^2 p^4+p^2 (-m_h^4+m_h^2 m^2+36 m^4)+18 m^6\big] \ln \left(\frac{m^2}{\mu ^2}\right) \nonumber \\
&-&3 p^2 (m_h^6+10 m_h^4 m^2+m_h^2 m^4+12 m^6)\Bigg\},
\eeq
and
\beq
C_{VV}(p^2)&=&\frac{1}{12 (4\pi)^2 }\int_0^1 dx\Bigg\{
(2  m_h^2+p^2) \ln \left(\frac{m_h^2 (1-x)+m^2 x+p^2 (1-x) x}{\mu ^2}\right)\Bigg\}.
\eeq
The resummed form factor \eqref{GAAA} is depicted in Figure \ref{propAAT1}, as well as the modified version \eqref{dv2} and their second derivative, which is imported for the spectral analysis in section \ref{IIII}.

		\begin{figure}[H]
	\center
	\includegraphics[width=12cm]{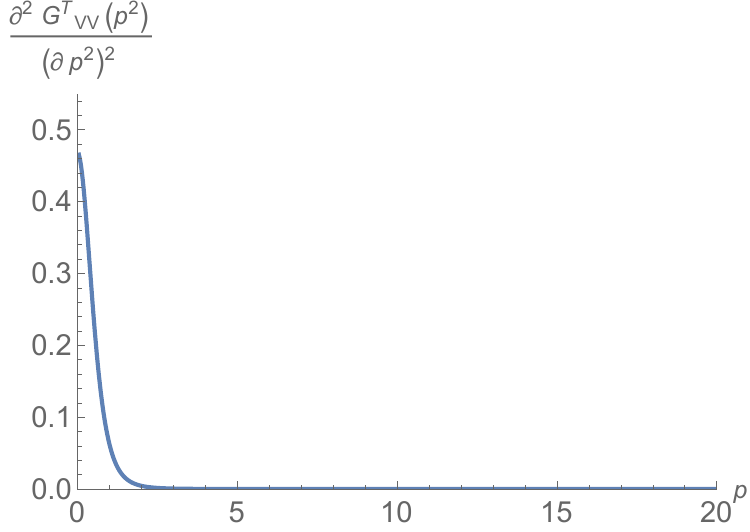}
	\caption{Second derivative of the form factor.
		The momentum $p$ is given in units of the energy scale $\mu$, for the parameter values $e=1$, $v=1 \, \mu$, $\lambda=\frac{1}{5}$.}
	\label{propAAT}
\end{figure}

For the longitudinal part of the propagator (see appendix \ref{ah} for details) , we find the divergent part

\beq
\langle V'(p)V'(-p)\rangle^L_{div}&\overset{\epsilon \rightarrow 0}{=}& -\frac{v^2 \left(3 e^4+\lambda ^2\right)}{(4 \pi) ^2 \lambda  \epsilon }
\eeq
and the total finite correction up to first order in $\hbar$ is given by

\beq
\langle V'(p)V'(-p)\rangle^L_{fin}&=& v^2	-\Bigg(\frac{m_h^4-m_h^4 \ln \left(\frac{m_h^2}{\mu ^2}\right)+m^4-3 m^4 \ln \left(\frac{m^2}{\mu ^2}\right)}{32 \pi ^2 m_h^2} \Bigg)\nonumber \\
\eeq
which means the longitudinal part of the propagator $\braket{V'_{\mu}(x),V'_{\nu}(y)}$ is not propagating.

\section{Spectral properties of the gauge-invariant local  operators $(V_{\mu}(x),O(x))$ \label{IIII}}
In this section, we will calculate the spectral properties associated with the correlation function obtained in the last section. In \ref{2a}, we will shortly review the techniques employed in the last chapter to obtain the pole mass, residue and spectral density up to first order in $\hbar$. In \ref{2c}, the spectral properties of the composite operators $(V_{\mu}(x),O(x))$ are discussed.

\subsection{Obtaining the spectral function \label{2a}}
For elementary fields we obtained the spectral density function by comparing the K\"all\'{e}n-Lehmann  spectral representation for the propagator of a generic field $\widetilde{O}(p)$
\beq
\braket{\widetilde{O}(p)\widetilde{O}(-p)}=G(p^2)=\int_0^{\infty} dt \frac{\rho (t)}{t+p^2},
\label{ltt}
\eeq
where $\rho(t)$ is the spectral density function and $G(p^2)$ stands for the  resummed propagator
\beq
G(p^2)&=& \frac{1}{p^2+m^2-\Pi(p^2)}.
\label{iuh}
\eeq
For higher-dimensional operators, the resummed propagator acquires an overall (dimensionful) factor identical to the one appearing in its tree level result, as we have seen in section \ref{fmsp}.
We also note that in the case of higher dimensional operators, the spectral representation,
eq.\eqref{ltt}, might require appropriate subtraction terms in order to ensure convergence. A standard way to cure this problem is to
subtract from $G(p^2)$ the first few terms of its Taylor expansion at $p=0$ \cite{colangelo2001qcd}, making the integral more and more convergent. These subtraction terms are directly related to the renormalization of the composite operators, and one can see that the modified Green's functions for the composite scalar field \eqref{dv1} and for the composite vector field \eqref{dv2} are in fact subtractions of the Taylor series to first and second order, respectively. In our theory we can make use of the subtracted equations at $p=0$ because all fields are massive in the $R_{\xi}$-gauge, so there are no divergences at zero momentum. Also, we stress that the spectral function $\rho(t)$ is not affected by the subtraction procedure. Moreover, we can see that these subtractions do not have an influence on the (second) derivative of the propagator. For the scalar composite operator

\beq
\frac{\partial (G_{OO}(p^2)-G_{OO}(0))
}{\partial p^2}= \frac{\partial G_{OO}(p^2)}{\partial p^2}=-\int_0^{\infty} dt \frac{\rho (t)}{(t+p^2)^2},
\eeq
which means that for a positive spectral function the first derivative of $G_{OO}$ is strictly negative, as is confirmed in Figure \ref{prophh3}. For the vector composite operator 
\beq
\frac{\partial^2 (G_{VV}(p^2)-G_{VV}(0)-p^2 G'_{VV}(0))
}{(\partial p^2)^2}= \frac{\partial^2 G_{VV}(p^2)}{(\partial p^2)^2}= 2\int_0^{\infty} dt \frac{\rho (t)}{(t+p^2)^3},
\eeq
which should be strictly positive for a positive spectral function, as is shown in Figure \ref{propAAT}. Remember, however, that we can also obtain the spectral function directly, by the methods developed in section \eqref{32}.
\subsection{Spectral properties of the gauge-invariant composite operators $V_{\mu}(x)$ and $O(x)$  \label{2c}}

\begin{figure}[H]
	\center
	\includegraphics[width=12cm]{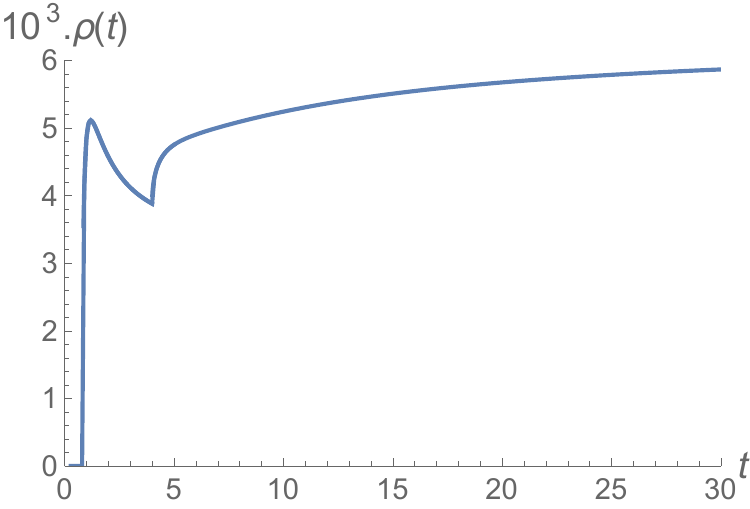}
	\caption{Spectral function for the reduced propagator of the scalar composite operator, $\langle {O}(p){O}(-p) \rangle$, with $t$ given in units of $\mu^2$, for the parameter values $e=1$, $v=1 \, \mu$, $\lambda=\frac{1}{5}$.}
	\label{Y44}
\end{figure}

For the scalar composite operator ${O}(x)$ with two-point function given by expression \eqref{dk34}, we find the first-order pole mass for our set of parameter values to be
\beq
m^2_{h,pole}=0.213472\, \mu^2,
\label{rar}
\eeq
which is exactly equal to the pole mass of the elementary Higgs field correlator. Following the steps from \ref{31}, we find the first-order residue to correct the tree-level result $Z_{\text{tree}}=v^2$ by $\sim 7\%$:
\beq
Z=v^2(1+\partial_{p^2}\Pi_{OO}(p^2)_{p^2=-m_h^2})=1.06577 v^2\,,
\eeq
while the first-order spectral function is shown in Figure \ref{Y44}. Similarly as for the spectral function of the Higgs field in Figure \ref{Y38}, one finds a two-particle threshold for Higgs pair production at $t=(m_h+m_h)^2= 0.8\, \mu^2$, and a two-photon state starting at $t=(m+m)^2=4 \,\mu^2$. The difference is that for this gauge-invariant correlation function we no longer have the unphysical Goldstone two-particle state. Due to the absence of this negative contribution, the spectral function is always positive. Therefore, this quantity is suitable for describing a physical Higgs excitation spectrum as opposed to the elementary propagator $\langle hh\rangle$. \\
\\
Finally, it is interesting to note that below the unphysical threshold the elementary correlator displays the same qualitative spectral properties as this gauge-invariant approach. This means that spectral description of the physical Higgs mode could in principle be successfully encoded in the elementary propagator in the unitary gauge, in which $\xi\to \infty $ and the Goldstone bosons are infinitely heavy. We shall make an explicit comparison in section \ref{unitarylimit}.

\begin{figure}[H]
	\center
	\includegraphics[width=10cm]{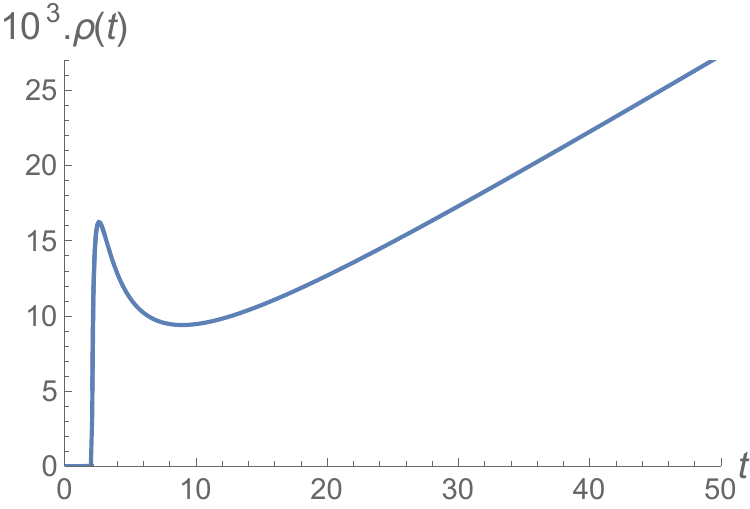}
	\caption{Spectral function for reduced transverse propagator of the vector composite operator $\langle {V}(p){V}(-p) \rangle^T$, with $t$ given in units of $\mu^2$, for the parameter values $e=1$, $v=1 \, \mu$, $\lambda=\frac{1}{5}$.}
	\label{W}
\end{figure}

For the transverse vector composite operator $V^T_{\m}(x)$, with our set of parameters we find the first-order pole mass
\beq
m^2_{pole}=1.05417 \mu^2,
\eeq
which is -- as expected from the Nielsen identities -- exactly the same as the pole mass of the transverse photon field correlator \eqref{polem3}. Furthermore, we find the first-order residue 
\beq
Z=\frac{e^2v^4}{4}(1+\partial_{p^2}\Pi_{VV}^T(p^2)_{p^2=-m^2})=1.09332 \frac{e^2v^4}{4}\,,
\eeq
and the first order spectral density for the reduced propagator is displayed in Figure \ref{W}. Like the photon spectral density in Figure \ref{Y4}, we find a photon-Higgs two-particle state at $t=(m_h+m)^2=2.09 \, \mu^2$, and the spectral density is positive for all values of $t$.
\subsection{Pole masses}
We can explain the fact that the pole mass of the elementary propagator equals that of its gauge-invariant extension in a qualitative way by looking at the definition of a first-order pole mass, eq.~\eqref{pol}. When calculating one-loop corrections to the two-point function of the composite operators $O(p)$ and $R_{\mu}^a(p)$, we find that
\beq
\Pi_{\rm composite}(p^2)&=& \Pi_{\rm elementary}(p^2)+ \Pi_{\rm 1-leg}(p^2) (p^2+m^2)+\Pi_{\rm 0-leg}(p^2)(p^2+m^2)^2,
\eeq
where $\Pi_{\rm 1-leg}(p^2)$ and $\Pi_{\rm 0-leg}(p^2)$ are the composite one-loop contributions to the correction of the composite field's two-point functions, with one external leg and zero external legs, respectively. From this, we see immediately that
\beq
\Pi_{\rm composite}(-m^2)&=& \Pi_{\rm elementary}(-m^2)
\eeq
and therefore, up to first order in $\hbar$, we find
\beq
m^2_{\rm pole, composite}&=&m^2_{\rm pole, elementary}
\label{zlf}
\eeq
which means that the elementary operators and their composite extensions share the same mass. This is an important feature, providing an alternative way to the Nielsen identities, to understand  why the pole masses of the elementary particles are gauge invariant  and not just gauge parameter independent.

\section{Unitary limit \label{unitarylimit}}
\begin{figure}[H]
	\center
	\includegraphics[width=15cm]{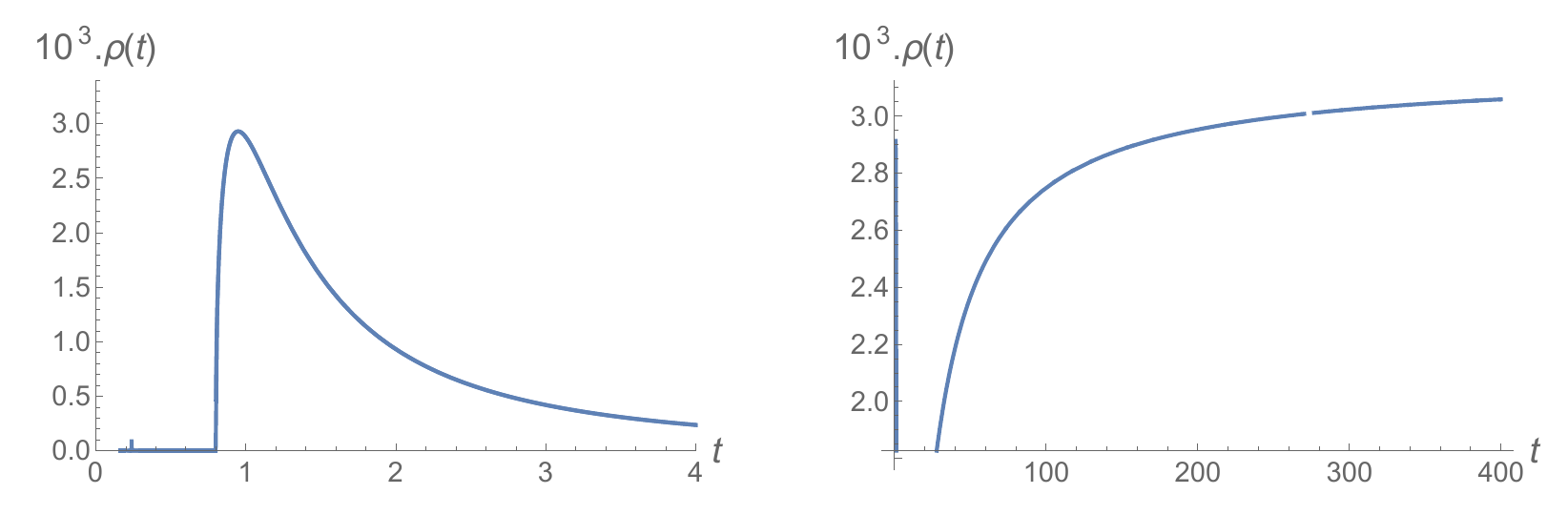}
	\caption{Spectral function for the reduced elementary propagator $\langle h(p) h(-p) \rangle$ for small values of $t$ (left) and large values of $t$ (right), with $t$ given in units of $\mu^2$, for $\xi=1000$ the parameter values $e=1$, $v=1 \, \mu$, $\lambda=\frac{1}{5}$.}
	\label{W3}
\end{figure}

\begin{figure}[H]
	\center
	\includegraphics[width=15cm]{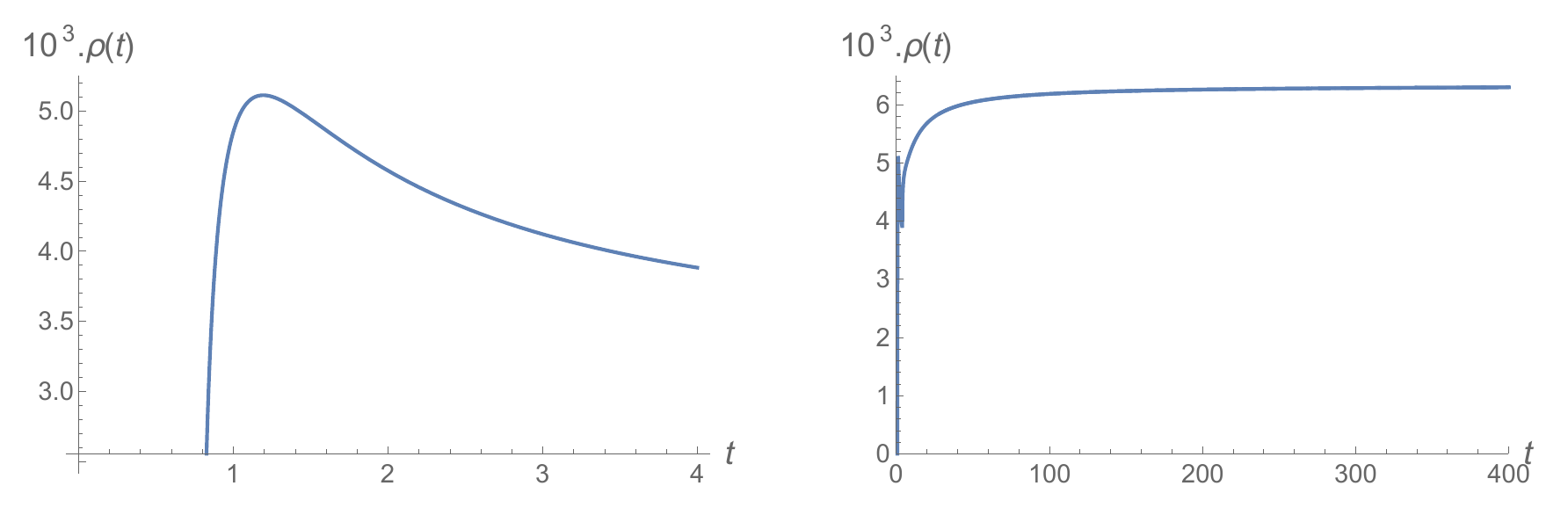}
	\caption{Spectral function for the reduced composite propagator $\langle O(p),O(-p) \rangle$ for small values of $t$ (left) and large values of $t$ (right), with $t$ given in units of $\mu^2$, for the parameter values $e=1$, $v=1 \, \mu$, $\lambda=\frac{1}{5}$.}
	\label{hkg}
\end{figure}

It is well-known \cite{Peskin:1995ev} that for the Higgs model, the unitary gauge represents the physical gauge, as it decouples the unphysical fields, i.e. the ghost field and the Goldstone field. The unitary gauge can be formally obtained in the $R_{\xi}$-gauges by taking $\xi \rightarrow \infty$. However, this gauge is non-renormalizable, as one can see by looking at this limit for the tree-level propagator of the photon field

\beq
\langle A_{\m}(p)A_{\n}(-p)\rangle_{\text{tree}} &\overset{\xi \rightarrow \infty}{=}& \frac{1}{p^2+ m^2}{\mathcal{P}}_{\m\n}+\frac{1}{m^2}\mathcal{L}_{\m\n}.
\eeq
Nonetheless, we can approximate the unitary gauge by taking large values of $\xi$. This is especially interesting when looking at the spectral function of the elementary Higgs field, which is $\xi$-dependent. In Figure \ref{W3} one finds the spectral function for $\xi= 1000$ for small and large ranges of $t$. In Figure \ref{hkg} we show the spectral function of the scalar composite field $O(x)$ for the same ranges of $t$. As one can see, the pictures are qualitatively very similar. This means that when approximating the unitary gauge, the spectral function of the gauge-dependent, elementary field $h(x)$ approximates that of its composite, gauge-invariant counterpart. 


\section{Conclusion\label{VI}}

In the present chapter, following the local gauge-invariant setup of \cite{hooft1980we,hooft2012nonperturbative,frohlich1980higgs, frohlich1981higgs}, we have evaluated at one-loop order the two-point correlation functions $\langle V'_{\mu}(x) V'_{\nu}(y) \rangle$, $\langle O(x) O(y) \rangle$ of the two local gauge-invariant operators ${V}'_{\mu}(x)= -i \varphi^{\dagger}(x) D_{\mu} \varphi(x)+i \partial_{\m} O(x)$ and ${O}(x)= \varphi^{\dagger}(x) \varphi(x)-\frac{v}{2}$ in the $U(1)$ Abelian Higgs model quantized in the $R_\xi$ gauge. Our results can be summarized as follows:
\begin{itemize}
	\item both $\langle V'_{\mu}(x) V'_{\nu}(y) \rangle$ and  $\langle O(x) O(y) \rangle$ do not depend on the gauge parameter $\xi$, as expected;
	\item the pole masses of $\langle V'_{\mu}(x) V'_{\nu}(y) \rangle^T$ and  $\langle O(x) O(y) \rangle$ are exactly the same as those of the correlation functions of the elementary fields $\langle A_\mu(x) A_\nu(y) \rangle^{T}$ and $\langle h(x) h(y) \rangle$, respectively, where $\langle \cdots \rangle_{T}$ stands for the transverse component of the corresponding propagator;
	\item the spectral densities of the K\"all\'{e}n-Lehmann representation of the correlation functions $\langle V'_{\mu}(x) V'_{\nu}(y) \rangle$ and  $\langle O(x) O(y) \rangle$ turn out to be always positive, in contrast to the one associated with the (gauge-dependent) elementary Higgs field.
	
\end{itemize}
These important features give us a fully gauge-invariant picture in order to describe the spectrum of elementary excitations of the model, i.e. the massive photon and the Higgs mode.

\newpage
\chapter{Spectral properties of local BRST invariant composite operators in the $SU(2)$ Yang-Mills-Higgs model \label{VIII}}

	In this chapter we will extend the techniques of chapter
 \ref{VII} and \ref{VV} to the more complex case of $SU(2)$ Higgs model with a single Higgs field in the fundamental representation. This model is equal to the electroweak model from chapter \ref{hew} with the coupling constant of the hypercharge interaction taken to zero, $g' \rightarrow 0$. As we have seen for the Abelian gauge theory, the direct use of non-gauge invariant fields displays several limitations in the spectral representation, and this will become more severe in the case of a non-Abelian gauge theory. For instance, in the case of the $U(1)$ Higgs model, the transverse component of the Abelian gauge field $A_\mu$ is gauge invariant, so that the two-point correlation function
	$\mathcal{P}_{\mu\nu}(p) \langle A_\mu(p) A_\nu(-p) \rangle$, where  $\mathcal{P}_{\mu\nu}(p) = (\delta_{\mu\nu} -\frac{p_\mu p_\nu}{p^2})$ is the transverse  projector, turns out to be independent from the gauge parameter $\xi$. However, this is no more true in the non-Abelian case, where both Higgs and gauge boson two-point functions, {\it i.e.}~$\braket{h(p) h(-p)}$ and $\mathcal{P}_{\mu\nu}(p) \langle A^a_\mu(p) A^b_\nu(-p) \rangle$, where $h$ stands for the Higgs field and $A^a_\mu$ for the gauge boson field,  exhibit a strong gauge dependence from $\xi$. As a consequence, the understanding of the two-point  correlation functions of both Higgs field $h$ and gauge vector boson $A^a_\mu$ in terms of the K\"all\'en-Lehmann (KL) spectral representation is completely jeopardized by an unphysical dependence from the gauge parameter $\xi$, obscuring a direct interpretation of the above mentioned correlation functions in terms of the elementary excitations of the physical spectrum, namely the Higgs and the vector gauge boson particles.\\
	\\
	 We also note here that from a lattice perspective, it is expected that the spectrum of a gauge (Higgs) theory should be describable in terms of local gauge invariant operator correlation functions, with concrete physical information hiding in the various (positive and gauge invariant) spectral functions, not only pole masses, decay widths, but also transport coefficients at finite temperature etc. Clearly, such information will not correctly be encoded in gauge variant, non-positive spectral functions.
	\\
	\\	
	 As we discussed in section \ref{cus}, besides the exact BRST invariance, the quantized theory exhibits a global $SU(2)$ symmetry known as custodial symmetry.  In this chapter we will show how the local composite BRST invariant operators corresponding to the gauge bosons  transform as a triplet under the custodial symmetry, a property which will imply useful relations for their two-point correlation functions.\\
\\
\\
This chapter is organized as follows. In section~\ref{I3}, we give a review of the $SU(2)$ HYM model with a single Higgs field in the fundamental representation, of the gauge fixing procedure and its ensuing BRST invariance.  In section~\ref{III} we calculate the two-point correlation functions of the elementary fields up to one-loop order. In section~\ref{IIIIt}, we define the BRST invariant local composite operators  $(O(x), R^a_\mu(x))$ corresponding to the BRST invariant extension of $(h,A^a_\mu)$ and calculate their one-loop correlation functions. In section~\ref{V}, we discuss the spectral properties of both  elementary and composite operators. In order to give a more general idea of the  behavior of the spectral functions, we shall be using two sets of parameters which we shall refer as to  Region I and Region II. To some extent, Region I can be associated to the perturbative weak coupling regime, while in Region II we keep the gauge coupling a little bit larger,  while decreasing the minimizing value $v$ of the Higgs field. Section~\ref{VI} is devoted to our conclusion. The technical details are all collected in the appendices.

\section{The action and its symmetries \label{I3}}
The YM action with a single Higgs field in the fundamental representation is given by
\begin{eqnarray}
S_0&=& \int d^4x \left\{ \frac{1}{4}F_{\mu\nu}^a F_{\mu\nu}^a+(D_{\mu}^{ij}\Phi^{\dagger j})(D_{\mu}^{ik}\Phi^k)+\frac{\lambda}{2}(\Phi^{\dagger i}\Phi^i-\frac{1}{2}v^2)^2\right\}\nonumber \\
&=& S_{\rm YM}+S_{\rm Higgs} \label{1}
\end{eqnarray}
with
\begin{eqnarray}
F_{\mu\nu}=\partial_{\mu}A_{\nu}^a-\partial_{\nu}A_{\mu}^a+g \epsilon^{abc}A_{\mu}^bA_{\nu}^c
\end{eqnarray}
and
\begin{eqnarray}
D_{\mu}^{ij}\Phi^j=\partial_{\mu}\Phi^i-\frac{i}{2}g(\tau^a)^{ij}A_{\mu}^a\Phi^j, \,\,\, (D_{\mu}^{ij}\Phi^j)^{\dagger}=\partial_{\mu}\Phi^{i\dagger}+\frac{i}{2}g\Phi^{j\dagger}(\tau^a)^{ji}A_{\mu}^a,
\end{eqnarray}
with the Pauli matrices $\tau^a (a=1,2,3)$ and the Levi-Civita tensor  $\epsilon^{abc}$ referring to the gauge symmetry group $SU(2)$. The scalar complex field $\Phi^i(x)$ is in the fundamental representation of $SU(2)$,
{\it i.e.}~$i,j=1,2$. Thus, $\Phi$ is an $SU(2)$-doublet of complex scalar fields that can be written as
\beq
\Phi&=&\frac{1}{\sqrt{2}}\begin{pmatrix}
	\phi^+ \\
	\phi^0
\end{pmatrix}=\frac{1}{\sqrt{2}}\begin{pmatrix}
	\phi_1+i \phi_2 \\
	\phi_3+i \phi_4
\end{pmatrix}.
\eeq
The configuration which minimizes the Higgs potential in the expression \eqref{1}  is
\begin{eqnarray}
\braket{\Phi}&=\frac{1}{\sqrt{2}}
\begin{pmatrix}
v \\
0 \end{pmatrix} \label{aa}
\end{eqnarray}
and we write down  $\Phi(x) $ as an expansion  around the configuration \eqref{aa}, so that
\begin{eqnarray}
\Phi= \frac{1}{\sqrt{2}}
\begin{pmatrix}
v+h+i\rho_3 \\
i\rho_1-\rho_2\end{pmatrix}, \label{expansion1}
\end{eqnarray}
where $h$ is the Higgs field and $\rho^a$, $a=1,2,3$, the would-be Goldstone bosons. We can use the  matrix notation\footnote{ This is of course possible thanks to the fact that $\Phi$ counts 3 Goldstone modes and that $SU(2)$ has three generators. This ``numerology'' is essentially what leads to a large custodial symmetry in the $SU(2)$ case. }:
\begin{eqnarray}
\Phi=\frac{1}{\sqrt{2}}((v+h)\textbf{1}+i\rho^a \tau^a) \cdot
\begin{pmatrix}
1\\
0
\end{pmatrix}, \label{expansion2}
\end{eqnarray}
so that the second term in eq.~\eqref{1} becomes
\begin{eqnarray}
(D^{ij}_{\mu}\Phi^j)^{\dagger}D_{\mu}^{ik}\Phi^k &=& \frac{1}{2} (1,0) \cdot \Bigg[\partial_{\mu}h \cdot \textbf{1}-i\partial_{\mu}\rho^a \tau^a+\frac{ig}{2}\tau^a A_{\mu}^a \Big((v+h)\textbf{1}\nonumber\\
&&-i\rho^b\tau^b\Big)\Bigg]\times\Bigg[\partial_{\mu}h \cdot \textbf{1}+i\partial_{\mu}\rho^c \tau^c-\frac{ig}{2} ((v+h)\textbf{1}\nonumber\\
&&-i\rho^d\tau^d\Big)\tau^c A_{\mu}^c\Bigg] \cdot \begin{pmatrix} 1\\
0 \end{pmatrix}\nonumber\\
&=& \tilde{\mathcal{L}_0}+ \tilde{\mathcal{L}_1}+ \tilde{\mathcal{L}_2},
\end{eqnarray}
with $\tilde{\mathcal{L}_i}$ the $i$th term in powers of $A_{\mu}$:
\begin{eqnarray}
\tilde{\mathcal{L}}_0&=&\frac{1}{2}\left((\partial_{\mu}h)^2+\partial_{\mu}\rho^a\partial_{\mu}\rho^a\right), \nonumber\\
\tilde{\mathcal{L}}_1&=&-\frac{1}{2}\left\{gv A_{\mu}^a\partial_{\mu} \rho^a-gA_{\mu}^a\rho^a\partial_{\mu}h+gA_{\mu}^a(\partial_{\mu}\rho^a)h+g \epsilon^{abc} \partial_{\mu} \rho^a \rho^b A_{\mu}^c \right\},\nonumber\\
\tilde{\mathcal{L}_2}&=&\frac{g^2}{8}A_{\mu}^a A_{\mu}^a \left[(v+h)^2+\rho^b\rho^b \right],
\end{eqnarray}
and we have the full action
\begin{eqnarray}
S_0&=&\int d^4x \frac{1}{2}\Bigg\{\frac{1}{2}F^a_{\mu\nu}F^a_{\mu\nu}+\frac{1}{4}v^2g^2A_{\mu}^aA_{\mu}^a+(\partial_{\mu}h)^2+\partial_{\mu}\rho^a\partial_{\mu}\rho^a-gvA_{\mu}^a\partial_{\mu}\rho^a+gA_{\mu}^a\rho^a\partial_{\mu}h-gA^a_{\mu}(\partial_{\mu}\rho^a)h\nonumber\\
&-&g \epsilon^{abc}\partial_{\mu}\rho^a\rho^b A_{\mu}^c+\frac{g^2}{4}A_{\mu}^a A_{\mu}^a\left[2vh+h^2+\rho^b\rho^b\right]+\lambda v^2h^2+\lambda v h (h^2+\rho^a \rho^a)\nonumber\\
&+&\frac{\lambda}{4}(h^2+\rho^a\rho^a)^2 \Bigg\}.
\end{eqnarray}
One sees that both gauge field $A^a_\mu$ and Higgs field $h$ have acquired a mass given, respectively, by
\beq
m^2= \frac{1}{4} g^2 v^2, \,\,\,\,\,\,\, m_h^2= \lambda v^2 \;.
\eeq
\subsection{Gauge fixing and BRST symmetry}
The action \eqref{1} is invariant under the local $\omega$-parametrized gauge transformations
\begin{equation}
\delta A_{\mu}^a=-D_{\mu}^{ab}\omega^b, \,\,\, \delta \Phi=-\frac{ig}{2}\omega^a\tau^a\Phi, \,\,\, \delta \Phi^{\dagger}=\frac{ig}{2}\omega^a \Phi^{\dagger}\tau^a,
\end{equation}
which, when written in terms of the fields $(h,\rho^a)$, become
\begin{equation}
\delta h = \frac{g}{2}\omega^a \rho^a, \,\,\, \delta \rho^a=-\frac{g}{2}(\omega^a(v+h)\textbf{1}-\epsilon^{abc}\omega^b\rho^c).
\end{equation}
As done in the $U(1)$ case, we shall be using the $R_{\xi}$-gauge. We add thus need the gauge fixing term
\beq
\mathcal{S}_{\rm gf}&=&s \int d^4 x  \Bigl\{-i \frac{\xi}{2}\bar{c}^a b^a+\bar{c}^a(\partial_{\m} A_{\m}^a-\xi m \rho^a)\Bigr\} \nonumber\\
&=& \frac{1}{2}\int d^4 x \Bigl\{ \xi b^a b^a+ 2 i b^a \partial_{\m}  A_{\m}^a+2\bar{c}^a \partial_{\m}D_{\m}^{ab}c^b-2i \xi m b^a \rho^a \nonumber\\
&-& 2 \xi \bar{c}^a m c^a - g \xi \bar{c}^a m h c^a - \xi g \epsilon^{abc} \bar{c}^a  c^b \rho^c\Bigr\} \;,
\eeq
so that the gauge fixed action $S_{\text{full}} = S_0 +\mathcal{S}_{\rm gf} $, namely
\begin{eqnarray}
S_{\text{full}}&=&\int d^4x \frac{1}{2}\Bigg\{\frac{1}{2}F^a_{\mu\nu}F^a_{\mu\nu}+\frac{1}{4}v^2g^2A_{\mu}^aA_{\mu}^a\nonumber\\&+&(\partial_{\mu}h)^2+\partial_{\mu}\rho^a\partial_{\mu}\rho^a-gvA_{\mu}^a\partial_{\mu}\rho^a+gA_{\mu}^a\rho^a\partial_{\mu}h-gA^a_{\mu}(\partial_{\mu}\rho^a)h\nonumber\\
&-&g \epsilon^{abc}\partial_{\mu}\rho^a\rho^b A_{\mu}^c+\frac{g^2}{4}A_{\mu}^a A_{\mu}^a\left[2vh+h^2+\rho^b\rho^b\right]+\lambda v^2h^2\nonumber\\
&+&\lambda v h (h^2+\rho^a \rho^a)+\frac{\lambda}{4}(h^2+\rho^a\rho^a)^2 +\xi b^a b^a+ 2 i b^a \partial_{\m}  A_{\m}^a+2\bar{c}^a \partial_{\m}D_{\m}^{ab}c^b-2i \xi m b^a \rho^a \nonumber\\
&-& 2 \xi \bar{c}^a m c^a - g \xi \bar{c}^a m h c^a - \xi g \epsilon^{abc} \bar{c}^a  c^b \rho^c \Bigg\}
\end{eqnarray}
turns out to be left  invariant by the BRST transformations
\begin{eqnarray}
sA_{\mu}^a&=&-D_{\mu}^{ab}c^b, \,\,\, s h = \frac{g}{2}c^a \rho^a, \,\,\, s \rho^a=-\frac{g}{2}(c^a(v+h)\textbf{1}-\epsilon^{abc}c^b\rho^c)\nonumber\\
sc^a&=&\frac{1}{2}g \epsilon^{abc}c^bc^c, \,\,\, s \bar{c}^a=ib^a, \,\,\, sb^a=0 \;,
\end{eqnarray}
\begin{equation}
s S_{\text{full}} = 0 \;. \label{brstgt}
\end{equation}
\subsection{Custodial symmetry \label{cust}}
As already mentioned, apart from the BRST symmetry, there is an extra global symmetry, which we shall refer to as the custodial symmetry:
\beq
\delta A^a_\mu &=& \epsilon^{abc} \beta^b A^c_\mu, \nonumber\\
\delta \rho^a &=& \epsilon^{abc} \beta^b \rho^c,                      \nonumber\\
\delta \overline{c}^a&=& \epsilon^{abc} \beta^b \overline{c}^c, \nonumber
\\
\delta c^a&=& \epsilon^{abc} \beta^b c^c,
\nonumber\\
\delta b^a&=& \epsilon^{abc} \beta^b b^c, \nonumber\\
\delta h&=&0 \;, \label{custodial}
\eeq
where $\beta^a$ is a constant parameter, $\partial_{\mu}\beta^a=0$,
\begin{equation}
\delta S_{\text{full}} = 0 \;. \label{beta}
\end{equation}
One notices that all fields carrying the index $a=1,2,3$, {\it i.e.}~$(A^a_\mu, b^a, c^a, {\bar c}^a, \rho^a)$, undergo a global transformation in the adjoint representation of $SU(2)$. The origin of this symmetry is an $SU(2)_{\rm gauge} \times SU(2)_{\rm global}$ symmetry of the action in the unbroken phase, see section \ref{cus}. The exception is the Higgs field $h$, which is left invariant, {\it i.e.}~it is a singlet. As we shall see in the following, this additional global symmetry will provide useful relationships for the two-point correlation functions of the BRST invariant composite operators.
\section{One-loop evaluation of the
	correlation function of the elementary fields \label{III}}
For the elementary fields $h(x)$ and $A^a_{\mu}$, the correlation functions are calculated in Appendices \ref{hp} and \ref{Ap}. For the Higgs field, for the one-loop propagator we get
\beq
\braket{h(p)h(-p)}&=&\frac{1}{p^2+m_h^2}+\frac{1}{(p^2+m_h^2)^2} \Pi_{hh}(p^2)+\mathcal{O}(\hbar^2)
\label{hh}
\eeq
where
\beq
\Pi_{hh}(p^2)&=&\frac{3g^2}{{8 (4 \pi )^2}}\int_0^1 dx \, \Bigg\{2  \xi  \left(m_h^2+p^2\right) \ln \left(\frac{m^2 \xi }{\mu ^2}\right)-2 \xi  m_h^2+2 \left(6 m^2-p^2\right) \ln \left(\frac{m^2}{\mu ^2}\right)\nonumber\\
&-&(12 m^2+\frac{p^4}{m^2}+4 p^2) \ln \left(\frac{m^2+p^2 (1-x) x}{\mu ^2}\right)+\left(\frac{p^4 }{m^2}-\frac{m_h^4}{m^2}\right)\ln \left(\frac{m^2 \xi +p^2 (1-x) x}{\mu ^2}\right)\nonumber\\
&-&12 m^2-2 \xi  p^2+2 p^2-\frac{ m_h^4}{m^2} \left(-2 \ln \left(\frac{m_h^2}{\mu ^2}\right)+3 \ln \left(\frac{m_h^2+p^2 (1-x) x}{\mu ^2}\right)+2\right)\Bigg\}.
\label{hhf}
\eeq
Before trying to resum the self-energy $\Pi_{hh}(p^2) $, we notice that this resummation is tacitly assuming that the
second term in \eqref{hh} is much smaller than the first term. However, we see that eq.~\eqref{hh} contains terms of the order of $ \frac{p^4}{(p^2+m_h^2)^2}\ln \left(\frac{m^2+p^2 (1-x) x}{\mu ^2}\right)$  which cannot be resummed for big values of $p^2$. We therefore proceed as in eq. \eqref{u1}-\eqref{jju2}  and use the identity
\beq
p^4=(p^2+m_h^2)^2-m_h^4-2 p^2 m_h^2
\eeq
to rewrite
\beq
\frac{p^4}{(p^2+m_h^2)^2} \ln \frac{p^2x(1-x)+m^2}{\mu^2} &=& \ln \frac{p^2x(1-x)+m^2}{\mu^2} -\underline{\frac{(m_h^4+2p^2m_h^2)}{(p^2+m_h^2)^2} \ln \frac{p^2x(1-x)+m^2}{\mu^2}}.
\label{jju}
\eeq
The term which has been underlined  in eq.~\eqref{jju} can be safely resummed, as it decays fast enough for large values of $p^2$.  We thence rewrite
\beq
\frac{{\Pi}_{hh}(p^2)}{(p^2+m_h^2)^2}&=& \frac{\hat{\Pi}_{hh}(p^2)}{(p^2+m_h^2)^2}+C_{hh}(p^2) \;,
\eeq
with
\beq
\hat{\Pi}_{hh}(p^2)&=&\frac{3g^2}{{8 (4 \pi )^2}}\int_0^1 dx \,  \Bigg\{2  \xi  \left(m_h^2+p^2\right) \ln \left(\frac{m^2 \xi }{\mu ^2}\right)-2 \xi  m_h^2+2 \left(6 m^2-p^2\right) \ln \left(\frac{m^2}{\mu ^2}\right)\nonumber\\
&-&(12 m^2-\frac{(m_h^4+2p^2 m_h^2)}{m^2}+4 p^2) \ln \left(\frac{m^2+p^2 (1-x) x}{\mu ^2}\right)-\frac{(2m_h^4+2p^2 m_h^2)}{m^2}\ln \left(\frac{m^2 \xi +p^2 (1-x) x}{\mu ^2}\right)\nonumber\\
&-&12 m^2-2 \xi  p^2+2 p^2-\frac{ m_h^4}{m^2} \Big(-2 \ln \left(\frac{m_h^2}{\mu ^2}\right)+3 \ln \left(\frac{m_h^2+p^2 (1-x) x}{\mu ^2}\right)+2\Big)\Bigg\}
\eeq
and
\beq
C_{hh}(p^2)= -\frac{3g^2}{{8 m^2(4 \pi )^2}}\int_0^1 dx \left( \ln \frac{p^2x(1-x)+m^2}{\mu^2} -\ln \frac{p^2x(1-x)+\xi m^2}{\mu^2} \right)  \;.
\eeq
Thus, for the one-loop Higgs propagator, we get
\beq
\braket{h(p)h(-p)}&=&\frac{1}{p^2+m_h^2- \hat{\Pi}_{hh}(p^2)} + C_{hh}(p^2) + \mathcal{O}(\hbar^2) \;.
\label{hhresum}
\eeq
For the gauge  field, we split the
two-point function into transverse and longitudinal parts in the usual way
\beq
\braket{A_{\mu}^a(p)A^b_{\nu}(-p)}= \braket{A_{\mu}^a(p)A^b_{\nu}(-p)}^T \mathcal{P}_{\mu \nu}(p)+ \braket{A_{\mu}^a(p)A^b_{\nu}(-p)}^L \mathcal{L}_{\mu \nu}(p),
\eeq
where we have introduced the transverse and longitudinal projectors, given respectively by
\beq
\mathcal{P}_{\mu\nu}(p)&=& \delta_{\mu\nu}-\frac{p_{\mu}p_{\nu}}{p^2}\,, \qquad	\mathcal{L}_{\mu\nu}(p)~=~ \frac{p_{\mu}p_{\nu}}{p^2}\,.
\eeq
We find
\beq
\braket{A_{\mu}^a(p)A_{\nu}^b(-p)}^T=\frac{\delta^{ab}}{p^2+m^2}+ \frac{\delta^{ab}}{(p^2+m^2)^2} \Pi_{AA^T}(p^2)+ \mathcal{O}(\hbar^2),
\label{piAA}
\eeq
\beq
\Pi_{AA^T}(p^2)&=& -\frac{ \delta^{ab} g^2 }{36 (4 \pi )^2 m^4 p^2 m_h^2}\int_0^1 dx \Bigg\{-27 m^4 p^2 m_h^4 \ln \left(\frac{m_h^2}{\mu ^2}\right)-27 m^6 \xi  p^2 m_h^2 \ln \left(\frac{m^2 \xi }{\mu ^2}\right)\nonumber\\
&+&3 m^4 m_h^4 \left(m_h^2-m^2+2 p^2\right) \ln \left(\frac{m_h^2}{\mu ^2}\right)+27 m^4 p^2 m_h^2 \left(m_h^2+m^2 \xi \right)\nonumber\\
&-&3 m^4 \xi  m_h^2 \left(2 m^4 (\xi -1)+m^2 (4 \xi +7) p^2+2 (\xi +9) p^4\right) \ln \left(\frac{m^2 \xi }{\mu ^2}\right)\nonumber\\
&+&3 m^4 \ln \left(\frac{m^2}{\mu ^2}\right) \left(-m^2 m_h^4+m_h^2 \left(m^4 (2 \xi -1)+m^2 (4 \xi +45) p^2+2 (\xi +9) p^4\right)-54 m^4 p^2\right)\nonumber\\
&+&m^4 \Big(6 m^2 m_h^4+m_h^2 \left(3 m^4 (2 (\xi -2) \xi +1)+3 m^2 (\xi -1) (4 \xi -1) p^2+2 (3 \xi  (\xi +4)-17) p^4\right)\nonumber\\
&-&3 m_h^6+54 m^4 p^2\Big)\nonumber\\
&-&3 m_h^2 \Big[m^4 \left(2 p^2 \left(m_h^2-5 m^2\right)+\left(m_h^2-m^2\right){}^2+p^4\right) \ln \left(\frac{p^2 (1-x) x+m_h^2(1-x) +m^2 x}{\mu ^2}\right)\nonumber\\
&-&2 \left(m^2+p^2\right)^2 \left(m^4 (\xi -1)^2+2 m^2 (\xi -5) p^2+p^4\right) \ln \left(\frac{p^2 (1-x) x+ \xi m^2 (1-x)+m^2 x}{\mu ^2}\right)\nonumber\\
&+&p^2 \left(p^4-m^4\right) \left(4 m^2 \xi +p^2\right) \ln \left(\frac{p^2 (1-x) x+ \xi m^2}{\mu ^2}\right)\nonumber\\
&+&p^2 \left(4 m^2+p^2\right) \left(12 m^4-20 m^2 p^2+p^4\right) \ln \left(\frac{p^2 (1-x) x+m^2}{\mu ^2}\right)\Big]\Bigg\}.
\label{AAf}
\eeq
We see that \eqref{piAA} contains again terms of the order $ \frac{p^4}{(p^2+m^2)^2}\ln \left(\frac{m^2+p^2 (1-x) x}{\mu ^2}\right)$ and $ \frac{p^6}{(p^2+m^2)^2}\ln \left(\frac{m^2+p^2 (1-x) x}{\mu ^2}\right)$, which cannot be resummed for big values of $p^2$. We use
\beq
\frac{p^4}{(p^2+m^2)^2} \ln \frac{p^2x(1-x)+m^2}{\mu^2} &=& \ln \frac{p^2x(1-x)+m^2}{\mu^2} -\underline{\frac{(m^4+2p^2m^2)}{(p^2+m^2)^2} \ln \frac{p^2x(1-x)+m^2}{\mu^2}}
\label{u3}
\eeq
and
\beq
\frac{p^6}{(p^2+m^2)^2} \ln \frac{p^2x(1-x)+m^2}{\mu^2} &=& (p^2-2m^2)\ln \frac{p^2x(1-x)+m^2}{\mu^2} \nonumber\\
&+&\underline{\frac{2m^6+3p^2m^4}{(p^2+m^2)^2} \ln \frac{p^2x(1-x)+m^2}{\mu^2}}.
\label{u2}
\eeq
The underlined terms in \eqref{u3} and \eqref{u2} can be safely resummed. We rewrite
\beq
\frac{{\Pi}_{AA^T}(p^2)}{(p^2+m^2)^2}&=& \frac{\hat{\Pi}_{AA^T}(p^2)}{(p^2+m_h^2)^2}+C_{AA^T}(p^2),
\eeq
with \allowdisplaybreaks
\beq
\hat{\Pi}_{AA^T}(p^2)&=& -\frac{ \delta^{ab} g^2 }{36 (4 \pi )^2 m^4 p^2 m_h^2}\int_0^1 dx \Bigg\{-27 m^4 p^2 m_h^4 \ln \left(\frac{m_h^2}{\mu ^2}\right)-27 m^6 \xi  p^2 m_h^2 \ln \left(\frac{m^2 \xi }{\mu ^2}\right)\nonumber\\
&+&3 m^4 m_h^4 \left(m_h^2-m^2+2 p^2\right) \ln \left(\frac{m_h^2}{\mu ^2}\right)+27 m^4 p^2 m_h^2 \left(m_h^2+m^2 \xi \right)\nonumber\\
&-&3 m^4 \xi  m_h^2 \left(2 m^4 (\xi -1)+m^2 (4 \xi +7) p^2+2 (\xi +9) p^4\right) \ln \left(\frac{m^2 \xi }{\mu ^2}\right)\nonumber\\
&+&3 m^4  \left(-m^2 m_h^4+m_h^2 \left(m^4 (2 \xi -1)+m^2 (4 \xi +45) p^2+2 (\xi +9) p^4\right)-54 m^4 p^2\right)\ln \left(\frac{m^2}{\mu ^2}\right)\nonumber\\
&+&m^4 \left(6 m^2 m_h^4+m_h^2 \left(3 m^4 (2 (\xi -2) \xi +1)+3 m^2 (\xi -1) (4 \xi -1) p^2+2 (3 \xi  (\xi +4)-17) p^4\right)-3 m_h^6+54 m^4 p^2\right)\nonumber\\
&-&3 m_h^2 \Big[m^4 \left(2 p^2 \left(m_h^2-5 m^2\right)+\left(m_h^2-m^2\right){}^2+p^4\right) \ln \left(\frac{p^2 (1-x) x+(1-x) m_h^2+m^2 x}{\mu ^2}\right)\nonumber\\
&-&2 m^4 (\xi -1)^2 \left(m^2+p^2\right)^2 \ln \left(\frac{p^2 (1-x) x+\xi  m^2 (1-x)+m^2 x}{\mu ^2}\right)\nonumber\\
&+&\left(-2 m^4 (4 \xi -1) p^2 \left(m^2+p^2\right)\right) \ln \left(\frac{p^2 (1-x) x+\xi m^2}{\mu ^2}\right)\nonumber\\
&+&(66 m^6 p^2-33 m^4 p^4) \ln \left(\frac{p^2 (1-x) x+m^2}{\mu ^2}\right)\Big]\Bigg\}.
\eeq
and
\beq
C_{AA^T}(p^2)&=& \frac{ \delta^{ab} g^2 }{12 (4 \pi )^2 m^4  }\int_0^1 dx \Bigg\{(-4 m^2 (\xi -5)-2p^2) \nonumber\\
&\times&\ln \left(\frac{p^2 (1-x) x+\xi m^2  (1-x)+m^2 x}{\mu ^2}\right)  \nonumber\\
&+&\left(4 \,\xi m^2   +p^2-2m^2\right) \ln \left(\frac{p^2 (1-x) x+\xi m^2}{\mu ^2}\right)\nonumber\\
&+&(-18 m^2 +p^2) \ln \left(\frac{p^2 (1-x) x+m^2}{\mu ^2}\right)\Bigg\}.
\eeq
Finally
\beq
\braket{A_{\mu}^a(p)A_{\nu}^b(-p)}^T=\delta^{ab}\left(  \frac{1}{p^2+m^2- \hat{\Pi}_{AA^T}(p^2)}+ C_{AA^T}(p^2)\right)  + \mathcal{O}(\hbar^2) \;.
\label{piAAll}
\eeq
\section{One-loop evaluation of the
	correlation function of the local BRST invariant  composite operators  \label{IIIIt}}

\subsection{Correlation function of the scalar BRST invariant composite operator $O(x)$}
The BRST invariant local scalar composite operator $O(x)$ is given by
\begin{equation}
O(x)= \Phi^{\dagger}\Phi -\frac{v^2}{2} \;,\qquad s\; O(x) =0 \;,  \label{OOO}
\end{equation}
which, after using the expansion \eqref{expansion2}, becomes
\beq
O(x)&=& \frac{1}{2} \Big[\begin{pmatrix}
	1 & 0
\end{pmatrix}((v+h)\textbf{1}-i\rho^a \tau^a))((v+h)\textbf{1}+i\rho^b \tau^b))\begin{pmatrix}
	1\\
	0
\end{pmatrix}\Big]-\frac{v^2}{2}\nonumber\\
&=&\frac{1}{2}\Big(h^2(x)+2 v h(x) + \rho^a (x) \rho^a(x)\Big) \;, \label{scopt}
\eeq
so that
\beq
\braket{{O}(x) {O}(y)} &=& v ^2 \braket{h(x) h(y)}+ v  \braket{h(x) \rho^b (y)\rho^b (y) }+ v  \braket{h(x) h(y)^2}+\frac{1}{4}\braket{h(x)^2 \rho^b (y)\rho^b (y)}+\frac{1}{4}\braket{h(x)^2 h(y)^2} \nonumber \\
&+&\frac{1}{4}\braket{\rho^a (x)\rho^a (x) \rho^b (y)\rho^b (y)}. \label{exp1}
\eeq
Looking at the tree level expression of eq.~\eqref{exp1}, one easily obtains
\beq
\braket{{O}(p) {O}(-p)}_{\rm tree} = v ^2 \braket{h(p) h(-p)}_{\rm tree}
= v^2 \frac{1}{p^2+m_h^2} \;, \label{treeoo}
\eeq
showing that the BRST invariant scalar operator $O(x)$ is directly linked to the Higgs propagator.

Concerning now the one-loop calculation of expression \eqref{exp1}, after evaluating each term, see Appendix \ref{OO} for details, we find that the two-point correlation function of the scalar composite operator $O(x)$ develops a geometric series in the same way as the elementary field $h(x)$. This allows us the make a resummed approximation. Using  dimensional regularization in the  $\overline{MS}bar$-scheme, we find
\beq
\braket{O(p)O(-p)}(p^2)&=& \frac{v^2}{p^2+m_h^2}+\frac{v^2 \,}{(p^2+m_h^2)^2}\Pi_{OO}(p^2)+\mathcal{O}(\hbar^2),
\label{ark}
\eeq
\beq
\Pi_{OO}(p^2)&=&\frac{1}{{32 v^2\pi ^2  m_h^2}}\int_0^1 dx \Bigg\{-24 m_h^2 m^4-6 m^2 p^2 \left(m_h^2+6 m^2\right) \ln \left(\frac{m^2}{\mu ^2}\right)\nonumber\\
&-&m_h^2\left(p^2-2 m_h^2\right)^2 \ln \left(\frac{m_h^2 +p^2 x(1-x)}{\mu ^2}\right)\nonumber\\
&-&3 m_h^2\left(12 m^4+4 m^2 p^2+p^4\right) \ln \left(\frac{m^2 +p^2 x(1-x)}{\mu ^2}\right)\nonumber\\
&+&6 p^2 \left(m_h^4+m_h^2 m^2+2 m^4\right)-6 m_h^4 p^2 \ln \left(\frac{m_h^2}{\mu ^2}\right)\Bigg\}.
\eeq
Since \eqref{ark} contains terms of the order of $\frac{p^4}{(p^2+m^2)^2}\ln (p^2)$,
we follow the steps \eqref{u1}-\eqref{jju2} to find the resummed correlation function in the one-loop approximation
\beq
G_{{O}{O}}(p^2)
&=&\frac{v^2}{p^2+m_h^2-\hat{\Pi}_{OO}(p^2)}+C_{OO}(p^2)
\label{dk3a}
\eeq
with
\beq
\hat{\Pi}_{OO}(p^2)&=&\frac{1}{{32 v^2\pi ^2  m_h^2}}\int_0^1 dx \Bigg\{-24 m_h^2 m^4-6 m^2 p^2 \left(m_h^2+6 m^2\right) \ln \left(\frac{m^2}{\mu ^2}\right)\nonumber\\
&-&m_h^2(3 m_h^4 -6 m_h^2p^2) \ln \left(\frac{m_h^2 +p^2 x(1-x)}{\mu ^2}\right)\nonumber\\
&-&3 m_h^2\left(12 m^4+4 m^2 p^2-m_h^4-2p^2m_h^2 \right) \ln \left(\frac{m^2 +p^2 x(1-x)}{\mu ^2}\right)\nonumber\\
&+&6 p^2 \left(m_h^4+m_h^2 m^2+2 m^4\right)-6 m_h^4 p^2 \ln \left(\frac{m_h^2}{\mu ^2}\right)\Bigg\}
\label{dk3b}
\eeq
and
\beq
C_{OO}(p^2)&=&-\frac{1}{32 \pi ^2  }
\int_0^1 dx \Bigg\{\ln \left(\frac{m_h^2+p^2x(1-x)}{\mu^2}\right)+3\ln \left( \frac{ m^2 + p^2 x (1-x) }{\mu^2}\right)\Bigg\}.
\label{grgra}
\eeq
Expressions \eqref{dk3a} and \eqref{grgra} show that, as expected, and unlike the Higgs propagator, eq.~\eqref{hhresum}, the correlator $\braket{O(p)O(-p)}$ is  independent from the gauge parameter $\xi$.

\subsection{A little digression on the unitary gauge}
\begin{figure}[H]
	\centering
	\includegraphics[width=15cm]{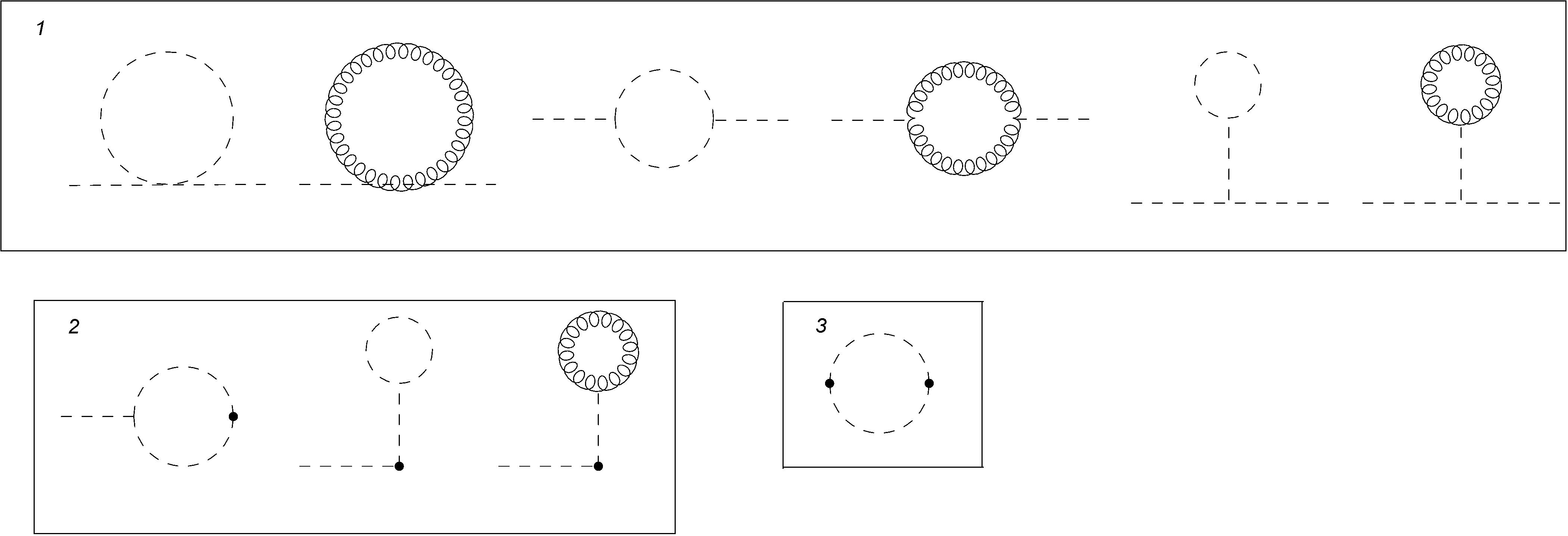}
	\caption{One-loop contributions for the propagator $\langle O(x)O(y) \rangle$ in the unitary gauge: $ \braket{h(x) h(y)}$ (first line),$ \braket{h(x) h(y)^2}$ (second line) and $\braket{h(x)^2 h(y)^2}$ (third line) . Wavy lines represent the gauge field, dashed lines the Higgs field, solid lines the Goldstone boson and double lines the ghost field. The $\bullet$  indicates the insertion of a composite operator. }.
	\label{OOu}
\end{figure}

Of course, since the composite operator $O(x)$ is  BRST invariant, any choice for the gauge parameter $\xi$ should give the same expression for the correlation function  $G_{OO}(p^2)$. One convenient choice is the so-called  unitary gauge, which is formally attained by taking  $\xi \rightarrow \infty$ at the end of the calculation. Though, we take here a different route and perform the same calculation done before for  $G_{OO}(p^2)$ by employing the tree level propagators and other Feynman rules which follow by taking the limit $\xi \rightarrow \infty$ at the beginning. In doing this, for the tree level propagators one finds
\beq
\langle A^a_{\m}(p)A^b_{\n}(-p)\rangle &=& \frac{\delta^{ab}}{p^2+ m^2} \mathcal{P}_{\m\n}(p)+\delta^{ab}\frac{1}{m^2}\mathcal{L}_{\m\n}(p)=  \frac{\delta^{ab}}{p^2+ m^2}\Big(\delta_{\mu \nu}+ \frac{p_{\m}p_{\n}}{m^2}\Big),\nonumber\\
\langle h(p)h(-p)\rangle &=&\frac{1}{p^2 + m_h^2}\,
\label{uuu}
\eeq
with all other propagators, {\it i.e.}~the Goldstone and Faddeev-Popov ghost propagators, vanishing.  Then, eq.~\eqref{exp1} simplifies to
\beq
\braket{{O}(x) {O}(y)}_{\rm unitary} &=& v ^2 \braket{h(x) h(y)}_{\rm unitary}+ v  \braket{h(x) h(y)^2}_{\rm unitary}+\frac{1}{4}\braket{h(x)^2 h(y)^2}_{\rm unitary},
\eeq
with the contributing diagrams shown in Figure \ref{OOu}. Making use of the dimensional regularization in the  $\overline{MS}bar$-scheme  and switching to momentum space, we get
\beq
v^2 \braket{h(p)h(-p)}_{\rm unitary}&=&\frac{3}{32 \pi^2  }\int_0^1 dx\Bigg\{\frac{1}{ \epsilon} \left(2m_h^4+12 m^2 p^2+2p^4\right)+2m_h^4  \ln \left(\frac{m_h^2}{\mu ^2}\right)+2 \left(6 m^4-m^2 p^2\right) \ln \left(\frac{m^2}{\mu ^2}\right)\nonumber\\
&-&3 m_h^4 \ln \left(\frac{p^2 (1-x) x+m_h^2}{\mu ^2}\right)- \left(12 m^4+4 m^2 p^2+p^4\right) \ln \left(\frac{p^2 (1-x) x+m^2}{\mu ^2}\right)\nonumber\\
&-&2m_h^4 +2 m^2 \left(p^2-6 m^2\right)\Bigg\}\frac{1}{\left(m_h^2+p^2\right){}^2},\\
v \braket{h(p) h(-p)^2}_{\rm unitary}&=& \frac{3}{16 \pi ^2 m_h^2 }\int_0^1 dx \Bigg\{\frac{12}{\epsilon}m^4-6 m^4\ln \left(\frac{m^2}{\mu ^2}\right)-m_h^4\ln \left(\frac{m_h^2}{\mu ^2}\right)\nonumber\\
&+&  m_h^4 \ln \left(\frac{m_h^2+p^2 (1-x) x}{\mu ^2}\right)+m_h^4+2m^4 \Bigg\}\frac{1}{\left(m_h^2+p^2\right)},\\
\frac{1}{4}\braket{h(p)^2 h(p)^2}_{\rm unitary}&=& \frac{1}{16 \pi ^2  }\int_0^1 dx \Bigg\{\frac{1}{\epsilon}-\frac{1}{2}\ln \left(\frac{m_h^2+p^2 (1-x) x}{\mu ^2}\right)\Bigg\}.
\eeq
Inserting now the unity
\beq
1= (p^2+m_h^2)/(p^2+m_h^2) = ((p^2+m_h^2)/(p^2+m_h^2))^2,
\label{2o}
\eeq
we find indeed that $\braket{O(x)O(y)}_{\rm unitary}= \braket{O(x)O(y)}$, showing that the same expression has been re-obtained by starting directly from the tree level propagators of the unitary gauge, $\xi \rightarrow \infty$, despite the fact that this gauge is known to be non-renormalizable. Though, at one-loop order, a simple explanation can be found for the previous result, which is due in part to the BRST  invariant nature of the correlation function  $\braket{O(x)O(y)}$ and to the fact that, at one-loop order, the handling of the overlapping divergences is not required. From two-loop onward these divergences will show up, requiring a fully renormalizable setup. More specifically, in this case, one would need to keep $\xi$ finite and use all Feynman rules of the $R_\xi$-gauge, while taking $\xi \rightarrow \infty$ only at the end. Nevertheless, having checked that the one-loop result for $G_{OO}(p^2)$ is the same by using both procedures, in the next section, we will use the same simplifying second trick to evaluate  the two-point function of the vectorial composite operator at one-loop order.

\subsection{Vectorial composite operators}
We identify three gauge invariant vector composite operators, following the definitions of 't Hooft in \cite{tHooft:1980xss}, namely
\beq
O_{\mu}^{3}&=&i\phi^{\dagger}D_{\mu}\phi,\nonumber\\
O_{\mu}^{+}&=&\phi^{T}\begin{pmatrix}0 & 1\nonumber\\
	-1 & 0
\end{pmatrix}D_{\mu}\phi,\\
O_{\mu}^{-}&=&(O_{\mu}^{+})^{\dagger}.
\eeq
The gauge invariance of $O_{\mu}^{3}$ is apparent. For $O^{+}_{\mu}$, we can show the gauge invariance by using  the following $2 \times 2$ matrix representation of a generic $SU(2)$ transformation,
\beq
U=\begin{pmatrix}a & -b^{\star}\\
	b & a^{\star}
\end{pmatrix}
\eeq
with determinant $\vert a \vert^2 +\vert b \vert^2=1$. Thus, we find that under a $SU(2)$ transformation
\beq
O_{\mu}^{+} &\rightarrow& (U\phi)^T \begin{pmatrix}0 & 1\\
	-1 & 0
\end{pmatrix} D_{\mu}(U\phi)\nonumber\\
&=& \phi^T U^T \begin{pmatrix}0 & 1\\
	-1 & 0
\end{pmatrix} U D_{\mu}\phi\nonumber\\
&=& \phi^T  \begin{pmatrix}0 & 1\\
	-1 & 0
\end{pmatrix} D_{\mu}\phi=O_{\mu}^{+},
\eeq
which shows the gauge invariance of $O_{\mu}^{+}$ and, subsequently, of $O_{\mu}^{-}$. After using the expansion \eqref{expansion2}, the first composite operator reads
\beq
O_{\mu}^{3}&=&i \phi^{\dagger}D_{\mu}\phi
\nonumber\\
&=&i \phi^{\dagger}\partial_{\mu}\phi+\frac{1}{2}g\phi^{\dagger}\tau^{a}A_{\mu}^{a}\phi\nonumber\\
&=&\frac{i}{2} \Bigg[(v+h)\partial_{\mu}h+i(v+h)\partial_{\mu}\rho^{3}-i\rho^{3}\partial_{\mu}h+\rho^{a}\partial_{\mu}\rho^{a}+i \rho^1 \partial_{\mu}\rho^2-i \rho^2 \partial_{\mu}\rho^1
-\frac{i}{2}g(v+h)^{2}A_{\mu}^{3}+ig(v+h)(A_{\mu}^{1}\rho^{2}-A_{\mu}^{2}\rho^{1})\nonumber
\\
&+&\frac{i}{2}g\rho^{a}A_{\mu}^{3}\rho^{a}-ig\rho^{3}A_{\mu}^{b}\rho^{b}\Bigg]\nonumber\\
&=& \frac{1}{2} \Bigg[-(v+h)\partial_{\mu}\rho^{3}+\rho^{3}\partial_{\mu}h- \rho^1 \partial_{\mu}\rho^2+ \rho^2 \partial_{\mu}\rho^1
+\frac{1}{2}g(v+h)^{2}A_{\mu}^{3}-g(v+h)(A_{\mu}^{1}\rho^{2}-A_{\mu}^{2}\rho^{1})\nonumber
\\
&-&\frac{1}{2}g\rho^{a}A_{\mu}^{3}\rho^{a}+g\rho^{3}A_{\mu}^{b}\rho^{b}\Bigg]+ \frac{i}{2} \partial_{\mu}O,
\eeq
and since the last term, {\it i.e.}~$\frac{i}{2} \partial_{\mu}O$, is  BRST invariant, the sum of the others terms has to be  BRST invariant too. Therefore, we can introduce the following three  ``reduced'' vector composite operators $R^a_\mu$ with  $a=1,2,3$ :
\beq
R_{\mu}^{1}&=&\frac{i}{2}\Big( O_{\mu}^+ - O^-_{\mu}\Big),\nonumber\\
R_{\mu}^{2}&=&\frac{1}{2}\Big( O_{\mu}^+ + O^-_{\mu}\Big),\nonumber\\
R_{\mu}^{3}&=&O_{\mu}^3-\frac{i}{2}\partial_{\mu}O_{\mu},  \label{threeop}
\eeq
so that
\beq
R_{\mu}^a &=& \frac{1}{2}\Bigg[-(v+h)\partial_{\mu}\rho^a+\rho^a \partial_{\mu} h - \varepsilon^{abc} \rho^b \partial_{\mu} \rho^c
+\frac{1}{2}g (v+h)^2 A_{\mu}^a-g(v+h)\varepsilon^{abc}(\rho^b A_{\mu}^c)\nonumber\\
&-& \frac{1}{2}g A_{\mu}^a \rho^m \rho^m +g \rho^a A_{\mu}^m \rho^m\Bigg]\;,  \label{threeop1}
\eeq
with
\begin{equation}
s R^a_\mu(x) = 0 \;. \label{BrstRR}
\end{equation}
Remarkably, the BRST  invariant operators $R_{\mu}^a$ transform like a triplet under the custodial symmetry \eqref{custodial}, namely
\beq
\delta R_{\mu}^a= \varepsilon^{abc}\beta^b R_{\mu}^c \;, \label{tripr}
\eeq
and since the only rank two invariant tensor is $\delta^{ab}$, we can write, moving to momentum space,
\beq
\braket{ R^a_{\mu}(p)\,R^b_{\nu}(-p)}= \delta^{ab} R_{\mu \nu}(p^2 )\,\, \rightarrow \,\, R_{\mu \nu}(p^2 )= \frac{1}{3} \braket{ R^a_{\mu}(p)\,R^a_{\nu}(-p)}, \label{cc1}
\eeq
as well as
\beq
R_{\mu \nu}(p^2)= R(p^2) \mathcal{P}_{\mu \nu} (p) + L (p^2) \mathcal{L}_{\mu \nu}(p), \label{cc2}
\eeq
so that  in $d$ dimensions,
\beq
R(p^2 )=\frac{1}{3}\frac{\mathcal{P}_{\mu \nu}(p)}{(d-1)}\braket{ R^a_{\mu}(p)\,R^a_{\nu}(-p)}, \label{cc3}
\eeq
and
\beq
L(p^2)= \frac{1}{3} \mathcal{L}_{\mu \nu}(p)\braket{ R^a_{\mu}(p)\,R^a_{\nu}(-p)}. \label{cc4}
\eeq
One recognizes that eqs.\eqref{cc1}-\eqref{cc4} display exactly the same structure of the gauge vector boson correlation function $\langle A^a_\mu(p) A_\nu(-p)\rangle$.

In the $R_{\xi}$-gauge, the non-vanishing contributions, up to first order in $\hbar$, to the correlation function $\braket{ R^a_{\mu}(p)\,R^b_{\nu}(-p)}$ are
\beq
\braket{R^a_{\mu}(p)\,R^a_{\nu}(-p)}&=&\frac{1}{16}g^{2}v^{4}\langle A_{\mu}^{a}(p)\, A_{\nu}^{a}(-p)\rangle-\langle(\rho^{a}\partial_{\mu}h)(p)\, (\partial_{\nu}\rho^{a}h)(-p)\rangle+\frac{1}{4}p_{\mu}p_{\nu}\langle(\rho^{a}h)(p) \,(\rho^{a}h)(-p)\rangle \nonumber\\
&+&\frac{1}{8}\partial_{\mu}\partial_{\nu}\langle(\rho^{a}\rho^{b})(p)\, (\rho^{a} \rho^{b})(-p)\rangle-\frac{1}{2}\langle(\rho^{a}\partial_{\mu}\rho^{b})(p)\,(\partial_{\nu}\rho^{a}\rho^{b})(-p)\rangle\nonumber\\
&+&\frac{1}{4}g^{2}v^{3}\langle A_{\mu}^{a}(p)\, (A_{\nu}^{a}h)(-p)\rangle-\frac{i}{4}gv^{3}p_{\nu}\langle A_{\mu}^{a}(p)\, \rho^{a}(-p)\rangle
\nonumber\\
&+&\frac{1}{4}v^{2}p_{\mu}p_{\nu}\langle\rho^{a}(x) \rho^{a}(y)\rangle+\frac{1}{6}g^{2}v^{2}\langle(\rho^{a}A_{\mu}^{b})(p)\, (\rho^{a}A_{\nu}^{b})(-p)\rangle\nonumber\\
&-&\frac{1}{24}g^{2}v^{2}\langle(\rho^{a}\rho^{a}A_{\mu}^{b})(p)\, A^{b}(-p)\rangle+\frac{1}{8}g^{2}v^{2}\langle (h^2 A_{\mu}^{a})(p) \,A_{\nu}^{a}(-p)\rangle
\nonumber\\
&+&\frac{1}{4}g^{2}v^{2}\langle (h A_{\mu}^{a})(p) \,(h A_{\nu}^{a})(-p)\rangle+\frac{i}{4}v^{2}gp_{\mu}\langle (h\rho^{a})(p)\,A_{\nu}^{a}(-p)\rangle
\nonumber\\
&+&\frac{1}{2}gv^{2}\langle(\partial_{\mu}h\rho^{a})(p)\,A_{\nu}^{a}(-p)\rangle+\frac{i}{2}gv^{2}p_{\mu}\langle\rho^{a}(p)\,(h A_{\nu}^{a})(-p)\rangle\nonumber\\
&-&\frac{1}{4}gv^{2}\varepsilon^{abc}\langle A^{a}(p)\,(\rho^{b}\partial_{\mu}\rho^{c})(-p)\rangle-\frac{ig v^{2}}{2}\varepsilon^{abc}p_{\mu}\langle\rho^{a}(p)\,(\rho^{b} A_{\nu}^{c})(-p)\rangle\nonumber
\\
&+&\frac{1}{2}vp_{\mu}p_{\nu}\langle (h\rho^{a})(p)\,\rho^{a}(-p)\rangle-ivp_{\nu}\langle(\partial_{\mu}h\rho^{a})(p)\,\rho^{a}(-p)\rangle
\label{OO+}
\eeq
where we have used the notation $\partial_{\mu}=\frac{\partial}{\partial x^{\mu}}$ and $\partial_{\nu}=\frac{\partial}{\partial y^{\nu}}$ \footnote{The derivative here is not expressed in momentum space because for composite operators, the derivative will bring down different momenta depending on the configuration of the Feynman diagram.}.  
The first term in expression \eqref{OO+} is the gauge field propagator $\braket{A_{\mu}^a (p)A_{\mu}^a(-p)} $, which means that $R^a_{\mu}$ can be thought as a kind of  BRST invariant extension of the elementary gauge field $A_{\mu}^a$. At tree-level, we find in fact
\beq
\braket{R^a_{\mu}(x)\,R^a_{\nu}(y)}_{tree}&=& \frac{1}{16}g^{2}v^{4}\langle A_{\mu}^{a}(p),A_{\nu}^{a}(-p)\rangle+\frac{1}{4}v^{2}\partial_{\mu}\partial_{\nu}\langle\rho^{a}(p),\rho^{a}(-p)\rangle\nonumber\\
&=& \frac{3}{16}g^2 v^4 \frac{1}{p^2+m^2}\mathcal{P}_{\mu\nu}(p)+\frac{3}{4}v^2 \mathcal{L}_{\mu\nu}(p),
\eeq
where we can see that, apart from the constant factor $\frac{3}{4}v^2$ appearing in the longitudinal sector, the transverse component reproduces exactly the transverse gauge tree-level propagator.

\begin{figure}[H]
	\centering
	\includegraphics[width=15cm]{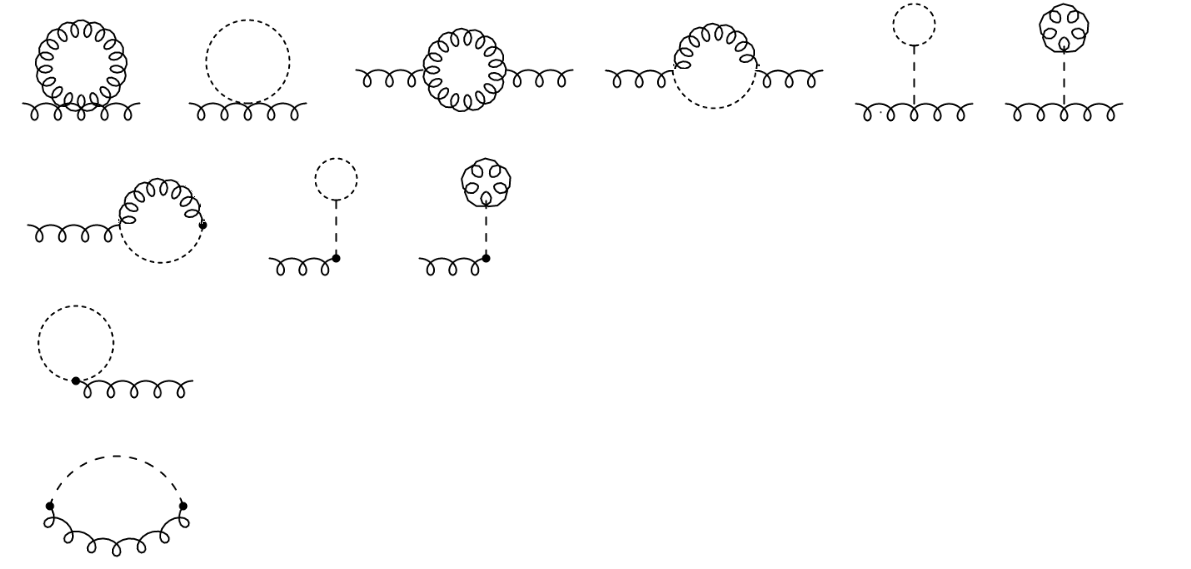}
	\caption{One-loop contributions for the correlation function $\langle OO \rangle$ in the unitary gauge: $ \braket{A_{\mu}^a(x) A_{\nu}^a(y)}$ (first two lines), $ \braket{A(x) (Ah)(y)}$ (third line), $\braket{h(x)^2 A_{\mu}^a}$ (fourth line) and $\braket{(A_{\mu}^a h)(x) (A_{\nu}^a h)(y)}$ . Wavy lines represent the gauge field, dashed lines the Higgs field, solid lines the Goldstone boson and double lines the ghost field. The $\bullet$  indicates the insertion of a composite operator. }.
	\label{OOuA}
\end{figure}

Since the correlation function $\braket{R^a_{\mu}(p),R^a_{\nu}(-p)}$ is independent from the gauge parameter $\xi$, due to the BRST invariant nature of the operator $R^a_{\mu}(x)$, we shall proceed as in the previous example by making use of  $\braket{R^a_{\mu}(p)\,R^a_{\nu}(-p)}_{\rm unitary}=\braket{R^a_{\mu}(p),R^a_{\nu}(-p)}$ and evaluating the correlator at the one-loop order  with the propagators given in \eqref{uuu}, so that
\beq
\braket{R^a_{\mu}(p)\,R^a_{\nu}(-p)}&=&\frac{1}{16}g^{2}v^{4}\langle A_{\mu}^{a}(p)\,A_{\nu}^{a}(-p)\rangle_{\rm unitary}+\frac{1}{4}g^{2}v^{3}\langle A_{\mu}^{a}(p)\,(A_{\nu}^{a}h)(-p)\rangle_{\rm unitary}\nonumber\\
&+&\frac{1}{8}g^{2}v^{2}\langle (h^2 A_{\mu}^{a})(p)\,A_{\nu}^{a}(-p)\rangle_{\rm unitary}+\frac{1}{4}g^{2}v^{2}\langle (h A_{\mu}^{a})(p)\,(h A_{\nu}^{a})(-p)\rangle_{\rm unitary}
\eeq
with the contributing diagrams shown in  Figure~\ref{OOuA}. Using dimensional regularization in the  $\overline{MS}bar$-scheme with $(d=4-\epsilon)$ and switching to momentum space, we find
\beq
\frac{1}{16}g^{2}v^{4}\langle A_{\mu}^{a}(p)\,A_{\nu}^{a}(-p)\rangle_{\rm unitary}&=& \frac{g^4 v^4}{32 (4\pi)^2} \int_0^1 dx\Bigg\{ -\frac{1}{\epsilon 6m^4 m_h^2}\left(9 m^4 m_h^4+m_h^2 \left(-9 m^6-83 m^4 p^2-14 m^2 p^4+p^6\right)+54 m^8\right)\nonumber\\
&+&\frac{m_h^2}{2 p^2} \left(-m_h^2+m^2+7 p^2\right) \ln \left(\frac{m_h^2}{\mu ^2}\right)\nonumber\\
&+&\frac{1}{2 m^2 p^2 m_h^2}\left(m^4 m_h^4-m_h^2 \left(m^6+47 m^4 p^2+16 m^2 p^4-2 p^6\right)+54 m^6 p^2\right) \ln \left(\frac{m^2}{\mu ^2}\right)\nonumber\\
&+&\frac{1}{2p^2}\left(-2 m_h^2 \left(m^2-p^2\right)+m_h^4+m^4-10 m^2 p^2+p^4\right)\ln \left(\frac{p^2 (1-x) x+(1-x) m_h^2+m^2 x}{\mu ^2}\right)\nonumber\\
&+&\frac{1}{2 m^4}\left(4 m^2+p^2\right) \left(12 m^4-20 m^2 p^2+p^4\right) \ln \left(\frac{m^2+p^2 (1-x) x}{\mu ^2}\right)\nonumber\\
&+&\frac{1}{6 m^4 p^2 m_h^2}\Big(3 m^4 m_h^6-3 m_h^4 \left(2 m^6+9 m^4 p^2\right)\nonumber\\
&+&m_h^2 \left(3 m^8-9 m^6 p^2-2 m^4 p^4-26 m^2 p^6-2 p^8\right)-54 m^8 p^2\Big)\Bigg\}\frac{\mathcal{P}_{\mu \nu}(p)}{(p^2+m^2)^2}\nonumber\\
&+& \frac{g^4 v^4}{32 (4\pi)^2m^4} \int_0^1 dx\Bigg\{\frac{1}{\epsilon}\frac{3}{ m_h^2} \left(m_h^2 \left(3 m^2+p^2\right)-3 m_h^4-18 m^4\right)\nonumber\\
&-&\frac{3 m^2}{2  p^2 m_h^2}\left(m_h^2 \left(p^2-m^2\right)+m_h^4-18 m^2 p^2\right) \ln \left(\frac{m^2}{\mu ^2}\right)\nonumber\\
&+&\frac{3 m_h^2}{2 p^2} \left(m_h^2-m^2+5 p^2\right) \ln \left(\frac{m_h^2}{\mu ^2}\right)\nonumber\\
&-&\frac{3}{2 p^2} \left(\left(m_h-m\right){}^2+p^2\right) \left(\left(m_h+m\right){}^2+p^2\right) \ln \left(\frac{p^2 (1-x) x+(1-x) m_h^2+m^2 x}{\mu ^2}\right)\nonumber\\
&-&\frac{3}{2p^2 m_h^2} \left(m_h^4 \left(5 p^2-2 m^2\right)+m_h^2 \left(m^4-m^2 p^2\right)+m_h^6+6 m^4 p^2\right)\Bigg\} \mathcal{L}_{\mu \nu}(p),\\
\frac{1}{4}g^2 v^3 \braket{A^a_{\mu}(p)\,(A^a h)(-p)}_{\rm unitary}&=&\frac{1}{16\pi ^2 }\int_0^1 dx \Bigg\{\frac{1}{\epsilon}\frac{m^2}{ m_h^2} \left(m_h^2 \left(p^2-9 m^2\right)+12 m_h^4+54 m^4\right)\nonumber\\
&-&\frac{m^2 m_h^2}{2 p^2} \left(-m_h^2+m^2+10 p^2\right) \ln \left(\frac{m_h^2}{\mu ^2}\right)\nonumber\\
&-&\frac{m^4}{2 p^2 m_h^2} \left(m_h^2 \left(p^2-m^2\right)+m_h^4+54 m^2 p^2\right) \ln \left(\frac{m^2}{\mu ^2}\right)\nonumber\\
&-&\frac{m^2}{2p^2} \left(2 p^2 \left(m_h^2-5 m^2\right)+\left(m^2-m_h^2\right){}^2+p^4\right) \ln \left(\frac{p^2(1-x)x+(1-x)m_h^2+x m^2}{\mu ^2}\right)\nonumber\\
&+&\frac{m^2}{6 p^2 m_h^2} \left(6 m_h^4 \left(m^2+6 p^2\right)+m_h^2 \left(-3 m^4+9 m^2 p^2+2 p^4\right)-3 m_h^6+54 m^4 p^2\right)\nonumber\Bigg\}\frac{\mathcal{P}_{\mu \nu}(p)}{\left(m^2+p^2\right)}\nonumber\\
&+&\frac{1}{16\pi ^2}\int_0^1 dx \Bigg\{
\frac{1}{\epsilon}\frac{3}{  m_h^2} \left(-m_h^2 \left(3 m^2+p^2\right)+m_h^4+18 m^4\right)\nonumber\\
&+&\frac{3 m^2}{2 p^2 m_h^2} \left(m_h^2 \left(p^2-m^2\right)+m_h^4-18 m^2 p^2\right) \ln \left(\frac{m^2}{\mu ^2}\right)\nonumber\\
&-&\frac{3 m_h^2}{2 p^2} \left(m_h^2-m^2+3 p^2\right) \ln \left(\frac{m_h^2}{\mu ^2}\right)\nonumber\\
&+&\frac{3}{2 p^2 m_h^2} \left(m_h^2 \left(\left(m-m_h\right){}^2+p^2\right) \left(\left(m_h+m\right){}^2+p^2\right)\right) \ln \left(\frac{p^2(1-x)x+(1-x)m_h^2+ x m^2}{\mu ^2}\right)\nonumber\\
&+&\frac{3}{2 p^2 m_h^2} \left(\left(m_h^3-m^2 m_h\right){}^2+p^2 \left(-m^2 m_h^2+3 m_h^4+6 m^4\right)\right)\nonumber\\
&&\Bigg\}\mathcal{L}_{\mu \nu }(p), 
\eeq
\beq
\frac{1}{8}g^{2}v^{2}\langle (h^2 A_{\mu}^{a})(p)\,A_{\nu}^{a}(-p)\rangle_{\rm unitary}&=& -\frac{3 m_h^2 m^2}{32\pi^2}\int_0^1 dx \Bigg\{\frac{2}{\epsilon }- \ln \left(\frac{m_h^2}{\mu ^2}\right)+1\Bigg\}\Big(\frac{1}{p^2+m^2}\mathcal{P}_{\mu \nu}(p)\nonumber\\
&+&\frac{1}{m^2}\mathcal{L}_{\mu\nu}(p)\Big),
\eeq
\beq
\frac{1}{4}g^{2}v^{2}\langle (h A_{\mu}^{a})(p)\,(h A_{\nu}^{a})(-p)\rangle_{\rm unitary}&=&\frac{1}{32\pi ^2} \int_0^1 dx \Bigg\{\frac{1}{ \epsilon }(-3 m_h^2+9 m^2-p^2)\nonumber\\
&+&\frac{m_h^2 }{2 p^2}\left(m^2+p^2-m_h^2\right) \ln \left(\frac{m_h^2}{\mu ^2}\right)+\frac{m^2 }{2 p^2} \left(m_h^2-m^2+p^2\right)\ln\left(\frac{m^2}{\mu ^2}\right)\nonumber\\
&+&\frac{1}{2p^2} \left(2 p^2 \left(m_h^2-5 m^2\right)+\left(m^2-m_h^2\right){}^2+p^4\right) \ln \left(\frac{p^2(1-x)x+(1-x)m_h^2+x m^2}{\mu ^2}\right)\nonumber\\
&+&\frac{1}{6p^2} \left(3 \left(m^2-m_h^2\right){}^2-9 p^2\left(m_h^2+m^2\right)-2 p^4\right)\Bigg\}\mathcal{P}_{\mu \nu}(p)\nonumber\\
&+& \frac{3}{32\pi ^2}\int_0^1 dx \Bigg\{\frac{1}{ \epsilon } \left(-m_h^2+3 m^2+p^2\right)\nonumber\\
&+&\frac{m^2}{2 p^2} \left(-m_h^2+m^2-p^2\right) \ln \left(\frac{m^2}{\mu ^2}\right)+\frac{m_h^2}{2 p^2} \left(m_h^2-m^2+3 p^2\right) \ln \left(\frac{m_h^2}{\mu ^2}\right)\nonumber\\
&-& \frac{1}{2p^2} \left(\left(m-m_h\right){}^2+p^2\right) \left(\left(m_h+m\right){}^2+p^2\right) \ln \left(\frac{p^2x(1-x)+(1-x)m_h^2+x m^2}{\mu ^2}\right)\nonumber\\
&-&\frac{1}{2p^2}\left(\left(m^2-m_h^2\right){}^2+3 p ^2m_h^2-p^2m^2\right)\Bigg\}\mathcal{L}_{\mu \nu}(p) \;.
\eeq
Using the unity \eqref{2o}, we find that the transverse part of the propagator is given by
\beq
R(p^2)&=& \frac{1}{16}g^2 v^4\Bigg(\frac{1}{p^2+m^2}+\frac{1}{(p^2+m^2)^2}\left( \Pi_R(p^2)+ \Pi_{\rm div}(p^2)\right) \Bigg)+ \mathcal{O}(\hbar^2)
\eeq
where the divergent part is, see also the comments in the  Appendix~\ref{un},
\beq
\Pi_{\rm div}(p^2)&=&\frac{g^2}{\epsilon \pi^2}\Big( -\frac{h^2 p^4}{32 m^4}+\frac{9 m^4}{16 h^2}+\frac{9 m^2 p^2}{8 h^2}+\frac{h^2 p^2}{8 m^2}+\frac{h^2}{16}-\frac{p^6}{48 m^4}+\frac{23 p^4}{96 m^2}+\frac{7 p^2}{8}\Big) \;,
\eeq
while for the finite part we get
\beq
\Pi_{R}(p^2)&=&\frac{3}{36 \pi^2 g^2 v^4 m_h^2}\int_0^1 dx \Bigg\{ 6 m^4 \left(m_h^4+3 m^4\right)-\frac{p^4 m_h^2}{3} \left(9 m_h^2+35 m^2+4 p^2\right)\nonumber\\
&+& p^2 \left(m^4 m_h^2+10 m^2 m_h^4+m_h^6+36 m^6\right)+m_h^4 \left(-p^2 \left(m_h^2+11 m^2\right)-6 m^4+p^4\right) \ln \left(\frac{m_h^2}{\mu ^2}\right)\nonumber\\
&+&m_h^2 \left(2 p^4 \left(m_h^2-5 m^2\right)+\left(m^2-m_h^2\right)^2 p^2+p^6\right) \ln \left(\frac{p^2(1-x)x+(1-x)m_h^2+x m^2}{\mu ^2}\right)\nonumber\\
&+&m_h^2\left(48 m^6-68 m^4 p^2-16 m^2 p^4+p^6\right) \ln \left(\frac{p^2 (1-x) x+m^2}{\mu ^2}\right)\nonumber\\
&+& m^2 \left(m_h^2 \left(-48 m^4-17 m^2 p^2+3 p^4\right)+p^2 m_h^4-54 \left(m^6+2 m^4 p^2\right)\right) \ln \left(\frac{m^2}{\mu ^2}\right)\Bigg\}.
\label{GG}
\eeq
Since \eqref{GG} contains terms of the order of $\frac{p^4}{(p^2+m^2)^2}\ln (p^2)$ and $\frac{p^6}{(p^2+m^2)^2}\ln (p^2)$,
we follow the steps \eqref{u1}-\eqref{jju2} to find the resummed   propagator in the one-loop approximation, namely
\beq
G_{R}(p^2)
&=&\frac{1}{16}g^2 v^4\Big(\frac{1}{p^2+m_h^2-\hat{\Pi}_{R}(p^2)}\Big)+C_{R}(p^2)
\label{dk3}
\eeq
with
\beq
\hat{\Pi}_{R}(p^2)&=&\frac{3}{36 \pi^2 g^2 v^4 m_h^2}\int_0^1 dx \Bigg\{ 6 m^4 \left(m_h^4+3 m^4\right)-\frac{p^4 m_h^2}{3} \left(9 m_h^2+35 m^2+4p^2\right)\nonumber\\
&+& p^2 \left(m^4 m_h^2+10 m^2 m_h^4+m_h^6+36 m^6\right)+m_h^4 \left(-p^2 \left(m_h^2+11 m^2\right)-6 m^4+p^4\right) \ln \left(\frac{m_h^2}{\mu ^2}\right)\nonumber\\
&+& m_h^2\left(m_h^4 p^2-2 m_h^2 m^4-6 m_h^2 m^2 p^2+12 m^6+24 m^4 p^2\right) \ln \left(\frac{x \left(m^2-p^2 (x-1)\right)-(x-1) m_h^2}{\mu ^2}\right)\nonumber\\
&+&33 m_h^2\left(2 m^6- m^4 p^2\right) \ln \left(\frac{m^2-p^2 (x-1) x}{\mu ^2}\right)\nonumber\\
&+& m^2 \left(m_h^2 \left(-48 m^4-17 m^2 p^2+3 p^4\right)+p^2 m_h^4-54 \left(m^6+2 m^4 p^2\right)\right) \ln \left(\frac{m^2}{\mu ^2}\right)\Bigg\}
\label{dk3}
\eeq
and
\beq
C_{R}(p^2)&=&\frac{1}{ 12(4\pi) ^2   }
\int_0^1 dx \Bigg\{\left(-18 m^2 +p^2\right) \ln \left(\frac{p^2 (1-x)x + m^2}{\mu ^2}\right)\nonumber\\
&+& \Big((2 \left(m_h^2-6 m^2\right)+p^2) \ln \left(\frac{p^2x(1-x)+(1-x)m_h^2+x m^2}{\mu ^2}\right)\Bigg\}.
\label{grgr}
\eeq
Looking now at the longitudinal part $L(p^2)$, it turns out to be
\beq
L(p^2)&=&\frac{1}{4} v^2	-\frac{1}{(4 \pi) ^2 }\Bigg(\frac{m_h^4-3m_h^4 \ln \left(\frac{m_h^2}{\mu ^2}\right)+9m^4-27 m^4 \ln \left(\frac{m^2}{\mu ^2}\right)}{2  m_h^2}  -\frac{1}{\epsilon} \left( m_h^2-9\frac{ m^4}{m_h^2}\right) \Bigg) \;.  \label{lexp}
\eeq
As it happens in the tree-level case, expression \eqref{lexp} is independent from the momentum $p^2$, meaning that it does not correspond to the propagation of some physical mode, a feature which is expected to persist at higher orders.

\section{Spectral properties \label{V}}
In this section, we will calculate the spectral properties associated with the correlation function obtained in the last section. For the employed techniques in obtaining the pole mass, residue and spectral density up to first order we refer to section \ref{2a}.  In  subsection \ref{IV}, we analyze the spectral properties of the elementary fields. In \ref{compp}, the spectral properties of the composite operators $O(x)$ and $R^a_{\mu}(x)$ are discussed.

\subsection{Spectral properties of the elementary fields \label{IV}}
\begin{figure}[H]
	\centering
	\includegraphics[width=10
	cm]{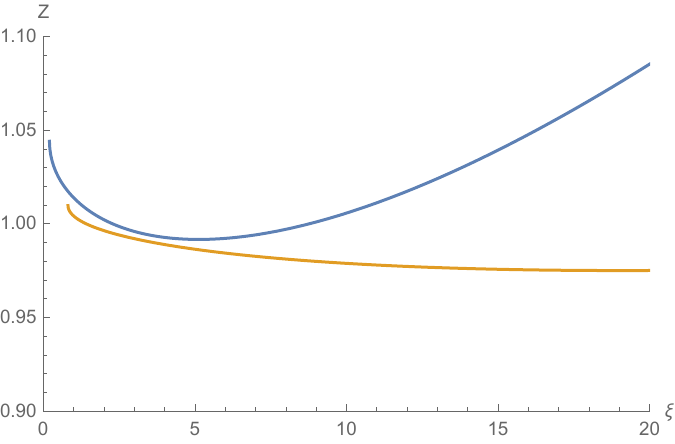}
	\caption{Dependence of the residue $Z$  for the Higgs field propagator on the gauge parameter $\xi$, for Region I (Blue), and Region II (Orange). }
	\label{ZZ}
\end{figure}

\begin{table}[h!]
	\center
	\begin{tabular}{|c|c|c|}
		\hline
		& \textbf{Region I} & \textbf{Region II} \\ \hline
		$~v~$  & 0.8 $\mu$                     & 1 $\mu$   \\ \hline
		$~g~$  & 1.2                          & 0.5           \\ \hline
		$~\lambda~$  & 0.3                          & 0.205    \\ \hline
	\end{tabular}		\caption{Parameter values used in the spectral density functions.}
	\label{tabel}
\end{table}
We first discuss the spectral properties of the elementary fields: the scalar Higgs field $h(x)$ and  the transverse part of the gauge field $A^a_{\m}(x)$. We will work with two sets of parameters, set out in Table \ref{tabel}. All values are given in units of the energy scale $\mu$. Also, we have that $m^2= \frac{1}{4}g^2 v^2$ and $m_h^2=\lambda v^2$, so that $m^2=0.23\, \mu^2$ and $m_h^2=0.192 \,\mu^2$ in Region I and    $m^2=0.625\, \mu^2$ and $m_h^2=0.205  \,\mu^2$ in Region II.
\\
\\
For the Higgs fields, following the steps from section \ref{2a}, we find the pole mass to first order in $\hbar$ to be: for Region I
\beq
m_{h,\rm pole}^2 &=& 0.207 \, \mu^2,
\label{higgsmass}
\eeq
and for Region II
\beq
m_{h,\rm pole}^2 &=& 0.206 \, \mu^2, \label{higgsmass2}
\eeq
for all values of the parameter $\xi$. This means that while the Higgs propagator \eqref{hhf} is gauge dependent, the pole mass is gauge independent. This is in full agreement  with the Nielsen identities of the $SU(2)$ Higgs model studied in \cite{gambino1999fermion}. The residue, however, is gauge dependent, as is depicted in  Figure~\ref{ZZ}.
For small values of $\xi$, including the Landau gauge $\xi=0$, the residue is not well-defined, and we cannot determine the spectral density function, as we will explain further in the next section.

\begin{figure}[H]
	\centering
	\includegraphics[width=18cm]{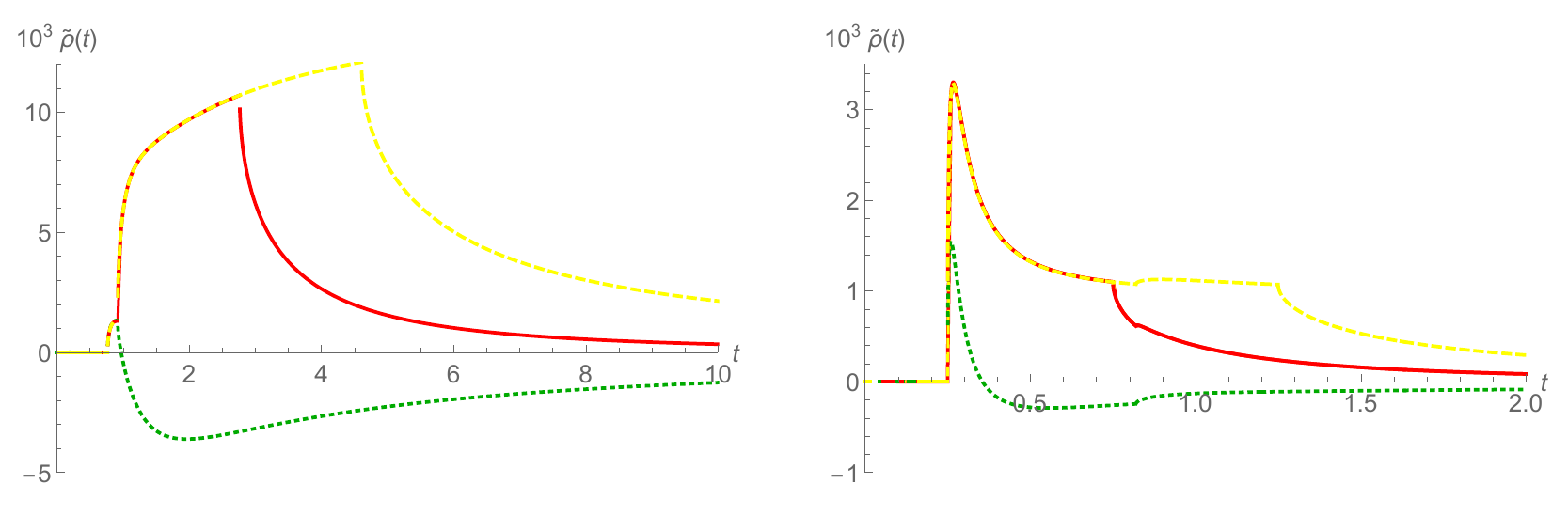}
	\caption{Spectral functions for the propagator $\langle h(p)h(-p) \rangle$, for $\xi = 1$ (Green, Dotted), $\xi = 3$ (Red, Solid),
		$\xi= 5$ (Yellow, Dashed),  with $t$ given in units of $\mu^2$, for Region I (left) and Region II (right) with the parameter values given in Table \ref{tabel}.}
	\label{Y33}
\end{figure}

In  Figure~\ref{Y33}, we find the spectral density functions both regions, for three values of $\xi: 1, 2, 5$. Looking at Region I, we see the first two-particle state appearing at $t=(m_h+m_h)^2=0.768\,  \mu^2$, followed by another two-particle state at $t=(m+m)^2=0.922 \, \mu^2$. Then, we see that there is a negative contribution, different for each diagram, at $t=(\sqrt{\xi} m+\sqrt{\xi} m)^2$. This corresponds to the (unphysical) two-particle state of two Goldstone bosons. For $\xi < 3$, this leads to a negative contribution for the spectral function, probably due to the large-momentum behaviour of the Higgs propagator \eqref{hhf}, for a detailed discussion see Appendix \ref{lot} . For Region II, we find essentially the same behaviour: a Higgs two-particle state at $t=(m_h+m_h)^2=0.81\,  \mu^2$, and a gauge field two-particle state at $t=(m+m)^2=0.25\, \mu^2$. We also see a negative contribution different for each diagram at $t=(\sqrt{\xi} m+\sqrt{\xi} m)^2$, corresponding to the (unphysical) two-particle state of two Goldstone bosons.

For the gauge field propagator, following the steps  from section~\ref{2a}, we find the first-order pole mass of the transverse gauge field to be: for Region I
\beq
m_{\rm pole}^2 &=& 0.274 \, \mu^2
\label{polem}
\eeq
and for Region II
\beq
m_{\rm pole}^2 &=& 0.065 \, \mu^2
\label{polema}
\eeq
for all values of the parameter $\xi$, so that the pole mass is gauge independent. The residue is, however,  gauge dependent as is depicted in Figure~\ref{Z}. For small values of $\xi$ the residue is not well-defined, as we will explain further in the next section.

\begin{figure}[H]
	\centering
	\includegraphics[width=10
	cm]{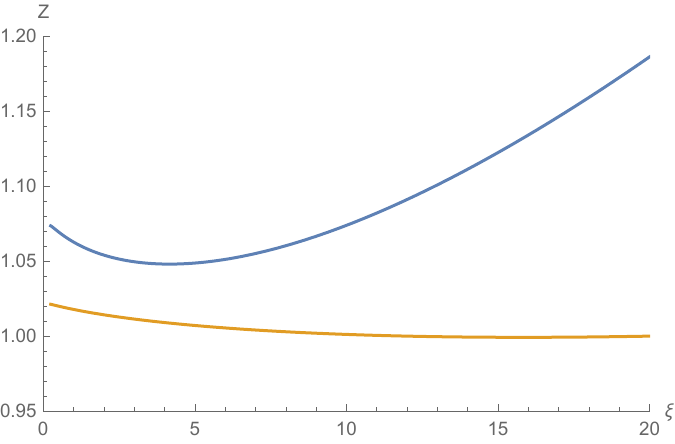}
	\caption{Dependence of the residue $Z$  for the gauge field propagator from  the gauge parameter $\xi$, for Region I (Blue), and Region II (Orange). }
	\label{Z}
\end{figure}

\begin{figure}[H]
	\centering
	\includegraphics[width=18cm]{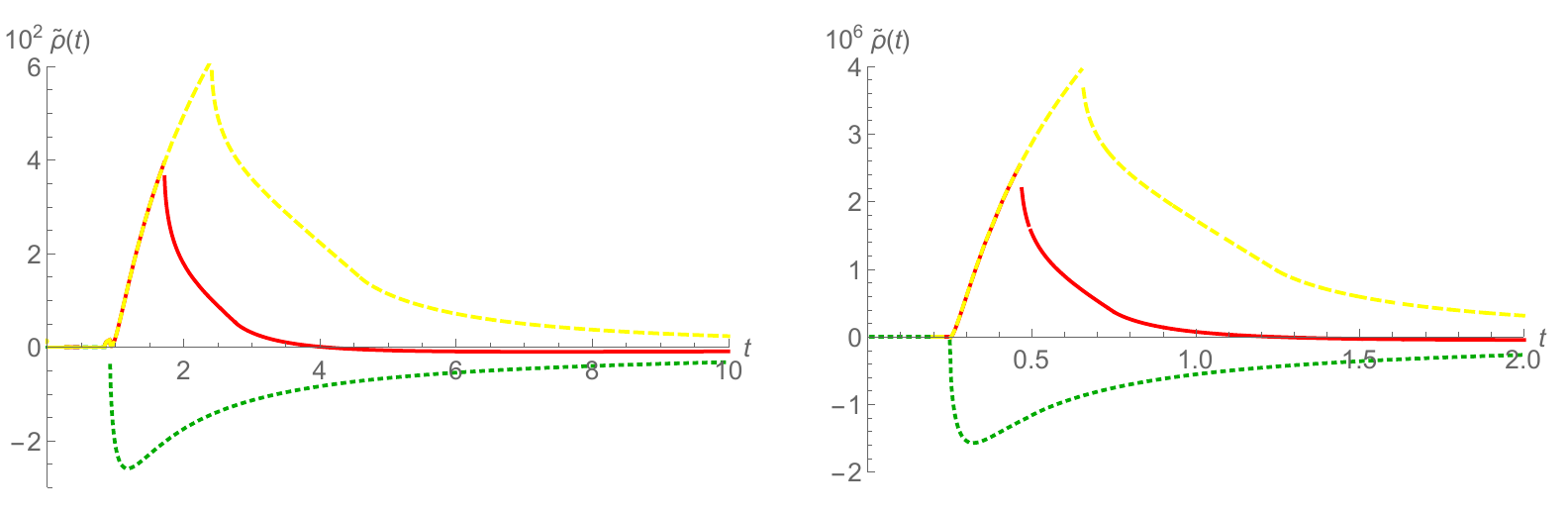}
	\caption{Spectral functions for the propagator $\langle A^a_{\mu}(p)A^b_{\nu}(-p) \rangle^T$, for $\xi = 1$ (Green, Dotted), $\xi = 3$ (Red, Solid),
		$\xi= 5$ (Yellow, Dashed),  with $t$ given in units of $\mu^2$, for Region I (left) and Region II (right) with the parameter values given in Table \ref{tabel}.}
	\label{Y3}
\end{figure}

In  Figure~\ref{Y3}, we find the spectral density functions for both regions, for three values of $\xi$: $1, 2, 5$. Looking at Region I, we see the first two-particle state appearing at $t=(m_h+m)^2=0.843\,  \mu^2$, followed by a two-particle state at $t=(m+m)^2=0.922 \, \mu^2$. Then, we see that there is a negative contribution, different for each diagram, at $t=( m+\sqrt{\xi} m)^2$. This corresponds to the (unphysical) two-particle state of a  gauge and Goldstone boson. For Region II, we find a gauge field two-particle state at $t=(m+m)^2=0.25 \, \mu^2$. We also see a negative contribution different for each diagram at $t=( m+\sqrt{\xi} m)^2$, corresponding to the (unphysical) two-particle state of two Goldstone bosons.

\subsection{Unphysical  threshold effects \label{stable}}

\begin{figure}[H]
	\centering
	\includegraphics[width=12cm]{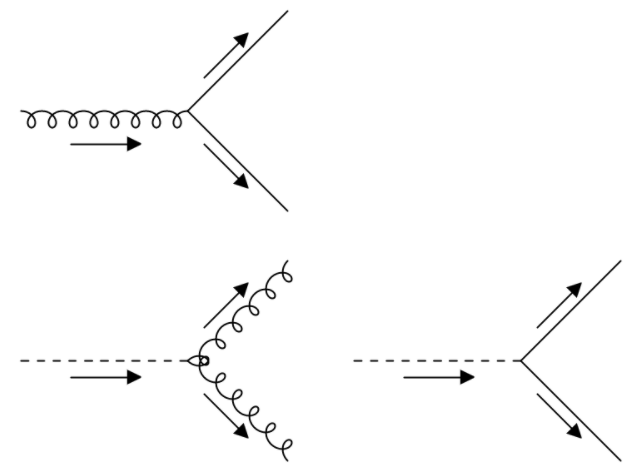}
	\caption{Possible decays of the  gauge boson (above) and the Higgs boson (below).}
	\label{yy}
\end{figure}
From the Feynman vertex rules given in Appendix~\ref{appA}, for certain values of the masses, unphysical  threshold effects can occur. These effects imply that for certain values of the (physical and unphysical) parameters, a ``decay'' occurs of a gauge and Higgs boson into two other particles, see Figure~\ref{yy}. We distinguish three cases:

\begin{itemize}
	\item[$(1.)$] Decay of a gauge vector boson in two Goldstone bosons: this happens when $m > 2  \sqrt{\xi} m$.
	\item[$(2.)$] Decay of a Higgs boson in two gauge vector bosons: this happens when $m_h > 2 m$.
	\item[$(3.)$] Decay of a Higgs boson in two Goldstone bosons: this happens when $m_h > 2 \sqrt{\xi} m$.
\end{itemize}
In order to guarantee the stability of the gauge boson, we therefore need from $(1.)$ that $\xi > \frac{1}{4}$. This means that for the Landau gauge $\xi =0$, the elementary gauge boson is not stable.
For the Higgs particle, to guarantee stability we need from $(2.)$ that $m_h < 2m$. Then, from $(3.)$ we find that $\xi >  \frac{m_h^2}{4m^2}$. This is the window in which we can work with a stable model.		We can have a look at what happens when we go outside of this window. For the Higgs particle, we see that for $m_h > 2m$, or $\lambda < g^2$, we will find a complex value for the first order pole mass, calculated through \eqref{ppp}. For $\lambda \geq g^2$, we will always find a real pole mass. Since the pole mass is gauge invariant, we find that this is true for all values of $\xi$. However, we do find that for $\xi >  \frac{m_h^2}{4m^2}$ and $ \lambda > g^2$, the real value of the pole mass is a real point inside the branch cut. This means that we cannot achieve the usual differentiation around this point. As a consequence, we cannot consistently construct the residue, so that we are unable to obtain a first-order spectral function. For the gauge field, we find the same problem when $\xi < \frac{1}{4}$.
\\
\\
The foregoing mathematically correct observations clearly show that there is something physically wrong with using the elementary fields' spectral functions.  Luckily, all of these shortcomings are surpassed by using the gauge invariant composite operators.

\subsection{Spectral properties for the composite fields \label{compp}}
\begin{figure}[H]
	\centering
	\includegraphics[width=18cm]{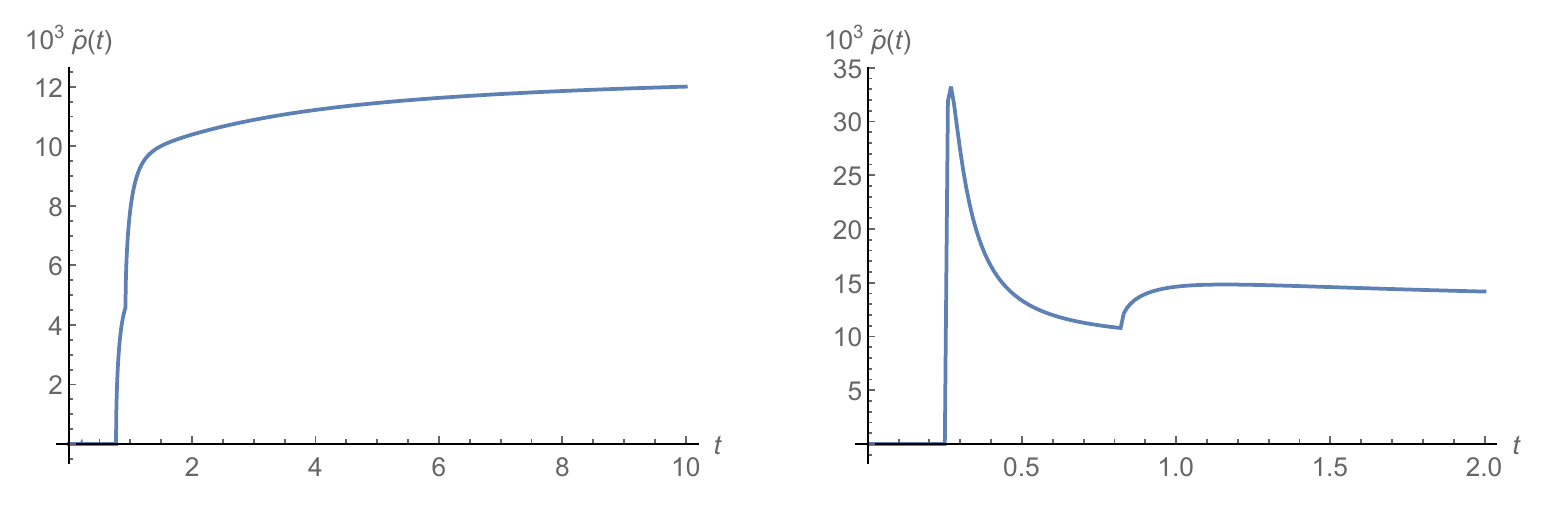}
	\caption{Spectral function for the two-point correlation function $\langle {O}(p){O}(-p) \rangle$, with $t$ given in units of $\mu^2$, for the Region I (left) and Region II (right) with parameter values given in Table \ref{tabel}. }
	\label{Y}
\end{figure}

For the scalar composite operator ${O}(x)$ we find the first-order pole mass for Region I
\beq
m_{OO, \rm pole}^2 &=& 0.207\, \mu^2, \label{OOa}
\eeq
and for Region II
\beq
m_{OO, \rm pole}^2 &=& 0.206 \, \mu^2, \label{OOb}
\eeq
which is equal to the pole mass of the elementary Higgs field in \eqref{higgsmass}, as we expect from eq.~\eqref{zlf}. Following the steps from section~\ref{2a}, we find the first-order residue
\beq
Z= 1.11 \, v^2
\eeq
for Region I and
\beq
Z=1.01 \,v^2
\eeq
for Region II. The first order spectral function for $\braket{O(x)\,O(y)}$ is shown in Figure~\ref{Y}. Comparing this result with that of the spectral function of the Higgs field in  Figure~\ref{Y3}, we see a two-particle state for the Higgs field at $t=(m_h+m_h)^2$, and a two-particle state for the gauge vector field, starting at $t=(m+m)^2$. The difference is that for the gauge invariant correlation function $\braket{O(x)\,O(y)}$ we no longer have the unphysical Goldstone two-particle state. Due to the absence of this negative contribution, the spectral function is positive throughout the spectrum. In fact, we see that for bigger values of $\xi$, we find that the spectral function of the elementary Higgs field resembles more and more the spectral function of the composite operator $O(x)$. This makes sense, since for $\xi \rightarrow \infty$, we are approaching the unitary gauge which has a more direct link with the physical spectrum of the elementary excitations. In Appendix~\ref{un}, one finds a detailed discussion about the unitary gauge limit $\xi \rightarrow \infty$ as well as the calculation of the spectral function.  The asymptotic (constant) behaviour is directly related to the (classical) dimension of the used composite operator.

\begin{figure}[H]
	\centering
	\includegraphics[width=18cm]{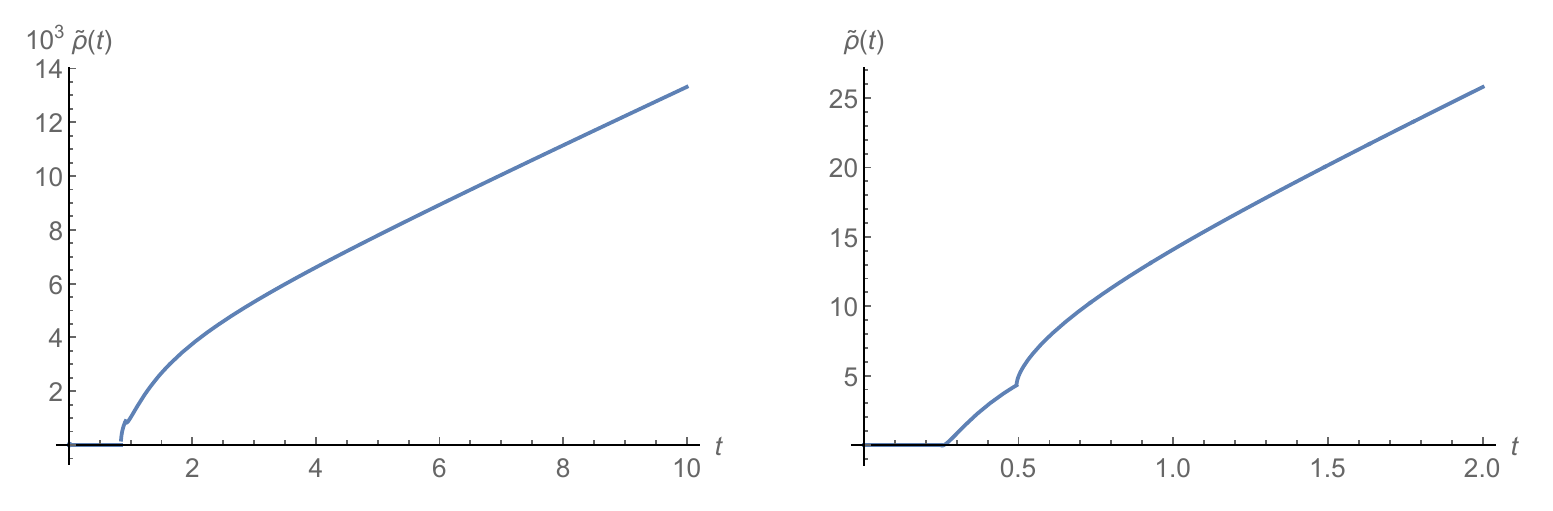}
	\caption{Spectral function for the transverse two-point correlation function  $ G_{R}(p^2)$, with $t$ given in units of $\mu^2$, for the Region I (left) and Region II (right) with parameter values given in Table \ref{tabel}. }
	\label{rtrt}
\end{figure}

For the transverse part of the two-point correlation function 	$ G_{R}(p^2)$, eq.~\eqref{dk3}, for our set of parameters we find the first-order pole mass: in Region I
\beq
m_{R, \rm pole}^2 &=& 0.274 \, \mu^2  \label{voa}
\eeq
and Region II
\beq
m_{R, \rm pole}^2 &=& 0.065 \, \mu^2  \label{vob}
\eeq
which is the same as the pole mass of the transverse gauge field, eq.~\eqref{polem}, in agreement with eq.~\eqref{zlf}. Following the steps from \ref{2a}, we find the first-order residue
\beq
Z=\frac{1}{16}g^2 v^4 \left(1.27\right)
\eeq
for Region I and
\beq
Z=\frac{1}{16}g^2 v^4 \left(1.05\right)
\eeq
for Region II. The first order spectral function for $ G_{R}(p^2)$  is shown in Figure~\ref{rtrt}. Comparing this to the spectral function of the gauge vector field in  Figure~\ref{Y3}, we see a two-particle state at $t=(m_h+m)^2$, and a two-particle state for the gauge field, starting at $t=(m+m)^2$. Again, as in the case of the two-point correlation function of the scalar operator $O(x)$, the difference is that for this gauge invariant correlation function we no longer have the unphysical Goldstone/gauge boson two-particle state. Due to the absence of this negative contribution, the spectral function is positive throughout the spectrum. In fact, we see that for bigger values of $\xi$, we find that the spectral function of the elementary gauge field resembles more and more the spectral function of the composite operator $R^a_{\mu}(x)$. As already mentioned previously, this relies on the fact that in the limit $\xi \rightarrow \infty$ we are approaching  the unitary gauge, see  Appendix \ref{un}.  Also here, the linear increase at large $t$ follows from the operator dimension.

\section{Conclusion \label{VI}}

This chapter is the natural extension of chapter \ref{VII} and \ref{VV}, where the Abelian $U(1)$ Higgs model has been scrutinized by employing two local composite BRST invariant operators whose two-point correlation functions provide a fully gauge independent description of the elementary excitations of the model, namely the Higgs and the massive gauge boson.\\\\
This formulation generalizes to the non-Abelian Higgs model as, for example, the $SU(2)$ YM theory with a single Higgs in the fundamental representation \cite{tHooft:1980xss,Frohlich:1980gj, Frohlich:1981yi}. This is the model which has been considered in the present analysis. The local BRST invariant composite operators $(O(x), R^a_\mu(x))$ which generalize their $U(1)$ counterparts are given in eq.~\eqref{scopt} and in eqs.~\eqref{threeop},\eqref{threeop1}.
The two-point correlation functions $\langle O(x) O(y) \rangle$ and
$\langle R^a_\mu(x) R^b_\nu(y) \rangle^T$, where the superscript $T$ stands for the transverse component, have been evaluated at one-loop order in the $R_\xi$-gauge and compared with the corresponding correlation functions of the elementary fields $\langle h(x) h(y) \rangle $ and $\langle A^a_\mu(x) A^b_\nu(y) \rangle^T$. It turns out that both $\langle O(x) O(y) \rangle$ and $\langle h(x) h(y) \rangle $ share the same gauge independent pole mass, eqs.\eqref{higgsmass},\eqref{higgsmass2},\eqref{OOa},\eqref{OOb}, in agreement with both Nielsen identities \cite{Nielsen:1975fs,Piguet:1984js,gambino1999fermion,gambino2000nielsen,Grassi:2000dz,Aitchison:1983ns,Andreassen:2014eha} and the BRST invariant nature of $O(x)$. Nevertheless, unlike the residue and spectral function of the elementary correlator $\langle h(x) h(y) \rangle $, which exhibit a strong unphysical dependence from the gauge parameter $\xi$, Figure \eqref{Y33}, the spectral density of $\langle O(x) O(y) \rangle$ turns out to be $\xi$-independent and positive over the whole $p^2$ axis, Figure \eqref{Y}.  The same features hold for $\langle A^a_\mu(x) A^b_\nu(y) \rangle^T$ and $\langle R^a_\mu(x) R^b_\nu(y) \rangle^T$. Again, both correlation function share have the same $\xi$-independent pole mass, eqs.~\eqref{polema},\eqref{polema},\eqref{voa},\eqref{vob}. Though, unlike the $\xi$-dependent spectral function associated to $\langle A^a_\mu(x) A^b_\nu(y) \rangle^T$, Figure \eqref{Y3}, that corresponding to $\langle R^a_\mu(x) R^b_\nu(y) \rangle^T$, Figure \eqref{rtrt},  turns out to be independent from the gauge parameter $\xi$ and positive. As such, the local composite operators $(O(x), R^a_\mu(x))$ provide a fully BRST consistent description of the observable scalar (Higgs) and vector boson particles.
\\
\\
It is worth mentioning here that, besides the BRST invariance of the gauge fixed action, the model exhibits an additional global custodial symmetry, eqs.\eqref{custodial},\eqref{beta}, according to which all fields carrying the index $a=1,2,3$, {\it i.e.}~$(A^a_\mu, b^a, c^a, {\bar c}^a, \rho^a)$, undergo a global transformation in the adjoint representation of $SU(2)$. The same feature holds for the composite operators $(O(x), R^a_\mu(x))$ which transform exactly as $h$ and $A^a_\mu$. More precisely, the operator $O(x)$ is a singlet under the custodial symmetry, while the operators $R^a_\mu$ transform like a triplet, eq.~\eqref{tripr}, so that the correlation function $\langle R^a_\mu(p) R^b_\nu(-p) \rangle$ displays the same $SU(2)$ structure of the elementary two-point function  $\langle A^a_\mu(p) A_\nu(-p)\rangle$, eqs.\eqref{cc1}-\eqref{cc4}. Although not being the aim of the present analysis, we expect that the existence of a global custodial symmetry will imply far-reaching consequences for the renormalization properties of the composite operators $(O(x), R^a_\mu(x))$ encoded in the corresponding Ward identities. The renormalization properties of the local gauge-invariant composite operators for the $U(1)$ Higgs model have been reported recently in \cite{Capri:2020ppe}.
\newpage

\chapter{Conclusion and Outlook \label{con}}
In this work, we have evaluated different BRST-invariant solutions for the introduction of a mass term in YM theories. More specifically, we have looked at gauge-invariant extensions of the elementary gauge field and Higgs boson, and we evaluated the two-point functions and spectral properties for these gauge-invariant local composite fields. While the proposed solutions differ in methods and levels at which the BRST invariance is introduced, they serve a common goal: to access the physical spectrum for the nuclear forces. \\
\\
 For the composite field $A^h(x)$, we have shown that a renormalizable gauge class can be developed for this field, which introduces a mass $\mu$ for the Stueckelberg-like field $\xi^a$, which is used in the localization procedure for the composite gauge field. This mass ensures the infrared safety of the propagator for this field. The gauge class, with $\mu$ as the gauge parameter, gives the opportunity to investigate physical objects related to the gauge field propagator outside of the Landau gauge. For example, in \cite{Mintz_2019} it was shown, by analyzing dimension two condensates, that the instability of the GZ action observed in the Landau gauge persists in this larger gauge class, suggesting a physical meaning for the refinement of the GZ action in the infrared regime. This indicates an important role for the RGZ action in the IR description of the YM theories. \\
\\
For the gauge-invariant local composite operators in the Higgs model, we have investigated spectral properties of the field propagators. The most important conclusion here is that for truly gauge-invariant objects, we have not observed any non-physical behavior. In particular, we have not observed any non-physical behaviour of the kind that is commonly associated with confinement, such as positivity violation of the spectral density function. We immediately add that this does not mean that positivity violation, as has been observed for the spectral density function of the elementary gauge field operator, should be solely ascribed to a lack of gauge-invariance. Even though we were able to make a direct comparison between gauge-invariant composite operators and the gauge-dependent elementary operators, and we concluded that the first always shows a positive spectral density function, while the second for some values of the gauge parameter shows a negative spectral density function, our methods are not suited to reach the non-perturbative region in which positivity violation has been observed in other studies. Since our study consisted of perturbative loop calculations in the Higgs model, we are far from the IR region where we expect this confinement-like behavior. It is therefore our hope that this work will encourage others to investigate the spectral properties of the  gauge-invariant composite local operators developed in this work with non-perturbative methods, such as lattice simulations. The composite gauge-invariant local operators have already led to some preliminary lattice results \cite{maas2019observable, maas2015field}. \\
\\
A direct comparison between the two methods of developing gauge-invariant composite fields discussed in this thesis can be made in the Landau gauge. In this gauge, the composite gauge field $A_{\mu}^h(x)$ is equal to the elementary field $A_{\mu}(x)$, and the action will be equal to the action of the massive YM model discussed in section \ref{massym}. The appearance of complex poles for the propagator of the gauge field, even at a perturbative level, means that there is no physical interpretation for these propagators. Thus, while BRST invariance is established by the configuration of $A^h_{\mu}(x)$, this does not seem to ensure the usual features of the definition of physical space that are associated with BRST invariance. Possibly, this is because the BRST invariance is introduced at a non-local level. In contrast, the local composite BRST invariant operators developed for the Higgs model always have a real pole mass. However, for the Landau gauge the residue is not well-defined, and we are unable to subtract a spectral density function of the gauge field for this gauge choice. Therefore, the Landau gauge might not be the most suitable gauge to work with for massive non-Abelian models.
\\
\\
 If we want to access the non-perturbative region to investigate further the local composite gauge-invariant operators in the Higgs model, the continuum offers some ways to access this region besides numerical methods. The BRST  invariant nature of $(O(x), R^a_\mu(x))$ makes them natural candidates to an attempt at facing the  challenge of investigating the infrared non-perturbative behaviour of the model, trying to make contact with the  analytical lattice predictions of Fradkin and Shenker \cite{Fradkin:1978dv}. \\\\
 For the $SU(2)$ HYM theory one may for example introduce a horizon term, in its BRST-invariant formulation encoded in the RGZ action (cf. \cite{Vandersickel:2012tz,Dudal:2008sp,Capri:2015ixa,Capri:2018ijg} and refs. therein) implementing the restriction to the Gribov region $\Omega$ \cite{gribov1978quantization} in order to take into account the existence of the Gribov copies plaguing the FP quantization procedure.
As a consequence, the gauge-invariant pole masses of the non-Abelian generalization of the correlation functions $\langle R^a_{\mu}(x) R^a_{\nu}(y) \rangle$ and  $\langle O(x) O(y) \rangle$ will now show an explicit dependence on the Gribov mass parameter as well as on the dimension-two condensates present in the RGZ action \cite{Vandersickel:2012tz,Dudal:2008sp,Capri:2015ixa,Capri:2018ijg}. Thus, extending the framework already outlined in \cite{Capri:2012ah}, the aforementioned  pole masses and further spectral properties could be employed as gauge-invariant probing quantities in order to extract non-perturbative information about the behaviour of the elementary excitations of HYM theories in the light of the Fradkin-Shenker results. \\
\\
The employment of an order parameter to distinguish between the confinement/deconfinement region can also be done in a YM model with finite temperature. In recent years, much valuable progress has been made towards the understanding of
non-abelian gauge theories at finite temperature using background field gauge (BFG) methods \cite{Abbott:1980hw, Abbott:1981ke} in the Landau-DeWitt gauge. BFG methods provide an efficient way to describe the confinement/deconfinement order parameter (the Polyakov loop or any of its proxies \cite{Braun:2007bx}) because the related center symmetry is explicit at the quantum level and is easily maintained in approximation schemes \cite{Herbst:2015ona,Reinosa:2015gxn,Reinosa_HDR}. Several models have been put forward in order to implement the BFG formalism in the Landau-deWitt gauge while restricting the number of Gribov copies. In \cite{Reinosa:2014ooa, Reinosa:2014zta,Reinosa:2015gxn,Reinosa:2015oua,Maelger:2017amh}, the formalism was used within the massive YM model to compute the background potential and Polyakov loop up to two-loop order, both in pure YM theories and in heavy-quark QCD. Recently, results on the BFG method in the Landau-DeWitt gauge for the GZ model have been reported \cite{Canfora:2015yia, Dudal:2017jfw, Kroff:2018ncl, egmond2020scalar}. It would be interesting to see if this method could be combined with the gauge-invariant composite operators developed in this work.
\\
\\
But even without accessing the non-perturbative region, the gauge-invariant composite operators offer several interesting lines of investigation. For example, we could compare different gauge classes that leave different remnant global gauges, such as the $R_{\xi}$ gauge and the Linear Covariant Gauges (LCG). The $R_{\xi}$ gauges break the global gauge symmetry of the model, in contrast to the LCG. We could  compare the pole masses of the elementary fields for these two gauge classes. Then, they can be compared with the pole mass of their gauge-invariant composite extensions, which should of course be the same in
every gauge class. This is a very interesting exercise as in a sense it will reveal which class of gauges is physical.
Another line of investigation is given by the gauge-invariant quadratic
Higgs condensate $\varphi \varphi^{\dagger}$, similar to the Coleman-Weinberg condensate \cite{Coleman:1973jx, knecht2001new}. In particular,
we are curious to see what is the influence of this condensate on the propagators of the
gauge-invariant operators, including their pole structure.
\\
\\
Finally, a most promising avenue to further explore is the $SU(2)\times U(1)$ setting of the electroweak theory. This sector is of course of special interest because of
the experimental data that are available for the masses of the W and Z bosons. We would be the first to ever calculate these masses analytically in a gauge-invariant way.
\newpage

\appendix
\chapter{Technical details of the gauge invariant operator $A^2_{\text{min}}$}
\section{Properties of the functional $f_{A}[u]$.} \label{apb}In this
Appendix we recall some useful properties of the functional
$f_{A}[u]$
\begin{equation}
f_{A}[u]\equiv \mathrm{Tr}\int d^{4}x\,A_{\mu }^{u}A_{\mu
}^{u}=\mathrm{Tr}\int d^{4}x\left( u^{\dagger }A_{\mu
}u+\frac{i}{g}u^{\dagger }\partial _{\mu }u\right) \left(
u^{\dagger }A_{\mu }u+\frac{i}{g}u^{\dagger }\partial _{\mu
}u\right) \;. \label{fa}
\end{equation}
For a given gauge field configuration $A_{\mu }$, $f_{A}[u]$ is a functional
defined on the gauge orbit of $A_{\mu }$. Let $\mathcal{A}$ be the space of
connections $A_{\mu }^{a}$ with finite Hilbert norm $||A||$, \textit{i.e.}
\begin{equation}
||A||^{2}=\mathrm{Tr}\int d^{4}x\,A_{\mu }A{_{\mu }=}\frac{1}{2}\int
d^{4}xA_{\mu }^{a}A_{\mu }^{a}<+\infty \;, \label{norm0}
\end{equation}
and let $\mathcal{U}$ be the space of local gauge transformations $u$ such
that the Hilbert norm $||u^{\dagger }\partial {u}||$ is finite too, namely
\begin{equation}
||u^{\dagger }\partial {u}||^{2}=\mathrm{Tr}\int d^{4}x\,\left(
u^{\dagger }\partial _{\mu }u\right) \left( u^{\dagger }\partial
_{\mu }u\right) <+\infty \;. \label{norm1}
\end{equation}
\noindent The following proposition holds
\cite{Zwanziger:1990tn,DellAntonio:1989wae,DellAntonio:1991mms,vanBaal:1991zw}
\begin{itemize}
	\item  \underline{Proposition}\newline
	The functional $f_{A}[u]$ achieves its absolute minimum on the gauge orbit
	of $A_{\mu }$.
\end{itemize}
\noindent This proposition means that there exists a $h\in
\mathcal{U}$ such that
\begin{eqnarray}
\delta f_{A}[h] &=&0\;,  \label{impl0} \\
\delta ^{2}f_{A}[h] &\ge &0\;,  \label{impl1} \\
f_{A}[h] &\le &f_{A}[u]\;,\;\;\;\;\;\;\;\forall \,u\in
\mathcal{U}\;. \label{impl2}
\end{eqnarray}
The operator $A_{\min }^{2}$ is thus given by
\begin{equation}
A_{\min }^{2}=\min_{\left\{ u\right\} }\mathrm{Tr}\int
d^{4}x\,A_{\mu }^{u}A_{\mu }^{u}=f_{A}[h]\;.  \label{a2min}
\end{equation}
Let us give a look at the two conditions (\ref{impl0}) and
(\ref{impl1}). To evaluate $\delta f_{A}[h]$ and $\delta
^{2}f_{A}[h]$ we set\footnote{The case of the gauge group $SU(N)$
	is considered here.}
\begin{equation}
v=he^{ig\omega }=he^{ig\omega ^{a}T^{a}}\;,  \label{set0}
\end{equation}
\begin{equation}
\left[ T^{a},T^{b}\right] =if^{abc}\;T^c\;,\;\;\;\;\;\mathrm{Tr}\left( T^{a}T^{b}\right) =%
\frac{1}{2}\delta ^{ab}\;,  \label{st000}
\end{equation}
where $\omega $ is an infinitesimal hermitian matrix and we
compute the linear and quadratic terms of the expansion of the
functional $f_{A}[v]$ in power series of $\omega $. Let us first
obtain an expression for $A_{\mu }^{v}$
\begin{eqnarray}
A_{\mu }^{v} &=&v^{\dagger }A_{\mu }v+\frac{i}{g}v^{\dagger }\partial _{\mu
}v  \nonumber \\
&=&e^{-ig\omega }h^{\dagger }A_{\mu }he^{ig\omega }+\frac{i}{g}e^{-ig\omega
}\left( h^{\dagger }\partial _{\mu }h\right) e^{ig\omega }+\frac{i}{g}%
e^{-ig\omega }\partial _{\mu }e^{ig\omega }  \nonumber \\
&=&e^{-ig\omega }A_{\mu }^{h}e^{ig\omega }+\frac{i}{g}e^{-ig\omega }\partial
_{\mu }e^{ig\omega }\;.  \label{orbit0}
\end{eqnarray}
Expanding up to the order $\omega ^{2}$, we get
\begin{eqnarray}
A_{\mu }^{v} &=&\left( 1-ig\omega -g^{2}\frac{\omega ^{2}}{2}\right) A_{\mu
}^{h}\left( 1+ig\omega -g^{2}\frac{\omega ^{2}}{2}\right) +\frac{i}{g}\left(
1-ig\omega -g^{2}\frac{\omega ^{2}}{2}\right) \partial _{\mu }\left(
1+ig\omega -g^{2}\frac{\omega ^{2}}{2}\right)  \nonumber \\
&=&\left( 1-ig\omega -g^{2}\frac{\omega ^{2}}{2}\right) \left( A_{\mu
}^{h}+igA_{\mu }^{h}\omega -g^{2}A_{\mu }^{h}\frac{\omega ^{2}}{2}\right) +
\nonumber \\
&+&\frac{i}{g}\left( 1-ig\omega -g^{2}\frac{\omega ^{2}}{2}\right) \left(
ig\partial _{\mu }\omega -\frac{g^{2}}{2}\left( \partial _{\mu }\omega
\right) \omega -\frac{g^{2}}{2}\omega \left( \partial _{\mu }\omega \right)
\right)  \nonumber \\
&=&A_{\mu }^{h}+igA_{\mu }^{h}\omega -\frac{g^{2}}{2}A_{\mu }^{h}\omega
^{2}-ig\omega A_{\mu }^{h}+g^{2}\omega A_{\mu }^{h}\omega -\frac{g^{2}}{2}%
\omega ^{2}A_{\mu }^{h}  \nonumber \\
&+&\frac{i}{g}\left( ig\partial _{\mu }\omega -\frac{g^{2}}{2}\left(
\partial _{\mu }\omega \right) \omega -\frac{g^{2}}{2}\omega \partial _{\mu
}\omega +g^{2}\omega \partial _{\mu }\omega \right) +O(\omega ^{3})\;,
\label{ex1}
\end{eqnarray}
from which it follows
\begin{equation}
A_{\mu }^{v}=A_{\mu }^{h}+ig[A_{\mu }^{h},\omega ]+\frac{g^{2}}{2}[[\omega
,A_{\mu }^{h}],\omega ]-\partial _{\mu }\omega +i\frac{g}{2}[\omega
,\partial _{\mu }\omega ]+O(\omega ^{3})\;,  \label{A0}
\end{equation}
We now evaluate
\begin{eqnarray}
f_{A}[v] &=&\mathrm{Tr}\int d^{4}xA_{\mu }^{u}A_{\mu }^{u}  \nonumber \\
&=&\mathrm{Tr}\int d^{4}x\,\left[ \left( A_{\mu }^{h}+ig[A_{\mu }^{h},\omega ]+\frac{%
	g^{2}}{2}[[\omega ,A_{\mu }^{h}],\omega ]-\partial _{\mu }\omega +i\frac{g}{2%
}[\omega ,\partial _{\mu }\omega ]+O(\omega ^{3})\right) \times
\right.
\nonumber \\
& &\left. \left( A_{\mu }^{h}+ig[A_{\mu }^{h},\omega ]+\frac{g^{2}}{2}[%
[\omega ,A_{\mu }^{h}],\omega ]-\partial _{\mu }\omega +i\frac{g}{2}[\omega
,\partial _{\mu }\omega ]+O(\omega ^{3})\right) \right]  \nonumber \\
&=&\mathrm{Tr}\int d^{4}x\,\left\{ A_{\mu }^{h}A_{\mu
}^{h}+igA_{\mu }^{h}[A_{\mu
}^{h},\omega ]+g^{2}A_{\mu }^{h}\omega {A}_{\mu }^{h}\omega -\frac{g^{2}}{2}%
A_{\mu }^{h}A_{\mu }^{h}\omega ^{2}-\frac{g^{2}}{2}A_{\mu }^{h}\omega
^{2}A_{\mu }^{h}-A_{\mu }^{h}\partial _{\mu }\omega \right.  \nonumber \\
&+&\left. i\frac{g}{2}A_{\mu }^{h}[\omega ,\partial _{\mu }\omega
]+ig[A_{\mu }^{h},\omega ]A_{\mu }^{h}-g^{2}[A_{\mu }^{h},\omega
][A_{\mu }^{h},\omega ]-ig[A_{\mu }^{h},\omega ]\partial _{\mu
}\omega +g^{2}\omega
A_{\mu }^{h}\omega A_{\mu }^{h}\right.  \nonumber \\
&-&\left. \frac{g^{2}}{2}A_{\mu }^{h}\omega ^{2}A_{\mu }^{h}-\frac{g^{2}}{2}%
\omega ^{2}A_{\mu }^{h}A_{\mu }^{h}-\partial _{\mu }\omega A_{\mu
}^{h}-ig\partial _{\mu }\omega [A_{\mu }^{h},\omega ]+\partial _{\mu }\omega
\partial _{\mu }\omega +i\frac{g}{2}[\omega ,\partial _{\mu }\omega ]A_{\mu
}^{h}\right\} +O(\omega ^{3})\nonumber\\
&=&f_{A}[h]-\mathrm{Tr}\int d^{4}x\,\left\{
A_{\mu }^{h},\partial _{\mu }\omega
\right\} +\mathrm{Tr}\int d^{4}x\,\left( g^{2}A_{\mu }^{h}\omega A_{\mu }^{h}\omega -%
\frac{g^{2}}{2}A_{\mu }^{h}A_{\mu }^{h}\omega ^{2}-\frac{g^{2}}{2}A_{\mu
}^{h}\omega ^{2}A_{\mu }^{h}\right.  \nonumber \\
&-&\left. g^{2}[A_{\mu }^{h},\omega ][A_{\mu }^{h},\omega
]+g^{2}\omega A_{\mu }^{h}\omega A_{\mu
}^{h}-\frac{g^{2}}{2}A_{\mu }^{h}\omega ^{2}A_{\mu
}^{h}-\frac{g^{2}}{2}\omega ^{2}A_{\mu }^{h}A_{\mu }^{h}\right)
+\mathrm{Tr}\int d^{4}x\,\left( \partial _{\mu }\omega \partial
_{\mu }\omega \right.
\nonumber \\
&+&\left. i\frac{g}{2}[\omega ,\partial _{\mu }\omega ]A_{\mu
}^{h}-ig\partial _{\mu }\omega [A_{\mu }^{h},\omega ]-ig[A_{\mu
}^{h},\omega ]\partial _{\mu }\omega +i\frac{g}{2}A_{\mu
}^{h}[\omega ,\partial _{\mu }\omega ]\right) +O(\omega ^{3})
\nonumber
\end{eqnarray}
\begin{eqnarray}
&=&f_{A}[h]+2\int {d^{4}x}\,tr\left( \omega \partial _{\mu
}{A}_{\mu }^{h}\right) +\int {d^{4}x}\,tr\left\{ 2g^{2}\omega
{A}_{\mu }^{h}\omega
A_{\mu }^{h}-2g^{2}A_{\mu }^{h}A_{\mu }^{h}\omega ^{2}\right.  \nonumber \\
&-&\left. g^{2}\left( A_{\mu }^{h}\omega -\omega {A}_{\mu }^{h}\right)
\left( A_{\mu }^{h}\omega -\omega {A}_{\mu }^{h}\right) \right\} +\int {%
	d^{4}x}\,tr\left( \partial _{\mu }\omega \partial _{\mu }\omega +i\frac{g}{2}%
\omega \partial _{\mu }\omega {A}_{\mu }^{h}-i\frac{g}{2}\partial _{\mu
}\omega \omega {A}_{\mu }^{h}\right.  \nonumber \\
&-&\left. ig\partial _{\mu }\omega {A}_{\mu }^{h}\omega +ig\partial _{\mu
}\omega \omega {A}_{\mu }^{h}-igA_{\mu }^{h}\omega \partial _{\mu }\omega
+ig\omega {A}_{\mu }^{h}\partial _{\mu }\omega +i\frac{g}{2}A_{\mu
}^{h}\omega \partial _{\mu }\omega -i\frac{g}{2}A_{\mu }^{h}\partial _{\mu
}\omega \omega \right) +O(\omega ^{3})  \nonumber \\
&=&f_{A}[h]+2\mathrm{Tr}\int d^{4}x\left( \,\omega \partial _{\mu
}A_{\mu }^{h}\right) +\mathrm{Tr}\int d^{4}x\,\left( \partial
_{\mu }\omega
\partial _{\mu }\omega +ig\omega \partial _{\mu }\omega {A}_{\mu
}^{h}-ig\partial _{\mu
}\omega \omega {A}_{\mu }^{h}\right.  \nonumber \\
&-&\left. 2ig\partial _{\mu }\omega A_{\mu }^{h}\omega +2ig\partial
_{\mu }\omega \omega A_{\mu }^{h}\right) +O(\omega ^{3})\;.
\end{eqnarray}
Thus
\begin{eqnarray}
f_{A}[v] &=&f_{A}[h]+2\mathrm{Tr}\int d^{4}x\,\left( \omega
\partial _{\mu }A_{\mu }^{h}\right) +\mathrm{Tr}\int d^{4}x\,\left(
\partial _{\mu }\omega \partial _{\mu }\omega +ig\omega \partial
_{\mu }\omega A_{\mu }^{h}-ig\partial _{\mu
}\omega \omega A_{\mu }^{h}\right.  \nonumber \\
&-&\left. ig\left( \partial _{\mu }\omega \right) A_{\mu }^{h}\omega
+ig\left( \partial _{\mu }\omega \right) \omega A_{\mu }^{h}\right)
+O(\omega ^{3})  \nonumber \\
&=&f_{A}[h]+2\mathrm{Tr}\int d^{4}x\,\left( \omega \partial _{\mu
}A_{\mu }^{h}\right) +\mathrm{Tr}\int d^{4}x\,\left\{ \partial
_{\mu }\omega \left(
\partial _{\mu }\omega -ig\left[ A_{\mu }^{h},\omega \right]
\right) \right\} +O(\omega ^{3})\;. \nonumber  \\ \label{f1}
\end{eqnarray}
Finally
\begin{equation}
f_{A}[v]=f_{A}[h]+2\mathrm{Tr}\int d^{4}x\,\left( \omega \partial
_{\mu }A_{\mu }^{h}\right) -\mathrm{Tr}\int d^{4}x\,\omega
\partial _{\mu }D_{\mu }(A^{h})\omega +O(\omega ^{3})\;,
\label{func2}
\end{equation}
so that
\begin{eqnarray}
\delta f_{A}[h] &=&0\;\;\;\Rightarrow \;\;\;\partial _{\mu }A_{\mu
}^{h}\;=\;0\;,  \nonumber \\
\delta ^{2}f_{A}[h] &>&0\;\;\;\Rightarrow \;\;\;-\partial _{\mu }D{_{\mu }(}%
A^{h}{)}\;>\;0\;.  \label{func3}
\end{eqnarray}
We see therefore that the set of field configurations fulfilling conditions (%
\ref{func3}), \textit{i.e.} defining relative minima of the functional $%
f_{A}[u]$, belong to the so called Gribov region $\Omega $, which is defined
as
\begin{equation}
\Omega =\left.\{A_{\mu }\right|\partial _{\mu }A_{\mu
}=0\;\mathrm{and}\;-\partial _{\mu }D_{\mu }(A)>0\}\;.
\label{gribov0}
\end{equation}
Let us proceed now by showing that the transversality condition,
$\partial
_{\mu }A_{\mu }^{h}=0$, can be solved for $h=h(A)$ as a power series in $%
A_{\mu }$. We start from
\begin{equation}
A_{\mu }^{h}=h^{\dagger }A_{\mu }h+\frac{i}{g}h^{\dagger }\partial _{\mu
}h\;,  \label{Ah0}
\end{equation}
with
\begin{equation}
h=e^{ig\phi }=e^{ig\phi ^{a}T^{a}}\;.  \label{h0}
\end{equation}
Let us expand $h$ in powers of $\phi $
\begin{equation}
h=1+ig\phi -\frac{g^{2}}{2}\phi ^{2}+O(\phi ^{3})\;.  \label{hh1}
\end{equation}
From equation (\ref{Ah0}) we have
\begin{equation}
A_{\mu }^{h}=A_{\mu }+ig[A_{\mu },\phi ]+g^{2}\phi A_{\mu }\phi -\frac{g^{2}%
}{2}A_{\mu }\phi ^{2}-\frac{g^{2}}{2}\phi ^{2}A_{\mu }-\partial _{\mu }\phi
+i\frac{g}{2}[\phi ,\partial _{\mu }]+O(\phi ^{3})\;.  \label{A1}
\end{equation}
Thus, condition $\partial _{\mu }A_{\mu }^{h}=0$, gives
\begin{eqnarray}
\partial ^{2}\phi &=&\partial _{\mu }A+ig[\partial _{\mu }A_{\mu },\phi
]+ig[A_{\mu },\partial _{\mu }\phi ]+g^{2}\partial _{\mu }\phi A_{\mu }\phi
+g^{2}\phi \partial _{\mu }A_{\mu }\phi +g^{2}\phi A_{\mu }\partial _{\mu
}\phi   \nonumber \\
&-&\frac{g^{2}}{2}\partial _{\mu }A_{\mu }\phi ^{2}-\frac{g^{2}}{2}A_{\mu
}\partial _{\mu }\phi \phi -\frac{g^{2}}{2}A_{\mu }\phi \partial _{\mu }\phi
-\frac{g^{2}}{2}\partial _{\mu }\phi \phi A_{\mu }-\frac{g^{2}}{2}\phi
\partial _{\mu }\phi A_{\mu }-\frac{g^{2}}{2}\phi ^{2}\partial _{\mu }A_{\mu
}  \nonumber \\
&+&i\frac{g}{2}[\phi ,\partial ^{2}\phi ]+O(\phi ^{3})\;.  \label{hh2}
\end{eqnarray}
This equation can be solved iteratively for $\phi $ as a power series in $%
A_{\mu }$, namely
\begin{equation}
\phi =\frac{1}{\partial ^{2}}\partial _{\mu }A_{\mu }+i\frac{g}{\partial ^{2}%
}\left[ \partial A,\frac{\partial A}{\partial ^{2}}\right] +i\frac{g}{%
	\partial ^{2}}\left[ A_{\mu },\partial _{\mu }\frac{\partial A}{\partial ^{2}%
}\right] +\frac{i}{2}\frac{g}{\partial ^{2}}\left[ \frac{\partial A}{%
	\partial ^{2}},\partial A\right] +O(A^{3})\;,  \label{phi0}
\end{equation}
so that
\begin{eqnarray}
A_{\mu }^{h} &=&A_{\mu }-\frac{1}{\partial ^{2}}\partial _{\mu }\partial A-ig%
\frac{\partial _{\mu }}{\partial ^{2}}\left[ A_{\nu },\partial _{\nu }\frac{%
	\partial A}{\partial ^{2}}\right] -i\frac{g}{2}\frac{\partial _{\mu }}{%
	\partial ^{2}}\left[ \partial A,\frac{1}{\partial ^{2}}\partial A\right]
\nonumber \\
&+&ig\left[ A_{\mu },\frac{1}{\partial ^{2}}\partial A\right] +i\frac{g}{2}%
\left[ \frac{1}{\partial ^{2}}\partial A,\frac{\partial _{\mu }}{\partial
	^{2}}\partial A\right] +O(A^{3})\;.  \label{minn2}
\end{eqnarray}
Expression (\ref{minn2}) can be written in a more useful way,
given in eq.(\ref{min0}). In fact
\begin{eqnarray}
A_{\mu }^{h} &=&\left( \delta _{\mu \nu }-\frac{\partial _{\mu }\partial
	_{\nu }}{\partial ^{2}}\right) \left( A_{\nu }-ig\left[ \frac{1}{\partial
	^{2}}\partial A,A_{\nu }\right] +\frac{ig}{2}\left[ \frac{1}{\partial ^{2}}%
\partial A,\partial _{\nu }\frac{1}{\partial ^{2}}\partial A\right] \right)
+O(A^{3})  \nonumber \\
&=&A_{\mu }-ig\left[ \frac{1}{\partial ^{2}}\partial A,A_{\mu }\right] +%
\frac{ig}{2}\left[ \frac{1}{\partial ^{2}}\partial A,\partial _{\mu }\frac{1%
}{\partial ^{2}}\partial A\right] -\frac{\partial _{\mu }}{\partial ^{2}}%
\partial A+ig\frac{\partial _{\mu }}{\partial ^{2}}\partial _{\nu }\left[
\frac{1}{\partial ^{2}}\partial A,A_{\nu }\right]   \nonumber \\
&-&i\frac{g}{2}\frac{\partial _{\mu }}{\partial ^{2}}\partial _{\nu }\left[
\frac{\partial A}{\partial ^{2}},\frac{\partial _{\nu }}{\partial ^{2}}%
\partial A\right] +O(A^{3})  \nonumber \\
&=&A_{\mu }-\frac{\partial _{\mu }}{\partial ^{2}}\partial A+ig\left[ A_{\mu
},\frac{1}{\partial ^{2}}\partial A\right] +\frac{ig}{2}\left[ \frac{1}{%
	\partial ^{2}}\partial A,\partial _{\mu }\frac{1}{\partial ^{2}}\partial
A\right] +ig\frac{\partial _{\mu }}{\partial ^{2}}\left[ \frac{\partial
	_{\nu }}{\partial ^{2}}\partial A,A_{\nu }\right]   \nonumber \\
&+&i\frac{g}{2}\frac{\partial _{\mu }}{\partial ^{2}}\left[ \frac{\partial A%
}{\partial ^{2}},\partial A\right] +O(A^{3})  \label{hhh3}
\end{eqnarray}
which is precisely expression (\ref{minn2}). The transverse field
given in eq.(\ref {min0}) enjoys the property of being gauge
invariant order by order in the
coupling constant $g$. Let us work out the transformation properties of $%
\phi _{\nu }$ under a gauge transformation
\begin{equation}
\delta A_{\mu }=-\partial _{\mu }\omega +ig[A_{\mu },\omega ]\;.
\label{gauge3}
\end{equation}
We have, up to the order $O(g^{2})$,
\begin{eqnarray}
\delta \phi _{\nu } &=&-\partial _{\nu }\omega +ig\left[ \frac{1}{\partial
	^{2}}\partial A,\partial _{\nu }\omega \right] -i\frac{g}{2}\left[ \omega
,\partial _{\nu }\frac{1}{\partial ^{2}}\partial A\right] -i\frac{g}{2}%
\left[ \frac{\partial A}{\partial ^{2}},\partial _{\nu }\omega \right]
+O(g^{2})  \nonumber \\
&=&-\partial _{\nu }\omega +i\frac{g}{2}\left[ \frac{1}{\partial ^{2}}%
\partial A,\partial _{\nu }\omega \right] +i\frac{g}{2}\left[ \partial _{\nu
}\frac{1}{\partial ^{2}}\partial A,\omega \right] +O(g^{2})\;.  \label{gg2}
\end{eqnarray}
Therefore
\begin{equation}
\delta \phi _{\nu }=-\partial _{\nu }\left( \omega -i\frac{g}{2}\left[ \frac{%
	\partial A}{\partial ^{2}},\omega \right] \right) +O(g^{2})\;,  \label{phi1}
\end{equation}
from which the gauge invariance of $A_{\mu }^{h}$ is established.\newline
\newline
Finally, let us work out the expression of $A_{\mathrm{min}}^{2}$ as a power
series in $A_{\mu }$.
\begin{eqnarray}
A_{\mathrm{min}}^{2} &=&\mathrm{Tr}\int d^{4}x\,A_{\mu }^{h}A_{\mu }^{h}  \nonumber \\
&=&\mathrm{Tr}\int d^{4}x\,\left[ \phi _{\mu }\left( \delta _{\mu \nu }-\frac{%
	\partial _{\mu }\partial _{\nu }}{\partial ^{2}}\right) \phi _{\nu }\right]
\nonumber \\
&=&\mathrm{Tr}\int d^{4}x\,\left[ \left( A_{\mu }-ig\left[ \frac{1}{\partial ^{2}}%
\partial A,A_{\mu }\right] +\frac{ig}{2}\left[ \frac{1}{\partial ^{2}}%
\partial A,\partial _{\mu }\frac{1}{\partial ^{2}}\partial A\right] \right)
\times \right.  \nonumber \\
&&\left.  \left( \delta _{\mu \nu }-\frac{\partial _{\mu }\partial
	_{\nu }}{\partial ^{2}}\right) \left( A_{\nu }-ig\left[
\frac{1}{\partial
	^{2}}\partial A,A_{\nu }\right] +\frac{ig}{2}\left[ \frac{1}{\partial ^{2}}%
\partial A,\partial _{\nu }\frac{1}{\partial ^{2}}\partial A\right] \right)
\right]  \nonumber \\
&=&\frac{1}{2}\int d^{4}x\left[ A_{\mu }^{a}\left( \delta _{\mu \nu }-\frac{%
	\partial _{\mu }\partial _{\nu }}{\partial ^{2}}\right) A_{\nu
}^{a}-2gf^{abc}\frac{\partial _{\nu }\partial A^{a}}{\partial ^{2}}\frac{%
	\partial A^{b}}{\partial ^{2}}A_{\nu }^{c}-gf^{abc}A_{\nu }^{a}\frac{%
	\partial A^{b}}{\partial ^{2}}\frac{\partial _{\nu }\partial A^{c}}{\partial
	^{2}}\right] +O(A^{4})\;.  \nonumber \\
&&  \label{a2em}
\end{eqnarray}
We conclude this Appendix by noting that, due to gauge invariance, $A_{%
	\mathrm{min}}^{2}$ can be rewritten in a manifestly invariant way
in terms of $F_{\mu \nu }$ and the covariant derivative $D_{\mu }$
\cite{Zwanziger:1990tn}.

\section{A generalised Slavnov-Taylor identity}  \label{apc}
In this Appendix we derive the Ward identities for the generalised gauge fixing  of eq.\eqref{gf12}. Since 
the quantity $\omega^a(\xi)$ is now a composite operator, {\it i.e.} a product of fields at the same space-time point, 
we need to define $\omega^a(\xi)$ by introducing it into the starting action though a suitable external source. In order to maintain BRST invariance, we 
make use of a BRST doublet of external sources $(Q^a, R^a)$, of dimension four and ghost number $(-1,0)$, 
\begin{equation}
sQ^a = R^a \;, \qquad sR^a = 0 \;, \label{qr} 
\end{equation} 
and introduce the term 
\begin{equation} 
\int d^4x \; s \left( Q^a \omega^a(\xi) \right) = \int d^4 x \left( R^a \omega^a(\xi) - Q^a \frac{\partial \omega^a}{\partial \xi^c} g^{cd}(\xi) c^d \right)  \;. \label{qr1} 
\end{equation}
We start thus with the complete classical action $\Sigma$ given now by 
\begin{eqnarray} 
\Sigma &=&  S_{inv} + \int d^4x \left( {\cal J}^a_\mu A^{ah}_\mu  + \Xi^a_\mu D^{ab}_\mu(A^h) \eta^b \right)  \nonumber \\
&+&   \int d^4x \left( i b^a \partial_\mu A^a_\mu +  \frac{\alpha}{2} b^a b^a 
- i M^{ab} b^a \omega^b(\xi) - N^{ab} {\bar c}^a \omega^b(\xi) + {\bar c}^a \partial_\mu D^{ab}_\mu c^b + M^{ab} {\bar c}^a \frac{\partial \omega^{b}(\xi)}{\partial \xi^c} g^{cd}(\xi) c^d  \right)  \nonumber \\
&+& \int d^4x \left( -\Omega^a_\mu D^{ab}_\mu c^b + L^a \frac{g f^{abc}}{2} c^b c^c + K^a g^{ab}(\xi) c^d + R^a \omega^a(\xi) - Q^a \frac{\partial \omega^a}{\partial \xi^c} g^{cd}(\xi) c^d \right) \;,  \label{ca1}
\end{eqnarray} 
with $S_{inv}$ given by expression \eqref{act1}. \\\\The action $\Sigma$, eq.\eqref{ca1},  obeys the following Ward identities:
\begin{itemize} 
	\item the Slavnov-Taylor identity
	\begin{equation}
	\int d^4x \left( \frac{\delta \Sigma}{\delta A^a_\mu} \frac{\delta \Sigma}{\delta \Omega^a_\mu}  +  \frac{\delta \Sigma}{\delta c^a} \frac{\delta \Sigma}{\delta L^a}  
	+ \frac{\delta \Sigma}{\delta \xi^a} \frac{\delta \Sigma}{\delta K^a} + ib^a \frac{\delta \Sigma}{\delta {\bar c}^a} + N^{ab} \frac{\delta \Sigma}{\delta M^{ab}}
	+ R^{a} \frac{\delta \Sigma}{\delta Q^{a}}\right)   = 0 \;, \label{stg}
	\end{equation}
	\item the equation of motion of the Lagrange multiplier $b^a$
	\begin{equation} 
	\frac{\delta \Sigma}{\delta b^a} = i \partial_\mu A^a_\mu + \alpha b^a - i M^{ab} \frac{\delta \Sigma}{\delta R^b} \;, \label{bg1} 
	\end{equation}  
	\item the anti-ghost equation 
	\begin{equation}
	\frac{\delta \Sigma}{\delta {\bar c}^a} + \partial_\mu \frac{\delta \Sigma}{\delta \Omega^a_\mu}  + M^{ab} \frac{\delta \Sigma}{\delta Q^b}  - N^{ab} \frac{\delta \Sigma}{\delta R^b} = 0 \;, \label{ghg1}
	\end{equation}
	\item the equation of  $\tau^{a}$
	\begin{eqnarray}
	\frac{\delta \Sigma}{\delta\tau^{a}}-\partial_{\mu}\frac{\delta \Sigma}{\delta\mathcal{J}_{\mu}^{a}} & = & 0 \;, \label{ap5} 
	\end{eqnarray}
	\item the equation of the ghost $\eta^{a}$
	\begin{eqnarray}
	\int d^{4}x\left(\frac{\delta \Sigma}{\delta\eta^{a}}+gf^{abc}\bar{\eta}^{b}\frac{\delta \Sigma}{\delta\tau^{c}}+gf^{abc}\Xi^{b}\frac{\delta \Sigma}{\delta\mathcal{J}_{\mu}^{c}}\right) & = & 0  \;, \label{ap6}
	\end{eqnarray}
	\item the equation of the antighost ${\bar \eta}^a$
	\begin{eqnarray}
	\frac{\delta \Sigma}{\delta\bar{\eta}^{a}}-\partial_{\mu}\frac{\delta \Sigma}{\delta\Xi_{\mu}^{a}} & = & 0 \;. \label{ap7}
	\end{eqnarray}
\end{itemize}
These Ward identities can be employed for the analysis of the algebraic renormalization when the generalised function $\omega^a(\xi)$ is explicitly present in the 
gauge-fixing. In this case, the general counterterm will be reabsorbed through a renormalization of $\omega^a(\xi)$, corresponding to a renormalization of the infinite set of unphysical gauge parameters $(a_1^{abc}, a_2^{abcd}, a_3^{abcde}, ..)$  of expression \eqref{rp}. \\\\Repeating the lengthy discussion of the previous sections, for the most general local invariant counterterm we find now

\begin{eqnarray}
\Sigma^{ct} & = & \int d^4 x \; \bigg\{ -a_{0}g^2\frac{\partial \Sigma}{\partial g^2}+d_{2}\left(\alpha\right)2\alpha\frac{\partial \Sigma}{\partial\alpha}+a_{7} m^{2}\frac{\partial \Sigma}{\partial m^{2} \Bigr. \nonumber }\\
&  & +a_{4}\left(\tau^{a}\frac{\delta \Sigma}{\delta\tau^{a}}+\mathcal{J}_{\mu}^{a}\frac{\delta \Sigma}{\delta\mathcal{J}_{\mu}^{a}}+\frac{1}{2}\bar{\eta}^{a}\frac{\delta \Sigma}{\delta\bar{\eta}^{a}}+\frac{1}{2}\eta^{a}\frac{\delta \Sigma}{\delta \eta^{a}}+\frac{1}{2}\Xi_{\mu}^{a}\frac{\delta \Sigma}{\delta\Xi_{\mu}^{a}}\right) \nonumber \\
& &+d_{2}\left(\alpha\right)A_{\mu}^{a}\frac{\delta \Sigma}{\delta A_{\mu}^{a}}-d_{2}\left(\alpha\right)\Omega_{\mu}^{a}\frac{\delta \Sigma}{\delta\Omega_{\mu}^{a}}-d_{1}\left(\alpha\right)c^{a}\frac{\delta \Sigma}{\delta c^{a}}+d_{1}\left(\alpha\right)L^{a}\frac{\delta \Sigma}{\delta L^{a}} \nonumber \\
&&+f_{1}^{ab}(\xi,\alpha)\xi^{a}\frac{\delta \Sigma}{\delta \xi^b }-\left(f_1^{ab}(\xi,\alpha)+\frac{\partial f_{1}^{kb}(\xi,\alpha)}{\partial\xi^a}\xi^k\right)K^b\frac{\delta \Sigma}{\delta K^a } \nonumber \\
& &-d_2(\alpha) \bar{c}\frac{\delta \Sigma}{\delta \bar{c}}+\left(d_2(\alpha)-f_2(0,\alpha)\right) M^{ab} \frac{\delta \Sigma}{\delta M^{ab}}+\left(d_2(\alpha)-f_2(0,\alpha)\right)N^{ab}\frac{\delta \Sigma}{\delta N^{ab}}\nonumber \\
& &-d_2(\alpha) b^a\frac{\delta \Sigma}{\delta b^a}-f_2(0,\alpha) Q^a \frac{\delta \Sigma}{\delta Q^a}-f_2(0,\alpha)R^a\frac{\delta \Sigma}{\delta R^a}\nonumber \\
& &+ \big[ \left(f_2(0,\alpha)a_1^{abc}+\tilde a_1^{abc}\right) \frac{\delta \Sigma}{\delta a_1^{abc}}+\left(f_2(0,\alpha)a_2^{abcd}+\tilde{a}_2^{abcd}\right)\frac{\delta \Sigma}{\delta a_2^{abcd}} \nonumber \\
& &\Bigl. +\left(f_2(0,\alpha)a_3^{abcde}+\tilde{a}_3^{abcde}\right)\frac{\delta \Sigma}{\delta a_3^{abcde}}+... \big] \bigg\} \;,   \label{cctt}
\end{eqnarray}
where the dots $ ... $ denote the reamaining, infinite set, of terms of the kind  
\begin{equation} 
\sum_{j} \left(f_2(0,\alpha)a_j^{abcde...}+\tilde{a}_j^{abcde...}\right)\frac{\delta \Sigma}{\delta a_j^{abcde...}}  \;, \qquad \ j=4, ..., \infty  \;, \label{jin}
\end{equation} 
The counterterm $\Sigma^{ct}$ in eq.\eqref{cctt}  can be rewritten  as 

\begin{eqnarray}
\Sigma^{ct} = \mathcal{R}\Sigma \;,  \label{rpp}
\end{eqnarray}
with
\begin{eqnarray}
\mathcal{R}&=& -a_{0}g^2\frac{\partial}{\partial g^2}+d_{2}\left(\alpha\right)2\alpha\frac{\partial}{\partial\alpha}+a_{7} m^{2}\frac{\partial }{\partial m^{2}} \nonumber \\
&  & +\int d^4x \,\bigg\{a_{4}\left(\tau^{a}\frac{\delta}{\delta\tau^{a}}+\mathcal{J}_{\mu}^{a}\frac{\delta }{\delta\mathcal{J}_{\mu}^{a}}+\frac{1}{2}\bar{\eta}^{a}\frac{\delta }{\delta\bar{\eta}^{a}}+\frac{1}{2}\eta^{a}\frac{\delta }{\delta \eta^{a}}+\frac{1}{2}\Xi_{\mu}^{a}\frac{\delta }{\delta\Xi_{\mu}^{a}}\right) \nonumber \\
& &+d_{2}\left(\alpha\right)A_{\mu}^{a}\frac{\delta}{\delta A_{\mu}^{a}}-d_{2}\left(\alpha\right)\Omega_{\mu}^{a}\frac{\delta}{\delta\Omega_{\mu}^{a}}-d_{1}\left(\alpha\right)c^{a}\frac{\delta}{\delta c^{a}}+d_{1}\left(\alpha\right)L^{a}\frac{\delta}{\delta L^{a}} \nonumber \\
&&+f_{1}^{ab}(\xi,\alpha)\xi^{a}\frac{\delta }{\delta \xi^b }-\left(f_1^{ab}(\xi,\alpha)+\frac{\partial f_{1}^{kb}(\xi,\alpha)}{\partial\xi^a}\xi^k\right)K^b\frac{\delta }{\delta K^a } \nonumber \\
& &-d_2(\alpha) \bar{c}\frac{\delta}{\delta \bar{c}}+\left(d_2(\alpha)-f_2(0,\alpha)\right) M^{ab} \frac{\delta }{\delta M^{ab}}+\left(d_2(\alpha)-f_2(0,\alpha)\right)N^{ab}\frac{\delta}{\delta N^{ab}} \nonumber \\
& &-d_2(\alpha) b^a\frac{\delta}{\delta b^a}-f_2(0,\alpha) Q^a \frac{\delta }{\delta Q^a}-f_2(0,\alpha)R^a\frac{\delta}{\delta R^a} \nonumber \\
& &+ \big[ \left(f_2(0,\alpha)a_1^{abc}+\tilde a_1^{abc}\right) \frac{\delta}{\delta a_1^{abc}}+\left(f_2(0,\alpha)a_2^{abcd}+\tilde{a}_2^{abcd}\right)\frac{\delta }{\delta a_2^{abcd}} \nonumber \\
& &+\left(f_2(0,\alpha)a_3^{abcde}+\tilde{a}_3^{abcde}\right)\frac{\delta }{\delta a_3^{abcde}}+... \big]\bigg\} \;. \label{rpp1}
\end{eqnarray}
For the renormalization factors, we have now 
\begin{equation} 
\Sigma(\Phi) +  \varepsilon  \Sigma^{ct}(\Phi)   =   \Sigma(\Phi) + \varepsilon \mathcal{R}\Sigma(\Phi) =  \Sigma(\Phi_0) + O(\varepsilon^2) \;, \label{rpp2}
\end{equation}
with
\begin{eqnarray}
\Phi_0=Z_{\Phi} \Phi= (1+\varepsilon \mathcal{R})\Phi + O(\varepsilon^2) \;. \label{rpp3} 
\end{eqnarray}
where
\begin{eqnarray}
A_0&=&Z_A^{1/2} A_{\mu}\;, \,\,\, b_0=Z_b^{1/2}\;, \,\,\, c_0=Z_c^{1/2}c\;, \,\,\, \bar{c}_0=Z_{\bar{c}}^{1/2}\bar{c}\;, \nonumber  \\
\xi_0^a& = & Z^{ab}_{\xi}(\xi)\xi^b, \,\,\, \tau_0=Z_{\tau}^{1/2} \tau \;, 
\Omega_0=Z_{\Omega} \Omega \;, \,\, \,L_0=Z_L L\;,  \nonumber \\
\,\,\, K_0^a&=&Z_K^{ab} (\xi) K^b\;, \,\,\, m_0^2=Z_{m^2} m^2\;, \,\,\,\mathcal{J}_0=Z_{\mathcal{J}}\mathcal{J} \;, \nonumber \\
g_0&=&Z_g\;, \,\,\, \alpha_0=Z_{\alpha} \alpha\;, \,\,\,\bar{\eta}_0=Z_{\bar{\eta}}^{1/2}\bar{\eta}\;, \,\,\, \eta_0=Z_{\eta}^{1/2} \eta\;, \nonumber \\
\Xi_0&=& Z_{\Xi} \Xi\;, \,\,\, M_0=Z_M M\;,  \nonumber \\
N_0&=&Z_N N, \,\,\, Q_0=Z_Q Q, \,\,\, R_0= Z_R R \;, \label{rpp4} 
\end{eqnarray}
and
\begin{eqnarray}
Z_g&&=1-\varepsilon \frac{a_0}{2}  \nonumber \\
Z^{1/2}_A&&=Z^{-1}_{\Omega}=Z^{-1/2}_{\bar{c}}=Z^{-1/2}_b=Z^{1/2}_{\alpha}=1+\varepsilon d_2(\alpha) \nonumber \\
Z_\xi^{ab}&&=\delta^{ab}+\varepsilon f_{1}^{ab}(\xi,\alpha) \nonumber \\
Z_L&&=Z^{-1/2}_c=1+\varepsilon d_1(\alpha) \nonumber \\
Z_{\bar{\eta}}&&=Z_{\eta}=Z^2_{\Xi}=Z^{1/2}_{\tau}=Z_{\mathcal{J}}=1+\varepsilon a_4 \nonumber \\
Z_{m^2}&&=1+\varepsilon a_7 \nonumber \\
Z_{M}&&=Z_{N}=1+\varepsilon (d_2-f_2(0,\alpha)) \nonumber \\
Z_Q&&=Z_R=1-\varepsilon (f_2(0,\alpha)) \nonumber \\
Z_{K}^{ab}&&=\delta^{ab}-\varepsilon\left(f_{1}^{ab}(\xi,\alpha)+\frac{\partial f_{1}^{kb}(\xi,\alpha)}{\partial \xi^a}\xi^{k}\right) \;,  \label{rpp5}
\end{eqnarray}
with the addition of a multiplicative renormalization of the infinite set of gauge parameters $(a_1^{abc}, a_2^{abcd}, a_3^{abcde}, ..)$  of expression \eqref{rp}, namely 
\begin{eqnarray}
(a_1^{abc})_0&=&(1+\varepsilon f_2(0,\alpha))a_1^{abc}+\varepsilon\tilde{a}_1^{abc} \nonumber \\
(a_2^{abcd})_0&=&(1+\varepsilon f_2(0,\alpha))a_2^{abcd}+\varepsilon\tilde{a}_2^{abcd}  \nonumber \\
(a_3^{abcde})_0&=&(1+\varepsilon f_2(0,\alpha))a_3^{abcde}+\varepsilon\tilde{a}_3^{abcde} \nonumber \\
&...& \;.  \label{rpp6} 
\end{eqnarray}
Equations \eqref{rpp5} and  \eqref{rpp6}  show that the inclusion of the ambiguity $\omega^a(\xi)$ in the generalised gauge fixing of eq.\eqref{gf12} gives rise to a standard renormalization of the fields, parameters and sources. Clearly, from eq.\eqref{rpp6}  one sees that the renormalization of $\omega^a(\xi)$ itself is now encoded into a multiplicative renormalization of the infinite set of the unphysical gauge parameters $(a_1^{abc}, a_2^{abcd}, a_3^{abcde}, ..)$.

\chapter{Technical details to the Abelian Higgs model\label{FR}}

\section{Field propagators of the Abelian Higgs model in the $R_{\xi}$ gauge}
The quadratic part of the action \eqref{fullaction} in the bosonic sector is given by
\beq
S_{bos}^{quad}&=&\ha\int d^4 x \Big\{A_{\m}(-\d_{\m\n}(\pa^2-m^2)+\pa_{\m}\pa_{\n})A_{\n}-\r \pa^2 \r  -h(\pa^2 - m_{h}^2)h +\bar{c}(\pa^2 -m^2 \e )c\nonumber\\
&+&2ib\pa_{\m}A_{\m}+\xi b^2+2im\xi b  \rho +2m A_{\m}\pa_{\m}\rho \Big\}.
\eeq
Putting this in a matrix form yields
\beq
S^{quad}_{bos}=\ha\int d^4 x \,\Y^{T}_{\m} {\mc O}_{\m\n} \Y_{\n},
\eeq
where
\beq
\Y^{T}_{\m}=\left( {\begin{array}{cccc}
		A_{\m} &
		b&
		\r&
		h
\end{array} } \right),\,\, \Y_{\n}=\left( {\begin{array}{cccc}
		A_{\n}\\
		b\\
		\r\\
		h
\end{array} } \right),
\eeq
and
\beq
{\mc O}=\left( {\begin{array}{cccc}
		(-\d_{\m\n}(\pa^2-m^2)+\pa_{\m}\pa_{\n})&-i\pa_{\mu}&m\pa_{\m}&0\\
		i \pa_{\n} &\xi &i m \xi &0\\
		-m \partial_{\n} &i m \xi&-\pa^2&0\\
		0&0&0&-(\pa^2 - m_{h}^2)
\end{array} } \right),
\eeq
the tree-level field propagators can be read off from the inverse of $\mathcal{O}$, leading to the following expressions
\beq
\langle A_{\m}(p)A_{\n}(-p)\rangle &=& \frac{1}{p^2+ m^2}{\mc P}_{\m\n}+\frac{\e}{p^2 + \e m^2}\mathcal{L}_{\m\n},\nonumber\\
\langle \r(p)\r(-p)\rangle &=&\frac{1}{p^2 +\e m^2},\nonumber\\
\langle h(p)h(-p)\rangle &=&\frac{1}{p^2 + m_h^2},\nonumber\\
\langle A_{\m}(p)b(-p)\rangle &=& \frac{p_{\m}}{p^2+\xi m^2},\nonumber\\
\langle b (p) \rho (-k) \rangle &=& \frac{-i m}{p^2+ \xi m^2},
\eeq
where $\mathcal{P}_{\m\n}=\delta_{\m\n}-\frac{p_{\m}p_{\n}}{p^2}$ and $\mathcal{L}_{\m\n}=\frac{p_{\m}p_{\n}}{p^2}$ are the transversal and longitudinal projectors, respectively. The ghost propagator is
\beq
\langle \bar{c}(p)c(-p)\rangle &=&\frac{1}{p^2 +\e m^2}.
\eeq

\section{Vertices of the Abelian Higgs model in the $R_{\xi}$ gauge}
From the action \eqref{fullaction}, we find the following vertices
\beq
\Gamma_{A_{\m}\r h}(-p_1,-p_2,-p_3)&=&ie(p_{\m,3}-p_{\m,2})\delta(p_1+p_2+p_3),\nonumber\\
\Gamma_{A_{\m}A_{\n}h}(-p_1,-p_2,-p_3)&=& -2e^2 v \delta_{\m\n}\delta(p_1+p_2+p_3),\nonumber\\
\Gamma_{A_{\m}A_{\n}hh}(-p_1,-p_2,-p_3,-p_4)&=&-2e^2 \delta_{\m\n}\delta(p_1+p_2+p_3+p_4),\nonumber\\
\Gamma_{A_{\m}A_{\n}\r\r}(-p_1,-p_2,-p_3,-p_4)&=&-2e^2 \delta_{\m\n}\delta(p_1+p_2+p_3+p_4),\nonumber\\
\Gamma_{hhhh}(-p_1,-p_2,-p_3,-p_4)&=&-3\l \, \delta(p_1+p_2+p_3+p_4),\nonumber \\
\Gamma_{hh\r \r}(-p_1,-p_2,-p_3,-p_4)&=&-\l \, \delta(p_1+p_2+p_3+p_4), \nonumber\\
\Gamma_{\r\r\r\r}(-p_1,-p_2,-p_3,-p_4)&=&-3\l \,\delta(p_1+p_2+p_3+p_4), \nonumber\\
\Gamma_{hhh}(-p_1,-p_2,-p_3)&=&-3\l v \,\delta(p_1+p_2+p_3),\nonumber \\
\Gamma_{h\r\r}(-p_1,-p_2,-p_3)&=&-\l v \,\delta(p_1+p_2+p_3), \nonumber\\
\Gamma_{\bar{c}h c}(-p_1,-p_2,-p_3)&=&-m\e e  \,\delta(p_1+p_2+p_3).
\eeq

\section{  Equivalence between including tadpole diagrams in the self-energies and shifting $\langle \vf \rangle$ \label{v}}
There is yet another way to come to \eqref{pop}. For this, we do not need to include the balloon type tadpoles  in the self-energies, but rather fix the expectation value of the Higgs field $\langle h \rangle =0$ by shifting the vacuum expectation value of the Higgs field to its proper one-loop value. The $h$ field one-point function has the following contributions at one-loop order :

\begin{itemize}
	\item the gluon contribution
	\beq
	-\frac{1}{m_h^2}\frac{2 e^2 v}{(4 \pi)^{d/2}}\frac{\Gamma(2-d/2)}{(2-d)}(m^{d-2}(d-1)+\xi (\xi m^2)^{d/2-1}),
	\eeq
	\item the Goldstone boson one
	\beq
	-\frac{1}{m_h^2}\lambda v  \frac{1}{(4\pi)^{d/2}} \frac{\Gamma(2-d/2)}{(2-d)}(\e m^2)^{d/2-1},
	\eeq
	\item the ghost loop
	\beq
	2 \frac{1}{m_h^2} \frac{ e^2 v \e}{(4\pi)^{d/2}} \frac{\Gamma(2-d/2)}{(2-d)} (\e m^2)^{d/2-1}
	\eeq
	\item the Higgs boson one
	\beq
	-3 \frac{1}{m_h^2} \frac{\lambda v}{(4\pi)^{d/2}} \frac{\Gamma(2-d/2)}{(2-d)}m_h^{d-2},
	\eeq
\end{itemize}
Together those four contributions yield
\beq
\Gamma_{\braket{h}}&=&\frac{1}{(4\pi)^{d/2}}\frac{\Gamma(2-d/2)}{(2-d)}\frac{1}{m_h^2}(-2e^2v m^{d-2}(d-1)-\lambda v (\xi m^2)^{d/2-1}-3\lambda v m_h^{d-2}),
\label{ll}
\eeq
that becomes, for $d=4-\epsilon$,
\beq
&=& -\frac{1}{2}\frac{1}{m_h^2}\frac{1}{(4\pi)^2}(\frac{2}{\epsilon}+1+\ln (\m^2))(-2e^2 v m^{2-\epsilon}(3-\epsilon)-\lambda v (\xi m^2) ^{1-\epsilon/2}-3\lambda v m_h^{2-\epsilon})\\
&=& -\frac{1}{2}\frac{1}{m_h^2}\frac{1}{(4\pi)^2}(\frac{2}{\epsilon}+1+\ln (\m^2))(-2 e^2 v m^2(1-\frac{\epsilon}{2}\ln m^2)(3-\epsilon)-\lambda v \xi m^2 (1-\frac{\epsilon}{2}\ln m^2)-3 \lambda v m_h^2(1-\frac{\epsilon}{2}\ln m_h^2)).\nonumber
\nonumber\\
\eeq
We can split this in a divergent part
\beq
\Gamma^{div}_{\braket{h}}= \frac{1}{\epsilon}\frac{1}{m_h^2}(6 e^2m^2 v + 3 m_h^2 v \lambda + \xi m^2 v),
\eeq
which we can cancel with the counterterms, and a finite part that reads
\beq
\Gamma_{\braket{h}}^{fin}= \frac{1}{m_h^2}\frac{e^2}{(4 \pi)^2} v \Big( m^2 (1-3 \ln \frac{m^2}{\m^2})\Big) + \frac{1}{m_h^2}\frac{\lambda}{(4 \pi)^2}\frac{v}{2}\Big(3 m_h^2 (1- \ln \frac{h^2}{\m^2})+\xi m^2 (1- \ln \frac{\xi m^2}{\m^2})\Big).
\label{finite<h>}
\eeq

Now, to see how this reflects on the propagator, we can rewrite our scalar field as
\beq
\vf=\frac{1}{\sqrt{2}}((\langle \vf \rangle +h)+i\r),
\eeq
where the vacuum expectation value of the Higgs field has tree-level and one-loop terms:
\beq
\langle \vf \rangle = v + \hbar v_1.
\eeq
Thus the ``classical'' potential part of the action becomes
\beq
\frac{\l}{2}\left(\vf^{\dagger}\vf-\frac{v^2}{2}\right)^2=\frac{\l}{8}\left(\langle \vf \rangle^2-v^2+2h \langle \vf \rangle +h^2+\rho^2\right)^2
\eeq
and expanding this, we find for the shifted tree level Higgs mass
\beq
m_h^2=\frac{1}{2}\lambda (3\langle \vf\rangle^2-v^2)=\lambda v^2+ 3\hbar \l v v_1,
\eeq
while the photon mass is
\beq
m^2= e^2 \langle \varphi \rangle^2= e^2v^2+2\hbar e^2 v v_1.
\eeq

As now per construction $ \langle h \rangle = 0$, we can fix the one-loop correction\footnote{This procedure is also equivalent to computing $\langle \vf \rangle$ via an effective potential minimization up to the same order.} to the Higgs minimizing value by requiring it to absorb the tadpole contributions:
\beq
v_1+ \Gamma^{fin}_{\braket{h}}=0,
\eeq
thus
\beq
v_1= - \frac{1}{m_h^2}\frac{e^2}{(4 \pi)^2} v \Big( m^2 (1-3 \ln \frac{m^2}{\m^2})\Big) - \frac{1}{m_h^2} \frac{\lambda}{(4 \pi)^2}\frac{v}{2}\Big(3 m_h^2 (1- \ln \frac{h^2}{\m^2})+\xi m^2 (1- \ln \frac{\xi m^2}{\m^2})\Big)\,.
\eeq

Implementing this in the transverse $\langle AA \rangle $-propagator, one gets
\beq
G^\perp_{AA}(p^2) = \frac{1}{p^2 + e^2 (v^2+2\hbar v_1 v ) - \Pi^\perp_{AA}(p^2)},
\label{r}
\eeq
where in the correction $\Pi_{AA}^\perp$, which is already of $\mathcal{O}(\hbar)$, we only  include the $\mathcal{O}(\hbar^0)$ part of $\langle\vf\rangle$, i.e. $v$.

We can now verify the $\xi$-independence of the transverse propagator $\langle AA \rangle$. The $\xi$-dependent part of $\Pi_{AA}^\perp(p^2)$ is
\beq
\Pi_{AA,\xi}^\perp(p^2)=\frac{-2e^2}{(4\pi)^{d/2}}\frac{\Gamma(2-d/2)}{2-d}(\e m^2)^{d/2-1},
\eeq
while we find the $\xi$-dependent part of $v_1$ to be (using \eqref{ll})
\beq
v_{1\xi}=\frac{1}{(4\pi)^{d/2}}\frac{\Gamma(2-d/2)}{(2-d)}\frac{1}{m_h^2}(\lambda v (\xi m^2)^{d/2-1}).
\eeq
In the denominator of \eqref{r} we now easily see that
\beq
2e^2 v_{1\xi} v_0-\Pi_{AA,\xi}^\perp(p^2)= 0,
\eeq
thereby establishing the gauge independence of the transverse photon propagator.

For the Higgs propagator, we similarly find
\beq
G_{hh}(p^2) = \frac{1}{p^2 + \lambda (v^2+3\hbar v_1 v ) - \Pi_{hh}(p^2)}.
\label{rr}
\eeq
Here we observe that the $\xi$-dependent part of $v_1$ has the same effect as the balloon tadpole of the Goldstone boson, consequently establishing the gauge parameter independence of the Higgs mass pole.
\section{Feynman integrals \label{apfeyn}}

\beq
\int_0^1 dx \ln \frac{K(m_1^2,m_2^2)}{\mu^2}&=&\frac{1}{2 p^2}\Bigg\{m_1^2 \ln(\frac{m_2^2}{m_1^2})+m_2^2 \ln(\frac{m_1^2}{m_2^2})+p^2 \ln(\frac{m_1^2 m_2^2}{\mu^4})\nonumber\\
&-&2 \sqrt{-m_1^4+2 m_1^2 m_2^2-2 m_1^2 p^2-m_2^4-2 m_2^2 p^2-p^4}\nonumber\\
&\times&\tan^{-1}\Big[\frac{-m_1^2+m_2^2-p^2}{\sqrt{-m_1^4+2 m_1^2 (m_2^2-p^2)-(m_2^2+p^2)^2}}\Big]\nonumber\\
&+&2 \sqrt{-m_1^4+2 m_1^2 m_2^2-2 m_1^2 p^2-m_2^4-2 m_2^2 p^2-p^4}\nonumber\\
&\times&\tan^{-1}\Big[\frac{-m_1^2+m_2^2+p^2}{\sqrt{-m_1^4+2 m_1^2 (m_2^2-p^2)-(m_2^2+p^2)^2}}\Big]\nonumber\\
&-&4 p^2\Bigg\}
\eeq
\section{Asymptotics of the Higgs propagator \label{lot}}
At one-loop, the Higgs propagator behaves like
\begin{equation}\label{dd1}
G_{hh}(p^2)= \frac{\mathcal{Z}}{p^2 \ln \frac{p^2}{\mu^2}} \qquad\text{for}\qquad p^2\to\infty.
\end{equation}
For $\mathcal{Z}>0$, this can only be compatible with
\begin{equation}\label{dd2}
G_{hh}(p^2)= \int_0^\infty \frac{\rho(t)dt}{t+p^2}
\end{equation}
if the superconvergence relation \cite{oehme1990superconvergence,cornwall2013positivity} $\int dt\rho(t)=0$ holds, which forbids a positive spectral function. Let us support this non-positivity of $\rho(t)$ by using \eqref{dd1} to show that $\rho(t)$ is certainly negative for very large $t$. This argument can also be found in the Appendix of \cite{Dudal:2019gvn}.

Since for a KL representation we have:
\beq
\rho(t)=\frac{1}{2\pi i}\lim_{\epsilon\to 0^+}\left(G(-t-i\epsilon)-G(-t+i\epsilon)\right),
\eeq
we find for $t\to+\infty$ and $\epsilon\to 0^+$,
\begin{eqnarray}
\rho(t) &=& \frac{\mathcal{Z}}{2\pi i}\left[\frac{\left(\ln\frac{-t-i\epsilon}{\mu^2}\right)^{-1}}{-t-i\epsilon}-\frac{\left(\ln\frac{-t+i\epsilon}{\mu^2}\right)^{-1}}{-t+i\epsilon}\right]\nonumber \\
&=&\frac{\mathcal{Z}}{2\pi i t} \left[-\left(\ln\frac{t}{\mu^2}-i\pi\right)^{-1}+\left(\ln\frac{t}{\mu^2}+i\pi\right)^{-1}\right]\nonumber\\
&=&\frac{\mathcal{Z}}{\pi t}\text{Im}\left[\left(\ln\frac{t}{\mu^2}+i\pi\right)^{-1}\right]\nonumber\\
&=&\frac{\mathcal{Z}}{\pi t}\left(\left(\ln\frac{t}{\mu^2}\right)^2+\pi^2\right)^{-1/2}\sin\left(-\arctan\frac{\pi}{\ln\frac{t}{\mu^2}}\right).
\end{eqnarray}
From the latter expression, we can indeed infer that $\rho(t)$ becomes negative for $t$ large. We find
\begin{eqnarray}
\rho(t) &\stackrel{t\to\infty}{=}& -\frac{\mathcal{Z} }{t}\left(\ln\frac{t}{\mu^2}\right)^{-2}<0
\end{eqnarray}
for $\mathcal{Z}>0$, and vice versa for $\mathcal{Z}<0$.

\section{Contributions to $\langle O(p) O(-p) \rangle$ \label{A}}
\begin{figure}[H]
	\includegraphics[width=15cm]{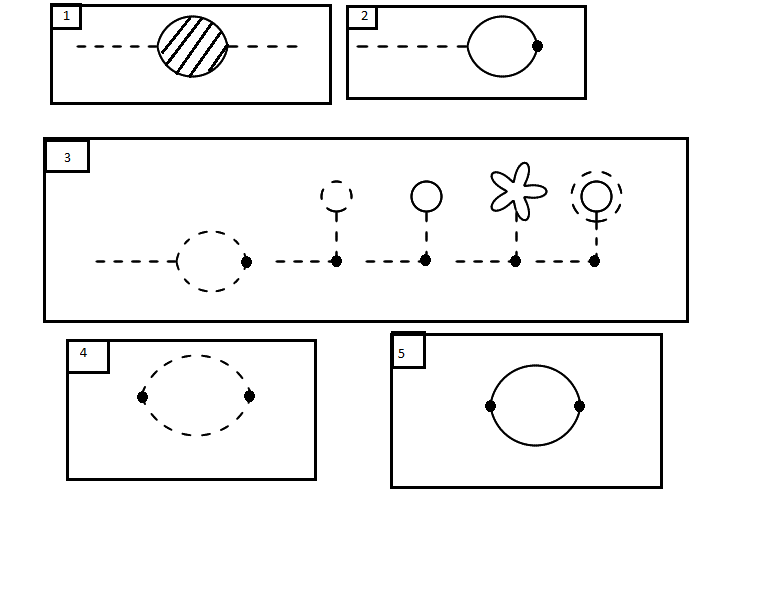}
	\caption{One-loop contributions to the propagator $\langle O(p)O(-p) \rangle$. Wavy lines represent the photon field, dashed lines the Higgs field, solid lines the Goldstone boson and double lines the ghost field.}
	
	\label{Yw}
\end{figure}
We consider each term in the two-point function $\braket{O(p)O(-p)}$, given by eq. \eqref{exp14}. The first term is the one-loop correction to the Higgs propagator $\langle h(p)h(-p) \rangle$ known from the last chapter, shown in frame $(1)$ in Figure \ref{Yw}, which gives

\beq
v^2\langle h(p) h(-p) \rangle &=&\frac{v^2}{m_h^2+p^2} + v^2 \Big(e^2 \eta \left(m^2,m^2\right) \left(2 (d-1) m^2+\frac{p^4}{2 m^2}+2 p^2\right)+\eta \left(m^2 \xi ,m^2 \xi \right) \left(\frac{m_h^2 \lambda }{2}-\frac{e^2 p^4}{2 m^2}\right)\nonumber\\
&+&\frac{9}{2} m_h^2 \lambda  \eta \left(m_h^2,m_h^2\right)+\frac{1}{2} e^2 \chi \left(m^2\right) \left(4 (d-1)-\frac{2 p^2}{m^2}\right)+\chi \left(m^2 \xi \right) \left(\frac{e^2 p^2}{m^2}+\lambda \right)\nonumber\\
&+&3 \lambda  \chi \left(m_h^2\right)\Big)\frac{1}{\left(m_h^2+p^2\right)^2}.
\eeq
The second term, the one-loop correction shown in frame (2) of Figure \ref{Yw}, gives

\beq
v \langle h(p) \rho(-p)^2 \rangle &=& -\frac{m_h^2 \eta \left(m^2 \xi ,m^2 \xi \right)}{m_h^2+p^2}.
\eeq
The third term, the one-loop correction shown in frame (3) of Figure \ref{Yw}, gives

\beq
v \langle h(p) h(-p)^2 \rangle &=&-\frac{3 m_h^2 \eta \left(m_h^2,m_h^2\right)}{m_h^2+p^2}-\frac{3 \chi \left(m_h^2\right)}{m_h^2+p^2}-\frac{\chi \left(m^2 \xi \right)}{m_h^2+p^2}-\frac{2 (d-1) m^2 \chi \left(m^2\right)}{m_h^2 \left(m_h^2+p^2\right)}-\frac{2 m^2 \xi  \chi \left(m^2 \xi \right)}{m_h^2 \left(m_h^2+p^2\right)}\nonumber\\
&+& \frac{2 m^2 \xi  \chi \left(m^2 \xi \right)}{m_h^2 \left(m_h^2+p^2\right)}.
\eeq
The fourth term has no 1-loop contributions.
The fifth term, the one-loop correction shown in frame (4) of Figure \ref{Yw}, gives

\beq
\frac{1}{4}\langle h(p)h(p) h(-p)h(-p) \rangle &=& \frac{1}{2}\eta \left(m_h^2, m_h^2\right).
\eeq
The sixth term, the one-loop correction shown in frame (5) of Figure \ref{Yw}, gives

\beq
\frac{1}{4}\langle \rho(p)\rho(p) \rho(-p)\rho(-p) \rangle &=& \frac{1}{2} \eta \left(\xi m^2,\xi m^2\right).
\eeq
Using the identity \eqref{2o} we are able to write the whole  one-loop correlation 
function $\braket{O(-p),O(p)}$, up to the order $\hbar$,  as 

\beq
\langle {O}(p) {O}(-p) \rangle &=& \frac{v^2}{p^2+m_h^2}+\frac{1}{(p^2+m_h^2)^2}\int_0^1 dx \Bigg(\frac{1}{2}\eta \left[m^2,m^2\right](4 (d-1) m^4+4 m^2 p^2+p^4)+\frac{1}{2}(p^2-2 m_h^2)^2\eta \left[m_h^2,m_h^2\right]\nonumber\\
&&-\frac{p^2 \chi [m^2] (2 (d-1) m^2+m_h^2)}{m_h^2}-3 p^2 \chi  [m_h^2]\Bigg)\,, \label{ch}
\eeq

\section{Contributions to $\langle V_{\mu}(x)V_{\nu}(y)\rangle$ \label{ah}}
\begin{figure}[H]
	\center
	\includegraphics[width=19cm]{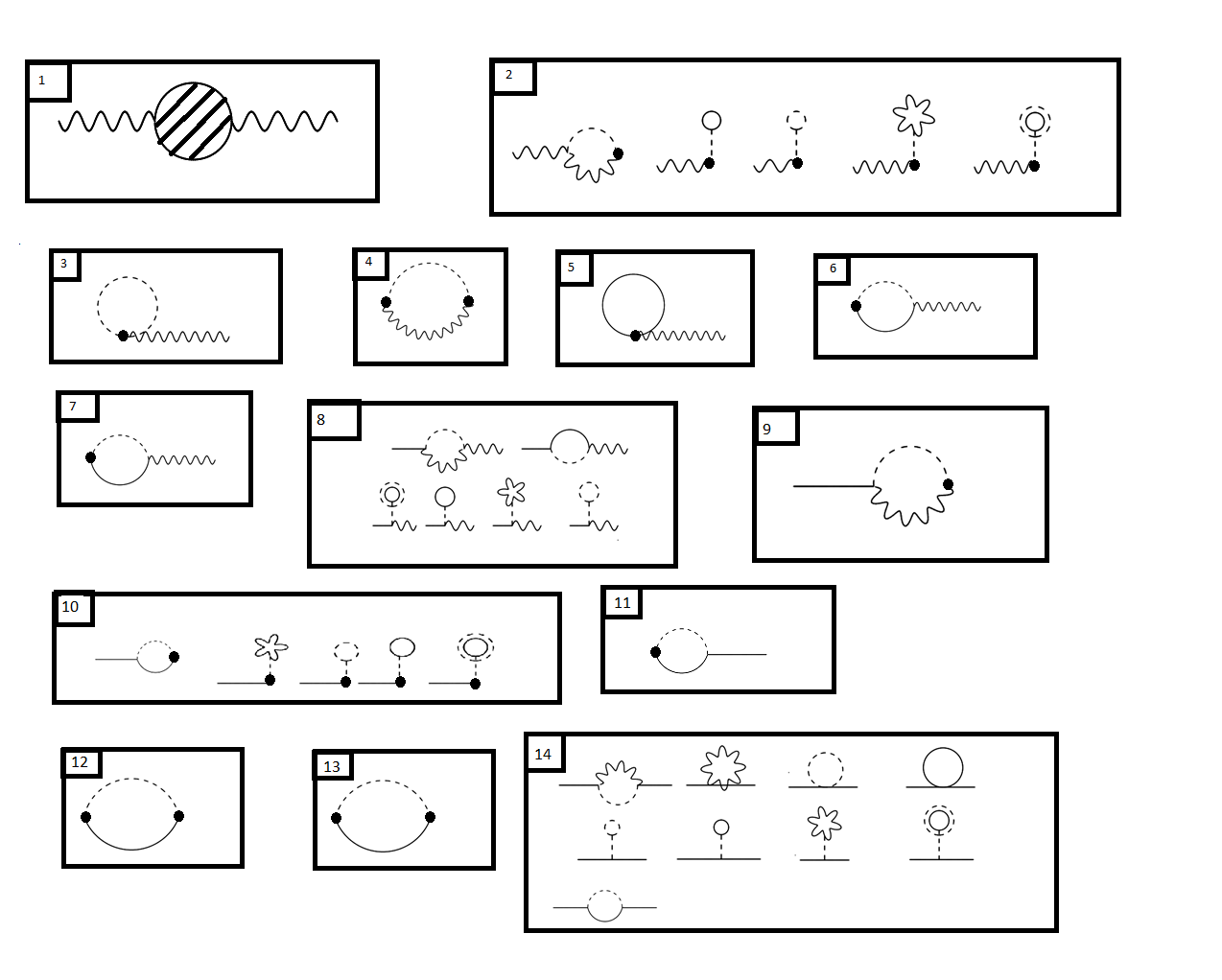}
	\caption{One-loop contributions for the propagator $\langle V_{\mu}(p)V_{\nu}(-p) \rangle$.  Wavy lines represent the photon field, dashed lines the Higgs field, solid lines the Goldstone boson and double lines the ghost field.}
	\label{Y22}
\end{figure}
We consider each term in the two-point function $\braket{V_{\m}(p)V_{\n}(-p)}$, given by eq. \eqref{exp2}. The first term is the one-loop correction to the photon propagator $\langle A_{\m}(p)A_{\n}(-p) \rangle$ known from the last chapter, shown in frame $(1)$ in Figure \ref{Y22}, which gives

\beq
\frac{1}{4}e^2 v^4  \langle  A_{\mu }(p)A_v(-p) \rangle &=&\frac{1}{4}e^2 v^4 \frac{\mathcal{P}_{\m\n}(p)}{p^2+m^2}+ \frac{1}{4}e^2 v^4 \Bigg[-\frac{m^2 \eta \left(m^2,m_h^2\right)\left( \left(m_h^2-m^2+p^2\right){}^2-4 ( d-2) m^2 p^2\right)}{( d-1) p^2 v^2}\nonumber\\
&+&\frac{m^2 \chi\left(m^2  \right) \left(2 ( d-1)^2 m^2 p^2+m_h^2 \left(p^2-m^2\right)+m_h^4\right)}{( d-1) p^2 v^2 m_h^2}\nonumber\\
&+&\frac{m^2 \chi \left(  m_h^2\right) \left((2  d-1) p^2-m_h^2+m^2\right)}{( d-1) p^2 v^2}\Bigg] \frac{\mathcal{P}_{\m\n}(p)}{(p^2+m^2)^2}\nonumber\\
&+&\frac{1}{4}e^2 v^4 \frac{\xi }{p^2+\xi m^2}\mathcal{L}_{\m\n}(p)+\frac{1}{4}e^2 v^4\Bigg[\frac{m^2 \xi ^2 \left(-2 m_h^2 \left(m^2-p^2\right)+m_h^4+\left(m^2+p^2\right)^2\right) \eta \left(m^2,m_h^2\right)}{p^2 v^2}\nonumber\\
&-&\frac{m^2 \xi ^2 \left(2 m_h^2-2 m^2 \xi +p^2\right) \eta \left(m_h^2,m^2 \xi \right)}{v^2}\nonumber\\
&+&\frac{m^2 \xi ^2 \chi \left(m^2\right) \left(2 ( d-1) m^2 p^2+m_h^2 \left(m^2-p^2\right)-m_h^4\right)}{p^2 v^2 m_h^2}\nonumber\\
&-&\frac{m^2 \xi ^2 \left(-m_h^2+m^2-3 p^2\right) \chi \left(m_h^2\right)}{p^2 v^2}+\frac{2 m^2 \xi ^2 \chi \left(m^2 \xi \right)}{v^2}\Bigg] \frac{\mathcal{L}_{\m\n}(p)}{(p^2+\xi m^2)^2}.
\eeq
The second term, the one-loop correction shown in frame (2) of Figure \ref{Y22}, gives
\beq
e^2 v^3  \langle h A_{\mu }(p)  A_v(-p)  \rangle &=& e^2 v^3  \Big[-\frac{e \eta \left(m_h^2,m^2 \xi \right) \left(2 e v m_h^4 \left(p^2-m^2 \xi \right)+e v m_h^2 \left(m^2 \xi +p^2\right)^2+e v m_h^6\right)}{2 (d-1) m^2 p^2 m_h^2}\nonumber\\
&-&\frac{e \eta \left(m^2,m_h^2\right) \left(e v m_h^2 \left(-2 (3-2 d) m^2 p^2-m^4-p^4\right)+2 e v m_h^4 \left(m^2-p^2\right)-e v m_h^6\right)}{2 (d-1) m^2 p^2 m_h^2}\nonumber\\
&-&\frac{e \chi \left(m^2 \xi \right) \left(d e p^2 v m_h^2+e m^2 \xi  v m_h^2-2 e p^2 v m_h^2-e v m_h^4\right)}{2 (d-1) m^2 p^2 m_h^2)}\nonumber\\
&-&\frac{e \chi \left(m_h^2\right) (3 d e p^2 v m_h^2-e m^2 \xi  v m_h^2+e m^2 v m_h^2-3 e p^2 v m_h^2}{2 (d-1) m^2 p^2 m_h^2}\nonumber\\
&-&\frac{e \chi \left(m^2\right) \left(2 (d-1)^2 e m^2 p^2 v-e m^2 v m_h^2+e p^2 v m_h^2+e v m_h^4\right)}{2 (d-1) m^2 p^2 m_h^2}\Big]\frac{\mathcal{P}_{\m\n}(p)}{(p^2+m^2)}\nonumber\\
&+&e^2 v^3 \Big[\frac{1}{2 p^2 v}\xi  \Big(\left(m_h^2+m^2 (-\xi )+p^2\right){}^2 \eta \left(m_h^2,m^2 \xi \right)\nonumber\\
&-&\left(-2 m_h^2 \left(m^2-p^2\right)+m_h^4+\left(m^2+p^2\right)^2\right) \eta \left(m^2,m_h^2\right)\nonumber\\
&+&\frac{\chi \left(m^2\right) \left(-2 (d-1) m^2 p^2+m_h^2 \left(p^2-m^2\right)+m_h^4\right)}{m_h^2}-\chi \left(m^2 \xi \right) \left(m_h^2+m^2 (-\xi )+2 p^2\right)\nonumber\\
&-&\chi \left(m_h^2\right) \left(m^2 (\xi -1)+3 p^2\right)\Big) \Big] \frac{\mathcal{L}_{\mu\nu}}{p^2+\xi m^2}.
\eeq
The third term, the one-loop correction shown in frame (3) of Figure \ref{Y22}, gives

\beq
\frac{1}{2} e^2 v^2  \langle h^2 A_{\mu }(p)  A_v(-p) \rangle =  \frac{1}{2} e^2 v^2 \Big[\frac{\chi \left(m_h^2\right)}{m^2+p^2}\Big] \mathcal{P}_{\m\n}+\frac{1}{2} e^2 v^2 \Big[\frac{\xi  \chi \left(m_h^2\right)}{m^2 \xi +p^2}\Big] \mathcal{L}_{\m\n}.
\eeq
The fourth term, the one-loop correction shown in frame (4) of Figure \ref{Y22}, gives

\beq
e^2 v^2  \langle h A_{\mu }(p) h  A_v(-p)  \rangle &=&e^2 v^2\Big[\frac{\left(\frac{\left(m_h^2+m^2 (-\xi )+p^2\right){}^2}{m^2 p^2}+4 \xi \right) \eta \left(m_h^2,m^2 \xi \right)}{4 (d-1)}\nonumber\\
&+&\frac{\eta \left(m^2,m_h^2\right) \left(4 (d-2)-\frac{\left(m_h^2-m^2+p^2\right){}^2}{m^2 p^2}\right)}{4 (d-1)}-\frac{\chi \left(m^2 \xi \right) \left(m_h^2+m^2 (-\xi )+p^2\right)}{4 (d-1) m^2 p^2}\nonumber\\
&+&\frac{\chi \left(m^2\right) \left(m_h^2-m^2+p^2\right)}{4 (d-1) m^2 p^2}-\frac{(\xi -1) \chi \left(m_h^2\right)}{4 (d-1) p^2}\Big]\mathcal{P}_{\m\n}\nonumber\\
&+&e^2 v^2\Big[\frac{1}{4 m^2 p^2}\Big(-\left(m_h^2+m^2 (-\xi )+p^2\right){}^2 \eta \left(m_h^2,m^2 \xi \right)\nonumber\\
&+&\left(\left(m_h^2-m^2+p^2\right){}^2+4 m^2 p^2\right) \eta \left(m^2,m_h^2\right)+m^2 (\xi -1) \chi \left(m_h^2\right)\nonumber\\
&+&\chi \left(m^2 \xi \right) \left(m_h^2+m^2 (-\xi )+p^2\right)+\chi \left(m^2\right) \left(-m_h^2+m^2-p^2\right)\Big)\Big]\mathcal{L}_{\m\n}.
\eeq
The fifth term, the one-loop correction shown in frame (5) of Figure \ref{Y22}, gives
\beq
\frac{1}{2} e^2 v^2 \langle \rho^2 A_{\mu }(p), A_v(-p) \rangle=\frac{1}{2} e^2 v^2\frac{\chi \left(m^2 \xi \right)}{m^2+p^2}\mathcal{P}_{\m\n}+ \frac{1}{2} e^2 v^2\frac{\xi  \chi \left(m^2 \xi \right)}{m^2 \xi +p^2} \mathcal{L}_{\m\n}.
\eeq
The sixth term, the one-loop correction shown in frame (6) of Figure \ref{Y22}, gives

\beq
-\frac{i}{2} e v^2 p_{\mu } \langle h  \rho (p)  A_v(-p) \rangle =-\frac{i}{2} e v^2  \Big[\frac{i e \xi  \left(m^2 \xi -m_h^2\right) \eta \left(m_h^2,m^2 \xi \right)}{m^2 \xi +p^2}-\frac{i e \xi  \chi \left(m_h^2\right)}{m^2 \xi +p^2}+\frac{i e \xi  \chi \left(m^2 \xi \right)}{m^2 \xi +p^2}\Big]\mathcal{L}_{\m\n}.
\eeq
The seventh term, the one-loop correction shown in frame (7) of Figure \ref{Y22}, gives

\beq
e v^2  \langle \partial ^x{}_{\mu } h\rho (p)  A_v(-p) \rangle&=&
e v^2\Big[-\frac{e \left(\left(-m_h^2+m^2 \xi +p^2\right){}^2+4 p^2 m_h^2\right) \eta \left(m_h^2,m^2 \xi \right)}{2 (d-1) p^2 \left(m^2+p^2\right)}+\frac{e \chi \left(m^2 \xi \right) \left(m_h^2+m^2 (-\xi )+p^2\right)}{2 (d-1) p^2 \left(m^2+p^2\right)}\nonumber\\
&+&\frac{e \chi \left(m_h^2\right) \left(-m_h^2+m^2 \xi +p^2\right)}{2 (d-1) p^2 \left(m^2+p^2\right)}\Big]\mathcal{P}_{\mu\nu}\nonumber\\
&+&e v^2\Big[\frac{e \xi  \left(-3 p^2 \left(m_h^2-m^2 \xi +p^2\right)+\left(m_h^2-m^2 \xi +p^2\right)^2+2 p^4\right) \eta \left(m^2 \xi ,m_h^2\right)}{2 p^2 \left(m^2 \xi +p^2\right)}\nonumber\\
&+&\frac{e \xi  \chi \left(m_h^2\right) \left(m_h^2-m^2 \xi \right)}{2 p^2 \left(m^2 \xi +p^2\right)}+\frac{e \xi  \chi \left(m^2 \xi \right) \left(-m_h^2+m^2 \xi +2 p^2\right)}{2 p^2 \left(m^2 \xi +p^2\right)}\Big]\mathcal{L}_{\m\n}.
\eeq

The eighth term, the one-loop correction shown in frame (11) of Figure \ref{Y22}, gives

\beq
-\frac{1}{2} i e v^3 p_{\mu } \langle \rho (p) A_v(-p) \rangle&=&
-\frac{1}{2} i e v^3 \Bigg[\frac{i e^3 \xi  v \left(\left(m_h^2-m^2+p^2\right){}^2+4 m^2 p^2\right) \eta \left(m^2,m_h^2\right)}{m^2 \left(m^2 \xi +p^2\right)^2}\nonumber\\
&+&\eta \left(m_h^2,m^2 \xi \right) \left(\frac{i e^3 \xi  v m_h^2 \left(m_h^2-m^2 \xi \right)}{m^2 \left(m^2 \xi +p^2\right)^2}-\frac{i e^3 \xi  v \left(m_h^2+p^2\right) \left(m_h^2+m^2 (-\xi )+p^2\right)}{m^2 \left(m^2 \xi +p^2\right)^2}\right)\nonumber\\
&+&\chi \left(m^2\right) \left(\frac{i (d-1) e^3 \xi  p^2 v}{m_h^2 \left(m^2 \xi +p^2\right)^2}-\frac{i e^3 \xi  v \left(m_h^2-m^2+p^2\right)}{m^2 \left(m^2 \xi +p^2\right)^2}\right)\nonumber\\
&+&\chi \left(m_h^2\right) \left(\frac{i e^3 \xi  v m_h^2}{m^2 \left(m^2 \xi +p^2\right)^2}-\frac{i e^3 \xi  v}{\left(m^2 \xi +p^2\right)^2}+\frac{3 i e^3 \xi  p^2 v}{2 m^2 \left(m^2 \xi +p^2\right)^2}\right)\nonumber\\
&+&\chi \left(m^2 \xi \right) \left(\frac{i e^3 \xi ^2 p^2 v}{m_h^2 \left(m^2 \xi +p^2\right)^2}+\frac{i e^3 \xi  v \left(m_h^2+p^2\right)}{m^2 \left(m^2 \xi +p^2\right)^2}\right.\nonumber\\
&-&\left.\frac{i e^3 \xi  v m_h^2}{m^2 \left(m^2 \xi +p^2\right)^2}+\frac{i e^3 \xi  p^2 v}{2 m^2 \left(m^2 \xi +p^2\right)^2}-\frac{i e^2 m \xi ^2 p^2}{m_h^2 \left(m^2 \xi +p^2\right)^2}\right)\Bigg]\mathcal{L}_{\m\n}.
\eeq
The ninth term, the one-loop correction shown in frame (12) of Figure \ref{Y22}, gives

\beq
- i e v^2 p_{\mu } \langle  \rho (p) h A_v(-p) \rangle &=&- i e v^2 \Bigg[-\frac{i e \left(m_h^2+p^2\right) \chi \left(m^2 \xi \right)}{2 m^2 \left(m^2 \xi +p^2\right)}\nonumber\\
&-&\frac{1}{2 m^2 \left(m^2 \xi +p^2\right)}\Big(i e \left(-\left(m_h^2+p^2\right) \left(m_h^2+m^2 (-\xi )+p^2\right) \eta \left(m^2 \xi ,m_h^2\right)\right.\nonumber\\
&+&\left.\left(\left(m_h^2-m^2+p^2\right){}^2+4 m^2 p^2\right) \eta \left(m^2,m_h^2\right)-m^2 \chi \left(m_h^2\right)\right.\nonumber\\
&-&\left.m_h^2 \chi \left(m^2\right)-p^2 \chi \left(m^2\right)+m^2 \chi \left(m^2\right)\right)\Big)\Bigg]
\mathcal{L}_{\m\n}.
\eeq
The tenth term, the one-loop correction shown in frame (14) of Figure \ref{Y22}, gives

\beq
-\frac{1}{2}v p_{\mu}p_{\nu}\langle h\rho(p) \rho(-p)\rangle&=& -\frac{1}{2}v \Bigg[p^2 \Big(-\frac{e^2 v m_h^2 \eta \left(m^2 \xi ,m_h^2\right)}{m^2 \left(m^2 \xi +p^2\right)}-\frac{(d-1) e^2 v \chi \left(m^2\right)}{m_h^2 \left(m^2 \xi +p^2\right)}\nonumber\\
&-&\frac{3 e^2 v \chi \left(m_h^2\right)}{2 m^2 \left(m^2 \xi +p^2\right)}-\frac{e^2 \xi  v \chi \left(m^2 \xi \right)}{m_h^2 \left(m^2 \xi +p^2\right)}-\frac{e^2 v \chi \left(m^2 \xi \right)}{2 m^2 \left(m^2 \xi +p^2\right)}+\frac{e m \xi  \chi \left(m^2 \xi \right)}{m_h^2 \left(m^2 \xi +p^2\right)}\Big)\Bigg]\mathcal{L}_{\m\n}.
\eeq
The eleventh term, the one-loop correction shown in frame (15) of Figure \ref{Y22}, gives
\beq
i v p_{\nu } \langle \partial ^x{}_{\mu }h \rho (p)\rho (-p) \rangle&=&i v\Big[\frac{i e^2 m_h^2 v \chi \left(m^2 \xi \right)}{2 m^4 \xi +2 m^2 p^2}-\frac{i e^2 m_h^2 v \left(\left(m_h^2-m^2 \xi -p^2\right) \eta \left(m^2 \xi ,m_h^2\right)+\chi \left(m_h^2\right)\right)}{2 m^2 \left(m^2 \xi +p^2\right)}\Big] \mathcal{L}_{\m\n}.
\eeq

The twelfth term, the one-loop correction shown in frame (16) of Figure \ref{Y22}, gives

\beq
\frac{1}{4}p_{\mu }p_{\nu } \langle h h(p) \rho \rho (-p) \rangle=p^2 \eta \left(m^2 \xi ,m_h^2\right)\mathcal{L}_{\m\n}.
\eeq
The thirteenth term, the one-loop correction shown in frame (17) of Figure \ref{Y22}, gives
\beq
-\langle\partial ^x{}_{\mu }h\rho (p) h \partial ^y{}_{\nu }\rho (-p) \rangle&=&-\Big[\frac{\left(\frac{\left(-m_h^2+m^2 \xi +p^2\right){}^2}{p^2}+4 m_h^2\right) \eta \left(m_h^2,m^2 \xi \right)}{4 (d-1)}+\frac{\chi \left(m^2 \xi \right) \left(-m_h^2+m^2 \xi -p^2\right)}{4 (d-1) p^2}\nonumber\\
&-&\frac{\chi \left(m_h^2\right) \left(-m_h^2+m^2 \xi +p^2\right)}{4 (d-1) p^2}\Big]\mathcal{P}_{\m\n}\nonumber\\
&-&\Big[-\frac{\left(m_h^2+m^2 (-\xi )-p^2\right) \left(m_h^2+m^2 (-\xi )+p^2\right) \eta \left(m^2 \xi ,m_h^2\right)}{4 p^2}\nonumber\\
&+&\frac{\chi \left(m^2 \xi \right) \left(m_h^2+m^2 (-\xi )-p^2\right)}{4 p^2}-\frac{\chi \left(m_h^2\right) \left(m_h^2+m^2 (-\xi )+p^2\right)}{4 p^2}\Big]\mathcal{L}_{\m\n}.
\eeq
The fourteenth term, the one-loop correction shown in frame (18) of Figure \ref{Y22}, gives

\beq
\frac{1}{4}v^2 p_{\mu }p_{\nu }  \langle\rho (p) \rho (-p)\rangle &=& \Big[ p^2 \big(\frac{e^4 v^2 m_h^4 \eta \left(m^2 \xi ,m_h^2\right)}{m^4 \left(m^2 \xi +p^2\right)^2}+\frac{e^2 \left(\left(m_h^2-m^2+p^2\right){}^2+4 m^2 p^2\right) \eta \left(m^2,m_h^2\right)}{m^2 \left(m^2 \xi +p^2\right)^2}\nonumber\\
&-&\frac{e^2 \left(m_h^2+p^2\right){}^2 \eta \left(m_h^2,m^2 \xi \right)}{m^2 \left(m^2 \xi +p^2\right)^2}+\frac{(d-1) e^4 v^2 \chi \left(m^2\right)}{m^2 \left(m^2 \xi +p^2\right)^2}-\frac{(d-1) e^2 \chi \left(m^2\right)}{\left(m^2 \xi +p^2\right)^2}+\frac{e^4 v^2 m_h^2 \chi \left(m^2 \xi \right)}{2 m^4 \left(m^2 \xi +p^2\right)^2}\nonumber\\
&+&\frac{3 e^4 v^2 m_h^2 \chi \left(m_h^2\right)}{2 m^4 \left(m^2 \xi +p^2\right)^2}+\frac{e^4 \xi  v^2 \chi \left(m^2 \xi \right)}{m^2 \left(m^2 \xi +p^2\right)^2}-\frac{e^3 \xi  v \chi \left(m^2 \xi \right)}{m \left(m^2 \xi +p^2\right)^2}+\frac{e^2 \chi \left(m^2 \xi \right) \left(m_h^2+m^2 \xi +p^2\right)}{m^2 \left(m^2 \xi +p^2\right)^2}\nonumber\\
&-&\frac{e^2 \chi \left(m_h^2\right)}{\left(m^2 \xi +p^2\right)^2}-\frac{e^2 \chi \left(m^2\right) \left(m_h^2-m^2+p^2\right)}{m^2 \left(m^2 \xi +p^2\right)^2}-\frac{3 e^2 m_h^2 \chi \left(m^2 \xi \right)}{2 m^2 \left(m^2 \xi +p^2\right)^2}-\frac{e^2 m_h^2 \chi \left(m_h^2\right)}{2 m^2 \left(m^2 \xi +p^2\right)^2}\nonumber\\
&-&\frac{e^2 \xi  \chi \left(m^2 \xi \right)}{\left(m^2 \xi +p^2\right)^2}\big)\Big] \mathcal{L}_{\m\n}.
\eeq
\chapter{Technical details to the $SU(2)$ Higgs model}
	\section{Propagators and vertices of the $SU(2)$ Higgs model in the $R_{\xi}$ gauge \label{appA}}
	The tree level elementary propagators of the fields are easily computed, being given by
	\beq
	\langle A^a_{\m}(p)A^b_{\n}(-p)\rangle &=& \frac{\delta^{ab}}{p^2+ m^2} \mathcal{P}_{\m\n}(p)+\delta^{ab}\frac{\xi}{p^2 + \xi m^2}\mathcal{L}_{\m\n}(p),\nonumber\\
	\langle \r^a(p)\r^b(-p)\rangle &=&\frac{\delta^{ab}}{p^2 +\xi m^2},\nonumber\\
	\langle h(p)h(-p)\rangle &=&\frac{1}{p^2 + m_h^2},\nonumber\\
	\langle A^a_{\m}(p)b^b(-p)\rangle &=&\delta^{ab} \frac{p_{\m}}{p^2+\xi m^2},\nonumber\\
	\langle b^a (p) \rho^b (-k) \rangle &=& \delta^{ab}\frac{im}{p^2+ \xi m^2}
	\eeq
	and
	\beq
	\langle \bar{c}^a(p)c^b(-p)\rangle &=&\frac{\delta^{ab}}{p^2 +\xi m^2}
	\eeq
	for the ghost propagator.
	For all vertices, adopting the convention that the momentum is flowing towards the vertex, we get
	\begin{itemize}
		\item The $AAh$-vertex:
		$\Gamma_{A_{\mu}^a A_{\nu}^b h}(-p_1,-p_2,-p_3)=-\frac{g^2 v}{2}\delta_{\mu\nu}\delta^{ab} \delta(p_1+p_2+p_3)$.
		\item The $\rho\rho A$-vertex: $\Gamma_{\rho^a \rho^b A_{\mu}^c} (-p_1,-p_2,-p_3)= \frac{g}{2}i \epsilon^{abc}(p_{\mu,1}-p_{\mu,2})\delta(p_1+p_2+p_3)$.
		\item The $A\rho h$-vertex: $\Gamma_{A_{\mu}^a \rho^b h}(-p_1,-p_2,-p_3)=i\frac{g}{2}\delta^{ab}(p_{\mu,3}-p_{\mu,2})\delta(p_1+p_2+p_3)$.
		\item The $hhh$ vertex: $\Gamma_{hhh}(-p_1,-p_2,-p_3)=- 3 \l v\,\,\delta(p_1+p_2+p_3)$.
		\item The $h\rho \rho $ vertex: $\Gamma_{h\rho^a \rho^b}(-p_1,-p_2,-p_3)=-\lambda v \delta^{ab}\,\,\delta(p_1+p_2+p_3)$.
		\item The $AAA$-vertex: $\Gamma_{A_{\m}^a A_{\n}^bA_{\s}^c}(-p_1,-p_2,-p_3)=-igf^{abc}\left[(p_1-p_3)_{\nu}\delta_{\s\m}+(p_3-p_2)_{\mu}\delta_{\n\s}+(p_2-p_1)_{\s}\delta_{\n\m}\right]\delta(p_1+p_2+p_3)$.
		\item The $\overline c A c$-vertex: $\Gamma_{\overline c^a A_{\m}^bc^c}(-p_1,-p_2,-p_3)=ig f^{abc}p_{1,\m}\delta(p_1+p_2+p_3)$.
		\item The $AAAA$-vertex: $\Gamma_{A_{\mu}^{a}A_{\nu}^{b}A_{\rho}^{c}A_{\sigma}^{d}}(-p_1,-p_2,-p_3,-p_4)=g^{2}[f^{eab}f^{ecd}(\delta_{\mu\sigma}\delta_{\nu\rho}-\delta_{\mu\rho}\delta_{\nu\sigma})+f^{eac}f^{ebd}(\delta_{\mu\sigma}\delta_{\nu\rho}-\delta_{\mu\nu}\delta_{\rho\sigma})+f^{ead}f^{ebc}(\delta_{\mu\rho}\delta_{\nu\sigma}-\delta_{\mu\nu}\delta_{\rho\sigma})]\delta(p_1+p_2+p_3+p_4)$.
		\item The $AAhh$-vertex: $\Gamma_{A_{\m}^aA_{\n}^b hh}(-p_1,-p_2,-p_3,-p_4)= -\frac{1}{2}g^2\delta^{ab}\delta_{\m\n}$.
		\item The $AA\r\r$-vertex: $\Gamma_{A_{\m}^aA_{\n}^b \rho^c \rho^d} (-p_1,-p_2,-p_3,-p_4)= - \frac{1}{2}g^2 \delta_{\m\n} \delta^{ab}\delta^{cd}$.
	\end{itemize}

	\section{Elementary propagators of the $SU(2)$ Higgs model in the $R_{\xi}$ gauge\label{appel}}
	Here we will calculate the one-loop corrections to the Higgs and gauge field propagator. This requires the calculation \footnote{We have used from \cite{passarino1979one} the technique of modifying integrals into ``master integrals'' without numerators.} of the Feynman diagrams as shown in Figures~\ref{more} and \ref{les}.  We will use the following definitions:
	\beq
	\eta(m_1,m_2)&\equiv& \frac{1}{(4\pi)^{d/2}}\Gamma(2-\frac{d}{2})\int_0^1 dx \left(p^2 x(1-x)+x m_1+(1-x) m_2\right)^{d/2-2},\nonumber\\
	\chi(m_1)&\equiv&  \frac{1}{(4\pi)^{d/2}} \Gamma(1-\frac{d}{2}) m_1^{d/2-1}.
	\eeq
	Notice that the last four diagrams for both particles are zero for $\langle h \rangle=0$. In fact, including these diagrams has the same effect as making a shift in the minimizing value of the scalar field $\Phi$  to demand $\langle h \rangle=0$, see the Appendix of \cite{Dudal:2019aew} for the technical details.  In the context of the FMS operators, we found it more convenient to expand around the (classical) $v$ that is gauge invariant, and thus to include the tadpoles.  Expanding the FMS operator around the quantum corrected vev would lead to cancellations in that quantum vev coming from the propagator loop corrections to render it gauge invariant again, indeed the minimum of the quantum corrected effective Higgs potential is not gauge invariant itself.
	
	\subsection{Higgs propagator \label{hp}}
	\begin{figure}[H]
		\centering
		\includegraphics[width=15cm]{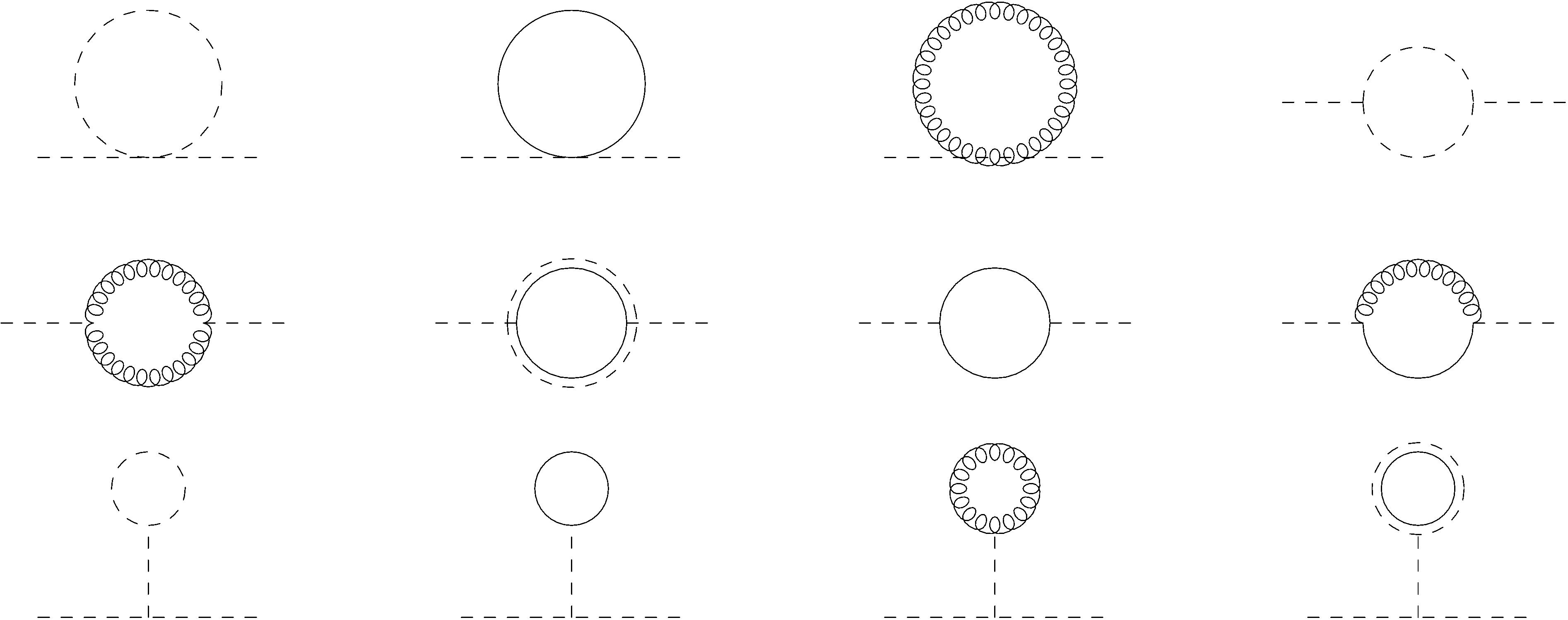}
		\caption{One-loop contributions to the propagator $\langle h(p)h(-p) \rangle$. Curly lines represent the gauge field, dashed lines the Higgs field, solid lines the Goldstone boson and double lines the ghost field.}
		\label{more}
	\end{figure}
	The first diagrams contributing to the Higgs self-energy are of the snail type, renormalizing the masses of the internal fields.
	The Higgs boson snail (first diagram in the first line of Figure \ref{more}):
	\beq
	\Gamma_{hh,1}(p^2)&=&-\frac{3 \lambda  \chi \left(m_h^2\right)}{2 \left(m_h^2+p^2\right)^2},
	\label{wat}
	\eeq
	the Goldstone boson snail  (second diagram in the first line of Figure \ref{more}):
	\beq
	\Gamma_{hh,2}(p^2) &=& -\frac{3 \lambda  \chi \left(m^2 \xi \right)}{2 \left(m_h^2+p^2\right)^2},
	\label{ii}
	\eeq
	and the gauge field  snail  (third diagram in the first line of Figure \ref{more}):
	\beq
	\Gamma_{hh,3}(p^2) &=& -\frac{3 (d-1) g^2 \chi \left(m^2\right)}{4 \left(m_h^2+p^2\right)^2}-\frac{3 g^2 \xi  \chi \left(m^2 \xi \right)}{4 \left(m_h^2+p^2\right)^2}.
	\eeq
	Next, we meet a couple of sunset diagrams. The Higgs boson sunset (fourth diagram in the first line of Figure \ref{more}):
	\beq
	\Gamma_{hh,4}(p^2)
	&=&\int_0^1 dx \Big\{\frac{9 \lambda ^2 v^2 \eta \left(m_h^2,m_h^2\right)}{2 \left(m_h^2+p^2\right)^2}\Big\},
	\eeq
	the gauge field sunset (first diagram in the second line of Figure \ref{more}):
	\beq
	\Gamma_{hh,5}(p^2)	&=& \int_0^1 dx \Big\{\frac{3 g^2 \eta \left(m^2,m^2\right) \left(4 (D-1) m^4+4 m^2 p^2+p^4\right)}{8 m^2 \left(m_h^2+p^2\right)^2}+\frac{3 g^2 \left(2 m^2 \xi +p^2\right)^2 \eta \left(m^2 \xi ,m^2 \xi \right)}{8 m^2 \left(m_h^2+p^2\right)^2}\nonumber\\
	&-&\frac{3 g^2 \left(m^4 (\xi -1)^2+2 m^2 \xi  p^2+2 m^2 p^2+p^4\right) \eta \left(m^2,m^2 \xi \right)}{4 m^2 \left(m_h^2+p^2\right)^2}+\frac{3 g^2 (\xi -1) \chi \left(m^2\right)}{4 \left(m_h^2+p^2\right)^2}\nonumber\\
	&-&\frac{3 g^2 (\xi -1) \chi \left(m^2 \xi \right)}{4 \left(m_h^2+p^2\right)^2}\Big\},
	\eeq
	the ghost sunset (second diagram in the second line of Figure \ref{more}):
	\beq
	\Gamma_{hh,6}(p^2)
	&=&-\int_0^1 dx \Big\{\frac{3 g^2 m^2 \xi ^2 \eta \left(m^2 \xi ,m^2 \xi \right)}{4 \left(m_h^2+p^2\right)^2}\Big\},
	\eeq
	the Goldstone boson sunset (third diagram in the second line of Figure \ref{more}):
	\beq
	\Gamma_{hh,7}(p^2)&=&\int_0^1 dx \Big\{\frac{3 \lambda ^2 v^2 \eta \left(m^2 \xi ,m^2 \xi \right)}{2 \left(m_h^2+p^2\right)^2}\Big\},
	\eeq
	and a mixed Goldstone-gauge sunset (fourth diagram in the second line of Figure \ref{more}):
	\beq
	\Gamma_{hh,8}(p^2)
	&=&\int_0^1 dx \Big\{\frac{3 g^2 \left(\left(m^2 (\xi -1)+p^2\right)^2+4 m^2 p^2\right) \eta \left(m^2,m^2 \xi \right)}{4 m^2 \left(m_h^2+p^2\right)^2}-\frac{3 g^2 \left(m^2 \xi +p^2\right)^2 \eta \left(m^2 \xi ,m^2 \xi \right)}{4 m^2 \left(m_h^2+p^2\right)^2}\nonumber\\
	&+&\frac{3 g^2 \chi \left(m^2 \xi \right) \left(m^2 (2 \xi -1)+p^2\right)}{4 m^2 \left(m_h^2+p^2\right)^2}-\frac{3 g^2 \chi \left(m^2\right) \left(m^2 (\xi -1)+p^2\right)}{4 m^2 \left(m_h^2+p^2\right)^2}\Big\}.
	\eeq
	Finally, we have the tadpole diagrams. The Higgs balloon (first diagram on the third line of  Figure~\ref{more}):
	\beq
	\Gamma_{hh,9}(p^2) &=&\frac{9 \lambda ^2 v^2 \chi \left(m_h^2\right)}{2 m_h^2 \left(m_h^2+p^2\right)^2},
	\eeq
	the gauge balloon (second diagram on the third line of  Figure~\ref{more}):
	\beq
	\Gamma_{hh,10}(p^2) &=& \frac{9 g \lambda  m v \left((d-1) \chi \left(m^2\right)+\xi  \chi \left(m^2 \xi \right)\right)}{2 m_h^2 \left(m_h^2+p^2\right)^2},
	\eeq
	the Goldstone boson balloon (third diagram on the third line of  Figure~\ref{more}):
	\beq
	\Gamma_{hh,11}(p^2) &=& \frac{9 \lambda ^2 v^2 \chi \left(m^2 \xi \right)}{2 m_h^2 \left(m_h^2+p^2\right)^2},
	\label{uu}
	\eeq
	the ghost balloon (fourth diagram on the third line of  Figure~\ref{more}):	
	\beq
	\Gamma_{hh,12}(p^2)&=&-\frac{9 g \lambda  m \xi  v \chi \left(m^2 \xi \right)}{2 m_h^2}.
	\label{poi}
	\eeq
	Putting together eqs.~\eqref{wat} to \eqref{poi} we find the Higgs propagator up to first order in $\hbar$
	\beq
	\braket{h(x)\,h(y)}&=& \frac{1}{p^2+m_h^2}+g^2\int_0^1 dx \Big\{ \frac{3  \left(4 (d-1) m^4+4 m^2 p^2+p^4\right)}{8 m^2}\eta \left(m^2,m^2\right)\nonumber\\
	&+&\frac{9 m_h^4}{8 m^2} \eta \left(m_h^2,m_h^2\right)+\frac{3 \left(m_h^4-p^4\right) }{8 m^2}\eta \left(m^2 \xi ,m^2 \xi \right)+\frac{ \left(6 (d-1) m^2-3 p^2\right)}{4 m^2}\chi \left(m^2\right)\nonumber\\
	&+&\frac{3 \left(m_h^2+p^2\right) }{4 m^2}\chi \left(m^2 \xi \right)+\frac{3 m_h^2}{4 m^2} \chi \left(m_h^2\right)\Big\} \frac{1}{(p^2+m_h^2)^2}.
	\eeq
	The resummed one-loop Higgs propagator can be now  approximated by
	\beq
	G^{-1}_{hh}(p^2)&=& p^2+m_h^2 - g^2\int_0^1 dx \Big\{ \frac{3  \left(4 (d-1) m^4+4 m^2 p^2+p^4\right)}{8 m^2}\eta \left(m^2,m^2\right)\nonumber\\
	&+&\frac{9 m_h^4}{8 m^2} \eta \left(m_h^2,m_h^2\right)+\frac{3 \left(m_h^4-p^4\right) }{8 m^2}\eta \left(m^2 \xi ,m^2 \xi \right)+\frac{ \left(6 (d-1) m^2-3 p^2\right)}{4 m^2}\chi \left(m^2\right)\nonumber\\
	&+&\frac{3 \left(m_h^2+p^2\right) }{4 m^2}\chi \left(m^2 \xi \right)+\frac{3 m_h^2}{4 m^2} \chi \left(m_h^2\right)\Big\}.
	\label{hhl}
	\eeq
	For $d=4$, the above expression, eq.~\eqref{hhl}, is divergent. Employing the procedure of dimensional regularization, {\it i.e.}~setting $d=4-\epsilon$, the  divergent part for $G_{hh}(p^2)$ is given by:
	\beq
	G_{hh, \rm div}(p^2)&=&\frac{g^2 \left(\frac{3 m_h^4}{m^2}-3 \xi  m_h^2-3 \xi  p^2+9 p^2\right)}{32 \pi ^2 \epsilon },
	\eeq
	which, following the  $\overline{MS}bar$-scheme, are re-absorbed by the introduction of suitable local  counterterms. We remain thus with the finite part of the Higgs propagator, given in  eq.~\eqref{hhf}.

	\subsection{Gauge field propagator \label{Ap}}
	
	\begin{figure}[H]
		\centering
		\includegraphics[width=\textwidth]{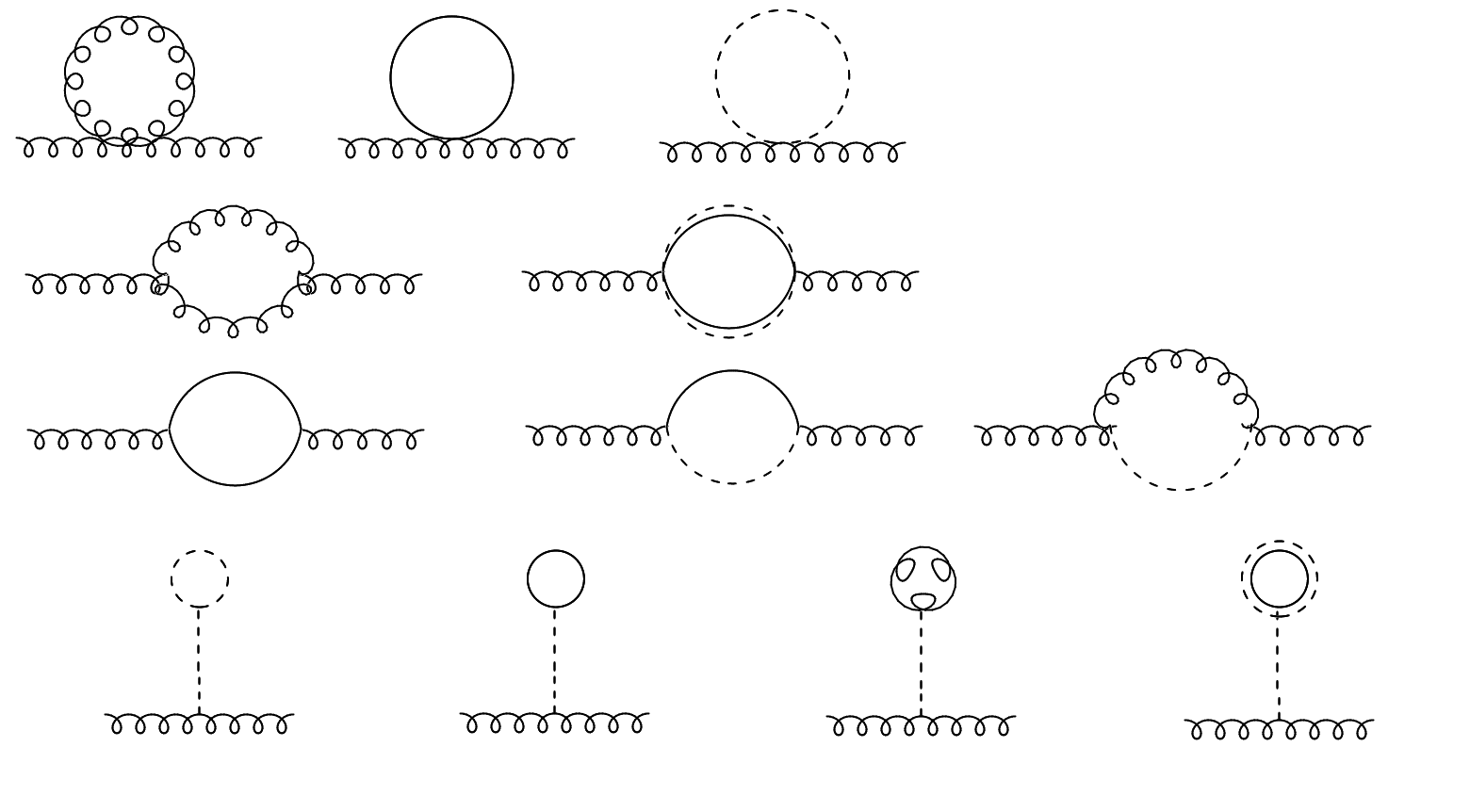}
		\caption{Contributions to the one-loop gauge field self-energy.}
		\label{les}
	\end{figure}
	The first diagram contributing to transverse part of the gauge field self-energy is the gauge field snail (first diagram in the first line of Figure \ref{les}) and gives a contribution:
	\beq
	\Pi_{AA^T, 1}(p^2)&=&\frac{2 g^2 \left(p^2-d \left(d^2-3 d+3\right) p^2\right) \chi \left(m^2\right)}{(d-1) d p^2 \left(m^2+p^2\right)^2}-\frac{2 g^2 \xi  \left((d-2) d p^2+p^2\right) \chi \left(m^2 \xi \right)}{(d-1) d p^2 \left(m^2+p^2\right)^2}.
	\label{besu}
	\eeq
	The second diagram is the Goldstone boson snail (second diagram in the first line of Figure \ref{les}):
	\beq
	\Pi_{AA^T, 2}(p^2)	&=&-\frac{3 g^2 \chi \left(m^2 \xi \right)}{4 \left(m^2+p^2\right)^2}.
	\eeq
	The third diagram is the Higgs boson snail (third diagram in the first line of Figure \ref{les}):
	\beq
	\Pi_{AA^T, 3}(p^2)	&=&-\frac{g^2 \chi \left(m_h^2\right)}{4 \left(m^2+p^2\right)^2}.
	\eeq
	The fourth diagram is the gauge field sunrise (first diagram in the second line of Figure \ref{les}):
	\beq
	\Pi_{AA^T, 4}(p^2) &=&g^2 \int_0^1 dx \Big\{\frac{\eta \left(m^2,m^2 \xi \right) \left(2 m^2 p^2 (-2 d+\xi +3)+m^4 (\xi -1)^2+p^4\right)}{2 (d-1) m^4 p^2}\nonumber\\
	&-&\frac{\left(4 m^2+p^2\right) \eta \left(m^2,m^2\right) \left(4 (d-1) m^4+4 (3-2 d) m^2 p^2+p^4\right)}{4 (d-1) m^4 \left(m^2+p^2\right)^2}\nonumber\\
	&-&\frac{\left(4 m^2 \xi  p^4+p^6\right) \eta \left(m^2 \xi ,m^2 \xi \right)}{4 (d-1) m^4 \left(m^2+p^2\right)^2}\nonumber\\
	&+&\frac{\chi \left(m^2 \xi \right) \left(4 d^2 \left(m^2 (\xi +1) p^2+p^4\right)+d \left(m^4 (\xi -1)-m^2 (6 \xi +7) p^2+(\xi -7) p^4\right)+4 m^2 \xi  p^2\right)}{2 (d-1) d m^2 p^2 \left(m^2+p^2\right)^2}\nonumber\\
	&-&\frac{\chi \left(m^2\right) \left(4 d^2 p^4+d \left(m^4 (\xi -1)+m^2 (2 \xi -5) p^2+(\xi -7) p^4\right)+4 m^2 p^2\right)}{2 (d-1) d m^2 p^2 \left(m^2+p^2\right)^2}\Big\}.
	\eeq
	The fifth diagram is the ghost sunrise (second diagram in the second line of Figure \ref{les}):
	\beq
	\Pi_{AA^T, 5}(p^2)
	&=&g^2 \int_0^1 dx\eta \left(m^2 \xi ,m^2 \xi \right) \left(\frac{2  m^2 \xi }{(d-1) \left(m^2+p^2\right)^2}+\frac{ p^2}{2 (d-1) \left(m^2+p^2\right)^2}\right)\nonumber\\
	&-&\frac{ \chi \left(m^2 \xi \right)}{(d-1) \left(m^2+p^2\right)^2}.
	\eeq
	The sixth diagram is the Goldstone sunrise (first diagram in the third line of Figure \ref{les}):
	\beq
	\Pi_{AA^T, 6}(p^2)
	&=&g^2 \int_0^1 dx \eta \left(m^2 \xi ,m^2 \xi \right) \left(-\frac{ m^2 \xi }{(d-1) \left(m^2+p^2\right)^2}-\frac{p^2}{4 (d-1) \left(m^2+p^2\right)^2}\right)\nonumber\\
	&+&\frac{\chi \left(m^2 \xi \right)}{2 (d-1) \left(m^2+p^2\right)^2}.
	\eeq
	The seventh diagram is the mixed Goldstone-Higgs sunrise (second diagram in the third line of Figure \ref{les}):
	\beq
	\Pi_{AA^T, 7}(p^2)
	&=&g^2 \int_0^1 dx -\frac{\left(\left(-m_h^2+m^2 \xi +p^2\right)^2+4 m_h^2 p^2\right) \eta \left(m_h^2,m^2 \xi \right)}{4 (d-1) p^2 \left(m^2+p^2\right)^2}+\frac{\chi \left(m_h^2\right) \left(-m_h^2+m^2 \xi +p^2\right)}{4 (d-1) p^2 \left(m^2+p^2\right)^2}\nonumber\\
	&+&\frac{\chi \left(m^2 \xi \right) \left(m_h^2-m^2 \xi +p^2\right)}{4 (d-1) p^2 \left(m^2+p^2\right)^2}.
	\eeq
	The eighth diagram is the mixed Goldstone-gauge field sunrise (third diagram in the third line of Figure \ref{les}):
	\beq
	\Pi_{AA^T, 8}(p^2)
	&=&g^2 \int_0^1 dx\frac{\left(\left(m_h^2-m^2 \xi +p^2\right)^2+4 m^2 \xi  p^2\right) \eta \left(m_h^2,m^2 \xi \right)}{4 (D-1) p^2 \left(m^2+p^2\right)^2}\nonumber\\
	&-&\frac{\eta \left(m^2,m_h^2\right) \left(\left(m_h^2-m^2+p^2\right)^2-4 (D-2) m^2 p^2\right)}{4 (D-1) p^2 \left(m^2+p^2\right)^2}-\frac{m^2 (\xi -1) \chi \left(m_h^2\right)}{4 (D-1) p^2 \left(m^2+p^2\right)^2}\nonumber\\
	&-&\frac{\chi \left(m^2 \xi \right) \left(m_h^2-m^2 \xi +p^2\right)}{4 (D-1) p^2 \left(m^2+p^2\right)^2}+\frac{\chi \left(m^2\right) \left(m_h^2-m^2+p^2\right)}{4 (D-1) p^2 \left(m^2+p^2\right)^2}.
	\eeq
	Finally, we have four tadpole (balloon) diagrams. The Higgs boson balloon (first diagram of the last line in  Figure~\ref{les}):
	\beq
	\Pi_{AA^T, 5}(p^2)
	&=& \frac{3 g m \chi \left(m_h^2\right)}{2 v \left(m^2+p^2\right)^2},
	\eeq
	the Goldstone boson balloon (second diagram of the last line in  Figure~\ref{les}):
	\beq
	\Pi_{AA^T, 6}(p^2)
	&=&\frac{3 g \lambda  m v \chi \left(m^2 \xi \right)}{2 m_h^2 \left(m^2+p^2\right)^2},
	\eeq
	The gauge field balloon (third diagram of the last line in  Figure~\ref{les}):
	\beq
	\Pi_{AA^T,7}(p^2)&=&\frac{3 (D-1) g^2 m^2 \chi \left(m^2\right)}{2 m_h^2 \left(m^2+p^2\right)^2}+\frac{3 g^2 m^2 \xi  \chi \left(m^2 \xi \right)}{2 m_h^2 \left(m^2+p^2\right)^2}
	\eeq
	and finally, the ghost balloon (fourth diagram of the last line in  Figure~\ref{les}):
	\beq
	\Gamma_{AA^T,8}(p^2)&=&-\frac{3 g^2 m^2 \xi  \chi \left(m^2 \xi \right)}{2 m_h^2 \left(m^2+p^2\right)^2}.
	\label{absu}
	\eeq
	Combining all these contributions \eqref{besu}-\eqref{absu}, we find the total one-loop correction to the gauge field self-energy
	\beq
	\braket{A^a_{\mu}(p) A^b_{\nu}(p)}^T&=&\frac{\delta^{ab}}{p^2+m^2}+\delta^{ab}g^2 \int_0^1 dx  \Big\{-\frac{ \left(2 (3-2 d) m^2 p^2+m_h^4-2 m_h^2 \left(m^2-p^2\right)+m^4+p^4\right)}{4 (d-1) p^2}\eta \left(m^2,m_h^2\right)\nonumber\\
	&+&\frac{\left(m^2+p^2\right)^2 \left(2 m^2 p^2 (-2 d+\xi +3)+m^4 (\xi -1)^2+p^4\right)}{2 (d-1) m^4 p^2}\eta \left(m^2,m^2 \xi \right) \nonumber\\
	&+&\frac{\left(m^4-p^4\right) \left(4 m^2 \xi +p^2\right) }{4 (d-1) m^4}\eta \left(m^2 \xi ,m^2 \xi \right)-\frac{\left(4 m^2+p^2\right)  \left(4 (d-1) m^4+4 (3-2 d) m^2 p^2+p^4\right)}{4 (d-1) m^4}\eta \left(m^2,m^2\right)\nonumber\\
	&+&\frac{\left(m_h^2 \left(-m^2 p^2 \left(8 d^2-24 d+4 \xi +13\right)-2 p^4 (4 d+\xi -7)+m^4 (1-2 \xi )\right)+6 (d-1)^2 m^4 p^2+m_h^4 m^2\right)}{4 (d-1) m_h^2 m^2 p^2}\chi \left(m^2\right) \nonumber\\
	&-&\frac{ \left((d-2) p^2+m_h^2-m^2\right)}{4 (d-1) p^2}\chi \left(m_h^2\right)+\frac{ \left(m^2 p^2 (5 d+4 \xi -13)+2 p^4 (4 d+\xi -7)+2 m^4 (\xi -1)\right)}{4 (d-1) m^2 p^2}\chi \left(m^2 \xi \right)\nonumber\\
	&+&\frac{3 }{4}\chi \left(m_h^2\right)+\frac{3}{4} \chi \left(m^2 \xi \right)\Big\}\frac{1}{(p^2+m^2)^2}
	\eeq
	and the resummed propagator for the transverse gauge field can be approximated, at one-loop order,  by
	\beq
	G^{-1}_{AA^T}&=& \delta^{ab}\Bigg(p^2+m^2 - g^2 \int_0^1 dx  \Big\{-\frac{ \left(2 (3-2 d) m^2 p^2+m_h^4-2 m_h^2 \left(m^2-p^2\right)+m^4+p^4\right)}{4 (d-1) p^2}\eta \left(m^2,m_h^2\right)\nonumber\\
	&+&\frac{\left(m^2+p^2\right)^2 \left(2 m^2 p^2 (-2 d+\xi +3)+m^4 (\xi -1)^2+p^4\right)}{2 (d-1) m^4 p^2}\eta \left(m^2,m^2 \xi \right) \nonumber\\
	&+&\frac{\left(m^4-p^4\right) \left(4 m^2 \xi +p^2\right) }{4 (d-1) m^4}\eta \left(m^2 \xi ,m^2 \xi \right)-\frac{\left(4 m^2+p^2\right)  \left(4 (d-1) m^4+4 (3-2 d) m^2 p^2+p^4\right)}{4 (d-1) m^4}\eta \left(m^2,m^2\right)\nonumber\\
	&+&\frac{\left(m_h^2 \left(-m^2 p^2 \left(8 d^2-24 d+4 \xi +13\right)-2 p^4 (4 d+\xi -7)+m^4 (1-2 \xi )\right)+6 (d-1)^2 m^4 p^2+m_h^4 m^2\right)}{4 (d-1) m_h^2 m^2 p^2}\chi \left(m^2\right) \nonumber\\
	&-&\frac{ \left((d-2) p^2+m_h^2-m^2\right)}{4 (d-1) p^2}\chi \left(m_h^2\right)+\frac{ \left(m^2 p^2 (5 d+4 \xi -13)+2 p^4 (4 d+\xi -7)+2 m^4 (\xi -1)\right)}{4 (d-1) m^2 p^2}\chi \left(m^2 \xi \right)\nonumber\\
	&+&\frac{3 }{4}\chi \left(m_h^2\right)+\frac{3}{4} \chi \left(m^2 \xi \right)\Big\}\Bigg).
	\eeq
	For $d=4-\epsilon$, following the procedure of dimensional regularization, we find that the divergent part for $G_{AA^T}(p^2)$ is given by:
	\beq
	G_{AA^T, \rm div}(p^2)&=&\frac{g^2}{\pi ^2 \epsilon } \left(-\frac{9 m^4}{16 m_h^2}-\frac{3 m_h^2}{32}-\frac{m^2 \xi }{8}-\frac{3 m^2}{32}-\frac{\xi  p^2}{8}+\frac{25 p^2}{48}\right),
	\eeq
	and these terms can be, following the  $\overline{MS}bar$-scheme, absorbed by means of appropriate counterterms. We remain with the finite part of the propagator, given in  eq.~\eqref{AAf}.
	
	\section{\label{IIII}Contributions to $\langle O(p) O(-p) \rangle$ \label{OO}}
	
\begin{figure}[H]
		\centering
		\includegraphics[width=15cm]{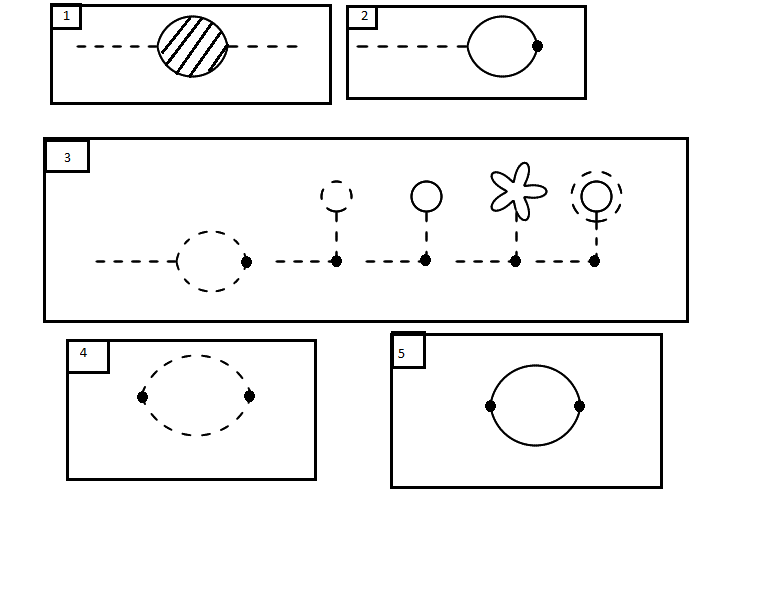}
		\caption{One-loop contributions to the correlation function $\langle OO \rangle$. Wavy lines represent the gauge field, dashed lines the Higgs field, solid lines the Goldstone boson and double lines the ghost field. The $\bullet$  indicates the insertion of a composite operator. }.  \label{hhhh}
	\end{figure}

	The diagrams which contribute to the correlation function $\langle O(p) O(-p) \rangle$ are depicted in Figure\eqref{hhhh}
	
	The first term is $v^2$ times the one-loop correction to the Higgs propagator, given in  eq.~\eqref{hhl}. The second term is
	\beq
	v \langle h(p) (\rho^a\rho^a)(-p) \rangle &=& -3\frac{m_h^2 \eta \left(m^2 \xi ,m^2 \xi \right)}{m_h^2+p^2}.
	\eeq
	The third term is
	\beq
	v \langle h(p) h^2(-p) \rangle &=&-\frac{3 m_h^2 \eta \left(m_h^2,m_h^2\right)}{m_h^2+p^2}-\frac{3 \chi \left(m_h^2\right)}{m_h^2+p^2}-\frac{\chi \left(m^2 \xi \right)}{m_h^2+p^2}-\frac{2 (D-1) m^2 \chi \left(m^2\right)}{m_h^2 \left(m_h^2+p^2\right)}-\frac{2 m^2 \xi  \chi \left(m^2 \xi \right)}{m_h^2 \left(m_h^2+p^2\right)}\nonumber\\
	&+& \frac{2 m^2 \xi  \chi \left(m^2 \xi \right)}{m_h^2 \left(m_h^2+p^2\right)}.
	\eeq
	The fourth term is
	\beq
	\langle m_h^2(p)m_h^2(-p) \rangle &=& \frac{1}{2}\eta \left(m_h^2,m_h^2\right).
	\eeq
	The fifth term is
	\beq
	\langle (\rho^a\rho^a)(p) (\rho^b\rho^b)(-p) \rangle &=& \frac{3}{2} \eta \left(\xi m^2,\xi m^2\right)
	\eeq
	and together these terms give the correlation function of the scalar composite operator $O$ up to first order in $\hbar$
	\beq
	\langle O(p) O(-p) \rangle &=& \frac{v^2}{p^2+m_h^2}+\int_0^1 dx \Big\{\frac{3}{2}\eta \left(m^2,m^2\right)(4 (d-1) m^4+4 m^2 p^2+p^4)+\frac{1}{2}(p^2-2 m_h^2)^2\eta \left(m_h^2,m_h^2\right)\nonumber\\
	&-&\frac{3 p^2 \chi (m^2) (2 (d-1) m^2+m_h^2)}{m_h^2}-3 p^2 \chi  (m_h^2)\Big\}\frac{1}{(p^2+m_h^2)^2}. \label{ch}
	\eeq
	Thus, for the one-loop resummed correlation function, we get
	\beq
	G^{-1}_{OO}(p^2)&=& \frac{p^2+m_h^2}{v^2}-\frac{1}{v^4} \int_0^1 dx \Big\{\frac{3}{2}\eta \left(m^2,m^2\right)(4 (d-1) m^4+4 m^2 p^2+p^4)+\frac{1}{2}(p^2-2 m_h^2)^2\eta \left(m_h^2,m_h^2\right)\nonumber\\
	&-&\frac{3 p^2 \chi (m^2) (2 (d-1) m^2+m_h^2)}{m_h^2}-3 p^2 \chi  (m_h^2)\Big\}\frac{1}{(p^2+m_h^2)^2}.
	\eeq
	Following the procedure of dimensional regularization for $d=4-\epsilon$, we find that the divergent part of the correlator is given by
	\beq
	G_{OO, \rm div}^{-1}&=&\frac{1}{4 v^4 \pi ^2 \epsilon }\Big(\frac{9 g^4 p^2 v^2}{16 \lambda }+\frac{9 g^4 v^4}{16}+\frac{9}{8} g^2 p^2 v^2+p^4+\frac{1}{2} \lambda  p^2 v^2+\lambda ^2 v^4\Big),
	\eeq
	which can be accounted for by appropriate counterterm, following the  $\overline{MS}bar$-scheme renormalization procedure. We remain with the finite part of the correlator, given in  eq.~\eqref{dk3}.

	\section{A few comments on the unitary gauge \label{un}}
	
\begin{figure}[H]
		\centering
		\includegraphics[width=18cm]{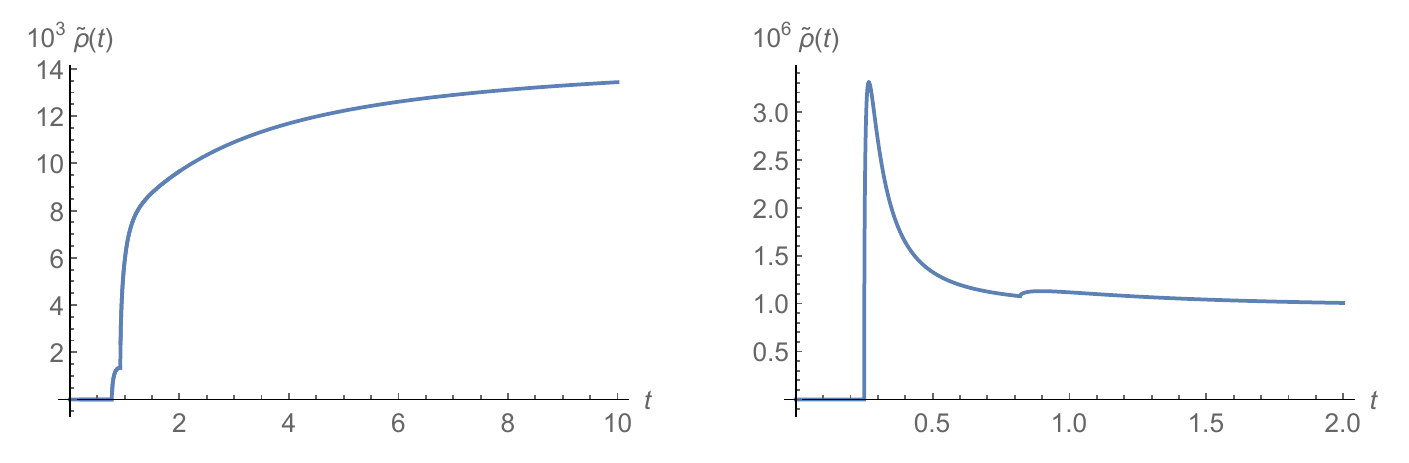}
		\caption{Spectral function for the propagator $\langle h(p) h(-p) \rangle$ in the unitary gauge, with $t$ given in unity of $\mu^2$, for the Region I (left) and Region II (right), with parameter values given in Table \ref{tabel}.}
		\label{Yhh}
	\end{figure}

	It is well-known that in the unitary gauge the unphysical fields, like the Goldstone and ghost fields, decouple, a feature which allows for a more direct link with the spectrum of the elementary excitations of the model.  However, this gauge is known to be  non-renormalizable.  In fact, working directly with the elementary tree level propagators taken already in the unitary limit, {\it i.e.}~$\xi \rightarrow \infty$, and following the steps of dimensional regularization, we find that the divergent part of the Higgs propagator reads
	\beq
	G^{-1}_{hh, \rm div}(p^2)&=&\frac{3 g^2 \left(m_h^4+6 m^2 p^2+p^4\right)}{64 \pi ^2 m^2 \epsilon }.   \label{divu}
	\eeq
	In expression \eqref{divu} we clearly see  the presence of the term  $\sim \frac{p^4}{\epsilon m^2}$, signalling the aforementioned issue of the non-renormalizability. Nevertheless, it is interesting to observe that, if we remove the divergent  part \eqref{divu} anyway,  we obtain the spectral function as shown in  Figure~\ref{Yhh}. This spectral function is almost identical to that obtained for the composite operator $O(x)$, see $ $ Figure~\ref{Y}.
	
\begin{figure}[H]
		\centering
		\includegraphics[width=18cm]{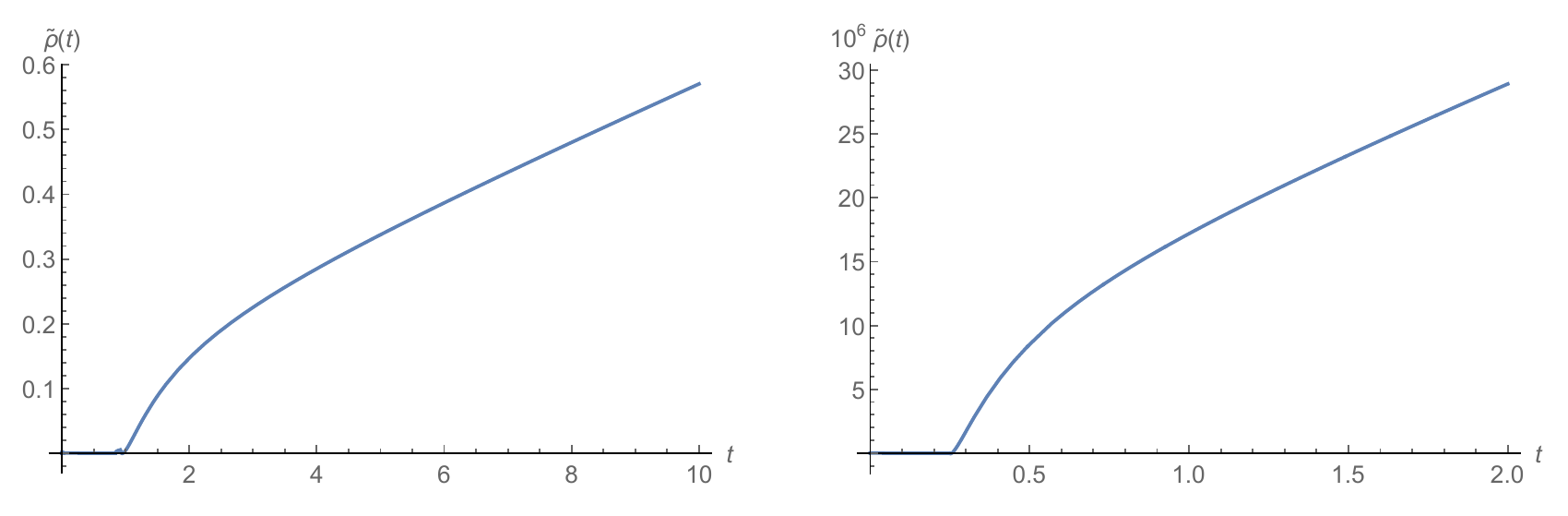}
		\caption{Spectral function for the propagator $\langle A^a_{\mu}(p) A^b_{\nu}(-p) \rangle$ in the unitary gauge, with $t$ given in unity of $\mu^2$, for  the Region I (left) and Region II (right), with parameter values given in Table \ref{tabel}.}
		\label{Yt}
	\end{figure}

	For the gauge field propagator, proceeding in the same way, we find the divergent part
	\beq
	G^{-1}_{AA, \rm div}(p^2)&=&\frac{1}{\epsilon }\Big(-\frac{9 g^2 m^4}{16 \pi ^2 m_h^2}-\frac{g^2 p^6}{96 \pi ^2 m^4}+\frac{7 g^2 p^4}{48 \pi ^2 m^2}+\frac{3 g^2 m^2}{32 \pi ^2}+\frac{83 g^2 p^2}{96 \pi ^2}-\frac{3 \lambda  m^2}{8 \pi ^2}\Big) \label{gdivu}
	\eeq
	which shows again the non-renormalizability of unitary gauge,  through the terms $\sim \frac{p^6}{\epsilon m^4}$ and $\sim \frac{p^4}{\epsilon m^2}$ . However, if we remove again those terms anyway,  we obtain the spectral function as shown in  Figure~\ref{Yt}. Nevertheless, as already remarked in the previous sections, this nice behaviour of the spectral densities for the Higgs and gauge field obtained by a direct use of the tree level propagators already taken in the unitary limit, $\xi \rightarrow \infty$, can be, to some extent, justified by the fact that we are working at the one-loop order in perturbation theory. Since overlapping divergences start from one-loop onward, we can easily figure out that the naive use of the elementary tree level propagators taken already in the unitary limit will run into  severe non-renormalizibility issues, making the removal of the  (overlapping) divergent parts \eqref{divu}, \eqref{gdivu} quite problematic.

\bibliographystyle{abntex2-alf}

\bibliography{ref}

\begin{thebibliography}{100}

\bibitem{maxwell1865viii}
J.~C. Maxwell, ``Viii. a dynamical theory of the electromagnetic field,'' {\em
  Philosophical transactions of the Royal Society of London}, no.~155,
  pp.~459--512, 1865.

\bibitem{planck2013theory}
M.~Planck, {\em The theory of heat radiation}.
\newblock Courier Corporation, 2013.

\bibitem{schrodinger1926undulatory}
E.~Schr{\"o}dinger, ``An undulatory theory of the mechanics of atoms and
  molecules,'' {\em Physical review}, vol.~28, no.~6, p.~1049, 1926.

\bibitem{einstein1905electrodynamics}
A.~Einstein {\em et~al.}, ``On the electrodynamics of moving bodies,'' {\em
  Annalen der physik}, vol.~17, no.~10, pp.~891--921, 1905.

\bibitem{dirac1928quantum}
P.~A.~M. Dirac, ``The quantum theory of the electron,'' {\em Proceedings of the
  Royal Society of London. Series A, Containing Papers of a Mathematical and
  Physical Character}, vol.~117, no.~778, pp.~610--624, 1928.

\bibitem{landau1955niels}
L.~Landau {\em et~al.}, ``Niels bohr and the development of physics,'' {\em W.
  Pauli with assistance of L. Rosenfeld and V. Weisskopf (eds.). Pergamon
  Press, London}, 1955.

\bibitem{einstein2019relativity}
A.~Einstein, {\em Relativity: The Special and the General Theory-100th
  Anniversary Edition}.
\newblock Princeton University Press, 2019.

\bibitem{Verlinde:2016toy}
E.~P. Verlinde, ``{Emergent Gravity and the Dark Universe},'' {\em SciPost
  Phys.}, vol.~2, no.~3, p.~016, 2017.

\bibitem{Yang:1954ek}
C.-N. Yang and R.~L. Mills, ``{Conservation of Isotopic Spin and Isotopic Gauge
  Invariance},'' {\em Phys. Rev.}, vol.~96, pp.~191--195, 1954.

\bibitem{gross1973ultraviolet}
D.~J. Gross and F.~Wilczek, ``Ultraviolet behavior of non-abelian gauge
  theories,'' {\em Physical Review Letters}, vol.~30, no.~26, p.~1343, 1973.

\bibitem{politzer1973reliable}
H.~D. Politzer, ``Reliable perturbative results for strong interactions?,''
  {\em Physical Review Letters}, vol.~30, no.~26, p.~1346, 1973.

\bibitem{Ripka:2003vv}
G.~Ripka, {\em {Dual superconductor models of color confinement}}, vol.~639.
\newblock 2004.

\bibitem{Jegerlehner:2001fb}
F.~Jegerlehner, M.~{\relax Yu}. Kalmykov, and O.~Veretin, ``{MS versus pole
  masses of gauge bosons: Electroweak bosonic two loop corrections},'' {\em
  Nucl. Phys.}, vol.~B641, pp.~285--326, 2002.

\bibitem{Jegerlehner:2002em}
F.~Jegerlehner, M.~{\relax Yu}. Kalmykov, and O.~Veretin, ``{MS-bar versus pole
  masses of gauge bosons. 2. Two loop electroweak fermion corrections},'' {\em
  Nucl. Phys.}, vol.~B658, pp.~49--112, 2003.

\bibitem{Martin:2015lxa}
S.~P. Martin, ``{Pole Mass of the W Boson at Two-Loop Order in the Pure
  $\overline {MS}$ Scheme},'' {\em Phys. Rev.}, vol.~D91, no.~11, p.~114003,
  2015.

\bibitem{Martin:2015rea}
S.~P. Martin, ``{$Z$-Boson Pole Mass at Two-Loop Order in the Pure
  $\overline{MS}$ Scheme},'' {\em Phys. Rev.}, vol.~D92, no.~1, p.~014026,
  2015.

\bibitem{Piguet:1995er}
O.~Piguet and S.~P. Sorella, ``{Algebraic renormalization: Perturbative
  renormalization, symmetries and anomalies},'' {\em Lect. Notes Phys.
  Monogr.}, vol.~28, pp.~1--134, 1995.

\bibitem{Kugo:1977yx}
T.~Kugo and I.~Ojima, ``{Manifestly Covariant Canonical Formulation of
  Yang-Mills Field Theories. 1. The Case of Yang-Mills Fields of Higgs-Kibble
  Type in Landau Gauge},'' {\em Prog. Theor. Phys.}, vol.~60, p.~1869, 1978.

\bibitem{Kugo:1977mk}
T.~Kugo and I.~Ojima, ``{Manifestly Covariant Canonical Formulation of
  Yang-Mills Field Theories. 2. The Case of Pure Yang-Mills Theories Without
  Spontaneous Symmetry Breaking in General Covariant Gauges},'' {\em Prog.
  Theor. Phys.}, vol.~61, p.~294, 1979.

\bibitem{zwanziger1989local}
D.~Zwanziger, ``{Local and Renormalizable Action From the Gribov Horizon},''
  {\em Nucl. Phys.}, vol.~B323, pp.~513--544, 1989.

\bibitem{zwanziger1993renormalizability}
D.~Zwanziger, ``{Renormalizability of the critical limit of lattice gauge
  theory by BRS invariance},'' {\em Nucl. Phys.}, vol.~B399, pp.~477--513,
  1993.

\bibitem{Dudal:2008sp}
D.~Dudal, J.~A. Gracey, S.~P. Sorella, N.~Vandersickel, and H.~Verschelde, ``{A
  Refinement of the Gribov-Zwanziger approach in the Landau gauge: Infrared
  propagators in harmony with the lattice results},'' {\em Phys. Rev.},
  vol.~D78, p.~065047, 2008.

\bibitem{Serreau:2012cg}
J.~Serreau and M.~Tissier, ``{Lifting the Gribov ambiguity in Yang-Mills
  theories},'' {\em Phys. Lett.}, vol.~B712, pp.~97--103, 2012.

\bibitem{Capri:2015nzw}
M.~A.~L. Capri, D.~Fiorentini, M.~S. Guimaraes, B.~W. Mintz, L.~F. Palhares,
  S.~P. Sorella, D.~Dudal, I.~F. Justo, A.~D. Pereira, and R.~F. Sobreiro,
  ``{More on the nonperturbative Gribov-Zwanziger quantization of linear
  covariant gauges},'' {\em Phys. Rev.}, vol.~D93, no.~6, p.~065019, 2016.

\bibitem{Zwanziger:2001kw}
D.~Zwanziger, ``{Nonperturbative Landau gauge and infrared critical exponents
  in QCD},'' {\em Phys. Rev.}, vol.~D65, p.~094039, 2002.

\bibitem{bowman2007scaling}
P.~O. Bowman, U.~M. Heller, D.~B. Leinweber, M.~B. Parappilly, A.~Sternbeck,
  L.~von Smekal, A.~G. Williams, and J.-b. Zhang, ``{Scaling behavior and
  positivity violation of the gluon propagator in full QCD},'' {\em Phys.
  Rev.}, vol.~D76, p.~094505, 2007.

\bibitem{Cucchieri:2004mf}
A.~Cucchieri, T.~Mendes, and A.~R. Taurines, ``{Positivity violation for the
  lattice Landau gluon propagator},'' {\em Phys. Rev.}, vol.~D71, p.~051902,
  2005.

\bibitem{Strauss:2012dg}
S.~Strauss, C.~S. Fischer, and C.~Kellermann, ``{Analytic structure of the
  Landau gauge gluon propagator},'' {\em Phys. Rev. Lett.}, vol.~109,
  p.~252001, 2012.

\bibitem{Dudal:2013yva}
D.~Dudal, O.~Oliveira, and P.~J. Silva, ``{K\"all\'{e}n-Lehmann spectroscopy
  for (un)physical degrees of freedom},'' {\em Phys. Rev.}, vol.~D89, no.~1,
  p.~014010, 2014.

\bibitem{Dudal:2019gvn}
D.~Dudal, O.~Oliveira, M.~Roelfs, and P.~Silva, ``{Spectral representation of
  lattice gluon and ghost propagators at zero temperature},'' {\em Nucl. Phys.
  B}, vol.~952, p.~114912, 2020.

\bibitem{cornwall2013positivity}
J.~M. Cornwall, ``{Positivity violations in QCD},'' {\em Mod. Phys. Lett.},
  vol.~A28, p.~1330035, 2013.

\bibitem{Krein:1990sf}
G.~Krein, C.~D. Roberts, and A.~G. Williams, ``{On the implications of
  confinement},'' {\em Int. J. Mod. Phys.}, vol.~A7, pp.~5607--5624, 1992.

\bibitem{Roberts:1994dr}
C.~D. Roberts and A.~G. Williams, ``{Dyson-Schwinger equations and their
  application to hadronic physics},'' {\em Prog. Part. Nucl. Phys.}, vol.~33,
  pp.~477--575, 1994.

\bibitem{Lowdon:2017gpp}
P.~Lowdon, ``{Non-perturbative constraints on the quark and ghost
  propagators},'' {\em Nucl. Phys.}, vol.~B935, pp.~242--255, 2018.

\bibitem{higgs1964broken}
P.~W. Higgs, ``Broken symmetries and the masses of gauge bosons,'' {\em
  Physical Review Letters}, vol.~13, no.~16, p.~508, 1964.

\bibitem{englert1964broken}
F.~Englert and R.~Brout, ``Broken symmetry and the mass of gauge vector
  mesons,'' {\em Physical Review Letters}, vol.~13, no.~9, p.~321, 1964.

\bibitem{guralnik1964global}
G.~S. Guralnik, C.~R. Hagen, and T.~W. Kibble, ``Global conservation laws and
  massless particles,'' {\em Physical Review Letters}, vol.~13, no.~20, p.~585,
  1964.

\bibitem{Glashow:1961tr}
S.~Glashow, ``{Partial Symmetries of Weak Interactions},'' {\em Nucl. Phys.},
  vol.~22, pp.~579--588, 1961.

\bibitem{Weinberg:1967tq}
S.~Weinberg, ``{A Model of Leptons},'' {\em Phys. Rev. Lett.}, vol.~19,
  pp.~1264--1266, 1967.

\bibitem{Salam:1968rm}
A.~Salam, ``{Weak and Electromagnetic Interactions},'' {\em Conf. Proc. C},
  vol.~680519, pp.~367--377, 1968.

\bibitem{Peskin:1995ev}
M.~E. Peskin and D.~V. Schroeder, {\em {An Introduction to quantum field
  theory}}.
\newblock Reading, USA: Addison-Wesley, 1995.

\bibitem{elitzur1975impossibility}
S.~Elitzur, ``Impossibility of spontaneously breaking local symmetries,'' {\em
  Physical Review D}, vol.~12, no.~12, p.~3978, 1975.

\bibitem{tHooft:1980xss}
G.~'t~Hooft, C.~Itzykson, A.~Jaffe, H.~Lehmann, P.~K. Mitter, I.~M. Singer, and
  R.~Stora, ``{Recent Developments in Gauge Theories. Proceedings, Nato
  Advanced Study Institute, Cargese, France, August 26 - September 8, 1979},''
  {\em NATO Sci. Ser. B}, vol.~59, pp.~pp.1--438, 1980.

\bibitem{Frohlich:1980gj}
J.~Frohlich, G.~Morchio, and F.~Strocchi, ``{Higgs phenomenon without a
  symmetry breaking order parameter},'' {\em Phys. Lett.}, vol.~97B,
  pp.~249--252, 1980.

\bibitem{Frohlich:1981yi}
J.~Frohlich, G.~Morchio, and F.~Strocchi, ``{Higgs phenomenon without symmetry
  breaking order parameter},'' {\em Nucl. Phys.}, vol.~B190, pp.~553--582,
  1981.

\bibitem{Nambu:1960tm}
Y.~Nambu, ``{Quasiparticles and Gauge Invariance in the Theory of
  Superconductivity},'' {\em Phys. Rev.}, vol.~117, pp.~648--663, 1960.

\bibitem{Goldstone:1961eq}
J.~Goldstone, ``{Field Theories with Superconductor Solutions},'' {\em Nuovo
  Cim.}, vol.~19, pp.~154--164, 1961.

\bibitem{itzykson1991statistical}
C.~Itzykson and J.-M. Drouffe, {\em Statistical field theory: volume 2, strong
  coupling, Monte Carlo methods, conformal field theory and random systems},
  vol.~2.
\newblock Cambridge University Press, 1991.

\bibitem{Caudy:2007sf}
W.~Caudy and J.~Greensite, ``{On the ambiguity of spontaneously broken gauge
  symmetry},'' {\em Phys. Rev.}, vol.~D78, p.~025018, 2008.

\bibitem{englertorigin}
F.~Englert, ``The origin and status of spontaneous symmetry breaking,''

\bibitem{Fradkin:1978dv}
E.~H. Fradkin and S.~H. Shenker, ``{Phase Diagrams of Lattice Gauge Theories
  with Higgs Fields},'' {\em Phys. Rev.}, vol.~D19, pp.~3682--3697, 1979.

\bibitem{v2011study}
N.~Vandersickel, ``A study of the gribov-zwanziger action: from propagators to
  glueballs,'' 2011.

\bibitem{maas2014two}
A.~Maas and T.~Mufti, ``{Two- and three-point functions in Landau gauge
  Yang-Mills-Higgs theory},'' {\em JHEP}, vol.~04, p.~006, 2014.

\bibitem{maas2015field}
A.~Maas, ``{Field theory as a tool to constrain new physics models},'' {\em
  Mod. Phys. Lett.}, vol.~A30, no.~29, p.~1550135, 2015.

\bibitem{Nielsen:1975fs}
N.~K. Nielsen, ``{On the Gauge Dependence of Spontaneous Symmetry Breaking in
  Gauge Theories},'' {\em Nucl. Phys.}, vol.~B101, pp.~173--188, 1975.

\bibitem{gambino1999fermion}
P.~Gambino, P.~A. Grassi, and F.~Madricardo, ``{Fermion mixing renormalization
  and gauge invariance},'' {\em Phys. Lett.}, vol.~B454, pp.~98--104, 1999.

\bibitem{gambino2000nielsen}
P.~Gambino and P.~A. Grassi, ``{The Nielsen identities of the SM and the
  definition of mass},'' {\em Phys. Rev.}, vol.~D62, p.~076002, 2000.

\bibitem{Grassi:2000dz}
P.~A. Grassi, B.~A. Kniehl, and A.~Sirlin, ``{Width and partial widths of
  unstable particles},'' {\em Phys. Rev. Lett.}, vol.~86, pp.~389--392, 2001.

\bibitem{Capri_2018}
M.~Capri, D.~van Egmond, G.~Peruzzo, M.~Guimaraes, O.~Holanda, S.~Sorella,
  R.~Terin, and H.~Toledo, ``On a renormalizable class of gauge fixings for the
  gauge invariant operatoramin2,'' {\em Annals of Physics}, vol.~390,
  p.~214–235, Mar 2018.

\bibitem{ozthesis}
O.~Holanda, {\em Renormalizabilidade de teorias de Yang-Mills massivas com um
  campo tipo Stueckelberg}.
\newblock PhD thesis, 2019.

\bibitem{Dudal:2019aew}
D.~Dudal, D.~van Egmond, M.~Guimarães, O.~Holanda, B.~Mintz, L.~Palhares,
  G.~Peruzzo, and S.~Sorella, ``{Some remarks on the spectral functions of the
  Abelian Higgs Model},'' {\em Phys. Rev. D}, vol.~100, no.~6, p.~065009, 2019.

\bibitem{Dudal:2019pyg}
D.~Dudal, D.~M. van Egmond, M.~S. Guimaraes, O.~Holanda, L.~F. Palhares,
  G.~Peruzzo, and S.~P. Sorella, ``{Gauge-invariant spectral description of the
  $U(1)$ Higgs model from local composite operators},'' {\em JHEP}, vol.~02,
  p.~188, 2020.

\bibitem{Dudal:2019pyg2}
D.~Dudal, D.~M. van Egmond, M.~S. Guimaraes, O.~Holanda, L.~F. Palhares,
  G.~Peruzzo, and S.~P. Sorella, ``{Gauge-invariant spectral description of the
  $SU(2)$ Higgs model from local composite operators},''

\bibitem{logan2014tasi}
H.~E. Logan, ``Tasi 2013 lectures on higgs physics within and beyond the
  standard model,'' 2014.

\bibitem{Aad:2012tfa}
G.~Aad {\em et~al.}, ``{Observation of a new particle in the search for the
  Standard Model Higgs boson with the ATLAS detector at the LHC},'' {\em Phys.
  Lett. B}, vol.~716, pp.~1--29, 2012.

\bibitem{Chatrchyan:2012ufa}
S.~Chatrchyan {\em et~al.}, ``{Observation of a New Boson at a Mass of 125 GeV
  with the CMS Experiment at the LHC},'' {\em Phys. Lett. B}, vol.~716,
  pp.~30--61, 2012.

\bibitem{anderson1962theory}
P.~Anderson, ``Theory of flux creep in hard superconductors,'' {\em Physical
  Review Letters}, vol.~9, no.~7, p.~309, 1962.

\bibitem{georgi1972unified}
H.~Georgi and S.~L. Glashow, ``Unified weak and electromagnetic interactions
  without neutral currents,'' {\em Physical Review Letters}, vol.~28, no.~22,
  p.~1494, 1972.

\bibitem{tHooft:1974kcl}
G.~'t~Hooft, ``{Magnetic Monopoles in Unified Gauge Theories},'' {\em Nucl.
  Phys. B}, vol.~79, pp.~276--284, 1974.

\bibitem{georgi1974unity}
H.~Georgi and S.~L. Glashow, ``Unity of all elementary-particle forces,'' {\em
  Physical Review Letters}, vol.~32, no.~8, p.~438, 1974.

\bibitem{Alkofer_2001}
R.~Alkofer, ``The infrared behaviour of qcd green’s functions confinement,
  dynamical symmetry breaking, and hadrons as relativistic bound states,'' {\em
  Physics Reports}, vol.~353, p.~281–465, Nov 2001.

\bibitem{gribov1978quantization}
V.~N. Gribov, ``{Quantization of Nonabelian Gauge Theories},'' {\em Nucl.
  Phys.}, vol.~B139, p.~1, 1978.
\newblock [,1(1977)].

\bibitem{Dudal:2007cw}
D.~Dudal, S.~Sorella, N.~Vandersickel, and H.~Verschelde, ``{New features of
  the gluon and ghost propagator in the infrared region from the
  Gribov-Zwanziger approach},'' {\em Phys. Rev. D}, vol.~77, p.~071501, 2008.

\bibitem{Dudal:2011gd}
D.~Dudal, S.~Sorella, and N.~Vandersickel, ``{The dynamical origin of the
  refinement of the Gribov-Zwanziger theory},'' {\em Phys. Rev. D}, vol.~84,
  p.~065039, 2011.

\bibitem{Capri:2015ixa}
M.~Capri, D.~Dudal, D.~Fiorentini, M.~Guimaraes, I.~Justo, A.~Pereira,
  B.~Mintz, L.~Palhares, R.~Sobreiro, and S.~Sorella, ``{Exact nilpotent
  nonperturbative BRST symmetry for the Gribov-Zwanziger action in the linear
  covariant gauge},'' {\em Phys. Rev. D}, vol.~92, no.~4, p.~045039, 2015.

\bibitem{Capri:2016aqq}
M.~Capri, D.~Dudal, D.~Fiorentini, M.~Guimaraes, I.~Justo, A.~Pereira,
  B.~Mintz, L.~Palhares, R.~Sobreiro, and S.~Sorella, ``{Local and
  BRST-invariant Yang-Mills theory within the Gribov horizon},'' {\em Phys.
  Rev. D}, vol.~94, no.~2, p.~025035, 2016.

\bibitem{Capri:2016gut}
M.~Capri, D.~Dudal, A.~Pereira, D.~Fiorentini, M.~Guimaraes, B.~Mintz,
  L.~Palhares, and S.~Sorella, ``{Nonperturbative aspects of Euclidean
  Yang-Mills theories in linear covariant gauges: Nielsen identities and a
  BRST-invariant two-point correlation function},'' {\em Phys. Rev. D},
  vol.~95, no.~4, p.~045011, 2017.

\bibitem{Capri:2017bfd}
M.~Capri, D.~Fiorentini, A.~Pereira, and S.~Sorella, ``{Renormalizability of
  the refined Gribov-Zwanziger action in linear covariant gauges},'' {\em Phys.
  Rev. D}, vol.~96, no.~5, p.~054022, 2017.

\bibitem{Vandersickel:2012tz}
N.~Vandersickel and D.~Zwanziger, ``{The Gribov problem and QCD dynamics},''
  {\em Phys. Rept.}, vol.~520, pp.~175--251, 2012.

\bibitem{cornwall1982dynamical}
J.~M. Cornwall, ``{Dynamical Mass Generation in Continuum QCD},'' {\em Phys.
  Rev.}, vol.~D26, p.~1453, 1982.

\bibitem{Parisi:1980jy}
G.~Parisi and R.~Petronzio, ``{On Low-Energy Tests of QCD},'' {\em Phys.
  Lett.}, vol.~94B, pp.~51--53, 1980.

\bibitem{Bernard:1981pg}
C.~W. Bernard, ``{Monte Carlo Evaluation of the Effective Gluon Mass},'' {\em
  Phys. Lett.}, vol.~108B, pp.~431--434, 1982.

\bibitem{papavassiliou2013effective}
D.~Binosi, D.~Ibanez, and J.~Papavassiliou, ``{The all-order equation of the
  effective gluon mass},'' {\em Phys. Rev.}, vol.~D86, p.~085033, 2012.

\bibitem{Aguilar:2014tka}
A.~C. Aguilar, D.~Binosi, and J.~Papavassiliou, ``{Renormalization group
  analysis of the gluon mass equation},'' {\em Phys. Rev.}, vol.~D89, no.~8,
  p.~085032, 2014.

\bibitem{Cyrol:2016tym}
A.~K. Cyrol, L.~Fister, M.~Mitter, J.~M. Pawlowski, and N.~Strodthoff,
  ``{Landau gauge Yang-Mills correlation functions},'' {\em Phys. Rev.},
  vol.~D94, no.~5, p.~054005, 2016.

\bibitem{Huber:2018ned}
M.~Q. Huber, {\em {Nonperturbative properties of Yang-Mills theories}}.
\newblock habilitation, Graz U., 2018.

\bibitem{Boucaud:2011ug}
P.~Boucaud, J.~P. Leroy, A.~L. Yaouanc, J.~Micheli, O.~Pene, and
  J.~Rodriguez-Quintero, ``{The Infrared Behaviour of the Pure Yang-Mills Green
  Functions},'' {\em Few Body Syst.}, vol.~53, pp.~387--436, 2012.

\bibitem{Cucchieri:2007md}
A.~Cucchieri and T.~Mendes, ``{What's up with IR gluon and ghost propagators in
  Landau gauge? A puzzling answer from huge lattices},'' {\em PoS},
  vol.~LATTICE2007, p.~297, 2007.

\bibitem{Cucchieri:2007rg}
A.~Cucchieri and T.~Mendes, ``{Constraints on the IR behavior of the gluon
  propagator in Yang-Mills theories},'' {\em Phys. Rev. Lett.}, vol.~100,
  p.~241601, 2008.

\bibitem{Bogolubsky:2009dc}
I.~L. Bogolubsky, E.~M. Ilgenfritz, M.~Muller-Preussker, and A.~Sternbeck,
  ``{Lattice gluodynamics computation of Landau gauge Green's functions in the
  deep infrared},'' {\em Phys. Lett.}, vol.~B676, pp.~69--73, 2009.

\bibitem{Maas:2008ri}
A.~Maas, ``{More on Gribov copies and propagators in Landau-gauge Yang-Mills
  theory},'' {\em Phys. Rev.}, vol.~D79, p.~014505, 2009.

\bibitem{Cucchieri:2009kk}
A.~Cucchieri, T.~Mendes, and E.~M. Santos, ``{Covariant gauge on the lattice: A
  New implementation},'' {\em Phys. Rev. Lett.}, vol.~103, p.~141602, 2009.

\bibitem{Cucchieri:2009zt}
A.~Cucchieri and T.~Mendes, ``{Landau-gauge propagators in Yang-Mills theories
  at beta = 0: Massive solution versus conformal scaling},'' {\em Phys. Rev.},
  vol.~D81, p.~016005, 2010.

\bibitem{Cucchieri:2010xr}
A.~Cucchieri and T.~Mendes, ``{Numerical test of the Gribov-Zwanziger scenario
  in Landau gauge},'' {\em PoS}, vol.~QCD-TNT09, p.~026, 2009.

\bibitem{Cucchieri:2011ig}
A.~Cucchieri, D.~Dudal, T.~Mendes, and N.~Vandersickel, ``{Modeling the Gluon
  Propagator in Landau Gauge: Lattice Estimates of Pole Masses and
  Dimension-Two Condensates},'' {\em Phys. Rev. D}, vol.~85, p.~094513, 2012.

\bibitem{Bornyakov:2009ug}
V.~G. Bornyakov, V.~K. Mitrjushkin, and M.~Muller-Preussker, ``{SU(2) lattice
  gluon propagator: Continuum limit, finite-volume effects and infrared mass
  scale m(IR)},'' {\em Phys. Rev.}, vol.~D81, p.~054503, 2010.

\bibitem{Oliveira:2012eh}
O.~Oliveira and P.~J. Silva, ``{The lattice Landau gauge gluon propagator:
  lattice spacing and volume dependence},'' {\em Phys. Rev.}, vol.~D86,
  p.~114513, 2012.

\bibitem{Bicudo:2015rma}
P.~Bicudo, D.~Binosi, N.~Cardoso, O.~Oliveira, and P.~Silva, ``{Lattice gluon
  propagator in renormalizable $\xi$ gauges},'' {\em Phys. Rev. D}, vol.~92,
  no.~11, p.~114514, 2015.

\bibitem{Cucchieri:2016jwg}
A.~Cucchieri, D.~Dudal, T.~Mendes, and N.~Vandersickel, ``{Modeling the
  Landau-gauge ghost propagator in 2, 3, and 4 spacetime dimensions},'' {\em
  Phys. Rev. D}, vol.~93, no.~9, p.~094513, 2016.

\bibitem{Duarte:2016iko}
A.~G. Duarte, O.~Oliveira, and P.~J. Silva, ``{Lattice Gluon and Ghost
  Propagators, and the Strong Coupling in Pure SU(3) Yang-Mills Theory: Finite
  Lattice Spacing and Volume Effects},'' {\em Phys. Rev.}, vol.~D94, no.~1,
  p.~014502, 2016.

\bibitem{Dudal:2018cli}
D.~Dudal, O.~Oliveira, and P.~J. Silva, ``{High precision statistical Landau
  gauge lattice gluon propagator computation vs. the Gribov–Zwanziger
  approach},'' {\em Annals Phys.}, vol.~397, pp.~351--364, 2018.

\bibitem{Boucaud:2018xup}
P.~Boucaud, F.~De~Soto, K.~Raya, J.~Rodríguez-Quintero, and S.~Zafeiropoulos,
  ``{Discretization effects on renormalized gauge-field Green’s functions,
  scale setting, and the gluon mass},'' {\em Phys. Rev.}, vol.~D98, no.~11,
  p.~114515, 2018.

\bibitem{Aguilar:2008xm}
A.~Aguilar, D.~Binosi, and J.~Papavassiliou, ``{Gluon and ghost propagators in
  the Landau gauge: Deriving lattice results from Schwinger-Dyson equations},''
  {\em Phys. Rev. D}, vol.~78, p.~025010, 2008.

\bibitem{Aguilar:2015bud}
A.~Aguilar, D.~Binosi, and J.~Papavassiliou, ``{The Gluon Mass Generation
  Mechanism: A Concise Primer},'' {\em Front. Phys. (Beijing)}, vol.~11, no.~2,
  p.~111203, 2016.

\bibitem{Fischer:2008uz}
C.~S. Fischer, A.~Maas, and J.~M. Pawlowski, ``{On the infrared behavior of
  Landau gauge Yang-Mills theory},'' {\em Annals Phys.}, vol.~324,
  pp.~2408--2437, 2009.

\bibitem{Aguilar:2015nqa}
A.~Aguilar, D.~Binosi, and J.~Papavassiliou, ``{Yang-Mills two-point functions
  in linear covariant gauges},'' {\em Phys. Rev. D}, vol.~91, no.~8, p.~085014,
  2015.

\bibitem{Huber:2015ria}
M.~Q. Huber, ``{Gluon and ghost propagators in linear covariant gauges},'' {\em
  Phys. Rev. D}, vol.~91, no.~8, p.~085018, 2015.

\bibitem{Fischer:2009tn}
C.~S. Fischer and J.~M. Pawlowski, ``{Uniqueness of infrared asymptotics in
  Landau gauge Yang-Mills theory II},'' {\em Phys. Rev. D}, vol.~80, p.~025023,
  2009.

\bibitem{Weber:2011nw}
A.~Weber, ``{Epsilon Expansion for Infrared Yang-Mills theory in Landau
  Gauge},'' {\em Phys. Rev. D}, vol.~85, p.~125005, 2012.

\bibitem{Frasca:2007uz}
M.~Frasca, ``{Infrared Gluon and Ghost Propagators},'' {\em Phys. Lett. B},
  vol.~670, pp.~73--77, 2008.

\bibitem{Siringo:2015wtx}
F.~Siringo, ``{Analytical study of Yang--Mills theory in the infrared from
  first principles},'' {\em Nucl. Phys. B}, vol.~907, pp.~572--596, 2016.

\bibitem{tissier2010infrared}
M.~Tissier and N.~Wschebor, ``{Infrared propagators of Yang-Mills theory from
  perturbation theory},'' {\em Phys. Rev.}, vol.~D82, p.~101701, 2010.

\bibitem{Curci:1976bt}
G.~Curci and R.~Ferrari, ``{On a Class of Lagrangian Models for Massive and
  Massless Yang-Mills Fields},'' {\em Nuovo Cim. A}, vol.~32, pp.~151--168,
  1976.

\bibitem{Tissier:2017fqf}
M.~Tissier, ``{Gribov copies, avalanches and dynamic generation of a gluon
  mass},'' {\em Phys. Lett. B}, vol.~784, pp.~146--150, 2018.

\bibitem{de_Boer_1996}
J.~de~Boer, K.~Skenderis, P.~van Nieuwenhuizen, and A.~Waldron, ``On the
  renormalizability and unitarity of the curci-ferrari model for massive vector
  bosons,'' {\em Physics Letters B}, vol.~367, p.~175–182, Jan 1996.

\bibitem{tissier2011infrared}
M.~Tissier and N.~Wschebor, ``{An Infrared Safe perturbative approach to
  Yang-Mills correlators},'' {\em Phys. Rev.}, vol.~D84, p.~045018, 2011.

\bibitem{Gracey:2019xom}
J.~A. Gracey, M.~Peláez, U.~Reinosa, and M.~Tissier, ``{Two loop calculation
  of Yang-Mills propagators in the Curci-Ferrari model},'' {\em Phys. Rev. D},
  vol.~100, no.~3, p.~034023, 2019.

\bibitem{Bowman:2007du}
P.~O. Bowman, U.~M. Heller, D.~B. Leinweber, M.~B. Parappilly, A.~Sternbeck,
  L.~von Smekal, A.~G. Williams, and J.-b. Zhang, ``{Scaling behavior and
  positivity violation of the gluon propagator in full QCD},'' {\em Phys. Rev.
  D}, vol.~76, p.~094505, 2007.

\bibitem{PhysRevLett.79.3591}
L.~von Smekal, A.~Hauck, and R.~Alkofer, ``Infrared behavior of gluon and ghost
  propagators in landau gauge qcd,'' {\em Phys. Rev. Lett.}, vol.~79,
  pp.~3591--3594, Nov 1997.

\bibitem{vonSmekal:1997ern}
L.~von Smekal, A.~Hauck, and R.~Alkofer, ``{A Solution to Coupled
  Dyson--Schwinger Equations for Gluons and Ghosts in Landau Gauge},'' {\em
  Annals Phys.}, vol.~267, pp.~1--60, 1998.
\newblock [Erratum: Annals Phys. 269, 182 (1998)].

\bibitem{Alkofer_2004}
R.~Alkofer, W.~Detmold, C.~S. Fischer, and P.~Maris, ``Analytic properties of
  the landau gauge gluon and quark propagators,'' {\em Physical Review D},
  vol.~70, Jul 2004.

\bibitem{Alkofer:2000wg}
R.~Alkofer and L.~von Smekal, ``{The Infrared behavior of QCD Green's
  functions: Confinement dynamical symmetry breaking, and hadrons as
  relativistic bound states},'' {\em Phys. Rept.}, vol.~353, p.~281, 2001.

\bibitem{kugo1979local}
T.~Kugo and I.~Ojima, ``{Local Covariant Operator Formalism of Nonabelian Gauge
  Theories and Quark Confinement Problem},'' {\em Prog. Theor. Phys. Suppl.},
  vol.~66, pp.~1--130, 1979.

\bibitem{Ojima:1981fs}
I.~Ojima, ``{Comments on Massive and Massless {Yang-Mills} Lagrangians With a
  Quartic Coupling of Faddeev-popov Ghosts},'' {\em Z. Phys.}, vol.~C13,
  p.~173, 1982.

\bibitem{deBoer:1995dh}
J.~de~Boer, K.~Skenderis, P.~van Nieuwenhuizen, and A.~Waldron, ``{On the
  renormalizability and unitarity of the Curci-Ferrari model for massive vector
  bosons},'' {\em Phys. Lett.}, vol.~B367, pp.~175--182, 1996.

\bibitem{Kondo:2019rpa}
K.-I. Kondo, M.~Watanabe, Y.~Hayashi, R.~Matsudo, and Y.~Suda, ``{Reflection
  positivity and complex analysis of the Yang-Mills theory from a viewpoint of
  gluon confinement},'' {\em Eur. Phys. J. C}, vol.~80, no.~2, p.~84, 2020.

\bibitem{Fischer:2020xnb}
C.~S. Fischer and M.~Q. Huber, ``{Landau gauge Yang-Mills propagators in the
  complex momentum plane},'' 7 2020.

\bibitem{Eden:1966dnq}
R.~J. Eden, P.~V. Landshoff, D.~I. Olive, and J.~C. Polkinghorne, {\em {The
  analytic S-matrix}}.
\newblock Cambridge: Cambridge Univ. Press, 1966.

\bibitem{Baulieu:2009ha}
L.~Baulieu, D.~Dudal, M.~S. Guimaraes, M.~Q. Huber, S.~P. Sorella,
  N.~Vandersickel, and D.~Zwanziger, ``{Gribov horizon and i-particles: About a
  toy model and the construction of physical operators},'' {\em Phys. Rev.},
  vol.~D82, p.~025021, 2010.

\bibitem{Windisch:2012sz}
A.~Windisch, M.~Q. Huber, and R.~Alkofer, ``{On the analytic structure of
  scalar glueball operators at the Born level},'' {\em Phys. Rev.}, vol.~D87,
  no.~6, p.~065005, 2013.

\bibitem{cucchieri2013crossing}
A.~Cucchieri and T.~Mendes, ``Crossing the gribov horizon: an unconventional
  study of geometric properties of gauge-configuration space in landau gauge,''
  {\em arXiv preprint arXiv:1311.4699}, 2013.

\bibitem{PhysRevD.81.074505}
D.~Dudal, O.~Oliveira, and N.~Vandersickel, ``Indirect lattice evidence for the
  refined gribov-zwanziger formalism and the gluon condensate $⟨{A}^{2}⟩$
  in the landau gauge,'' {\em Phys. Rev. D}, vol.~81, p.~074505, Apr 2010.

\bibitem{Cucchieri_2012}
A.~Cucchieri, D.~Dudal, T.~Mendes, and N.~Vandersickel, ``Modeling the gluon
  propagator in landau gauge: Lattice estimates of pole masses and
  dimension-two condensates,'' {\em Physical Review D}, vol.~85, May 2012.

\bibitem{Cucchieri_2012b}
A.~Cucchieri, D.~Dudal, and N.~Vandersickel, ``No-pole condition in landau
  gauge: Properties of the gribov ghost form factor and a constraint on
  the2dgluon propagator,'' {\em Physical Review D}, vol.~85, Apr 2012.

\bibitem{Cornwall:2009ud}
J.~M. Cornwall, ``{Positivity issues for the pinch-technique gluon propagator
  and their resolution},'' {\em Phys. Rev. D}, vol.~80, p.~096001, 2009.

\bibitem{Fiorentini:2016rwx}
M.~Capri, D.~Fiorentini, M.~Guimaraes, B.~Mintz, L.~Palhares, and S.~Sorella,
  ``{Local and renormalizable framework for the gauge-invariant operator
  $A^2_{\min}$ in Euclidean Yang-Mills theories in linear covariant gauges},''
  {\em Phys. Rev. D}, vol.~94, no.~6, p.~065009, 2016.

\bibitem{Zwanziger:1990tn}
D.~Zwanziger, ``{Quantization of Gauge Fields, Classical Gauge Invariance and
  Gluon Confinement},'' {\em Nucl. Phys. B}, vol.~345, pp.~461--471, 1990.

\bibitem{DellAntonio:1989wae}
G.~Dell'Antonio and D.~Zwanziger, ``{Ellipsoidal Bound on the Gribov Horizon
  Contradicts the Perturbative Renormalization Group},'' {\em Nucl. Phys. B},
  vol.~326, pp.~333--350, 1989.

\bibitem{DellAntonio:1991mms}
G.~Dell'Antonio and D.~Zwanziger, ``{Every gauge orbit passes inside the Gribov
  horizon},'' {\em Commun. Math. Phys.}, vol.~138, pp.~291--299, 1991.

\bibitem{vanBaal:1991zw}
P.~van Baal, ``{More (thoughts on) Gribov copies},'' {\em Nucl. Phys. B},
  vol.~369, pp.~259--275, 1992.

\bibitem{Lavelle:1995ty}
M.~Lavelle and D.~McMullan, ``{Constituent quarks from QCD},'' {\em Phys.
  Rept.}, vol.~279, pp.~1--65, 1997.

\bibitem{Capri_2018b}
M.~A.~L. Capri, D.~M. van Egmond, M.~S. Guimaraes, O.~Holanda, S.~P. Sorella,
  R.~C. Terin, and H.~C. Toledo, ``Renormalizability of n=1 super yang–mills
  theory in landau gauge with a stueckelberg-like field,'' {\em The European
  Physical Journal C}, vol.~78, Oct 2018.

\bibitem{Dudal:2002pq}
D.~Dudal, H.~Verschelde, and S.~Sorella, ``{The Anomalous dimension of the
  composite operator A**2 in the Landau gauge},'' {\em Phys. Lett. B},
  vol.~555, pp.~126--131, 2003.

\bibitem{Gracey:2002yt}
J.~Gracey, ``{Three loop MS-bar renormalization of the Curci-Ferrari model and
  the dimension two BRST invariant composite operator in QCD},'' {\em Phys.
  Lett. B}, vol.~552, pp.~101--110, 2003.

\bibitem{Cucchieri:2011aa}
A.~Cucchieri, T.~Mendes, G.~M. Nakamura, and E.~M. Santos, ``{Feynman gauge on
  the lattice: New results and perspectives},'' {\em AIP Conf. Proc.},
  vol.~1354, no.~1, pp.~45--50, 2011.

\bibitem{Dragon:1996tk}
N.~Dragon, T.~Hurth, and P.~van Nieuwenhuizen, ``{Polynomial form of the
  Stuckelberg model},'' {\em Nucl. Phys. B Proc. Suppl.}, vol.~56B,
  pp.~318--321, 1997.

\bibitem{Blasi:1988sh}
A.~Blasi, F.~Delduc, and S.~Sorella, ``{The Background Quantum Split Symmetry
  in Two-dimensional $\sigma$ Models: A Regularization Independent Proof of Its
  Renormalizability},'' {\em Nucl. Phys. B}, vol.~314, pp.~409--424, 1989.

\bibitem{Becchi:1988nh}
C.~Becchi and O.~Piguet, ``{On the Renormalization of Two-dimensional Chiral
  Models},'' {\em Nucl. Phys. B}, vol.~315, pp.~153--165, 1989.

\bibitem{Piguet:1981fb}
O.~Piguet and K.~Sibold, ``{Renormalization of $N=1$ Supersymmetrical
  {Yang-Mills} Theories. 1. The Classical Theory},'' {\em Nucl. Phys. B},
  vol.~197, pp.~257--271, 1982.

\bibitem{Piguet:1981hh}
O.~Piguet and K.~Sibold, ``{Renormalization of $N=1$ Supersymmetrical
  {Yang-Mills} Theories. 2. The Radiative Corrections},'' {\em Nucl. Phys. B},
  vol.~197, pp.~272--289, 1982.

\bibitem{Ruegg:2003ps}
H.~Ruegg and M.~Ruiz-Altaba, ``{The Stueckelberg field},'' {\em Int. J. Mod.
  Phys. A}, vol.~19, pp.~3265--3348, 2004.

\bibitem{Ferrari:2004pd}
R.~Ferrari and A.~Quadri, ``{Physical unitarity for massive non-Abelian gauge
  theories in the Landau gauge: Stueckelberg and Higgs},'' {\em JHEP}, vol.~11,
  p.~019, 2004.

\bibitem{Oehme:1979ai}
R.~Oehme and W.~Zimmermann, ``{Quark and Gluon Propagators in Quantum
  Chromodynamics},'' {\em Phys. Rev.}, vol.~D21, p.~471, 1980.

\bibitem{gieres1997symmetries}
F.~Gieres, {\em {About symmetries in physics}}.
\newblock 1997.

\bibitem{t1981recent}
G.~'t~Hooft, C.~Itzykson, A.~Jaffe, H.~Lehmann, P.~K. Mitter, I.~M. Singer, and
  R.~Stora, ``{Recent Developments in Gauge Theories. Proceedings, Nato
  Advanced Study Institute, Cargese, France, August 26 - September 8, 1979},''
  {\em NATO Sci. Ser. B}, vol.~59, pp.~pp.1--438, 1980.

\bibitem{Becchi:1974md}
C.~Becchi, A.~Rouet, and R.~Stora, ``{Renormalization of the Abelian
  Higgs-Kibble Model},'' {\em Commun. Math. Phys.}, vol.~42, pp.~127--162,
  1975.

\bibitem{Haussling:1996rq}
R.~Haussling and E.~Kraus, ``{Gauge parameter dependence and gauge invariance
  in the Abelian Higgs model},'' {\em Z. Phys.}, vol.~C75, pp.~739--750, 1997.

\bibitem{Piguet:1984js}
O.~Piguet and K.~Sibold, ``{Gauge Independence in Ordinary {Yang-Mills}
  Theories},'' {\em Nucl. Phys.}, vol.~B253, pp.~517--540, 1985.

\bibitem{Becchi:1974xu}
C.~Becchi, A.~Rouet, and R.~Stora, ``{The Abelian Higgs-Kibble Model. Unitarity
  of the S Operator},'' {\em Phys. Lett.}, vol.~52B, pp.~344--346, 1974.

\bibitem{passarino1979one}
G.~Passarino and M.~J.~G. Veltman, ``{One Loop Corrections for e+ e-
  Annihilation Into mu+ mu- in the Weinberg Model},'' {\em Nucl. Phys.},
  vol.~B160, pp.~151--207, 1979.

\bibitem{irges2017renormalization}
N.~Irges and F.~Koutroulis, ``{Renormalization of the Abelian–Higgs model in
  the $R_\xi$ and Unitary gauges and the physicality of its scalar
  potential},'' {\em Nucl. Phys.}, vol.~B924, pp.~178--278, 2017.
\newblock [Erratum: Nucl. Phys.B938,957(2019)].

\bibitem{oehme1990superconvergence}
R.~Oehme, ``{On superconvergence relations in quantum chromodynamics},'' {\em
  Phys. Lett.}, vol.~B252, pp.~641--646, 1990.

\bibitem{curci1976slavnov}
G.~Curci and R.~Ferrari, ``{Slavnov Transformations and Supersymmetry},'' {\em
  Phys. Lett.}, vol.~63B, pp.~91--94, 1976.

\bibitem{curci1976class}
G.~Curci and R.~Ferrari, ``{On a Class of Lagrangian Models for Massive and
  Massless Yang-Mills Fields},'' {\em Nuovo Cim.}, vol.~A32, pp.~151--168,
  1976.

\bibitem{Delduc:1989uc}
F.~Delduc and S.~P. Sorella, ``{A Note on Some Nonlinear Covariant Gauges in
  {Yang-Mills} Theory},'' {\em Phys. Lett.}, vol.~B231, pp.~408--410, 1989.

\bibitem{hayashi2018complex}
Y.~Hayashi and K.-I. Kondo, ``{Complex poles and spectral function of
  Yang-Mills theory},'' {\em Phys. Rev.}, vol.~D99, no.~7, p.~074001, 2019.

\bibitem{Binosi:2009qm}
D.~Binosi and J.~Papavassiliou, ``{Pinch Technique: Theory and Applications},''
  {\em Phys. Rept.}, vol.~479, pp.~1--152, 2009.

\bibitem{Grassi:2001bz}
P.~A. Grassi, B.~A. Kniehl, and A.~Sirlin, ``{Width and partial widths of
  unstable particles in the light of the Nielsen identities},'' {\em Phys.
  Rev.}, vol.~D65, p.~085001, 2002.

\bibitem{Maas:2017xzh}
A.~Maas, R.~Sondenheimer, and P.~Torek, ``{On the observable spectrum of
  theories with a Brout–Englert–Higgs effect},'' {\em Annals Phys.},
  vol.~402, pp.~18--44, 2019.

\bibitem{Maas:2017wzi}
A.~Maas, ``{Brout-Englert-Higgs physics: From foundations to phenomenology},''
  {\em Prog. Part. Nucl. Phys.}, vol.~106, pp.~132--209, 2019.

\bibitem{Binosi:2019ecz}
D.~Binosi and R.-A. Tripolt, ``{Spectral functions of confined particles},''
  {\em Phys. Lett. B}, vol.~801, p.~135171, 2020.

\bibitem{Sanchis-Alepuz:2015hma}
H.~Sanchis-Alepuz, C.~S. Fischer, C.~Kellermann, and L.~von Smekal,
  ``{Glueballs from the Bethe-Salpeter equation},'' {\em Phys. Rev.}, vol.~D92,
  p.~034001, 2015.

\bibitem{maas2019observable}
A.~Maas, R.~Sondenheimer, and P.~T{\"o}rek, ``On the observable spectrum of
  theories with a brout--englert--higgs effect,'' {\em Annals of Physics},
  vol.~402, pp.~18--44, 2019.

\bibitem{hooft1980we}
G.~Hooft, ``Why do we need local gauge invariance in theories with vector
  particles? an introduction,'' {\em Recent Developments in Gauge Theories},
  pp.~101--115, 1980.

\bibitem{hooft2012nonperturbative}
G.~Hooft, A.~Jaffe, G.~Mack, P.~Mitter, and R.~Stora, {\em Nonperturbative
  quantum field theory}, vol.~185.
\newblock Springer Science \& Business Media, 2012.

\bibitem{frohlich1980higgs}
J.~Fr{\"o}hlich, G.~Morchio, and F.~Strocchi, ``Higgs phenomenon without a
  symmetry breaking order parameter,'' {\em Physics Letters B}, vol.~97, no.~2,
  pp.~249--252, 1980.

\bibitem{frohlich1981higgs}
J.~Fr{\"o}hlich, G.~Morchio, and F.~Strocchi, ``Higgs phenomenon without
  symmetry breaking order parameter,'' {\em Nuclear Physics B}, vol.~190,
  no.~3, pp.~553--582, 1981.

\bibitem{colangelo2001qcd}
P.~Colangelo and A.~Khodjamirian, ``Qcd sum rules, a modern perspective,'' in
  {\em At The Frontier of Particle Physics: Handbook of QCD (In 3 Volumes)},
  pp.~1495--1576, World Scientific, 2001.

\bibitem{Aitchison:1983ns}
I.~Aitchison and C.~Fraser, ``{Gauge Invariance and the Effective Potential},''
  {\em Annals Phys.}, vol.~156, p.~1, 1984.

\bibitem{Andreassen:2014eha}
A.~Andreassen, W.~Frost, and M.~D. Schwartz, ``{Consistent Use of Effective
  Potentials},'' {\em Phys. Rev. D}, vol.~91, no.~1, p.~016009, 2015.

\bibitem{Capri:2020ppe}
M.~Capri, I.~Justo, L.~Palhares, G.~Peruzzo, and S.~Sorella, ``{Study of the
  renormalization of BRST invariant local composite operators in the $U(1)$
  Higgs model},'' 7 2020.

\bibitem{Mintz_2019}
B.~W. Mintz, L.~F. Palhares, G.~Peruzzo, and S.~P. Sorella, ``Infrared massive
  gluon propagator from a brst-invariant gribov horizon in a family of
  covariant gauges,'' {\em Physical Review D}, vol.~99, Feb 2019.

\bibitem{Capri:2018ijg}
M.~Capri, D.~Dudal, M.~Guimaraes, A.~Pereira, B.~Mintz, L.~Palhares, and
  S.~Sorella, ``{The universal character of Zwanziger's horizon function in
  Euclidean Yang--Mills theories},'' {\em Phys. Lett. B}, vol.~781, pp.~48--54,
  2018.

\bibitem{Capri:2012ah}
M.~A.~L. Capri, D.~Dudal, A.~J. Gomez, M.~S. Guimaraes, I.~F. Justo, S.~P.
  Sorella, and D.~Vercauteren, ``{Semiclassical analysis of the phases of 4d
  SU(2) Higgs gauge systems with cutoff at the Gribov horizon},'' {\em Phys.
  Rev.}, vol.~D88, p.~085022, 2013.

\bibitem{Abbott:1980hw}
L.~Abbott, ``{The Background Field Method Beyond One Loop},'' {\em Nucl. Phys.
  B}, vol.~185, pp.~189--203, 1981.

\bibitem{Abbott:1981ke}
L.~Abbott, ``{Introduction to the Background Field Method},'' {\em Acta Phys.
  Polon. B}, vol.~13, p.~33, 1982.

\bibitem{Braun:2007bx}
J.~Braun, H.~Gies, and J.~M. Pawlowski, ``{Quark Confinement from Color
  Confinement},'' {\em Phys. Lett. B}, vol.~684, pp.~262--267, 2010.

\bibitem{Herbst:2015ona}
T.~K. Herbst, J.~Luecker, and J.~M. Pawlowski, ``{Confinement order parameters
  and fluctuations},'' 10 2015.

\bibitem{Reinosa:2015gxn}
U.~Reinosa, J.~Serreau, M.~Tissier, and N.~Wschebor, ``{Two-loop study of the
  deconfinement transition in Yang-Mills theories: SU(3) and beyond},'' {\em
  Phys. Rev. D}, vol.~93, no.~10, p.~105002, 2016.

\bibitem{Reinosa_HDR}
U.~Reinosa, ``{Perturbative aspects of the deconfinement transition -
  \textit{Physics beyond the Faddeev-Popov model}},'' Habilitation thesis,
  2019.

\bibitem{Reinosa:2014ooa}
U.~Reinosa, J.~Serreau, M.~Tissier, and N.~Wschebor, ``{Deconfinement
  transition in SU($N$) theories from perturbation theory},'' {\em Phys. Lett.
  B}, vol.~742, pp.~61--68, 2015.

\bibitem{Reinosa:2014zta}
U.~Reinosa, J.~Serreau, M.~Tissier, and N.~Wschebor, ``{Deconfinement
  transition in SU(2) Yang-Mills theory: A two-loop study},'' {\em Phys. Rev.},
  vol.~D91, p.~045035, 2015.

\bibitem{Reinosa:2015oua}
U.~Reinosa, J.~Serreau, and M.~Tissier, ``{Perturbative study of the QCD phase
  diagram for heavy quarks at nonzero chemical potential},'' {\em Phys. Rev.
  D}, vol.~92, p.~025021, 2015.

\bibitem{Maelger:2017amh}
J.~Maelger, U.~Reinosa, and J.~Serreau, ``{Perturbative study of the QCD phase
  diagram for heavy quarks at nonzero chemical potential: Two-loop
  corrections},'' {\em Phys. Rev. D}, vol.~97, no.~7, p.~074027, 2018.

\bibitem{Canfora:2015yia}
F.~Canfora, D.~Dudal, I.~Justo, P.~Pais, L.~Rosa, and D.~Vercauteren, ``{Effect
  of the Gribov horizon on the Polyakov loop and vice versa},'' {\em Eur. Phys.
  J. C}, vol.~75, no.~7, p.~326, 2015.

\bibitem{Dudal:2017jfw}
D.~Dudal and D.~Vercauteren, ``{Gauge copies in the Landau--DeWitt gauge: A
  background invariant restriction},'' {\em Phys. Lett. B}, vol.~779,
  pp.~275--282, 2018.

\bibitem{Kroff:2018ncl}
D.~Kroff and U.~Reinosa, ``{Gribov-Zwanziger type model action invariant under
  background gauge transformations},'' {\em Phys. Rev. D}, vol.~98, no.~3,
  p.~034029, 2018.

\bibitem{egmond2020scalar}
D.~M. van Egmond and U.~Reinosa, ``The scalar sunset diagram at finite
  temperature with imaginary square masses,'' 2020.

\bibitem{Coleman:1973jx}
S.~R. Coleman and E.~J. Weinberg, ``{Radiative Corrections as the Origin of
  Spontaneous Symmetry Breaking},'' {\em Phys. Rev. D}, vol.~7, pp.~1888--1910,
  1973.

\bibitem{knecht2001new}
K.~Knecht and H.~Verschelde, ``New start for local composite operators,'' {\em
  Physical Review D}, vol.~64, no.~8, p.~085006, 2001.

\end{thebibliography}

\end{document}